\documentclass{report}
\usepackage{color}
\usepackage{graphicx}
\usepackage{epsfig}
\usepackage{amsfonts}
\usepackage{amssymb}
\usepackage[spanish]{babel}
\usepackage[latin1]{inputenc}
\usepackage{pstricks}
\usepackage[latin1]{inputenc}
\usepackage[T1]{fontenc}
\usepackage{amsmath}
\usepackage{latexsym}
\newcommand{\be}{\begin{equation}}
\newcommand{\ee}{\end{equation}}
\newcommand{\beq}{\begin{equation}}
\newcommand{\eeq}{\end{equation}}
\newcommand{\bea}{\begin{eqnarray}}
\newcommand{\eea}{\end{eqnarray}}
\newcommand{\nn}{\nonumber}
\def\be{\begin{equation}}
\def\ee{\end{equation}}
\def\ba{\begin{eqnarray}}
\def\ea{\end{eqnarray}}

\begin{document}
\thispagestyle{empty}
\begin {center}
\vspace{3.5cm}
\textbf{\Large AdS/CFT and its Applications}\\
\begin{center}
(in spanish)
\end{center}
\vspace{3cm}
A thesis submitted to
Universidad Nacional de La Plata
for the
Ph.D. Degree in Physics\\
\vspace{1cm}
\textbf{Raúl E. Arias}\\
\vspace{5.5cm}
\medskip
UNIVERSIDAD NACIONAL DE LA PLATA\\
Facultad de Ciencias Exactas\\
Departamento de Física\\
\vspace{1.5cm}
\noindent 2013
\end {center}
\pagebreak
\pagenumbering{arabic}
\newpage
\pagestyle{headings}
\begin{center}
\subsection*{Abstract}
\end{center}

In this thesis we study some applications of the gauge/gravity correspondence. In chapter \ref{int} we introduce the gauge/gravity
conjecture focusing on the aspects that will be relevant in the rest of the work, in chapter \ref{Wormhole} we extend the
Gubser-Klebanov-Polyakov-Witten (GKPW) prescription generalizing it to gravity backgrounds with two boundaries.

In chapter \ref{Wh} we discuss the Maldacena-Rey prescription to compute the quark-antiquark (and monopole-antimonopole) potential in some gravity backgrounds
duals to conformal and non-conformal field theories. In particular, we analyze the stability of the classical solutions under small fluctuations.

In chapter \ref{SC} we study the thermodynamic properties and the entanglement entropy of the 2+1 dimensional $p$ and $p+ip$ superconductors through their
dual gravity backgrounds. We analyze the limit where the number of charged degrees of freedom is comparable with the total number of degrees of freedom, which
means that the backreaction on the geometry due to the non-Abelian gauge field is relevant.

\newpage

\begin{center}
\subsection*{Agradecimientos}
\end{center}

En primer lugar quiero agradecer a mi director, colega y amigo Guillermo Silva. A mis amigos de la facultad Ale, Blai,
Carlos L, Carlos C, Colo, Diego C, Diego M, Diego R, Gastón, Gianni, Juan, Marcelo, Mariela, Mauricio, Nacho, Nico, Pablo P, Pablo RP y Walter.
Mención especial para mi compañero de oficina Diego T por los gratos momentos compartidos.  También agradezco a mi familia por el apoyo constante
y especialmente a Romi, a quien le dedico esta tesis.

\tableofcontents

\def\baselinestretch{1}
\chapter{Introducción}\label{int}
\def\baselinestretch{1.66}

\section{Correspondencia Gauge/Gravedad}

En 1997 Maldacena \cite{malda} conjetur\'o la
realizaci\'on de una antigua idea de 't Hooft \cite{thooft} en un
modelo concreto: la teor\'\i a de cuerdas IIB formulada sobre
$AdS_5\times S^5$ es exactamente dual al límite $N\gg1$ de la fase conforme de la teor\'\i a
supersim\'etrica de Yang-Mills (SYM) ${\cal N}=4$ con grupo de gauge
$SU(N)$. Esta idea se conoce como conjetura de dualidad
$AdS/CFT$ puesto que la teor\'\i a de gauge es conforme. Si bien la
conjetura no ha sido probada, la dualidad ha superado un n\'umero
importante de chequeos (ver \cite{magoo}). En particular, la
conjetura establece que la teor{\'\i}a de gauge dual a un cierto
fondo gravitatorio está definida sobre el borde asint\'otico de dicho
espacio-tiempo \cite{GKP,Ewitten}, ver sección \ref{funcco}. La dualidad entre estas dos teor\'\i
as tan dis\'\i miles se encontr\'o analizando las descripciones
complementarias en t\'erminos de cuerdas abiertas y cerradas de un
sistema de $N$ $D3$-branas en el l\'\i mite de horizonte cercano, ver sección \ref{corresp}.
Puesto que la teor\'\i a de bajas energ\'\i as de la teor\'\i a de
cuerdas es supergravedad, gran parte de las
aplicaciones se centran en el estudio de la
supergravedad IIB sobre el fondo $AdS_5\times S^5$. Más concretamente, la
supergravedad IIB describe el r\'egimen "planar"/no perturbativo de la teoría ${\cal N}=4$ SYM. Siguiendo
estas ideas, muchos trabajos se han dedicado a la extensi\'on a
teor\'\i as con menor n\'umero de supersimetr\'\i as
\cite{mn,ks} y la introducci\'on de quarks
(materia en la fundamental) \cite{Karch}. Por otro lado la conjetura
es sorprendente desde el punto de vista que relaciona dos
teor\'ias totalmente diferentes definidas sobre espacio-tiempos de
dimensiones distintas.

?` Por qu\'e es importante esta conjetura? La respuesta est\'a
relacionada con la siguiente propiedad: la dualidad relaciona el
regimen de acoplamiento fuerte de una teor\'ia de gauge con el
regimen semicl\'asico de la teor\'ia de cuerdas IIB. Esto
significa que usando la conjetura podemos derivar propiedades de la teor\'ia de gauge
en el regimen de acoplamiento fuerte (el cual es un l\'imite
dif\'icil de estudiar con técnicas usuales de teorías de campos) estudiando una teor\'ia de
cuerdas semicl\'asica en su regimen de acoplamiento d\'ebil.

En este capítulo se introducirán los conceptos fundamentales de la correspondencia AdS/CFT señalando más
detenidamente aquellos aspectos que se desarrollarán a lo largo de esta tesis.

\subsection{La Correspondencia}\label{corresp}

La conjetura de Maldacena, o AdS/CFT, fue propuesta en \cite{malda} y establece la existencia de una equivalencia entre una teoría de gauge y
una teoría de cuerdas \cite{GSW1,GSW2}. La correspondencia se infiere a partir de dos descripciones alternativas de un sistema de D-branas \cite{Polchinski}.

Consideremos una pila de $N$ D3-branas (coincidentes) en un espacio-tiempo de
Minkowski 10-dimensional, es decir, no consideramos la backreaction de las branas sobre el espacio-tiempo, lo que equivale a tomar $g_sN\ll1$, donde $g_s$ es
la constante de acoplamiento de la cuerda.
El espectro de bajas energías de las cuerdas abiertas que terminan sobre el volumen 4-dimensional de las branas es descripto por partículas
puntuales no masivas que componen un multiplete supersimétrico ${\cal{N}}=4$ super Yang-Mills y que transforman bajo el grupo de gauge $SU(N)$.
Esta teoría de campos goza de invarianza conforme aún a nivel cuántico. Esquemáticamente la acción de bajas energías que describe el sistema es
\be
S_{open}=S_{brane}+S_{bulk}+S_{int}\approx S_{{\cal{N}}=4}+S_{bulk\, flat}+S_{int},~~~~~~~~~~g_sN\ll1\nn
\ee
Aquí $S_{bulk}$ es la acción de supergravedad en 10 dimensiones e involucra modos no masivos de cuerdas cerradas que viven en el bulk (espacio plano).
$S_{brane}$ da cuenta de los modos no masivos del espectro de cuerdas abiertas y su límite de bajas energías es el Lagrangiano de ${\cal N}=4$ SYM.
El termino $S_{int}$ representa el acoplamiento entre modos de las cuerdas abiertas y las cerradas, a bajas energías se obtiene que
$S_{int}\rightarrow 0$ \cite{Klebanovabsorp}.
En el límite en que $\lambda=g_sN\gg1$ debemos considerar la deformación al espacio-tiempo debida a las branas. La dinámica
del sistema queda descripta simplemente
por cuerdas cerradas en la geometría
\bea
ds^2&=&H(r)^{-1/2}(-dt^2+dx_3^2)+H(r)^{1/2}(dr^2+r^2d\Omega_5^2)\label{brane}\\
&&H(r)=1+\frac{N G_N}{r^4},~~~~~~G_N=g_s\alpha'\nn
\eea
La acción que describe este sistema es
\be
S_{close}=S_{bulk\, throat}+S_{bulk\, flat}+S_{int},~~~~~~~~~~g_sN\gg1.\nn
\ee
en donde  $S_{bulk\, flat}$ da cuenta de los modos no masivos de cuerdas cerradas en la región $r\rightarrow\infty$ (espacio plano) de \eqref{brane} y $S_{bulk\, throat}$ denota la acción para todos los modos (masivos y no masivos) de cuerdas
en la región descripta por $r\rightarrow0$ en la geometría \eqref{brane}, es decir, en AdS$_5\times S^5$.
Notemos que debido al corrimiento al rojo gravitatorio modos masivos de cuerdas en la región que interpola entre Mikowski en 10 dimensiones y AdS$_5\times S^5$
(throat) son de baja energía con respecto a un observador en el infinito, y deben ser tenidos en cuenta al tomar el límite de bajas energías.
Por lo tanto $S_{bulk\, throat}\equiv S_{strings\,AdS_5\times S^5}$. $S_{int}$ representa el acoplo entre modos que viven en el throat y modos que viven
en el espacio plano. De nuevo, a bajas energías $S_{int}\rightarrow0$.
Pero ahora, nótese que hay dos descripciones para el mismo sistema y por lo tanto $S_{open}$ y $S_{close}$ deben ser equivalentes. Luego, comparando $S_{open}$ y $S_{close}$
se obtiene que
\be
S_{{\cal{N}}=4}= S_{strings\,AdS_5\times S^5}.\nn
\ee

En resumen, la correspondencia AdS/CFT conjetura la equivalencia entre dos teorías:
\begin{itemize}
\item ${\cal{N}}=4$ super Yang-Mills con grupo de gauge $SU(N)$ (generada por los modos no masivos de cuerdas abiertas) para $N\gg1$

\item teoría de supercuerdas IIB sobre un fondo AdS$_5\times S^5$ con un número entero del flujo de la 5-forma de Ramond-Ramond, $N=\int_{S^5}F_5$,
(generada por toda la torre de estados de la teoría de cuerdas sobre AdS).

\end{itemize}

El diccionario de la correspondencia relaciona los dos parámetros de la teoría de campos, es decir la constante de acoplamiento $g_{YM}$ y el rango del grupo de
gauge $N$, con la constante de acoplamiento de la teoría de cuerdas $g_s$ y el tamaño de la cuerda $l_s=\sqrt{\alpha'}$ mediante \cite{malda}
\be
g_{YM}^2=g_s,~~~~~~\left(\frac{R}{l_s}\right)^4=4\pi g_{YM}^2N=4\pi\lambda\label{dictionary}
\ee
en donde $R$ es el tamaño de AdS y $\lambda=g_{YM}^2N$ es la constante de ´t Hooft. Estas ecuaciones indican que dependiendo de la elección
de los parámetros podemos estudiar el régimen de acoplamiento débil de la teoría de gauge o de la teoría de cuerdas, pero nunca será simultáneo en ambas teorías.
Se observa de (\ref{dictionary}) que en el límite en el cual $\lambda\rightarrow\infty$ (el tamaño de la cuerda es despreciable frente al tamaño de AdS),
corresponde al regimen semiclásico de cuerdas, es decir una teoría de supergravedad. La tensión efectiva de la cuerda $T_{ef}=\sqrt\lambda\rightarrow\infty$ y las cuerdas son puntuales. Nótese que en este límite la teoría
de gauge se encuentra fuertemente acoplada y por lo tanto no es posible estudiarla perturbativamente.
El régimen opuesto (cuando el tamaño de la cuerda es comparable con el tamaño del espacio) nos dice que debemos analizar la teoría cuántica de cuerdas
en AdS a fin de hacer predicciones sobre el límite de acoplamiento débil de la teoría de campos.
Luego, la correspondencia puede ser aplicada en dos direcciones, podemos usar la teoría de cuerdas en su límite semiclásico para aprender sobre las propiedades
de una teoría de gauge fuertemente acoplada ó podemos estudiar una teoría de campos en su regimen perturbativo para entender la teoría cuántica de
cuerdas sobre fondos no triviales como lo es por ejemplo AdS.

Un primer chequeo de la conjetura es la correspondencia entre el grupo de isometrías de la teoría de cuerdas y el grupo conforme de ${\cal N}=4$.
En particular, el grupo de isometrías de AdS$_5$ coincide con el grupo conforme en 4 dimensiones $SO(4,2)$. Por otro lado, la teoría ${\cal N}=4$
tiene una simetría global adicional, además de las de Poincaré y supersimetría, dada por el grupo $SU(4)_{\cal R}\sim SO(6)$ y denominada simetría ${\cal R}$. No es difícil ver entonces que el origen del espacio 10-dimensional en donde vive la teoría de cuerdas
debe ser el producto de un espacio AdS y una esfera 5-dimensional $S^5$. En otras palabras, cada vez que la teoría de gauge sea una teoría conforme
(a nivel cuántico) debemos tener
un espacio AdS en su dual gravitatorio. Por otra parte cualquier grupo de simetría adicional en la teoría SYM se relaciona con la simetría del
espacio interno ${\cal M}$, la teoría de cuerdas dual queda definida en AdS$\times {\cal M}$.

 La conjetura AdS/CFT implica una identificaci\'on entre la
coordenada radial $r$ en la soluci\'on de supergravedad y la escala
de energ\'\i a de la teor\'\i a de campos dual a ella. Esta
identificaci\'on surge de  analizar c\'omo se encuentra realizado el
operador de dilataciones $D\subset Conf({\mathbb R}^{3,1})=so(2,4)$
en la soluci{\'o}n de gravedad. El resultado es que una dilataci{\'o}n
$x_i\to\lambda x_i$ en la teor\'\i a de gauge se corresponde en AdS:
$ds^2=dr^2+e^{2r/R} (dx_{i})^2$ con la isometr\'\i a
\be
\begin{array}{c}
  x_i\to \Lambda x_i \\
    \mathrm{teoria~de~gauge}
                \end{array}
:
\begin{array}{c}
  ~~~~~x_i\to \lambda x_i,~~~r\to r-R\log \lambda \\
  \mathrm{gravedad }
\end{array}
\ee
El borde del AdS ($r\to\infty$) se asocia con el regimen ultravioleta (UV) de la teoría de campos conforme
(CFT), mientras que el horizonte del AdS ($r\to 0$) se corresponde con
el r\'egimen infrarrojo (IR). Esta asociaci\'on es más que una identificaci\'on
formal como veremos al calcular lazos de Wilson: c\'alculos de valores de expectación
de vacío (VEV) de operadores en la
teor\'\i a de gauge con una escala caracter\'\i stica
$\mu$, se encuentran dominados a nivel semiclásico por contribuciones provenientes de la
regi\'on $r=R\log\mu$ de la geometría (ver \cite{magoo} para mas ejemplos y referencias).

Un cambio de escala en una CFT es irrelevante f\'\i sicamente. Sin
embargo, en una teor\'\i a no conforme las constantes de
acoplamiento cambian con la escala. La no invarianza conforme de la teoría cuántica de campos se manifiesta en el lado gravitatorio en la ausencia de una simetría que relacione diferentes valores de la dirección radial. Esto sugiere interpretar la
dependencia de los acoplamientos con la escala en t{\'e}rminos de una
dependencia espec\'\i fica para los campos de la soluci\'on de
gravedad. Esta interpretaci\'on en la teoría de campos funciona razonablemente a nivel
cualitativo, sin embargo, la identificaci{\'o}n cuantitativa del radio
con la escala de energ\'\i a puede ser dif\'\i cil de hallar.
 Una posibilidad para identificar la relaci\'on
radio/energ\'\i a es considerar el factor de enrollamiento $Z(r)$ que
multiplica el hiperplano 3+1 en la m\'etrica
$ds^2=Z(r)dx_{i}dx^{i}+....$, puesto que $Z(r)$ es el factor que
relaciona las energ\'\i as de dos observadores ubicados en puntos
distintos del espacio transverso mediante:
$Z(r')^{-1/2}E'=Z(r)^{-1/2}E$ \cite{magoo}.

\subsubsection{Extensión a teorías no conformes y menos supersimétricas}

Hasta acá se discutió la conjetura originalmente postulada por Maldacena, en donde la teoría de campos es invariante conforme y supersimétrica.
Una pregunta natural es cómo encontrar duales gravitatorios para teorías de gauge más realistas, es decir, con menos (o ninguna) supersimetrías y que no gocen de invarianza conforme. Como veremos a continuación, resulta problemático describir teorías realistas mediante supergravedad (ver \cite{Imeroni}).

Las razones son las siguientes: existen conceptualmente dos maneras
de romper la simetr\'\i a conforme: (i) comenzar con una teor\'\i a
conforme para la cual conocemos su dual gravitatorio y deformarla
mediante operadores relevantes o marginales que rompen supersimetría e
invarianza conforme, técnicamente esto corresponde a encender campos en la solución de gravedad,
ó (ii) comenzar con una configuraci\'on donde
la supersimetr\'\i a y la invarianza conforme se encuentran rotas
(las soluciones discutidas en el capítulo \ref{Wh} pertenecen a esta segunda
categor\'\i a). En el primer caso, una manera posible de obtener "SYM" pura sería
agregar una deformaci\'on con par\'ametro  de masa $M$ a los campos escalares de ${\cal N}=4$. Ahora bien, el par\'ametro de masa $M$
inducirá una escala dimensional $\Lambda\sim M e^{-1/g^2_{YM}N}$. El
l\'\i mite en el cual la deformación da origen a SYM ($\Lambda$ finito) es $M\to\infty$, $\lambda=g^2_{YM}N\ll1$ con
$\lambda$ fijo. Sin embargo, como discutimos en la introducci\'on,
la aproximaci\'on de supergravedad es confiable en el l\'\i mite
opuesto $\lambda\gg1$. De manera que, en términos generales, la descripci\'on de bajas
energ\'\i as de SYM pura requiere el conocimiento completo del espectro de la
teor\'\i a de cuerdas. El segundo caso (ii) enumerado arriba corresponde al de
enrollamiento de branas. La caracter\'\i stica principal y recurrente de este tipo
de soluciones es que los grados de libertad de la teor\'\i a de
gauge se encuentran entrelazados con los modos de Kaluza-Klein
(modos KK) de la variedad transversa y no es posible desacoplarlos.
Los modos KK entran en juego a una  escala de energ\'\i a que es
inversamente proporcional  al tama{\~n}o de la variedad, y el
inconveniente en este tipo de soluciones es que este tama{\~n}o de la
variedad compacta es comparable con la escala en la que los
fen\'omenos no perturbativos son relevantes (confinamiento,
ruptura de simetría quiral,...).
En resumen, soluciones cl\'asicas de supergravedad describen
apropiadamente teor\'\i as que no son puramente teorías de Yang-Mills, sino que contienen
un n\'umero infinito de campos adicionales. Se espera que
el dual holográfico en el límite de acoplamiento débil de QCD sea un modelo de cuerdas fuertemente acoplado.

\subsection{Funciones de Correlación}\label{funcco}

La conjetura fue definida más precisamente luego de los trabajos de Witten \cite{Ewitten} y Polyakov et. al. \cite{GKP}, en donde
se propuso una técnica de cálculo para la relación entre valores de expectación de operadores de la CFT y la función de partición de la teoría de cuerdas. Es el propósito de esta sección
analizar en detalle dicha relación.

Para definir la correspondencia necesitamos una relación unívoca entre los observables de ambas teorías y una prescripción para comparar cantidades físicas.
Desde el punto de vista de la teoría 4-dimensional, todo operador ${\cal O}$ invariante de gauge puede asociarse a una fuente $h$
\be
L_{CFT}+\int d^4x\, h\,{\cal O},
\ee
en donde $L_{CFT}$ es el Lagrangiano de la teoría 4-dimensional. Usualmente se define la funcional generatriz, $W(h)$, de funciones de correlación
del operador ${\cal O}$ a partir de
\be
e^{-W(h)}=\langle e^{-\int h {\cal O}}\rangle_{QFT}
\ee
y el valor de expectación de los operadores se escribe como
\be
\langle {\cal O}\ldots {\cal O}\rangle_c=(-1)^n\left[\frac{\delta^nW}{\delta h^n}\right]_{h=0}.
\ee
Un aspecto básico de la conjetura es la siguiente afirmación: todo operador invariante de gauge de la teoría cuántica se relaciona con un campo $h$ en AdS (ambos deben tener los mismos números cuánticos).
La correspondencia propone identificar a la fuente del operador ($h(x)$) con el valor en el borde de un campo 5-dimensional $h(x,z)$ ($z$ es la coordenada radial de AdS). Para cada posible fuente externa en la
CFT hay un campo 5-dimensional $h(x,z)$. Con esto podemos
entender la ecuación fundamental de la conjetura AdS/CFT
\be
e^{W(h)}=\langle e^{\int h O}\rangle_{QFT}= {\cal Z}_{strings\,\, en\, AdS}[h(x,z=borde)=h_0].\label{GKPW1}
\ee
En el lado izquierdo de esta ecuación tenemos una funcional que depende de una fuente arbitraria 4-dimensional (QFT en presencia de fuente externa) y
del lado derecho tenemos la función de partición de la teoría de cuerdas con particulares condiciones de contorno para los campos definidos en el espacio AdS.
La función de partición es un objeto complicado de calcular. En el límite semiclásico de cuerdas ($\lambda\gg1$) se aproxima como
\be
{\cal Z}_{strings\,\, en\, AdS}[h(x,z=borde)]\sim e^{S_{sugra}[h_0]},~~~~~~\lambda\gg1\label{GKPW2}
\ee
En el lado derecho la acción de supergravedad 5-dimensional se encuentra evaluada sobre la solución de sus ecuaciones
de movimiento que tiene por valor de borde $h_0(x)$. En resumen, (\ref{GKPW1}) y (\ref{GKPW2}) prescriben que el valor de borde de
campos de supergravedad actúa como una fuente para operadores de la teoría de gauge dual.

Algunos comentarios adicionales al respecto de las ecuaciones (\ref{GKPW1}) y (\ref{GKPW2}):

\begin{itemize}

\item Las ecuaciones de movimiento en AdS son ecuaciones diferenciales de segundo orden y por lo tanto vamos a necesitar estudiar las condiciones de borde para
determinar una solución única.  La asociación no es tan simple como $h(x,borde)=h_0(x)$, veremos más adelante que la condición adecuada es pedir
$h(x,z)\sim f(z)h(x)$. Por otro lado, la segunda condición de borde depende fuertemente de la signatura de AdS. Si se está trabajando con AdS
Euclídeo pedir regularidad en el interior del espacio determina la solución unívocamente. Se verá en el capitulo 2 que en signatura Lorentziana
pedir regularidad en el horizonte no es suficiente para tener una solución única, se requiere información adicional en el bulk relativa a los
estados entrantes y salientes en las tapas del cilindro de AdS \cite{bala, bala2,balagid,skenderis}. 

\item El campo $h$ que se acopla al operador ${\cal O}$ usualmente puede identificarse mediante simetrías debido a que ambos deben tener
los mismos números cuánticos.
En particular, para el caso de corrientes conservadas los acoplamientos son conocidos
\be
L_{CFT}+\int d^4x \sqrt{g}\left(g_{\mu\nu}T^{\mu\nu}+J_\mu A^\mu+\phi\, Tr\,(F_{\mu\nu}F^{\mu\nu})+\ldots\right).
\ee
El operador asociado al gravitón se corresponde con el tensor de energía-momento y el asociado con un campo de gauge en AdS a una corriente (global) conservada.
Nótese que la conservación de $T_{\mu\nu}$ ó de $J_\mu$ se asocia con una invariancia de gauge al nivel de las fuentes (esto significa que $W(g_{\mu\nu},A_\mu)$
es una funcional invariante de gauge frente a $A_\mu$ y $g_{\mu\nu}$). Este es un hecho general de AdS/CFT, simetrías de gauge en el bulk son simetrías globales en
la teoría del borde.

\end{itemize}

En lo que sigue de la sección voy a reproducir un ejemplo canónico con el objetivo de entender el poder de la expresión (\ref{GKPW2}).
Consideremos el caso de un campo escalar masivo en AdS$_{d+1}$ \cite{GKP, Ewitten}, el cual es dual a un operador ${\cal O}$ en la CFT dual.
La métrica de AdS$_{d+1}$ en coordenadas de Poincaré se escribe
\be
ds^2=R^2\frac{dx^\mu dx_\mu+dz^2}{z^2}\equiv g_{AB}dx^Adx^B,~~~~~~A=0,\ldots,d,~~~~~~x^A=(z,x^\mu)\label{AdSpoincare}
\ee
En estas coordenadas el borde de AdS se sitúa en $z=0$ y la geometría del mismo es la del espacio de Minkowski.


Dado un campo escalar $\phi(z,x^\mu)$, su acción en un espacio curvo se escribe
\be
S\sim\int d^{d+1}x\sqrt{g}(g^{AB}\partial_A\phi\partial_B\phi+m^2\phi^2).\label{accionscalar}
\ee
Integrando por partes y usando el teorema de Gauss tenemos
\be
S\sim\int_{\partial AdS} d^dx\sqrt{g}g^{zB}\phi\partial_B\phi+\int d^{d+1}x\,\sqrt{g}\phi(-\square+m^2)\phi,\label{accionescalar2}
\ee
en donde  $\square\phi=\frac{1}{\sqrt g}\partial_A\left(\sqrt{g}g^{AB}\partial_B\right)\phi$. Nótese que el segundo término es
la ecuación de Klein-Gordon para un campo escalar masivo en un espacio curvo y, por lo tanto, la acción on-shell está dada para el caso libre simplemente por el primer término de \eqref{accionescalar2}.

Nuestra intención ahora es resolver la ecuación de movimiento para
$\phi$, $(\Box+m^2)\phi=0$ con condición de borde
"$\phi(0,x)=\phi_0(x)$". Descomponiendo al campo escalar en modos
de Fourier \be \phi(z,x^\mu)=e^{i k_\mu x^\mu}f_k(z),~~~~~~k_\mu
x^\mu=-\omega t+\vec{k}.\vec{x} \ee y usando el hecho de que la
métrica solo depende de $z$, la ecuación de movimiento para el
campo escalar queda \be
0=\frac{1}{R^2}\left(z^2k^2-z^{d+1}\partial_z(z^{-d+1}\partial_z)+m^2R^2\right)f_k(z),~~~~~~k^2=\vec{k}^2-\omega^2.\label{scalarKG}
\ee En el límite $z\rightarrow0$ la solución asintótica de la
ecuación de movimiento es $f_k\sim z^\Delta$, en donde $\Delta$
queda definida por \be m^2R^2=\Delta(\Delta-d)\label{confdim}. \ee
Esta ecuación tiene dos soluciones, genéricamente una positiva
$\Delta_+$ y otra negativa $\Delta_-$. $\Delta_+$ se identifica
con la dimensión conforme del campo dual ${\cal O}$. Las
coordenadas de Poincaré no cubren toda (la variedad) AdS en el
caso Lorentziano, solo una parte. 
De manera que al definir el problema de $\phi$ en el interior debemos
dar condiciones de contorno no solo en $z=0$, sino también en el
horizonte ($z\rightarrow\infty$). Las dos soluciones linealmente
independientes de la ecuación diferencial \eqref{scalarKG} cerca
del borde son \be \phi\sim\phi_+
z^{\Delta_+}+\phi_-z^{\Delta_-}\label{phieucnb}, \ee siendo
$\Delta_{\pm}=\frac{d}{2}\pm\sqrt{\left(\frac{d}{2}\right)^2+m^2R^2}=\frac{d}{2}\pm\nu$
las raíces de la ecuación (\ref{confdim}) y $\phi_{\pm}(k)$ dos
constantes. El primer término $\phi_+(k)$ da origen a una solución
normalizable en el sentido de que la expresión \be \int
dz\sqrt{g}|\phi|^2<\infty. \ee El coeficiente $\phi_-(k)$
corresponde a una solución no normalizable.

Debido al carácter singular del borde de AdS ($z=0$) es conveniente introducir una regularización en $z=\epsilon$ y tomar $\epsilon\rightarrow0$ al final de los cálculos como veremos a continuación. La condición de contorno es entonces
\be
\phi(z,x^\mu)|_{z=\epsilon}\rightarrow \epsilon^{\Delta_-}\phi_-^{ren}(x^\mu).\label{phiren}
\ee
La función $\phi_-^{ren}$ se identifica con la fuente del operador ${\cal O}$.

Hay varias observaciones que mencionar con respecto a la ecuación (\ref{confdim}):

\begin{itemize}

\item Heurísticamente puede demostrarse que ${\cal O}(\epsilon,x)\sim \epsilon^{\Delta_+}{\cal O}^{ren}(x)$ y por lo tanto, la dimensión de escala de ${\cal O}$ es $\Delta_+$. Supongamos que

\be
S_{borde}\ni\int_{z=\epsilon}d^dx\sqrt{g_\epsilon}\phi(x,\epsilon){\cal O}(x,\epsilon)=\int d^dx\left(\frac{R}{\epsilon}\right)^d\epsilon^{\Delta_-}\phi_-^{ren}(x){\cal O}(x,\epsilon)
\ee
en donde $g_\epsilon$ es la métrica de AdS evaluada en $z=\epsilon$. Pidiendo que esta expresión sea finita cuando $\epsilon\rightarrow0$ se tiene
\be
{\cal O}(x,\epsilon)\sim\epsilon^{d-\Delta_-}{\cal O}^{ren}(x)=\epsilon^{\Delta_+}{\cal O}^{ren}(x)
\ee
y por lo tanto la dimensión de escala de ${\cal O}^{ren}$ es $\Delta_+$.

\item Si  $m^2>0$ entonces $\Delta_+>d$ y el operador ${\cal O}$ es irrelevante. Esto significa que si perturbamos la CFT con este operador su coeficiente tendrá una escala de masa negativa y por lo tanto sus efectos no serán importantes en el IR ($z$ grande).

\item El operador ${\cal O}$ es marginal si $m^2=0$.

\item Si $m^2<0$ entonces $\Delta_+<d$ y el operador ${\cal O}$ es relevante en el IR. Hay que destacar que en AdS, masas negativas están permitidas siempre y cuando se satisfaga la cota de Breitenlohner-Freedman \cite{BF} $m^2>-|m_{BF}|^2=-\left(\frac{d}{2R}\right)^2$. Valores de $m^2$ dentro de esta cota aseguran energía positiva y por lo tanto una teoría bien definida.

\end{itemize}

Volvamos a la ecuación diferencial (\ref{scalarKG}). Para valores de $k^\mu$ tipo espacio ($k^2>0$) la solución es
\be
f_k(z)=A(k) z^{d/2}K_{\nu}(k z)+B(k) z^{d/2}I_\nu(k z),
\ee
en donde $A(k), B(k)$ son constantes arbitrarias y $K_\nu$, $I_\nu$ son funciones de Bessel. Incidentalmente el caso $k^2>0$ coincide con la ecuación diferencial que se obtiene en el caso de signatura Euclídea. Regularidad en el interior de AdS ($z=\infty$) requiere $B(k)=0$ y la solución de (\ref{scalarKG}) es entonces única si fijamos los coeficientes $A(k)$. Dichos coeficientes $A(k)=\phi_+(k)$ se relacionan con la transformada de Fourier de la condición de contorno $\phi_+(x)$ en $z=\epsilon$ de la siguiente manera: normalizando la función de onda para que $f_k(z=\epsilon)=1$
\be
f_{k}(z)=\frac{z^{d/2}K_\nu(k z)}{\epsilon^{d/2}K_\nu(k \epsilon)}.
\ee
La solución de \eqref{scalarKG} con condición de contorno \eqref{phiren} es
\be
\phi(x,z)=\int d^dk e^{ikx}f_k(z)\phi(k,\epsilon),
\ee
con $\phi(k,\epsilon)=\int dx e^{-i k x}\phi(x,\epsilon)$. La acción on-shell resulta
\be
S(\phi)=2R^{d-1}\int d^dk\,\phi(k,\epsilon)\phi(-k,\epsilon)\epsilon^{-d+1}\partial_z\left(\frac{z^{d/2}K_\nu(k z)}{\epsilon^{d/2}K_\nu(k \epsilon)}\right)_{z=\epsilon}.
\ee
Expandiendo en serie la función de Bessel vemos que tenemos dos tipos de contribuciones a la acción. Por un lado se obtienen términos con potencias enteras de $k$, es decir analíticos en $k$, estos son términos de contacto y pueden ser substraídos tomando contratérminos en la acción (\ref{accionscalar}). La contribución física, no analítica en $k$ tiene la forma,
\be
-\frac{(-1)^{\nu-1}2^{1-2\nu}}{\Gamma(\nu)^2}k^{2\nu}\ln k\epsilon.\epsilon^{2\nu-d}+\ldots
\ee
La transformada de Fourier de este término y la apropiada condición de borde (\ref{phiren}) permiten obtener la función de dos puntos (finita) para el operador ${\cal O}$, el resultado es
\be
\langle {\cal O}(x_1){\cal O}(x_2)\rangle=-\frac{\delta}{\delta\phi_-^{ren}(x_1)}\frac{\delta}{\delta\phi_-^{ren}(x_2)}S=\frac{2\Gamma(\Delta_+)}{\pi^{d/2}\Gamma(\Delta_+-d/2)}
\frac{1}{(x_1-x_2)^{2\Delta_+}}.
\ee

En el caso de $k^\mu$ tipo tiempo, es decir $k^2<0$, las dos soluciones linealmente independientes tienen un comportamiento asintótico similar
en $z\rightarrow\infty$
\be
z^{d/2}K_{\pm\nu}(iqz)\sim e^{\pm -iqz}.
\ee
En este caso regularidad en el interior no elimina una de las soluciones. Esta ambigüedad tiene su origen en que existen diferentes elecciones para
la función de Green en AdS$_{d+1}$. Por ejemplo, una posible elección es la función de Green retardada, la cual da cuenta de la respuesta causal del sistema
ante una perturbación.

En el capítulo \ref{Wormhole} se extenderá en detalle la discusión hecha aquí arriba para AdS$_{d+1}$ en coordenadas globales y signatura Lorentziana.
Además se utilizará la prescripción GKPW para calcular funciones de dos puntos en espacios con dos bordes. En particular se estudiará la función de
correlación en un agujero de gusano que resuelve las ecuaciones de movimiento de una acción de Einstein-Gauss-Bonnet (EGB) cuyos bordes son
asintóticamente AdS. Además se revisará la prescripción de Gubser-Klebanov-Polyakov-Witten (GKPW) \cite{GKP,Ewitten} para extraer
correladores de la QFT a partir de cálculos de gravedad y se discutirá su aplicación a la solución de
agujero de gusano Lorentziana encontrada en \cite{Dotti}. A lo largo del camino se mencionarán similitudes y diferencias
con el caso de AdS$_2$ (ver \cite{diego,ali} para otros estudios sobre el agujero de gusano analizado en esta tesis).
Recordemos que la prescripción GKPW en signatura Lorentziana involucra no solo conocer datos en el borde del espacio-tiempo
sino también el conocimiento de los estados inicial y final, veremos cómo estos estados aparecen en el cálculo
(ver \cite{Herzog,bala,bala2,balagid,marolf,Skenderis} para discusiones sobre temas relacionados a la prescripción
GKPW en signatura Lorentziana).

\subsection{Lazos de Wilson}\label{Wilint}

El confinamiento de quarks en teorías de gauge no abelianas
aún no se ha demostrado debido al hecho de que este fenómeno ocurre
en el regimen de acoplamiento fuerte, lo cual nos impide tratarlo con
técnicas perturbativas de teoría cuántica de campos. Sin embargo,
Wilson propuso \cite{wilson} un criterio para
establecer cuándo una teoría es confinante y cuándo no lo es.
Dicho criterio establece que el confinamiento del flujo
cromoeléctrico (cuerda de QCD) ocurre en teorías para las
cuales el valor de expectación del lazo de Wilson es proporcional
al área encerrada por él. En particular, esta ley de área para un
contorno rectangular en el espacio-tiempo (banda infinita)
correspondiente a un par quark-antiquark estático, corresponde a un
potencial confinante lineal en la distancia de separación entre quarks.

En esta sección vamos a estudiar cómo la conjetura AdS/CFT permite estudiar valores de expectación de operadores no locales como el lazo de Wilson y cómo el resultado obtenido estudiando teoría de cuerdas concuerda con los resultados esperados para la teoría de campos.

\subsubsection{Teoría de Gauge}

Esencialmente un lazo de Wilson es un factor de fase en teorías de gauge tanto Abelianas como no Abelianas. Estos operadores son invariantes de gauge
y por lo tanto son observables de la teoría cuántica. En QCD permite medir el potencial de interacción entre un par de quarks.

La linea de Wilson se define como
\be
U_{{\cal C}}(y,z)={\cal P}e^{i\int_{{\cal C}}dx^\mu A_\mu^a T^a}\label{U}.
\ee
donde ${\cal C}$ es una curva que conecta $(y,z)$. El operador $U_{{\cal C}}(y,z)$ nos provee del valor de un campo $\varphi(y)$ transportando paralelamente el valor de $\varphi(z)$ a lo largo de la curva ${\cal C}$ como
\be
\varphi_{{\cal C}}(y)=U_{{\cal C}}(y,z)\varphi(z).
\ee
$U_{{\cal C}}(y,z)$ satisface $D_\mu U_{{\cal C}}(y,z)=0$ donde $D_\mu=\partial_\mu+i A_\mu$ es la derivada covariante.
El símbolo ${\cal P}$ en \eqref{U} nos indica que debemos tomar el orden de caminos, es decir, como en una teoría no Abeliana el campo de gauge $A_\mu(x)$ no
conmuta en puntos diferentes debemos ordenar el camino de tal forma que en la expansión en serie de la exponencial (\ref{U}) campos evaluados en
 valores mayores del parámetro $t$ que describe la curva ${\cal C}$ queden a la izquierda. $T^a$ son los generadores del grupo $G$ en alguna de
 sus representaciones \cite{Peskin}. El lazo de Wilson se define como la traza de una linea de Wilson a lo largo de una curva cerrada
\be
W_{{\cal C}}({\cal R})=Tr {\cal P}e^{i\oint_{{\cal C}}dx^\mu A_\mu^a T_{{\cal R}}^a}\label{W}.
\ee
El operador así definido depende solamente de la representación ${\cal R}$ del grupo de gauge y del camino ${\cal C}$. Este operador invariante de gauge sirve para expandir
cualquier función de $A_\mu$ invariante de gauge. También se utiliza para distinguir entre las fases de confinamiento y deconfinamiento
de una teoría de gauge. Veamos este última propiedad con más detalle: Físicamente el lazo de Wilson describe la contribución del campo de Yang-Mills a la amplitud de
propagación de una partícula muy masiva, externa y cargada ante la representación ${\cal R}$ del grupo que se mueve a lo largo de ${\cal C}$. Si
${\cal C}$ es un camino rectangular que tiene una longitud $T$ a lo largo del eje temporal y una separación espacial $l$ ($T\gg l$),
su valor de expectación se comporta como
\be
\langle W_{{\cal C}}(\square) \rangle\sim e^{-T V(l)}\label{qcdwilson},
\ee
en donde $V$ es el potencial quark-antiquark estático y $\square$ denota la representación fundamental de $SU(N)$. La teoría es linealmente confinante
 si $V(l)\sim l$.
Nótese que para un potencial confinante lineal el argumento de la exponencial en (\ref{qcdwilson}) es el área encerrada por la curva que define
${\cal C}$. Esta es la denominada ley de área. Por otro lado, para QED con quarks externos infinitamente masivos se encuentra que $V(l)\sim \frac{1}{l}$, es decir
es de tipo Coulomb, el comportamiento esperado para una teoría conforme.

\subsubsection{Teoría de Cuerdas}

Veremos ahora cuál es la prescripción en el lado de cuerdas de la conjetura que nos permite calcular el valor de expectación del lazo de Wilson.
En particular voy a enfocarme con detalle en el caso en que el camino ${\cal C}$ es un rectángulo y la representación ${\cal R}$ es la fundamental.
Más adelante en esta sección comentaré cómo obtener el valor de expectación en otras representaciones del grupo de gauge.

\begin{itemize}
\item {\bf Representación Fundamental}
\end{itemize}

En su construcción la conjetura goza de una simetría de gauge $U(N)$ cuyo origen son las cuerdas cuyos extremos terminan en cualquiera de las $N$ $D3$-branas que se apilan
paralelamente unas a otras y no guardan distancia perpendicular entre ellas.
Las cuerdas cuyos extremos se ubican sobre diferentes branas son no masivas debido a que no hay
una separación física entre las mismas. Consideremos $N+1$ D3-branas ubicadas en el mismo punto del espacio, generando un grupo de gauge $U(N+1)$. Tomemos
una de ellas y separémosla del resto, la cuantización del sistema muestra que este procedimiento es la realización del mecanismo de Higgs donde se rompe la simetría de gauge $U(N+1)\rightarrow U(N)\times U(1)$. La cuerda que tiene un extremo
sobre cualquiera de las $N$ branas y el otro extremo en la brana alejada será masiva y su masa será proporcional a la distancia de separación $\rho$ entre las branas, asimismo
transformará en la representación fundamental del grupo $U(N)$. El extremo de la cuerda es fuente para campos de gauge $U(N)$, y debe ser interpretado como
un quark en la representación fundamental. Obtendremos un quark (externo) infinitamente masivo tomando $\rho\rightarrow\infty$. Si tenemos en cuenta la backreaction de las $N D3-$branas, la cuerda se extenderá a lo largo de la dirección radial de AdS y su hoja de mundo formará una superficie suave.

Como el dual gravitatorio de ${\cal N}=4$ es AdS$_5\times S^5$, debemos también especificar cómo se extiende la cuerda en la 5-esfera parametrizada
por los ángulos $\theta^I$.  Debido a esto, se encuentra que la hoja de mundo descripta anteriormente
no es fuente para el lazo de Wilson usual sino que lo es para uno supersimétrico, acoplándose también a los campos escalares $X^I, I=1,\ldots,6$ de ${\cal N}=4$ SYM
\be
W_{{\cal C},\,{\cal C}_{int}}(\square)=Tr {\cal P}e^{\oint (iA_\mu\dot{x}^\mu+\theta^IX^I(x^\mu)\sqrt{\dot{x}^2})d\tau}\label{halfwilson}.
\ee
En esta expresión $x^\mu(\tau)$ parametriza el camino ${\cal C}$, $\theta^I(\tau)$ denota el acoplo a los escalares $X^I$. ${\cal C}_{int}$ denota la curva sobre el espacio interno. En esta tesis voy a considerar el caso $\theta^I=constante$, es decir, la cuerda se sitúa en un punto sobre la
$S^5$.

\begin{figure}[htbp]
\centering
\includegraphics[width=0.5\textwidth]{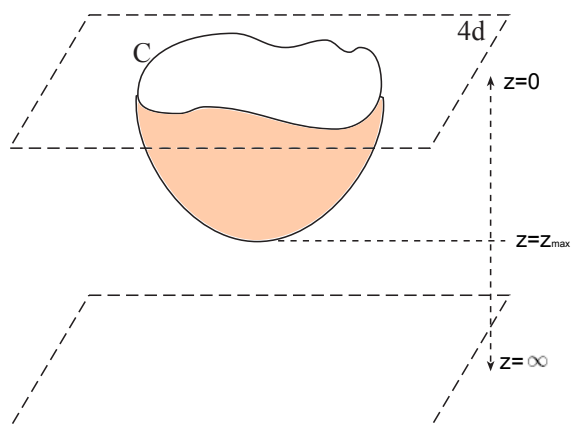}
\caption{Hoja de mundo de la cuerda que tiene a $\cal C$ como borde. La teor\'\i a de gauge vive en el borde del espacio ($z=0$).}
\label{surface}
\end{figure}

La prescripción para obtener el valor de expectación del lazo de Wilson \cite{rey, maldawilson}, dice que debemos
calcular la función de partición de la cuerda cuya hoja de mundo termina sobre la curva ${\cal C}$
\be
\langle W_{{\cal C},\,{\cal C}_{int}}(\square)\rangle={\cal Z}_{string}[{\cal C},{\cal{C}}_{int},\square].
\ee
En el límite de supergravedad $g_s\rightarrow 0, \lambda\rightarrow\infty$ (y fijo) se tiene
\be
{\cal Z}_{string}[{\cal C},{\cal{C}}_{int},\square]\cong e^{-S_{string}^{on\, shell}[{\cal C},{\cal{C}}_{int}]},\label{GKPWwilson}
\ee
en donde $S_{string}$ es la acción de Nambu-Goto evaluada en la solución clásica para la cuerda cuyos extremos se ubican sobre la curva ${\cal C}$.
La mayor contribución a la integral funcional se obtendrá para la/las configuraciones de área mínima. Antes de pasar a calcular el potencial
quark-antiquark explícitamente, debo hacer notar que dicha área mínima es divergente debido a que el volumen de mundo de la cuerda se extiende hasta el borde de AdS. Esta divergencia es interpretada
de la siguiente manera: lo que en realidad estamos calculando es la energía del par quark-antiquark incluyendo la auto-energía debida a su masa infinita.
Dicha masa debe ser substraída, la misma es equivalente a la hoja de mundo recta que se extiende desde el origen de AdS hasta infinito.
En resumen, el valor de expectación del lazo de Wilson se calcula a partir del área mínima de la hoja de mundo que termina sobre el camino ${\cal C}$
en el borde (ver figura \ref{surface}) convenientemente renormalizada.

\subsubsection{Potencial quark-antiquark}

\begin{figure}[htbp]
\centering
\includegraphics[width=0.36\textwidth]{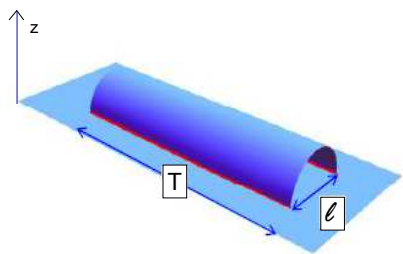}
\caption{Hoja de mundo de la cuerda que tiene al rectángulo de lados $T$ y $l$ como borde. La teor\'\i a de gauge vive en el borde del espacio ($z=0$).}
\label{rectangularwl}
\end{figure}

Estudiemos ahora el potencial quark-antiquark en ${\cal N}=4$ a partir de la correspondencia AdS/CFT calculando el valor de expectación del lazo de Wilson
por medio de la ecuación (\ref{GKPWwilson}) \cite{rey, maldawilson}. Para hallar dicho potencial la curva ${\cal C}$ es un rectángulo de lados $l,T$ con $T\gg l$
(ver figura \ref{rectangularwl}). Primero debemos parametrizar la hoja de mundo de la cuerda $X^\mu(\tau,\sigma)$ con $, \mu=0,\ldots,9$ y
decidir cómo está inmersa en AdS. La inmersión es la siguiente
\be
\tau=t,~~~~\sigma=x,~~~~z=z(x)\label{embeddingwilson}
\ee
Las condiciones de contorno para nuestra configuración son $\tau\in(-\frac{T}{2},\frac{T}{2}),\sigma\in(-\frac{l}{2},\frac{l}{2})$ y $z(\pm\frac{l}{2})=0$, $l$
debe ser interpretado como la distancia de separación entre quarks en la teoría de gauge.

~~

~

\begin{figure}[h]
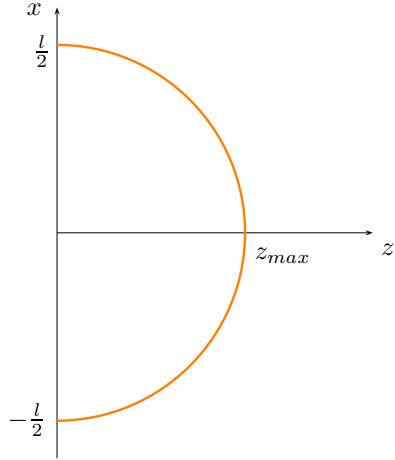

\vspace{3.2cm}
\hspace{7cm}
\psline[linewidth=.3pt]{->}(0,-3)(0,3)
\psline[linewidth=.3pt]{->}(0,0)(4.2,0)
\psarc[linewidth=.9pt,linecolor=orange](0,0){2.5}{270}{90}
\rput[bI](-0.3,2.9){$x$}
\rput[bI](4.4,-0.3){$z$}
\rput[bI](3,-0.4){$z_{max}$}
\rput[bI](-.2,2.2){$\frac l2$}
\rput[bI](-.4,-2.7){$-\frac l2$}
\vspace{3cm}
\caption{Plano $x-z$. La curva naranja muestra el perfil $z(x)$ de la cuerda en AdS, alcanzando un valor máximo  $z_{max}$.}
\label{will}
\end{figure}

La métrica inducida sobre la hoja de mundo $g_{\alpha\beta}=\partial_\alpha X^\mu\partial_\beta X^\nu G_{\mu\nu}(X)$ para el embedding \eqref{embeddingwilson} queda
\be
ds^2=\frac{R^2}{z^2}\left[-dt^2+dx^2\left(1+z'^2\right)\right]=\gamma_{ab}dx^adx^b,
\ee
en donde $z'=\frac{dz}{dx}$.
La acción de Nambu-Goto para la cuerda se escribe
\bea
S_{NG}&=&\frac{1}{2\pi\alpha'}\int_{-T/2}^{T/2}dt\int_{-l/2}^{l/2}dx\sqrt{-det(\gamma)}\nn\\
&=&\sqrt{4\pi\lambda}T\int_{-l/2}^{l/2}dx\frac{\sqrt{1+z'^2}}{z^2}.
\eea
Encontrar la superficie mínima en este caso es análogo a un problema de mecánica clásica en una variable. El Lagrangiano tiene una simetría traslacional
en $x$ y por lo tanto existe una corriente conservada ${\cal H}$ asociada a la misma dada por
\bea
{\cal H}&=&{\cal L}-z'\frac{\partial {\cal L}}{\partial z'}\nn\\
&=&\frac{1}{z^2\sqrt{1+z'^2}}=\frac{1}{z_{max}^2} ,\label{h}
\eea
en donde ${\cal L}$ es el Lagrangiano de Nambu-Goto y $z_{max}$ el máximo valor alcanzado por la hoja de mundo en el interior de AdS (ver figura \ref{will}). Esta ecuación puede reescribirse como un problema unidimensional en un potencial efectivo $V_{ef}(z)$ ($z_{max}$ es el punto de retorno)
\be
z'^2+V_{ef}(z)=0,~~~~ V_{ef}(z)=1-\left(\frac{z_{max}}{z}\right)^4\label{Vef}
\ee
A partir de \eqref{h} podemos despejar $x'(z)$ y calcular la separación entre quarks $l$ como función de $z_{max}$ obteniendo
\be
l=\int_{-l/2}^{l/2} dx=2\int_0^{z_{max}} dz\,\frac{dx}{dz}=2z_{max}\int_0^1dy\frac{y^2}{\sqrt{1-y^4}}=2z_{max}\frac{\sqrt{2\pi^3}}{\Gamma^2(1/4)},
\ee

Nótese que al aumentar la separación $l$ entre quarks en el borde, la cuerda explora regiones más interiores del espacio AdS.
Sustituyendo la expresión para $z'$ obtenida de \eqref{Vef} en la acción de Nambu-Goto queda
\be
S_{NG}=2\sqrt{4\pi\lambda}T\int_\epsilon^{z_{max}}dz\frac{z_{max}^2}{z^2\sqrt{z_{max}^4-z^4}}.
\ee
La divergencia para $z\rightarrow0$ en esta expresión se debe como mencionamos anteriormente a que la hoja de mundo se extiende hasta el borde de AdS.
Para regularizarla, sustraemos la auto-energía de los quarks restando el área de dos superficies planas que terminan sobre dos líneas rectas y
se extienden a lo largo de todo el espacio AdS
\be
m_{q}=\sqrt{4\pi\lambda}T\int_{\epsilon}^{z_{max}}\frac{dz}{z^2}.
\ee
Luego
\be
\ln\langle W_{{\cal C}}(\square)\rangle=S_{NG}-2m_{q}=-\frac{2\sqrt{4\pi\lambda}T}{z_{max}}\int_{\epsilon/z_{max}}^1\frac{dy}{y^2}\left(\frac{1}
{\sqrt{1-y^4}}-1\right),
\ee
y el potencial quark-antiquark queda
\be
V(l)=-\frac{\sqrt{\lambda}}{l}\frac{4\pi^2}{\Gamma^4(1/4)}.
\ee
Nótese que la energía se comporta como $1/l$ y es proporcional a $\sqrt{\lambda}$. Lo primero era esperable ya que ${\cal N}=4$ es invariante de escala.
Con respecto al coeficiente de proporcionalidad se debe decir que está predicción de AdS/CFT sugiere un apantallamiento en el régimen de acoplamiento fuerte.

\subsubsection{Lazo de Wilson circular}

\begin{figure}[htbp]
\centering
\includegraphics[width=0.38\textwidth]{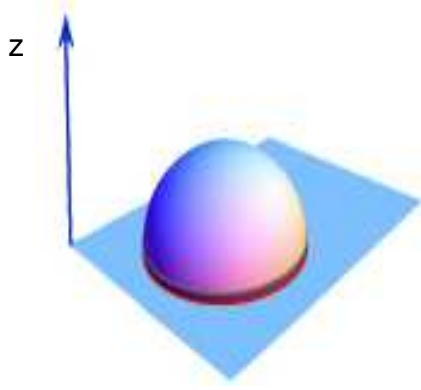}
\caption{Hoja de mundo de la cuerda que tiene al circulo de radio $a$ como borde.}
\label{circularWL}
\end{figure}

Para estudiar el lazo de Wilson definido sobre una curva circular spacelike \cite{corrado, Drukker} es conveniente escribir la métrica de AdS \eqref{AdSpoincare} (en signatura Euclídea) en la forma
\be
ds^2=\frac{1}{z^2}\left(dr^2+r^2d\phi^2+dz^2+dx_i^2\right).
\ee
En este caso la solución para la hoja de mundo de la cuerda (ver figura \ref{circularWL}) es
\be
z=\sqrt{a^2-r^2},~~~~0\leq r\leq a,~~~~0\leq \phi\leq 2\pi
\ee
en donde $a$ denota el radio del círculo en el borde de AdS. Calculando el área mínima de la superficie que termina sobre el cárculo (propiamente regularizada)
se encuentra que
\be
\ln\langle W_{{\cal C}}(\square)\rangle=-\sqrt{\lambda}
\ee

Tanto para la línea de Wilson como para el lazo circular se demostró en \cite{Zarembo, Semenoff} que el valor de expectación obtenido mediante
AdS/CFT coincide exactamente con lo obtenido en ${\cal N}=4$ SYM a todo orden en teoría de perturbaciones.
\begin{itemize}
\item {\bf Representaciones simétrica y antisimétrica, líneas y círculos}
\end{itemize}

La descripción gravitatoria dual para líneas de Wilson supersimétrico en ${\cal N}=4$ en la representación simétrica o antisimétrica del grupo
fue propuesta en \cite{drukker} y \cite{Yamaguchi} respectivamente. En \cite{drukker}
se argumento que soluciones de $D3$-branas enroscadas de forma tal que su volumen de mundo sea AdS$_2\times S^2\subset$ AdS$_5 $, con
campo eléctrico encendido y ubicadas en
un punto de $S^5$ son duales al valor de expectación del lazo de Wilson en la representación simétrica. El flujo del campo eléctrico caracteriza el
número de bloques de la representación. Por otro lado, la representación antisimétrica se obtiene enroscando $D5$ branas \cite{Yamaguchi}
(también eléctricamente cargadas) con el objetivo de obtener un volumen de mundo AdS$_2\times S^4$ donde AdS$_2\subset$ AdS$_5$ y $S^4\subset S^5$.
En ambos casos, la guía para obtener estas a priori caprichosas formas de embeber las branas en AdS$_5\times S^5$ fue la simetría.
 El lazo de Wilson 1/2-BPS descripto en la ecuación \eqref{halfwilson} para una curva tipo tiempo ${\cal C}$ recta o un circulo espacial
 es invariante ante el grupo $Osp(4^*|4)\approx SU(1,1)\times SU(2)\times SO(5)$ \cite{gomis}. La conjetura nos dice que debemos realizar estas
simetrías en el lado de gravedad y la forma natural de hacerlo es enroscando las $D3$ o $D5$ branas como ha sido explicado mas arriba.
 La presencia de un campo eléctrico da cuenta de la carga \cite{cm, Douglas}.
Al igual que en la representación fundamental, en el límite semiclásico debemos minimizar la acción de Dirac-Born-Infeld (DBI) para la D-brana
y calcular su valor evaluándola en la solución de las ecuaciones de movimiento para el embedding. En la representación simétrica el resultado obtenido
en el caso en que ${\cal C}$ es un circulo tipo espacio es \cite{drukker}
\be
\ln\langle W_{\cal C}(S)\rangle=2N\left(\kappa\sqrt{1+\kappa^2}+\sinh^{-1}\kappa\right),
\ee
en donde $\kappa$ es la constante de proporcionalidad del campo eléctrico. En la representación antisimétrica se obtiene \cite{Yamaguchi}
\be
\ln\langle W_{\cal C}(A)\rangle=-\frac{2N\sqrt{\lambda}}{3\pi}\sin^3\theta_k,
\ee
siendo $\theta_k$ la posición de la D5-brana en la $S^5$. El subíndice $k$ da cuenta de que dicha posición depende del campo eléctrico y determina el orden de la representación antisimétrica (número de bloques en el diagrama de Young).

En \cite{Kumar} se confirmaron estos resultados desde el punto de vista de la teoría conforme ${\cal N}=4$ utilizando técnicas de localización
y modelo de matrices.

\section{Aplicaciones a materia condensada}

En los último años la conjetura de Maldacena ha sido utilizada para modelar la física de materia condensada.
En particular la idea es usar la correspondencia AdS/CFT para estudiar materiales no convencionales que desafían los paradigmas tradicionales de la física
de materia condensada como, por ejemplo, materiales que involucran electrones fuertemente correlacionados.
Esta aplicación de la correspondencia ha sido llamada AdS/CMT \cite{lecturesadssmt}.

La técnica convencional para estudiar sistemas de materia condensada se basa en dos pilares fundamentales: la teoría del líquido de Fermi, la cual estudia
propiedades de electrones en sólidos como pequeñas perturbaciones de estado fundamental, y
la teoría de Landau, que afirma que al producirse una transición de fase debe haber una ruptura de simetría.

Recientemente diferentes experimentos con materiales no convencionales mostraron que en muchas situaciones de interés la teoría de Fermi no es una descripción
adecuada del problema. Una famoso indicativo de esto fue que la resistividad eléctrica observada era proporcional a la temperatura $T$ y no a $T^2$ como
predice la teoría de Fermi.

Lo que sucede en realidad es que las hipótesis de estas teorías no son validas, esto sucede por ejemplo en sistemas fuertemente correlacionados tales
como superconductores con alta temperatura crítica, compuestos de fermiones pesados, líquidos con efecto Hall cuántico fraccionario en gases de electrones
2-dimensionales
y líquidos de L\"{u}ttinger en una dimensión. En estos sistemas ocurren transiciones de fase que no se relacionan con la ruptura de ninguna simetría
 y asimismo presentan propiedades que no pueden ser descriptas con la teoría de perturbaciones usual de un líquido de Fermi.

Al estar en un regimen de acoplamiento fuerte la correspondencia AdS/CFT postula un dual gravitatorio para dichos sistemas.
 La idea en esta extensión de la conjetura es dar cuenta cualitativamente de fenómenos que no cuentan con una descripción adecuada.
 La aplicación de la conjetura AdS/CFT a estos sistemas persigue el objetivo de dar un respuesta desde el punto de vista gravitatorio
  a la siguiente pregunta que nos podemos hacer desde el punto de vista de la teoría de campos: ¿Cómo podemos clasificar todas
  las posibles fases gapless de la materia a baja temperatura? Las teorías de campos libres exhiben dos fases a bajas temperatura.
  Un bosón cargado libre será sometido a una condensación de Bose-Einstein, rompiendo espontáneamente la simetría de carga. El grado de
  libertad gapless es en consecuencia un bosón de Goldstone. Por otro lado, los fermiones libres cargados pueden formar una superficie de Fermi y
  los grados de libertad gapless son las excitaciones de dicha superficie. La dinámica de los bosones de Goldstone y de las excitaciones de la superficie de
  Fermi está estrechamente condicionada por la cinemática y está bien entendida. Para teorías fuertemente interactuantes la pregunta sobre las posibles fases
   de la materia se torna más complicada de responder, aquí es en donde la perspectiva gravitacional es útil.

En general, los fondos gravitatorios duales a un sistema de materia condensada involucran geometrías de agujeros negros cargados, la dualidad
prescribe que su horizonte y su carga eléctrica son mapeados a la temperatura y el potencial químico de la teoría de campos. Las técnicas desarrolladas
permiten estudiar, por ejemplo, conductividad Hall \cite{HartnollHall}, efecto Nernst \cite{HartnollNernst} y la ley de Ohm \cite{HartnollOhm}
en un régimen de acoplamiento fuerte del sistema de materia condensada.

La mayoría de los modelos AdS/CMT son del tipo bottom-up, es decir, son efectivos y en general no se derivan de la teoría de cuerdas. El objetivo de
estos modelos es dar una descripción macroscópica/cualitativa de ciertos fenómenos que ocurren en sistemas de materia condensada.

Hay muchos trabajos con aplicaciones de la conjetura al estudio de sistemas superfluídos y superconductores \cite{Herzog, Gubserbh, Hartnoll, albash,
 Horowitz}, líquidos de Fermi (FL) \cite{Cubrovic,Faulk}, líquidos que no cumplen con la teoría de Fermi (NFL) \cite{Liu, Faulkner, Vegh},
líquidos cuánticos semi-locales \cite{Iqbal}, puntos críticos cuánticos y aislantes topológicos \cite{Hoyos, AmmonTI} entre otras interesantes teorías de
materia condensada.

En el capítulo \ref{SC} se estudiará en detalle
cómo funciona AdS/CMT aplicada al estudio de un superconductor tipo $p$.

\newpage

Los resultados obtenidos en el desarrollo de esta tesis fueron
presentados en las siguientes reuniones científicas
\begin{itemize}
    \item XXXVIII reunión de la red Strings@ar, IAFE, Buenos Aires. Septiembre 2009.
    \item XLV reunión de la red Strings@ar, IAFE, Buenos Aires. Marzo 2011.
    \item Encuentro de Becarios del IFLP, La Plata. Diciembre 2010 y Diciembre 2012.
    \item International School on String Theory and Fundamental Physics,
DESY, Hamburg, Germany. Julio 2012.

\end{itemize}
y han sido documentados en las siguientes publicaciones científicas
internacionales con referato:
\begin{itemize}
    \item \begin{center}Wilson loop stability in the gauge/gravity correspondence.\\ Raúl Arias y Guillermo Silva.\\ JHEP 1001 (2010) 023;
    \\ arXiv:0911.0662 [hep-th]\end{center}
    \item\begin{center} Lorentzian AdS, Wormholes and Holography.\\ Raúl Arias, Marcelo Botta Cantcheff y Guillermo Silva.\\
    Phys.Rev. D83 (2011) 066015; \\arXiv:1012.4478 [hep-th]\end{center}
    \item \begin{center} Backreacting $p$-wave superconductors \\ Raúl Arias e Ignacio Salazar Landea.\\JHEP 1301 (2013) 157;\\arXiv:1210.6823 [hep-th]\end{center}
\end{itemize}
\newpage
\cleardoublepage
\def\baselinestretch{1}
\chapter{$AdS$ Lorentziano, Agujeros de Gusano y Holografía }\label{Wormhole}

En este capítulo se investigará la estructura de funciones de dos puntos para la QFT dual a un agujero de gusano Lorentziano.
La geometría en el volumen es una solución 5-dimensional de la gravedad de Einstein-Gauss-Bonnet y conecta causalmente dos
espacio-tiempos asintóticamente AdS. Se resumirá la prescripción de GKPW para calcular funciones de correlación de dos puntos para los
operadores duales $\cal O$ en signatura Lorentziana y se propondrá expresar los campos en el volumen en términos de su valor en los bordes
asintóticamente AdS independientes $\phi_0^\pm$, a lo largo del camino se mostrará cómo la ambigüedad de agregar modos normalizables en el volumen se
relaciona con los estados inicial y final sobre los que se calcula el correlador. Los valores en el borde son interpretados como fuentes para los operadores
duales $\cal O^\pm$ y se argumentará que, además de la posibilidad de entrelazamiento, existe un acoplamiento entre los grados de libertad definidos en
cada uno de los bordes. Se discutirá también la geometría AdS$_{1+1}$ debido a sus estructuras de borde similar. Basado en el análisis,
se propondrá un criterio geométrico que permitirá distinguir efectos de entrelazamiento y efectos de acoplamiento entre los dos
conjuntos de grados de libertad asociados a cada uno de los bordes desconectados.

Las geometrías asintóticamente AdS cumplen un rol destacado en la correspondencia AdS/CFT \cite{malda, GKP, Ewitten}
ya que estas proveen duales gravitatorios para teorías cuánticas de campos (QFT) con puntos fijos conformes en el UV.
Existe un consenso general, basado en un gran número de comprobaciones, para la interpretación dual de diferentes soluciones
asintóticamente AdS: una solución de agujero negro se supone describe una QFT térmica \cite{wittenT}, una solución que
interpola entre un horizonte de AdS (correspondiente a un punto fijo IR) y una geometría de AdS en infinito con
diferente radio realiza el flujo del grupo de renormalización entre dos puntos conformes \cite{pw}.
Una tercera posibilidad son ciertos solitones (cargados) en AdS que se interpretan como estados coherentes excitados de la QFT \cite{llm}.

Se discutirá la problemática que se genera cuando la solución de gravedad es un agujero de gusano que conecta
causalmente dos bordes asintóticamente AdS$_{d+1}$ (Lorentzianos). La holografía y la correspondencia AdS/CFT en presencia de
múltiples bordes aún no están bien comprendidas. La implementación del paradigma AdS/CFT en tales casos sugiere que la teoría
de campos dual vive en la unión de dos bordes disjuntos, y entonces es el producto de las teorías de campos de los diferentes bordes
(ver \cite{deboer}). Voy a rever esta afirmación y discutir cuándo las dos teorías duales son independientes, están acopladas o
son desacopladas.

Para signatura Lorentziana, las geometrías de agujero de gusano están descartadas (en $d\geq2$ dimensiones) como soluciones de
la acción de Einstein-Hilbert satisfaciendo condiciones naturales de causalidad: bordes desconectados deben estar separados
por horizontes \cite{galloway} (\cite{svr} es un trabajo reciente en 2+1 dimensiones). Por lo tanto, los estudios de agujeros
de gusano en teoría de cuerdas y en el contexto de la correspondencia AdS/CFT se han concentrado en espacios de signatura
Euclídea, particularmente motivados por sus aplicaciones a cosmología (ver referencias de \cite{maldamaoz,polchi}).
En el contexto Euclídeo existe un teorema que establece que los espacios con bordes desconectados (de curvatura escalar positiva)
también están descartados como soluciones de Einstein sobre variedades de curvatura negativa \cite{WittenYau} (ver también \cite{Cai}).
Además, \cite{WittenYau} prueba que para bordes de curvatura negativa la teoría de campos dual definida sobre ellos debe ser inestable
para $d\geq3$ (ver también \cite{buchel}). Los agujeros de gusano estudiados en \cite{maldamaoz} evitan el teorema de
\cite{WittenYau} debido a que fueron construidos como rebanadas hiperbólicas de AdS soportados con campos de supergravedad extras.

El ejemplo canónico de un espacio con dos bordes, con signatura Lorentziana, separados por un horizonte es el
agujero negro eterno. En \cite{Eternal} se hace contacto entre esta geometría y la formulación dinámica de campos térmicos (TFD)
de una QFT a temperatura finita \cite{TFD}: los dos bordes desconectados dan lugar a dos copias desacopladas ${\cal H}_{\pm}$ de la
teoría de campos dual y los correladores no nulos $\langle {\cal O}^+({\mathbf x})\, {\cal O}^-({\mathbf x'})\rangle$
se interpretan como promedios sobre un estado entrelazado que codifica la información estadística/térmica de la geometría
(ver también \cite{Azeyanagi} y \cite{vaman}). Un segundo ejemplo Lorentziano de interés con dos bordes desconectados fue
construido en \cite{balasu} trabajando con un orbifolio no singular de AdS$_3$. El resultado de la construcción condujo a dos bordes
cilíndricos causalmente conectados y su teoría de campos dual involucra el límite DLCQ de la teoría conforme que vive en una configuración de
D1-D5 branas, pero el acoplamiento entre los grados de libertad de los diferentes bordes no ha sido clarificado.
La diferencia principal entre estos dos ejemplos es que en el último caso existe contacto causal entre los bordes conformes.
El teorema \cite{galloway} se elude en el segundo caso debido a que el orbifold tiene una dirección compacta ($S^1$ en este caso) fibrada
sobre AdS$_2$, en donde el mencionado teorema no aplica.

Los teoremas \cite{galloway} que prohíben agujeros de gusano Lorentzianos no son válidos cuando se trabaja
con gravedades de orden más alto. Por otro lado, correcciones en la curvatura de orden más alto a la gravedad estándar
de Einstein son esperables en una teoría cuántica de la gravedad. Sin embargo, no se conoce mucho acerca de la
forma precisa de las correcciones en derivadas de orden superior más que en unos pocos casos máximamente supersimétricos.
Como desde el punto de vista puramente gravitatorio la teoría más general que involucra ecuaciones de campo de segundo orden
para la métrica es del tipo de Lovelock \cite{love}, voy a elegir trabajar con la más simple de ellas conocida como
teoría de Einstein-Gauss-Bonnet. La acción para esta teoría solo contiene términos hasta orden cuadrático en la curvatura
y mi interés en la solución de agujero de gusano, encontrada en \cite{Dotti}, es que permite un tratamiento analítico.
La geometría corresponde a un agujero de gusano estático que conecta dos regiones localmente asintóticas AdS con
variedad de base $\tilde\Sigma$, la cual en $d+1=5$ toma la forma
$\tilde\Sigma= H^3$ o $S^1\times H^2$, en donde
$H^2$ y $H^3$ son espacios hiperbólicos (cocientados) dos y tres- dimensionales. La geometría resultante es suave,
no contiene horizontes, y las dos regiones asintóticas están causalmente conectadas. En el caso de la solución
en 5 dimensiones \cite{Dotti} se realizó un análisis de estabilidad en \cite{diego}.

 Es comúnmente aceptado que la QFT dual a una geometría de agujero de gusano debe corresponder
a dos teorías de gauge independientes definidas sobre cada uno de los bordes y la geometría del agujero de gusano codifica
la información de un estado entrelazado entre ellos. Por otro lado, la conexión causal entre los bordes da lugar a un acoplamiento no
 trivial entre las dos teorías duales \cite{balasu}. Se va a argumentar, continuando analíticamente a una sección Euclídea
 del espacio-tiempo, que el resultado no nulo obtenido para el correlador $\langle {\cal O}^+({\mathbf x})\, {\cal O}^-({\mathbf x'})\rangle$
 entre operadores localizados en bordes opuestos señala la existencia de un acoplamiento entre los campos asociados a cada uno
 de los bordes.

El capítulo se organizará de la siguiente manera: en la sección \ref{GKPW} se revisará la prescripción GKPW para calcular
correladores de la QFT a partir de la teoría de gravedad mencionando las peculiaridades de la signatura Lorentziana.
En la sección \ref{gkpww} se extenderá la prescripción GKPW para el caso de datos sobre dos bordes asintóticos independientes.
Se aplicará esto a AdS$_2$, reproduciendo los resultados que aparecen en la literatura, y al agujero de gusano
\cite{Dotti} mostrando sus similitudes. En la sección \ref{holentanglement} se discutirán varios argumentos
relacionados con la posibilidad de entrelazamiento y/o interacción entre las dos QFT duales.

\section{ Prescripción GKPW con un solo borde asintótico}
\label{GKPW}

En esta sección se reformulará lo estudiado en la sección \ref{funcco} en términos de funciones de Green, con el objeto de facilitar la extensión de la prescripción GKPW a espacios con más de un borde que se realizará en la siguiente sección.

La prescripción de GKPW \cite{GKP,Ewitten} iguala la función de partición de gravedad
para un espacio-tiempo ${\cal M}$ asintóticamente AdS$_{d+1}$, entendida como una funcional
de los datos de borde, a la funcional generatriz de funciones de correlación para una teoría de
campos conforme (CFT) definida sobre el borde conforme del espacio-tiempo $\partial\cal M$.
Explícitamente la prescripción es
\be
\mathcal{Z}_{gravity}\left[\phi(\phi_0)\right]=\langle
e^{i\int_{\partial\cal
M}d^d{\mathbf{x}}\;\phi_0(\mathbf{x})\mathcal{O(\mathbf{x})}}\rangle\;.
\label{partition}
\ee
En el lado izquierdo $\phi_0=\phi_0(\mathbf{x})$ es el valor en el borde del campo
$\phi$, el lado derecho es la funcional generatriz de correladores en la CFT para operadores
$\mathcal{O}$ duales al campo $\phi$. En este capítulo se trabajará en el límite semiclásico del
espacio-tiempo (límite de $N$ grande en la CFT) y por lo tanto el lado izquierdo en
(\ref{partition}) podrá ser aproximado por la acción on-shell del campo $\phi$, el cual por simplicidad
será tomado como un campo escalar de masa $m$. Se centrará el interés en geometrías de tiempo real
(Lorentzianas) debido a que la prescripción (\ref{partition}) en este caso se encuentra incompleta,
esto se debe a la necesidad de especificar los estados inicial y final $\psi_{\mathrm {i,f}}$ sobre los cuales se
calcula los correladores del lado derecho.

Para fijar la notación se resumirá la prescripción para campos escalares masivos resaltando los puntos
importantes para nuestros argumentos. La métrica de $AdS_{d+1}$ en coordenadas de Poincaré es
\be
 ds^2=\frac{R^2}{z^2}(d\mathbf{x}^2+ dz^2)\,,
 \label{adsp}
\ee
en donde el termino $d\mathbf{x}^2$ significa $-dt^2+d\vec{x}^2$. El borde (conforme) de AdS se
ubica en $z=0$ y existe un horizonte en $z=\infty$\footnote{ En el caso Euclídeo $z=\infty$ es solo un punto,
lo que conduce a un semiplano $z\ge0$ en (\ref{adsp}) compactificado a una esfera.}. La solución para la
ecuación de movimiento para el campo $\phi$ sujeto a datos de borde $\phi_0$ se escribe comúnmente
\be
 \phi(\mathbf{x},z)=\int_{\partial\cal M}d\mathbf{y}\,{\sf K}(\mathbf{x},z\mid\mathbf{y})\,\phi_0(\mathbf{y})\,.
\label{phi}
\ee
En el límite de campo libre, la ecuación de Klein-Gordon (KG) muestra que el comportamiento asintótico para $\phi$ es
\be
\phi(\mathbf{x},z)\sim z^{\Delta_\pm}\phi_0(\mathbf x),~~~~z\to0
\label{asympfi}
\ee
en donde\footnote{Campos escalares de masa negativa están permitidos en AdS siempre que $\mu\ge0$. El mínimo
permitido para la masa del campo escalar en AdS$_{d+1}$ esta dado por la llamada cota de Breitenlohner-Freedman (BF)
$\mu_{BF}=0$, o equivalentemente $m_{BF}^2=-d^2/4$ \cite{BF}.}
\be
\Delta_\pm=\frac{d}{2}\pm\mu,~~~~~~\mu=\sqrt{\frac{d^2}{4}+m^2R^2}\,.
\label{lambda}
\ee
El propagador volumen-borde $\sf K$ en (\ref{phi}) debe satisfacer
\cite{GKP,Ewitten}
\be
(\Box-m^2)\,{\sf K}(\mathbf{x},z\mid \mathbf{y})=0\label{box}
\ee
con condiciones de borde
\be
 {\sf K}(\mathbf{x},z\mid \mathbf{y})\sim z^{\Delta_-}\delta(\mathbf{x}-\mathbf{y})\,,~~~~z\to0\,.
 \label{K boundary condition}
\ee
Finalmente, el propagador volumen-borde $\sf K$ puede ser relacionado con la función de Green (Dirichlet)
volumen-volumen ${\sf G}(\mathbf{x},z|\mathbf{y},z')$ a través de la segunda identidad de Green, resulta entonces
que ${\sf K}$ puede ser obtenido como la derivada normal de ${\sf G}$ evaluada en el borde del espacio-tiempo
(ver \cite{fdh,mcg}) como
\be
 {\sf K}(\mathbf{x},z\mid \mathbf{y})=\lim_{z'\to0}\sqrt{-g}g^{z'\!z'}\partial_{z'}{\sf
G}\,(\mathbf{x},z\mid\mathbf{y},z')\,.
\label{KfromG}
\ee
Dos comentarios vienen al caso: i) la solución en el volumen del espacio para unas dadas condiciones de borde
$\phi_0$ calculadas a partir de (\ref{phi}) no es única ya que espacios AdS Lorentzianos admiten soluciones
normalizables $\varphi(\mathbf{x},z)$ que pueden sumarse a (\ref{phi}) sin alterar el comportamiento en el borde
(\ref{K boundary condition}), explícitamente
\be
 \phi(\mathbf{x},z)=\int_{\partial\cal M}d\mathbf{y}\,{\sf K}(\mathbf{x},z\mid\mathbf{y})\,
 \phi_0(\mathbf{y})+\varphi(\mathbf{x},z)\,.
 \label{homo}
\ee
La consecuencia de su inclusión en la CFT se interpreta como que se están fijando los estados inicial y final
$|\psi_{\mathrm {i,f}}\rangle$ sobre los cuales se calculan los valores de expectación en el lado derecho de
(\ref{partition}). La segunda observación es ii) en signatura Lorentziana la superficie $z=\infty$ es un horizonte
de Killing y por lo tanto se necesita especificar el campo sobre un borde adicional para obtener un problema
de Dirichlet bien definido (ver fig.\ref{contributions}(a)). Estas dos observaciones están relacionadas al echo de
se requiere una segunda condición para fijar completamente el propagador volumen-borde {\sf K} (recordar que en espacio
Euclídeo pedir regularidad en el volumen implica $K\to0$ cuando $z\to\infty$). La condición faltante sobre
{\sf K}  impuesta en el horizonte ($z=\infty$) se expresa mejor en términos de modos de Fourier de ondas puramente entrantes
(funciones que decaen exponencialmente) para momento tipo tiempo (espacio), este es un problema bien conocido de las teorías
cuánticas de campos en espacio-tiempos curvos y se reduce a la elección de un vacío. La incorporación de
modos normalizables (tipo tiempo) induce una componente saliente del horizonte, la cual se interpreta naturalmente como
una excitación (ver \cite{GKP,bala,bala2,balagid}
y \cite{vaman,kv,star,herson} para trabajos relacionados).

Me interesa calcular la función de correlación de dos puntos en la teoría de campos dual. Con este fin se necesita
la acción evaluada en la solución para un campo escalar a orden cuadrático
\be
S=-\frac12\int d\mathbf{x}dz
\sqrt{-g}\left(g^{\mu\nu}\partial_\mu\phi\,\partial_\nu\phi+m^2\phi^2\right)\,.
\label{freefi}
\ee
integrando por partes y evaluando en la solución, la contribución del borde conforme esta dado por (ver \cite{star} para una
discusión sobre posibles contribuciones del horizonte)
\be
S[\phi_0]=\frac12\int
d\mathbf{x}\left[\sqrt{-g}g^{zz}\phi(\mathbf{x},z)\,\partial_z\phi(\mathbf{x},z)\right]_{z=0}
\label{action scalar}\,.
\ee
Insertando (\ref{phi}) en esta expresión obtenemos la acción evaluada en la solución como una funcional
del valor del campo en el borde  $\phi_0$
\be
S[\phi_0]=\frac12\int d\mathbf{y}d\mathbf{y'} \phi_0(\mathbf{y})\,\Delta(\mathbf{y},\mathbf{y'})\,\phi_0(\mathbf{y'})
\label{onshact}
\ee
en donde
\be
\Delta(\mathbf{y},\mathbf{y'})=\int
d\mathbf{x}\left[\sqrt{-g}g^{zz}{\sf K}(\mathbf{x},z\mid
\mathbf{y})\partial_z{\sf K}(\mathbf{x},z\mid
\mathbf{y'})\right]_{z=0}\,.\label{Delta}
\ee
Teniendo en cuenta (\ref{K boundary condition}) y (\ref{KfromG}) en (\ref{Delta}) se obtiene
\bea
\label{Delta-fromG}
\Delta(\mathbf{y},\mathbf{y'})&\sim&\left[\sqrt{-g}g^{zz}\partial_z{\sf
K}(\mathbf{y},z\mid\mathbf{y'})\right] _{z=0}\\
&\sim&\lim_{z,z'\rightarrow0}(\sqrt{-g}g^{zz})(\sqrt{-g}g^{z'\!z'})\frac{\partial^2}{\partial z\,\partial{z'}}{\sf
G}(\mathbf{y},z\mid\mathbf{y'},z')\,.\label{DeltafromG}
\eea
Esta relación ha sido usada para relacionar, en el límite semiclásico, la función de dos puntos con las
geodésicas de la geometría \cite{shenker}.

Resumiendo, la función de dos puntos para un operador $\cal O$ dual al campo $\phi$ se obtiene a partir
de la acción evaluada en la solución como
\be
\langle \psi_{\mathrm f}| \mathcal{O(\mathbf{y})}\mathcal{O(\mathbf{y'})}|\psi_{\mathrm i}\rangle=-i\frac{\delta^2
S[\phi_0]}{\delta\phi_0(\mathbf{y})\,\delta\phi_0(\mathbf{y'})}=-i\Delta^{\mathrm {i,f}}(\mathbf{y},\mathbf{y'})\,.
\label{2ptf}
\ee
Los estados inicial y final $\psi_{\mathrm {i,f}}$ del lado izquierdo involucran la ambigüedad ante la adición
de una solución normalizable a (\ref{phi}), en la siguiente sección se verá explícitamente como se manifiestan estos estados en
(\ref{Delta}).

\subsection*{AdS en coordenadas globales}

La receta para obtener correladores de una teoría de campos a partir de cálculos de gravedad involucra
evaluar cantidades, definidas en el volumen del espacio-tiempo, en el borde conforme, como puede sospecharse
de (\ref{asympfi}), (\ref{K boundary condition}) y (\ref{Delta}) la evaluación conduce a singularidades y por lo tanto
se necesita una regularización. En lo que sigue se discutirá cómo se hace ésto en AdS expresado en coordenadas globales
debido a que el caso del agujero de gusano discutido después coincide en su región asintótica con este sistema de coordenadas
y será por lo tanto regularizado de la misma manera. En el camino se mostrará cómo aparece en los cálculos la especificación de
los estados inicial y final $\psi_{\mathrm {i,f}}$.

La regularización de (\ref{2ptf}) se realiza imponiendo la condición de borde en alguna distancia finita en el volumen y tomando
el límite al borde al final de los cálculos (ver \cite{rastelli} para una sutileza al tomar este límite). La variedad
AdS$_{d+1}$ esta totalmente cubierta por las llamadas coordenadas globales, en las cuales la métrica toma la forma
\be
ds^2=R^2\left[-\frac{dt^2}{1-x^2}+\frac{dx^2}{(1-x^2)^2}+\frac{x^2}{1-x^2}d\Omega^2_{d-1}\right]
\ee
en donde se ha cambiado la variable $\rho$ a $x=\tanh\rho$ para mapear el borde conforme a $x=1$.

Se imponen los datos en el borde a una distancia finita $x_\epsilon=1-\epsilon$, luego, por consistencia se necesita que
\be
 \lim_{x\to x_\epsilon}{\sf K}(t,{\Omega},x\mid t',{\Omega'},x_\epsilon)=\frac{\delta(t-t')
 \delta({\Omega}-{\Omega'})}{\sqrt{g_{_\Omega}}}\, .
 \label{Kbc}
\ee
y {\sf K} regular en el interior. El propagador volumen-borde $\sf K$ que satisface (\ref{Kbc}) puede ser
obtenido a partir de la solución de la ecuaciones de Klein-Gordon ${\sf \phi}(t,{\Omega},x)=e^{-i\omega
t}Y_{lm}({\Omega})f_{l\omega}(x) $ como
\be
{\sf K}(t,{\Omega},x\mid t',{ \Omega'},x_\epsilon)=\int_{-\infty}^{\infty}
\frac{d\omega}{2\pi}\sum_{lm}e^{-i\omega
(t-t')}Y_{lm}({\Omega})Y_{lm}^\ast({\Omega'})f_{l\omega}(x)
\label{propfeyn}
\ee
si se normaliza\footnote{Los esféricos armónicos $Y_{lm}$ sobre la esfera $d-1$ satisfacen $\nabla^2 Y_{lm}=-q^2Y_{lm}$ con
$q^2=l(l+d-2),\,l=0,1,\ldots$} $f_{l\omega}(x_\epsilon)=1$. La ecuación diferencial que satisface $f_{l\omega}(x)$ es
\be
(1-x^2)\frac{d^2 f_{l\omega} }{dx^2}+\frac{d-1-x^2}{x}\frac{d
f_{l\omega}}{dx}+\left(\omega^2-\frac{q^2}{x^2}-
\frac{m^2R^2}{1-x^2}\right)f_{l\omega}=0\,.
\label{eqdiff2}
\ee
La solución de esta ecuación es una combinación lineal de dos funciones hipergeométricas, pero una de ellas diverge
cuando $x\rightarrow0$ y entonces pedir regularidad en el volumen del espacio demanda anularla. La solución regular propiamente
normalizada es
\be
f_{l\omega}(x)=\left(\frac{x^{-\frac{d}{2}+\nu +1}\left(1-x^2\right)^{\frac{1}{2}\Delta_+}}
{(1-\epsilon)^{-\frac{d}{2}+\nu+1} ((2-\epsilon ) \epsilon )^{\frac{1}{2}\Delta_+}}\right)
   \frac{ \, _2F_1\left(\frac{1}{2} (\mu +\nu -\omega +1),\frac{1}{2} (\mu +\nu +\omega +1);\nu +1;x^2\right)}
   {\, _2F_1\left(\frac{1}{2} (\mu +\nu -\omega
   +1),\frac{1}{2} (\mu +\nu +\omega +1);\nu +1;(1-\epsilon)^2\right)}
\label{solglobal}
\ee
aquí $_2F_1$ es la función hipergeométrica de Gauss con $\mu$ dado por (\ref{lambda}) y $\nu=\sqrt{(\frac{d-2}{2})^2+q^2}$,
la simetría de la función hipergeométrica en sus primeros dos argumentos implica que $ f_{l\omega}(x)=f_{l\,-\omega}(x)$.
El comportamiento asintótico de la solución (\ref{solglobal}) cerca del borde va como
\be
f_{l\omega}(x)\sim C_+\, (1-x)^{\frac12{\Delta_+}}+C_-\,(1-x)^{\frac12\Delta_-}\,,
\label{asympglobal}
\ee
en donde $\Delta_\pm$ esta dado por (\ref{lambda}) y $C_{\pm}=C_{\pm}(\mu,\nu,\omega)$. En signatura Lorentziana
el operador de KG posee soluciones normalizables, estas aparecen para valores particulares de $\omega$ dados por \cite{BF,bala,Avis}
\bea
 \omega_{nl}&=&\pm(2n+\nu+\mu+1)\nn\\
  &=&\pm(2n+l+\Delta_+),~~~~~~n,l=0,1,2\ldots\,,
\label{polos}
\eea
o dicho de otra manera, estas son las frecuencias para las cuales\footnote{ver \cite{BF} para una condición de cuantización alternativa
para $0\le\mu\le1$ y \cite{wittenDT} para su interpretación en el contexto de AdS/CFT.} $C_-=0$. La discretitud del espectro
manifiesta el carácter tipo "\,caja" de AdS y desde la perspectiva dual proviene del echo de que $S^3$ es una geometría compacta.
La cuantización de los estados (\ref{polos}) en el volumen se interpreta como dual a los estados definidos sobre el borde conforme
$S^3\times \mathbb R$ de AdS.
\begin{figure}
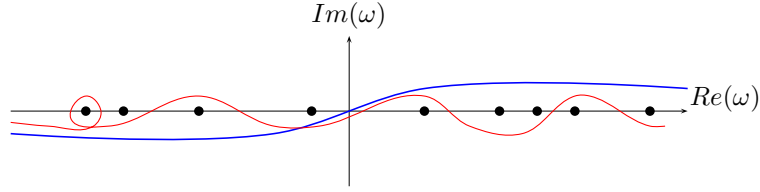

\vspace{2cm}
\hspace{8cm}
\psline[linewidth=.2pt]{->}(0,-1)(0,1)
\psline[linewidth=.2pt]{->}(-4.5,0)(4.5,0)
\psdots*(-3.5,0)(-3,0)(-2,0)(-0.5,0)(1,0)(2,0)(2.5,0)(3,0)(4,0)
\pscurve[linewidth=.6pt,linecolor=blue](-4.5,-0.3)(-1,-0.3)(1,0.3)(4.5,0.3)
\rput[bI](0,1.1){$Im(\omega)$}
\rput[bI](5,0){$Re (\omega)$}
\pscurve[linewidth=.4pt,linecolor=red](-4.5,-0.15)(-4,-0.2)(-3.4,-0.2)(-3.3,0)(-3.5,0.2)(-3.7,-0.1)(-3.2,-0.2)
(-2,0.2)(-1,-0.2)(-0.4,-0.2)(1,0.2)(1.6,-0.2)(2.3,-0.3)(3,0.2)(3.2,0.2)(4,-0.2)(4.2,-0.2)
\vspace{1cm}
\caption{{\bf Contornos en el plano complejo $\bf \omega$ :} al hacer la integración sobre $\omega$ en (\ref{deltatt})
cualquier contorno arbitrario (dibujado en rojo) puede ser deformado al contorno de Feynman (dibujado en azul) más
contribuciones provenientes de rodear los polos (\ref{polos}). El rodeo de polos positivos (negativos) fijan los estados
inicial (final) $\psi_{\mathrm {i,f}}$ en (\ref{2ptf}).}
\label{Contour}
\end{figure}

La función de dos puntos para los operadores de la QFT dual se obtienen de insertar
\be
\phi(t,\Omega,x)=\int dt'd\Omega'\sqrt{g_{_{\Omega'}}}\,{\sf K}(t,\Omega,x\mid
t',\Omega',x_\epsilon)\,\phi_0(t',\Omega') \label{identity}
\ee
en la acción (\ref{freefi}), nótese que no se incluyeron soluciones normalizables a (\ref{identity}) (ver próximo párrafo).
La acción evaluada en la solución conduce a un termino de borde evaluado en $x_\epsilon$ (ver (\ref{action scalar})-(\ref{Delta})) y
la expresión regularizada para $\Delta(t,\Omega\mid t',\Omega')$ en (\ref{onshact}) se escribe como\footnote{Siguiendo \cite{rastelli},
cuando se calcula (\ref{delta glob}) primero se calcula la derivada en $x$ y después se toma el límite $\epsilon\to 0$.}
\bea
  \Delta_{\sf reg}(t,\Omega|t',\Omega')&=&-\frac{1}{\sqrt{g_{_{\Omega}}}}\left[\sqrt{-g}g^{xx}\,\partial_x{\sf K}
  (t,\Omega,x\mid t',\Omega',x_\epsilon)\right] _{x=x_\epsilon}\nn\\
  &=&-\int\frac{d\omega}{2\pi}\,   e^{-i \omega (t- t')}
  \sum_{lm}Y_{lm}(\Omega)Y_{lm}^\ast(\Omega')
  \left[\frac{x^{d-1}}{(1-x^2)^{\frac{d-2}2}} \partial_x f_{l\omega}(x)\right]_{x=x_\epsilon}\nn\\
 &=& -\sum_{lm}Y_{lm}(\Omega)Y_{lm}^\ast(\Omega')\int \frac{d\omega}{2\pi}\,
 e^{-i \omega (t- t')}\left[\frac{x^{d-1}}{(1-x^2)^{\frac{d-2}2}} \partial_x f_{l\omega}(x)\right]_{x=x_\epsilon}\,,
 \label{delta glob}
\eea
en donde la primera linea proviene de (\ref{Delta}) teniendo en cuenta (\ref{Kbc}).

Algunos comentarios sobre (\ref{delta glob}): cuando se toma el límite $\epsilon\to 0$, la expresión en la última linea es ambigua
debido a la existencia de polos simples ubicados a lo largo del contorno de integración en el plano $\omega$ en (\ref{polos})
\footnote{Su origen se debe a haber pedido $f_{l\omega}(x_\epsilon)=1$.}.
Estos polos manifiestan la existencia de soluciones normalizables en el volumen de la geometría (ver (\ref{homo})) y entonces, para
definir el contorno de integración en $\omega$ se necesita alguna prescripción para evitar los polos.
La elección del contorno se entiende tradicionalmente como la elección entre las funciones de Green avanzada/retardada/Feynman, en lo que sigue
se trabajará con la de Feynman. Quiero llamar la atención acerca de la observación (ver \cite{Skenderis}) sobre la relación entre contornos
en el plano complejo $\omega$ y la elección de soluciones normalizables. La observación es simple, cualquier elección particular de contornos es
equivalente por deformación a elegir el contorno de Feynman más contribuciones que provendrán de rodear los polos (\ref{polos}).
Luego, la ambigüedad en la expresión  (\ref{identity}) proveniente de la adición de modos normalizables arbitrarios se translada
en una elección del contorno en el plano complejo $\omega$ (ver fig. \ref{Contour}). El contorno de Feynman naturalmente conduce
a correladores ordenados temporalmente, y el rodeo de modos normalizables positivos (negativos) fija el estado inicial (final)
$\psi_{\mathrm {i,f}}$ en el lado izquierdo de (\ref{2ptf}). La elección del contorno retardado da lugar a funciones respuesta en lugar de
funciones de correlación.
En resumen, los estados son interpretados como creados a partir de un fundamental $|\psi_0\rangle$ asociado al contorno de integración de
referencia elegido.

El límite $\epsilon\to0$ de la expresión dentro de los corchetes en (\ref{delta glob}) también presenta polos (analíticos y
no analíticos) en $\epsilon$. El resultado físico se obtiene renormalizando los datos en el borde teniendo en cuenta el comportamiento
asintótico en la dirección radial (ver (\ref{asympglobal})), en el presente caso esto lleva a reescalear $\phi_0$ como
(ver \cite{Ewitten,rastelli,mcg})
\be
\phi_0(t,\Omega)=\epsilon^{\frac12\Delta_-}\phi_{\sf ren}(t,\Omega)\,.
\label{renorm}
\ee
Además, como eventualmente estamos interesados en funciones de correlación para puntos separados, no se deben tener en cuenta
términos de contacto proporcionales a potencias enteras positivas de $q^2$. El término finito en el límite $\epsilon\to0$ es
\bea
\Delta_{\sf ren}(t,\Omega| t',\Omega')&\equiv&\lim_{\epsilon \to0}\epsilon^{\Delta_-}\,\Delta_{\sf reg}(t,\Omega| t',\Omega')\nn\\&=&\sum_{lm}Y_{lm}(\Omega)Y_{lm}^\ast(\Omega')
\int \frac{d\omega}{2\pi}\,e^{-i \omega (t-t')}\nn\\
 &&\times\frac{\Delta_+}{2^{\Delta_-}}\frac{\Gamma (1-\mu )}{\Gamma (1+\mu)}
 \frac{\Gamma \left(\frac{1}{2}(-\omega+\nu +\mu +1)\right)\Gamma \left(\frac{1}{2} (\omega+\nu+\mu +1)\right)}
{\Gamma \left(\frac{1}{2} (-\omega+\nu -\mu +1)\right)
 \Gamma \left(\frac{1}{2} (\omega+\nu -\mu +1)\right) }\,.
 \label{deltatt}
\eea
El numerador de esta última expresión muestra explícitamente la aparición de polos a lo largo del contorno de integración precisamente
a las frecuencias $\omega_{nl}$ dadas por (\ref{polos}). La especificación de un contorno $\cal C$ en el plano complejo $\omega$
fija los estados inicial y final $\psi_{\mathrm {i,f}}$ cuando se compara con el camino de Feynman estándar (ver fig. \ref{Contour}).

A partir de la discusión que sigue a (\ref{delta glob}) debe quedar claro que las funciones de correlación calculadas en el
estado de vacío de la QFT se obtienen de elegir el contorno de Feynman para la integración en  $\omega$ en (\ref{deltatt}).
Integrando sobre $\omega$ se obtiene\footnote{Generalmente (\ref{polos}) son las únicas divergencias en (\ref{deltatt}), se debe tener
un cuidado especial para valores enteros de $\mu$. No se discutirán los detalles en esta tesis debido a que la experiencia con la correspondencia
AdS/CFT ha mostrado que la función de correlación no cambia cualitativamente en este límite.}
\bea
\Delta_{\sf ren}^F(t,\Omega|t',\Omega')&=&2i\frac{\Delta_+\,\Gamma(1-\mu)}{2^{\Delta_-}\Gamma(1+\mu)}\sum_{lm}Y_{lm}(\Omega)Y_{lm}^\ast(\Omega')\nn\\
&&\times\left[\sum_{n=0}^\infty\frac{(-1)^{n}}{n!}\frac{\Gamma(n+l+\Delta_+)}{\Gamma(n+l+\frac d2)\Gamma(-(n+\mu))}
e^{-i|t-t'|(2n+l+\Delta_+)}\right]
\eea
La suma sobre los residuos conduce a
\bea
\langle 0|T{\cal O}(t,\Omega){\cal O}(t',\Omega')|0\rangle&=&-i\Delta_{\sf ren}^F(t,\Omega|t',\Omega')=
-\frac{2\Delta_+}{2^{\Delta_-}\Gamma(\mu)}
 \sum_{lm}Y_{lm}(\Omega)Y_{lm}^\ast(\Omega')\frac{\Gamma(l+\Delta_+)}{\Gamma(l+\frac d2)}\nn\\
 &&\times\,\, e^{-i|t-t'|(l+\Delta_+)}\,_2F_1\left(1+\mu,l+\Delta_+,l+\frac d2;e^{-2i|t-t'|}\right)\,. \label{2ptglobal}
\eea

\section{Prescripción GKPW para agujeros de gusano Lorentzianos}
\label{gkpww}

El gol en esta sección será extender la prescripción de GKPW al caso de múltiples bordes tipo tiempo, por simplicidad se discutirá el caso con dos bordes. Sobre fondos gravitatorios generales la correspondencia AdS/CFT sugiere que la presencia de dos bordes tipo tiempo se debe asociar con la
existencia de dos conjuntos de operadores duales ${\cal O}^\pm$ correspondientes a las dos condiciones de borde independientes $\phi^\pm_0$
que deben ser impuestas al campo $\phi$ cuando se resuelve la ecuación de onda\footnote{Algunos autores asumen que las dos teorías de campos
duales están desacopladas debido a la estructura disconexa del borde \cite{Azeyanagi}.}.

Se considerarán agujeros de gusano con topología en el borde conforme de la forma ${\mathbb R}\times \Sigma $, con $\mathbb R$ representando
el tiempo y $\Sigma=\Sigma_++\Sigma_-$ la unión de las dos copias (espaciales) compactas y disjuntas $\Sigma_\pm$.
Los agujeros de gusano pueden cubrirse por un sistema de coordenadas $(x,t,\theta)$ en donde $x$ es la coordenada holográfica radial
en el volumen y $(x_\pm,t,\theta)$ las coordenadas que parametrizan los dos bordes $\mathbb
R\times\Sigma_\pm$.
\begin{figure}
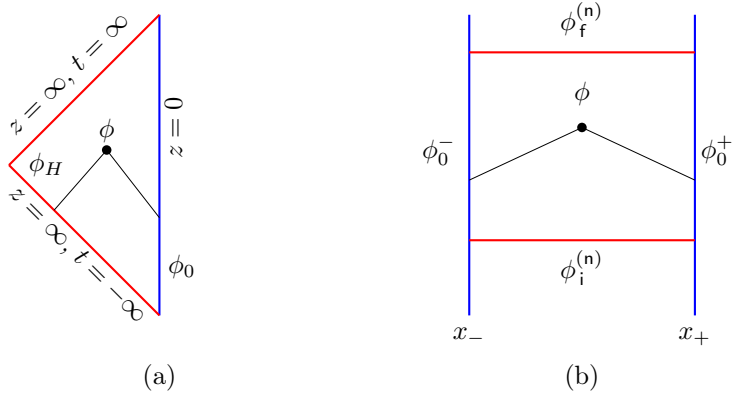

\vspace{2cm}
\hspace{5cm}
\psline[linewidth=.8pt,linecolor=blue](0,-2)(0,2)
\psline[linewidth=.8pt,linecolor=red](0,2)(-2,0)
\psline[linewidth=.8pt,linecolor=red](-2,0)(0,-2)
\psdots*(-0.7,0.2)
\psline[linewidth=.2pt](-0.7,0.2)(-1.4,-0.6)
\psline[linewidth=.2pt](-0.7,0.2)(0,-0.7)
\rput[bI](0,-3){(a)}
\rput[bI]{90}(0.3,0.5){$z=0$}
\rput[bI](0.3,-1.5){$\phi_0$}
\rput[bI](-1.5,-.15){$\phi_H$}
\rput[bI](-.7,.30){$\phi$}
\rput[bI]{45}(-1.1,1.1){$z=\infty$, $t=\infty$}
\rput[bI]{-45}(-1.2,-1.3){$z=\infty$, $t=-\infty$}
\hspace{4cm}
\psline[linewidth=.8pt,linecolor=blue](0,-2)(0,2)
\psline[linewidth=.8pt,linecolor=blue](3,-2)(3,2)
\psdots*(1.5,.5)
\psline[linewidth=.2pt](1.5,.5)(3,-0.2)
\psline[linewidth=.2pt](1.5,.5)(0,-0.2)
\psline[linewidth=.8pt,linecolor=red](0,-1)(3,-1)
\psline[linewidth=.8pt,linecolor=red](0,1.5)(3,1.5)
\rput[bI](1.5,-3){(b)}
\rput[bI](3,-2.4){$x_+$}
\rput[bI](0,-2.4){$x_-$}
\rput[bI](3.3,0){$\phi^+_0$}
\rput[bI](-.4,0){$\phi^-_0$}
\rput[bI](1.5,.8){$\phi$}
\rput[bI](1.5,-1.6){$\phi^{(\sf n)}_{\,\sf i}$}
\rput[bI](1.5,1.7){$\phi^{(\sf n)}_{\,\sf f}$}
\vspace{3cm}
\caption{(a) {\sf Coordenadas de Poincare para AdS Lorentziano}: el valor del campo escalar $\phi$ en cualquier punto del volumen no solo depende del valor en el borde $\phi_0$ sino que también depende del valor del campo en los horizontes pasado y futuro $\phi_H$. ~~~~~~(b)
{\sf Geometría de Agujero de Gusano Lorentziano}: el valor del campo escalar en el volumen no solo depende de los dos valores de borde $\phi^\pm_0$ (azul) sino que también depende de los modos normalizables en el volumen $\phi^{(\sf n)}$ (rojo). La dependencia, en ambos gráficos, sobre los modos normalizables (líneas rojas) se correlata con la elección de los estados inicial y final $|\psi_{\mathrm {i,f}}\rangle$ y se manifiesta
como la elección de un contorno de integración en el plano complejo $\omega$ cuando se calculan los correladores (\ref{deltatt}).}
\label{contributions}
\end{figure}
En presencia de dos bordes conformes desconectados se propone escribir al campo en el volumen en términos de los datos
$\phi_0^{\pm}(\mathbf{y})$ sobre cada uno de los bordes como (ver figura \ref{contributions})
\bea
\phi(\mathbf{y},x)&=&\int d\mathbf{y}' \,
{\sf K}^{i}(\mathbf{y},x|\mathbf{y}'  )\,\phi^{i}_0(\mathbf{y}' )\nn\\
&=&\int d\mathbf{y}' \,
\left[{\sf K}^{+}(\mathbf{y},x|\mathbf{y}'  )\,\phi^{+}_0(\mathbf{y}' )+ {\sf
K}^{-}(\mathbf{y},x|\mathbf{y}')\,\phi^{-}_0(\mathbf{y}')\right]
\,.
\label{phi2bdy}
\eea
aquí $\mathbf{y}=(t,\theta)$, y nótese que la solución para dadas condiciones de borde no es única debido a que en signatura Lorentziana
pueden agregarse soluciones normalizables a (\ref{phi2bdy}). Esta ambigüedad se resuelve, como se discutió en la sección \ref{GKPW}, cuando
se elige un contorno en el espacio de frecuencias $\omega$ del núcleo $\sf K$. Mi método difiere de la propuesta desarrollada en \cite{balasu}:
en aquel trabajo solo se discutió el propagador volumen-borde $K^+$ y su forma fue determinada pidiendo ausencia de cortes cuando se extiende la coordenada radial a valores complejos. La prescripción condujo a la conclusión de que los $\phi_0^\pm$ no eran independientes.

Consistencia con \eqref{phi2bdy} demanda que los propagadores volumen-borde ${\sf K}^{\pm}(\mathbf{y},x|\mathbf{y}')$ sean solución de la ecuación de Klein-Gordon
(\ref{box}) con las siguientes condiciones de contorno
\bea
\left.{\sf K}^+(\mathbf{y},x|\mathbf{y}')\right|_{x=x_+}&=&\delta(\mathbf{y}-\mathbf{y}'),~~~~~~~~
\left.{\sf K}^+(\mathbf{y},x|\mathbf{y}')\right|_{x=x_-}=0\nonumber\\
\left.{\sf K}^-(\mathbf{y},x|\mathbf{y}')\right|_{x=x_-}&=&\delta(\mathbf{y}-\mathbf{y}'),~~~~~~~~
\left.{\sf K}^-(\mathbf{y},x|\mathbf{y}')\right|_{x=x_+}=0
\label{condition2boundaries}\,,
\eea
Estas expresiones determinan completamente los propagadores volumen-borde $\sf K^\pm$. La acción evaluada en la solución (\ref{freefi})
resulta en dos términos provenientes de los bordes y toma la forma
\be
S=-\frac12\int d\mathbf{y}\left(\left[\sqrt{-g}g^{xx}\phi(\mathbf{y},x)\partial_x\phi(\mathbf{y},x)\right]_{x=x_+}
-\left[\sqrt{-g}g^{xx}\phi(\mathbf{y},x)\partial_x\phi(\mathbf{y},x)\right]_{x=x_-}\right)
\label{general two bdy action}
\ee
Insertando la solución (\ref{phi2bdy}) en (\ref{general two bdy action}) se obtiene
\bea
S[\phi_0]=-\frac12\int d\mathbf{y}\,\, d\mathbf{y}'\phi_0^i(\mathbf{y})\Delta_{ij}(\mathbf{y},\mathbf{y}')\phi_0^j(\mathbf{y}')
\label{two bdy on shell action}
\eea
con $i,j=+,-$ denotando los dos bordes y $\Delta_{ij}$ la generalización de (\ref{Delta}). Su forma explícita es
\be
\Delta_{+i}(\mathbf{y},\mathbf{y}')=  \left[\sqrt{-g}g^{xx} \partial_x{\sf K}^i(\mathbf{y},x|\mathbf{y}')\right]_{x=x_+}\,,~~~~~
\Delta_{-i}(\mathbf{y},\mathbf{y}')=  -\left[\sqrt{-g}g^{xx} \partial_x{\sf K}^i(\mathbf{y},x|\mathbf{y}')\right]_{x=x_-}
\label{two bdy deltas}
\ee
Como en la sección \ref{GKPW} la función de dos puntos de operadores en el mismo borde resulta
\be
\langle\psi_{\mathrm f}| {\cal O}^\pm(\mathbf{y}){\cal O}^\pm(\mathbf{y'})|\psi_{\mathrm i}\rangle\sim
-i\Delta_{\pm\pm}(\mathbf{y},\mathbf{y'})\label{samebdy}
\ee
y los correladores entre operadores sobre bordes opuestos es
\be
\langle\psi_{\mathrm f}| {\cal O}^\pm(\mathbf{y}){\cal O}^\mp(\mathbf{y'})|\psi_{\mathrm i}\rangle\sim-i\Delta_{\pm\mp}(\mathbf{y},\mathbf{y'})\,.
\label{opposbdy}
\ee
La generalización de las expresiones (\ref{KfromG}) y (\ref{DeltafromG}) a fondos de gravedad con dos bordes toma la forma
\be\label{K-fromG-2bound}
{\sf K}^i(\mathbf{y},x\mid\mathbf{y}')=
\lim_{x'\rightarrow x^i}\sqrt{-g}\, g^{x'x'}\partial_{x'}{\sf G}(\mathbf{y},x\mid\mathbf{y}',x'),
\ee
lo cual da
\be\label{Delta-fromG-2bound}
\Delta_{ij}(\mathbf{y},\mathbf{y}')\sim\lim_{x\rightarrow x^i,~ x'\rightarrow x^j
}(\sqrt{-g}g^{xx})
(\sqrt{-g}g^{x'\!x'})\frac{\partial^2}{\partial_x\partial_{x'}}{\sf G}(\mathbf{y},x\mid\mathbf{y}',x')\,.
\ee
Nótese que la función de correlación  $\Delta_{\pm\mp}$ involucra, en el límite semiclásico, una geodésica a través del volumen que conecta
los puntos sobre bordes opuestos.

\subsection*{Banda Lorentziana de AdS$_2$}

En esta sección se aplicará la prescripción desarrollada en la subsección anterior para AdS$_2$ Lorentziano, reobteniendo los resultados previos \cite{Azeyanagi}.
La métrica Lorentziana de AdS$_2$ puede escribirse como
\be
 ds^2=R^2\left[-\frac{dt^2}{1-x^2}+\frac{dx^2}{(1-x^2)^2}\right]\,,
\label{ads2}
\ee
el borde tipo tiempo esta ubicado en $x=\pm1$ y el sistema de coordenada $(t,x)$ cubre todo el espacio-tiempo. Para encontrar el propagador
volumen-borde $\mathsf K^{\pm}$ en (\ref{phi2bdy}) se propone
\be
{\sf K^{\pm}}(t,x)=\int_{-\infty}^{\infty}\frac{d\omega}{2\pi}e^{-i\omega t}f_\omega^{\pm}(x)\,.
\ee
Insertándolo dentro de la ecuación de KG (\ref{box}) se obtiene la ecuación diferencial para $f_\omega$
\be
 (1-x^2)\frac{d^2 f_\omega^{\pm}(x)}{dx^2}-{x}\frac{d
 f_\omega^{\pm}(x)}{dx}+\left(\omega^2-
 \frac{m^2R^2}{1-x^2}\right)f_\omega^{\pm}(x)=0\,.
 \label{efads2}
\ee
La solución a (\ref{efads2}) puede escribirse en términos de polinomios generalizados de Legendre como
\be
 f_{\omega}^{\pm}(x)=(1-x^2)^{\frac{1}{4}}[a_{\omega}^{\pm}\,P^\mu_\nu(x)+b_{\omega}^{\pm}\,Q^\mu_\nu(x)]
 \label{fm}
\ee
con $\mu=\sqrt{\frac1{4}+m^2R^2}$,  $\nu=\omega-\frac12$ y $a_\omega^{\pm}, b_\omega^{\pm}$ constantes arbitrarias que quedarán fijadas
cuando se impongan las condiciones de contorno (\ref{condition2boundaries}). Las condiciones se trasladan en\footnote{Como se discutió en la
sección \ref{GKPW} el dato en el borde se impone a una distancia finita $x=\pm x_\epsilon$ donde $x_\epsilon=1-\epsilon$.}
\be
f_{\omega}^\pm(\pm x_\epsilon)=1\,,~~~~~~~~
f_{\omega}^\pm(\mp x_\epsilon)=0\,.
\label{norm}
\ee
Las soluciones a (\ref{norm}) son
\bea
f_{\omega}^{+}(x)=\left(\frac{1-x^2}{1-x_\epsilon^2}\right)^\frac{1}{4}\frac{Q^\mu_\nu(x)P^\mu_\nu(-x_\epsilon)-Q^\mu_\nu(-x_\epsilon)P^\mu_\nu(x)}
{Q^\mu_\nu(x_\epsilon)P^\mu_\nu(-x_\epsilon)-Q^\mu_\nu(-x_\epsilon)P^\mu_\nu(x_\epsilon)}\\
f_{\omega}^{-}(x)=\left(\frac{1-x^2}{1-x_\epsilon^2}\right)^{\frac{1}{4}}\frac{Q^\mu_\nu(x)P^\mu_\nu(x_\epsilon)-Q^\mu_\nu(x_\epsilon)P^\mu_\nu(x)}
{Q^\mu_\nu(-x_\epsilon)P^\mu_\nu(x_\epsilon)-Q^\mu_\nu(x_\epsilon)P^\mu_\nu(-x_\epsilon)}\,.
\eea
Analizando el comportamiento asintótico cerca del borde en (\ref{fm}) se encuentran los modos normalizables para
\be
\omega_n=\pm\left(n+\mu+\frac12\right),~~~~~~n=0,1,2\ldots ~~~
 \mathrm {y}~~~~\frac{b_{\omega Q}}{a_{\omega Q}}=-\frac{2\tan\pi\mu}{\pi}  \label{polosads2}
\ee
Las funciones renormalizadas $\Delta_{ij}$ (sin tener en cuenta términos de contacto) resultan
\bea
\Delta_{\mathsf {ren}_{\pm\pm}}(t,t')&=&\mp\frac{2^{\Delta_-}}{2\pi}\frac{\Gamma (1-\mu )}{\Gamma (1+\mu )}\int
\frac{d\omega}{2\pi} e^{-i\omega(t-t')}
 \Gamma \left(\frac12+\mu-\omega \right) \Gamma \left(\frac12+\mu+\omega\right) \cos (\pi  \omega )
 \label{act}\\
\Delta_{\mathsf
{ren}_{\pm\mp}}(t,t')&=&\mp\,\frac{2^{\Delta_-}}{\Gamma (\mu
)^2}\int \frac{d\omega}{2\pi}e^{-i\omega(t-t')} \Gamma
\left(\frac12+\mu-\omega\right) \Gamma \left(\frac12+\mu+\omega\right) \,, \label{actionterms}
\eea
como antes, los integrandos en estas expresiones muestran polos en $\omega$ para los valores dados por (\ref{polosads2}).

Las integrales  (\ref{act})-(\ref{actionterms}) puede calcularse usando el teorema de los residuos una vez que se haya elegido un contorno en el plano complejo. Para el contorno de Feynman se obtiene
\bea
\Delta_{\mathsf
{ren}_{\pm\pm}}^F(t,t')&=&\mp\left(\frac{i^{\Delta_-
-\Delta_+}}{8^{\Delta_+}\pi^{\frac12}}\right)
\frac{\Gamma(\frac12+\mu)}{\Gamma(\mu)
\sin^{2\Delta_+} \left(\frac{t-t'}{2}\right)}\nn\\
\Delta_{\mathsf {ren}_{\pm\mp}}^F(t,t')&=&\mp\left(\frac{8^{\Delta_-}i}{4}\right)
\frac{\Gamma(1+2\mu)}
{\Gamma(\mu)^2\cos^{2\Delta_+} \left(\frac{t-t'}{2}\right)}
\eea
El valor de expectación de vacío entre operadores sobre el mismo y opuestos bordes resultan  (confrontar con \cite{Azeyanagi})
\bea
\langle 0|T
{\mathcal O}^\pm(t){\mathcal O}^\pm(t')|0\rangle&=&\pm\left(\frac{4^{\Delta_- }i^{2\Delta_-}}{8^{\Delta_+}}\right)\frac{\Gamma(2\mu)}
{\Gamma(\mu)^2\sin^{2\Delta_+}(\frac{t-t'}{2})},\\
\langle 0| T{\mathcal O}^\pm(t){\mathcal
O}^\mp(t')|0\rangle&=&\mp\left(\frac{8^{\Delta_-}}{4}\right)\frac{\Gamma(1+2\mu)}
{\Gamma(\mu)^2\cos^{2\Delta_+}(\frac{t-t'}{2})}. \label{ads2++}
\eea
La primera linea da el resultado para operadores sobre el mismo borde y tiene el comportamiento conforme esperado  $|
t-t'|^{-2\Delta_+}$ cuando los operadores se acercan uno al otro. La segunda linea corresponde a operadores localizados sobre bordes diferentes y
son singulares para $t=t'+(2n+1)\pi$, $n\in\mathbb Z$, esta singularidad refleja la existencia de curvas causales (nulas) que conectan los dos bordes
y se a argumentado que su existencia es una señal de interacción entre los dos conjuntos de grados de libertad
${\cal O}^\pm$ \cite{balasu}\footnote{\label{spatial} Esta interpretación asume que las foliaciones espaciales del borde $\Sigma_+^{(t)}$ y $\Sigma_-^{(t')}$
deben ser identificada como la misma superficie de Cauchy $\tilde\Sigma$ de una única variedad de base en donde se define la teoría de campos.}.
La periodicidad observada en el tiempo se relaciona con una propiedad particular de AdS, esta es la convergencia de geodésicas nulas cuando se pasa
al cubrimiento universal y puede ser entendidas como una consecuencia de que los modos propios (\ref{polosads2}) están equiespaciados (ver \cite{Avis}-
\cite{hawkingellis}).

En el caso sin masa ($\mu=\frac12$) las funciones de dos puntos toman la forma
\bea
\langle 0|T
{\mathcal O}^\pm(t){\mathcal O}^\pm(t')|0\rangle=\pm\frac{1}{8\pi\sin^2\left(\frac{t-t'}{2}\right)}\,,~~~
~\langle 0|T {\mathcal O}^\pm(t){\mathcal O}^\mp(t')|0\rangle=\mp\frac{1}{4\pi\cos^2\left(
\frac{t-t'}{2}\right)}
\eea

\subsection*{Agujero de Gusano}

Voy ahora a analizar las funciones de dos puntos en un fondo gravitatorio de agujero de gusano, esto es,
una geometría de espacio-tiempo con dos bordes conformes conectados a través del volumen. Se trabajará con una agujero de gusano que sirve como
 modelo de juguete, el cual permite un tratamiento analítico, que consiste de una geometría estática que conecta dos regiones asintóticas
 localmente AdS con variedad de base de la forma
$H^3$ o $S^1\times H^2$, con $H^n$ un espacio hiperbólico cocientado $n$- dimensional.
La geometría no contiene horizontes, y las dos regiones asintóticas están causalmente conectadas. El espacio-tiempo fue encontrado como
una solución de la gravedad de Einstein-Gauss-Bonnet, la cual en $d+1=5$ dimensiones toma la forma
\be
S_5=\kappa\int\epsilon_{abcde}\left(R^{ab}R^{cd}+\frac{2}{3l^2}R^{ab}e^ce^d+\frac{1}{5l^4}e^ae^be^ce^d\right)e^e\nn\label{EGB}\,,
\ee
aquí $R^{ab}=d\omega^{ab}+\omega^a_f\omega^{fb}$ es la dos forma de curvatura para la conección de spin $\omega^{ab}$, y $e^a$ es el vielbein.
La métrica $d+1$-dimensional encontrada en \cite{Dotti} es
\bea
  ds^2&=&R^2\left[-\cosh^2\!\rho\,{dt^2}+{d\rho^2}+\cosh^2\!\rho\,{d\tilde\Sigma^2_{d-1}}\right]\nn\\
 &=&R^2\left[-\frac{dt^2}{1-x^2}+\frac{dx^2}{(1-x^2)^2}+\frac{d\tilde\Sigma^2_{d-1}}{1-x^2}\right]
 \label{wh}
\eea
en donde $d\tilde\Sigma^2_{d-1}$ es una métrica constante de curvatura negativa sobre la variedad de base $\tilde\Sigma_{d-1}$. Nótese que los dos bordes conformes desconectados se ubican en $x=\pm1$.

Para construir los propagadores del borde al volumen ${\sf K}^\pm$ discutidas antes se propone
\be {\sf K}^\pm(t,x,\theta|t' ,\theta')=
\int_{-\infty}^{\infty} \frac{d\omega}{2\pi}\sum_{Q}e^{-i\omega
(t-t')}\,Y_{Q}(\theta)Y_{Q}^\ast(\theta')f_{ \omega Q}^\pm(x)
\label{ansatz2}
\ee
aquí $Y_Q(\theta)$ son funciones armónicas\footnote{Estas funciones satisfacen $\nabla_\Sigma ^2Y_Q=-Q^2\,Y_Q$ y en una variedad conforme con bordes se tiene $Q^2\ge0$. Los modos propios y autovalores para el Laplaciano sobre una variedad compacta {\it suave} hiperbólica no pueden expresarse en una forma cerrada analítica y depende de un subgrupo discreto de $SO(d-1,1)$ elegido para hacer el cociente. Ver \cite{Cornish} y las referencias citadas en él para tener un tratamiento numérico del problema. } sobre $\tilde \Sigma_{d-1}$.

Insertando (\ref{ansatz2}) en las ecuaciones de KG (\ref{box}) se encuentra que $f_{ \omega Q}$ satisface
\be
 (1-x^2)\,\frac{d^2 f_{\omega Q }^{\pm}(x)}{dx^2} +(d-2)\,x \frac{d f_{\omega Q }^{\pm}(x)}{dx}+
\left[ (\omega^2-Q^2)-\frac{m^2R^2}{1-x^2}\right]f_{\omega Q
}^{\pm}(x)=0 \label{eqdiffworm}\,.
\ee
Las soluciones a (\ref{eqdiffworm}) pueden ser escritas en términos de polinomios generalizados de Legendre como \cite{diego}
\be
f_{\omega Q
}^{\pm}(x)=(1-x^2)^{\frac{d}{4}}[a_{\omega Q}^{\pm}\,P^\mu_\nu(x)+b_{\omega
Q}^{\pm}\,Q^\mu_\nu(x)] \label{solution}
\ee
en donde $\mu$ está dado por (\ref{lambda}) y
\bea
\nu=\varpi-\frac12=\sqrt{\left(\frac{d-1}{2}\right)^2+\omega^2-Q^2}-\frac12\,.
\label{nu}
\eea
$a_{\omega Q}^{\pm},b_{\omega Q}^{\pm}$ en (\ref{solution}) son coeficientes constantes que quedan fijados cuando imponemos las condiciones
(\ref{condition2boundaries}), estos son
\be
f_{\omega Q}^\pm(\pm x_\epsilon)=1\,,~~~~~~~~
f_{\omega Q}^\pm(\mp x_\epsilon)=0\,.
\label{normalization}
\ee
Las soluciones para (\ref{solution}) que satisfacen (\ref{normalization}) son
\bea
f^+_{\omega Q}(x)&=&\left(\frac{1-x^2} {1-x_\epsilon ^2}\right)^{\frac{d}{4}}
\frac{ P_{\nu }^{\mu }(x)Q_{\nu }^{\mu }(-x_\epsilon)-P_{\nu }^{\mu }(-x_\epsilon ) \,Q_{\nu }^{\mu
   }(x)}{P_{\nu }^{\mu }(x_\epsilon )Q_{\nu }^{\mu
   }(-x_\epsilon)-P_{\nu }^{\mu }(-x_\epsilon) Q_{\nu }^{\mu }(x_\epsilon )}\nn\\
f^{-}_{\omega Q}(x)&=&\left(\frac{1-x^2} {1-x_\epsilon ^2}\right)^{\frac{d}{4}}
\frac{ P_{\nu }^{\mu }(x)Q_{\nu }^{\mu }(x_\epsilon )-P_{\nu }^{\mu }(x_\epsilon ) Q_{\nu }^{\mu
   }(x)}{{P_{\nu }^{\mu }(-x_\epsilon)Q_{\nu }^{\mu
   }(x_\epsilon )-P_{\nu }^{\mu }(x_\epsilon ) Q_{\nu }^{\mu }(-x_\epsilon)}}\label{solworm}\,,
\eea
La existencia de dos propagadores volumen-borde independientes proviene del echo de que la variedad de base $\tilde\Sigma$ nunca se reduce
a tamaño cero dentro del volumen (ver (\ref{wh})) y entonces la regularidad en el interior no impone ninguna condición sobre las soluciones
(\ref{solution}). Los modos normalizables aparecen para
\be
 \omega_{nQ}=\pm
\sqrt{\left(\mu+\frac12+n\right)^2+Q^2-\left(\frac{d-1}{2}\right)^2}\,,~~~~~n=0,1,\ldots,~~~
 \mathrm {y}~~~~\frac{b_{\omega Q}}{a_{\omega Q}}=-\frac{2\tan\pi\mu}{\pi} \label{poloswh}
\ee
En estas frecuencias el índice $\nu$ en (\ref{nu}) toma el valor $\nu_n=\mu+n$ y la solución resultante se hace normalizable.
Las funciones de dos puntos (\ref{samebdy})-(\ref{opposbdy}) entre operadores ${\cal O}_{\pm}$ sobre el mismo y opuestos bordes toman la forma
\bea
\langle \psi_{\mathrm {f}}|{  T}{\mathcal O}^\pm(t,\theta){\mathcal O}^\pm(t',\theta')|\psi_{\mathrm {i}}\rangle
&=&\pm\,i \frac{2^{\Delta_-}d}{\pi2^d}\frac{\Gamma(1-\mu)}{\Gamma(1+\mu)}\,\sum_{ Q}Y_Q(\theta)Y_{Q}^\ast(\theta')\nn\\
&&\quad\quad\times\int\frac{d\omega}{2\pi}e^{-i\omega(t-t')}\Gamma(\frac12+\mu-\varpi)\Gamma(\frac12+\mu+\varpi) \cos(\pi\varpi)\nn\\
\label{w++}\\
\langle \psi_{\mathrm {f}}|T{\mathcal O}^\pm(t,\theta){\mathcal O}^\mp(t',\theta')|\psi_{\mathrm {i}}\rangle
&=&\pm\,i \frac{2^{\Delta_-}}{2^{d-1}}\frac{1}{\Gamma (\mu )^2}\,\sum_{Q}Y_Q(\theta)Y_{Q}^\ast(\theta')\nn\\
&&\quad\quad\times\int\frac{d\omega}{2\pi}e^{-i\omega(t-t')} \Gamma(\frac12+\mu-\varpi)\Gamma(\frac12+\mu+\varpi).
\label{w+-}
\eea
Unos pocos comentarios sobre esta expresión: (i) en el límite de frecuencias grandes $\varpi\sim \omega$ y los integrandos en (\ref{w++})-(\ref{w+-})
coinciden con aquellos de AdS$_2$ (confrontar con  (\ref{act})-(\ref{actionterms})), (ii) cuando se elige el contorno de Feynman, debe entenderse el lado derecho de  (\ref{w++})-(\ref{w+-}) como ordenado temporalmente y (iii) aunque las funciones Gamma $\Gamma(\frac12+\mu\pm\varpi)$ presentan
dos cortes en $\omega=\pm\sqrt{Q^2-\left(\frac{d-1}{2}\right)^2}$ (ver (\ref{nu})), el producto de los integrandos (\ref{w++})-(\ref{w+-}) está libre de ellos.

La correlación entre operadores insertados sobre bordes opuestos es no nula, y este resultado ha sido explicado de diferentes maneras dependiendo del contexto: (i) como resultado de calcular el correlador  (\ref{w+-}) sobre un estado entrelazado de dos teorías de borde no interactuantes (contexto de agujero negro \cite{Eternal}) o (ii) como debido a una interacción entre teorías definidas sobre cada uno de los bordes (orbifold D1/D5 en \cite{balasu}). El punto crucial en ambos argumentos fue la ausencia/existencia de conexión causal entre las regiones asintóticas.

\section{Entrelazamiento vs. Acoplamiento}
\label{holentanglement}

En esta sección se van a revisar la interpretación de los resultados
(\ref{ads2++}) y (\ref{w+-}). Se intenta responder cuándo son consecuencia de: (i)
interacción entre las dos QFT duales o (ii) correladores evaluados
en un estado entrelazado o (iii) ambas situaciones.

\subsection*{Entropía de Entrelazamiento}

La entropía de entrelazamiento $\mathsf S_{_{\mathrm A}}$ es una
cantidad no local que mide cuan correlacionados están dos sistemas,
digamos A y B. Para una QFT en $d$ dimensiones se define como la
entropía de von Neumann de la matriz densidad reducida $\rho_A$
obtenida cuando se toma la traza sobre los grados de libertad
correspondientes a una sub-variedad tipo espacio de $d-1$
dimensiones, llamada B, que es el complemento de A (ver \cite{rt}
para un resumen).

En \cite{Ryu, takaya} se propuso una formula holográfica para la
entropía de entrelazamiento de una CFT$_d$ dual a una geometría
AdS$_{d+1}$, esta es
\be
 \mathsf S_{_{\mathrm
A}}=\frac{\mathrm{Area}(\gamma_{_{\mathrm A}})}{4G_N^{(d+1)}}\,,
\label{entang}
\ee
en donde $\gamma_{_{\mathrm A}}$ es un area mínima $(d-1)$ dimensional
en AdS$_{d+1}$ cuyo borde $\cal S$,  ubicado en el infinito de AdS, coincide con el de A, esto es ${\cal S}=\partial\gamma_{_{\mathrm A}}=\partial {\mathrm A}$
y $G_N^{(d+1)}$ es la constante de Newton en AdS$_{d+1}$. Esta fórmula, que supone la aproximación de supergravedad de la teoría de cuerdas completa,
a sido aplicada en el caso de AdS$_3$/CFT$_2$ mostrando perfecto acuerdo con los resultados conocidos de la 2D CFT \cite{rt}.

Podemos aplicar una generalización de (\ref{entang}) a la geometría de agujero de gusano (\ref{wh}) como sigue
\footnote{ Ver \cite{Hubeny} para obtener una generalización de (\ref{entang}) para el agujero de gusano Euclídeo construido en \cite{maldamaoz}.}:
el agujero de gusano presenta dos regiones espaciales desconectadas $\Sigma_\pm$ en las cuales se supone que viven dos grados de libertad idénticos.
Imaginemos que construimos una superficie cerrada de codimension 1 que es el borde, $\cal S$, de una {\it pequeña} región B sobre $\Sigma_-$,
experiencia en lazos de Wilson y embeddings de branas muestra que mientras B permanezca chico, la superficie mínima se ubicará cerca de $x=-1$.
A medida que vamos incrementando el tamaño de B, la superficie mínima extendida sobre $\Sigma_-$ explora regiones más profundas del volumen del espacio
y en el límite cuando B$\rightarrow \Sigma_-$ el borde $\cal S$ colapsa y la superficie mínima se localiza en la garganta $x=0$ de la geometría (ver figura \ref{wh}),
ofreciendo un resultado no nulo
\be
\mathsf S_{_{\tilde\Sigma}}=\frac{\mathrm{Area}(\tilde\Sigma)}{4G_N^{(d+1)}}\,.
\label{entro}
\ee
Este resultado debe entenderse como un indicador de que el estado cuántico en la QFT dual descripto por la geometría de agujero de gusano
es no separable, o dicho de otra manera, el resultado (\ref{w+-}) puede ser atribuido a que la geometría de agujero de gusano realiza un estado
entrelazado en el espacio de Hilbert producto $\cal H=\cal H_+\otimes\,\cal H_-$ con la entropía (\ref{entro}) resultante de integrar los
grados de libertad sobre $\cal H_-$.

\begin{figure}
\begin{center}
\includegraphics[trim= 0mm 90mm 0mm 90mm,scale=0.5]{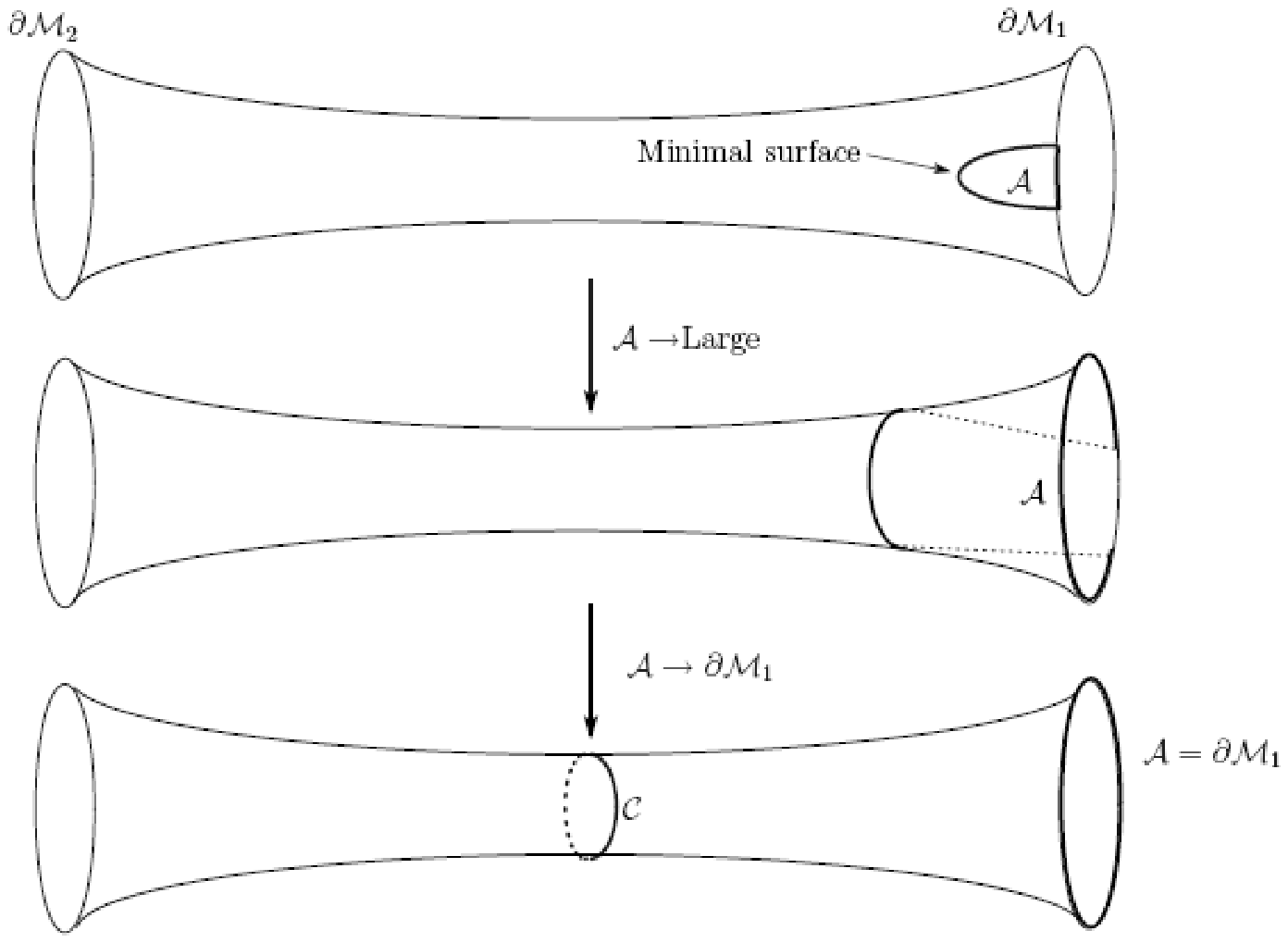}
\caption{Área mínima de la superficie cuyos extremos se encuentran sobre el borde de la región ${\cal A}$. A medida que ${\cal A}$ va cubriendo todos los
 grados de libertad de una de las teorías de campos, la superficie mínima explora el interior del agujero de gusano. Cuando ${\cal A}$ cubre todos los grados
  de libertad (${\cal A}=\partial {\cal M}_1$) la superficie mínima es el área del cuello. }
\label{wh}
\end{center}
\end{figure}

\subsection*{Interacción entre copias duales}
\label{double boundary geos}

Se argumentará que además del entrelazamiento discutido anteriormente, el contacto causal entre los bordes asintóticos de la geometría de agujero de
gusano conduce a un término de interacción entre los dos conjuntos de grados de libertad que viven en cada uno de los bordes. En particular, se concluirá
que existe un acoplamiento entra las teorías de campos cuando las regiones asintóticas se encuentren en contacto causal.

Comenzaré escribiendo la función de partición en el lado de gravedad para la geometría de agujero de gusano en el límite semiclásico, su forma es
\be
\mathcal{Z}_{gravity}\left[\,\phi(\phi_0^+,\phi_0^-,{\cal C})\right]\sim\, e^{-\frac{i}{2}
\int d{\mathbf{y}}\,\, d\mathbf {y}'\phi_0^i(\mathbf{y})\Delta_{ij}(\mathbf{y},\mathbf{y}')\phi_0^j(\mathbf{y}')}
\label{Z-2Bs}
\ee
aquí ${\bf y}=(t,\theta)$ denota los puntos en el borde y $i,j=+,-$ refieren a los bordes asintóticos en donde el dato $\phi_0^\pm$ es dado,
el contorno $\cal C$ fija las soluciones normalizables en el volumen del espacio y la expresión para $\Delta_{ij}$ esta dada por (\ref{two bdy deltas}).
De acuerdo a la prescripción GKPW la función de partición (\ref{Z-2Bs}) es la funcional generatriz para correladores ${\cal O}^\pm$, esto es
\be
\label{witten-gen-pm}
\mathcal{Z}_{gravity}\left[\phi(\phi_0^+,\phi_0^-,\cal C)\right]=\left\langle\,\psi_{\mathrm f}
\left|\,T
e^{i\int
d{\bf y}\,\phi_0^+({\bf y})
{\cal O^+({\bf y})} + i\int
d {\bf y}\,\phi_0^-({\bf y}){\cal O}^-({\bf y})}
\right|\,\psi_{\mathrm i}\right\rangle_{_{QFT}}\,,
\ee
en donde los observables ${\cal O}^\pm$ deben ser construidos como funcionales locales de los campos fundamentales $\Psi_\pm$ definidos en
cada uno de los bordes. Se asume que estos campos describen grados de libertad independientes: $[{\cal O}^+ , {\cal O}^-]=0$ sobre la misma
rebanada espacial (ver pie de página \ref{spatial}).

Consideremos la situación más simple correspondiente a elegir a $\cal C$ como el contorno de Feynman, esto es, se están calculando
las amplitudes de transición vacío-vacío en la teoría de campos. El lado derecho en (\ref{witten-gen-pm}) puede escribirse como
\be
\left\langle \psi_0 \left|\,T e^{i\int
d {\bf y}\,\phi_0^+({\bf y}){\cal O^+({\bf y})}\,+\,i \int
d {\bf y}\,
\phi_0^-({\bf y}){\cal O}^-({\bf y})}\right|\psi_0\right\rangle_{_{QFT}}=\mathsf {Tr}\left[\,\rho_{_{\psi_0}}\,T
e^{i\int
d {\bf y}\,\phi_0^+({\bf y}){\cal O^+({\bf y})} \,+\, i\int
d {\bf y}\,\phi_0^-({\bf y}){\cal O}^-({\bf y})}\right]
\label{whcorrel},
\ee
en donde ya se a realizado la operación de traza sobre un conjunto completo de estados del espacio de Hilbert de la teoría de campos dual
y $\rho_{_{\psi_0}}=|\psi_0\rangle\langle \psi_0 |$ es la matriz densidad asociada al estado de vacío de agujero de gusano.
Este estado de vacío pertenece al espacio de Hilbert ${\cal H} = {\cal H}_+ \otimes {\cal H}_-$ y de acuerdo con los argumentos resumidos en la
subsección anterior, no es separable como un único producto tensorial $|\psi_0\rangle\ne|\psi_+\rangle\otimes|\psi_-\rangle$.

Para analizar la posibilidad de interacción entre los campos definidos en cada uno de los bordes consideremos el sistema a temperatura finita
$T=\beta^{-1}$. La ausencia de singularidades en la continuación Euclídea implica que el agujero de gusano esta en equilibrio con un
reservorio térmico de temperatura arbitraria, o dicho de otra manera, la periodicidad en el tiempo Euclídeo es arbitraria. En equilibrio
térmico un estado de la teoría de campos queda descrito por la distribución de Boltzmann $\rho_{_\beta}= e^{-\beta H}$, con $H=H_+[\Psi_+] + H_-[\Psi_-] +
H_{int}[\Psi_+, \Psi_-]$ el Hamiltoniano de la teoría de campos dual, y $H_{int}$ un posible término de acoplamiento entre los conjuntos idénticos
de grados de libertad $\Psi_\pm$.

Ahora veremos que el término de interacción $H_{int}$ debe estar presente en el Hamiltoniano para evitar una contradicción.
El argumento es el siguiente: las funciones de correlación a temperatura finita sobre la teoría de campos son obtenidas a partir de
una rotación Euclídea de (\ref{whcorrel}),  con lo cual
\be
\left\langle  e^{- \int
d {\bf y}\,\phi_0^+({\bf y}){\cal O^+({\bf y})}\, -\, \int
d {\bf y}\,
\phi_0^-({\bf y}){\cal O}^-({\bf y}) }\right\rangle_{{\beta
}} =
\mathsf{Tr}\left[\,\rho_{_\beta}\, e^{-\int d {\bf y}\, \phi_0^+({\bf y}) {\cal O}^+({\bf y})\,
 -\, \int d {\bf y}\,\phi_0^-({\bf y}){\cal O}^-({\bf y}) }\,\right]\,,\label{euclqft}
\ee
la teoría de campos esta definida sobre una variedad Euclídea con topología $S^1_\beta\times \tilde\Sigma$ (ver pie de página \ref{spatial}).
La propuesta de AdS/CFT iguala esta expresión con la continuación Euclídea del lado izquierdo de (\ref{witten-gen-pm}), el resultado es
\bea
\label{thermalAdSCFT}
\mathsf{Tr} \,\left[\, e^{-\beta H}\, \,e^{-\int d {\bf y}\,\phi_0^+({\bf y}) {\cal O^+({\bf y})} -
\int d {\bf y}\,\phi_0^-({\bf y}){\cal O}^-({\bf y})} \,\right] &=&
\mathcal{Z}_{Egravity}\left[\phi(\phi_0^+,\,\phi_0^-)\right]\sim e^{ -S_E[\phi_0^+ ,\,\phi_0^-]}\,,
\label{eucZ}
\eea
sobre el lado derecho $S_E[\phi_0^+ ,\,\phi_0^-]$ corresponde a la acción para el campo escalar sobre el fondo gravitatorio del agujero de gusano
evaluada en la solución de las ecuaciones de movimiento para dicho campo con el tiempo compactificado sobre un circulo de radio $\beta$. Luego de
la rotación Euclídea y la compactificación temporal, la geometría resultante es un cilindro con topología $S_\beta^1 \times I  \times \tilde\Sigma$ y
los dos bordes sobre los cuales se imponen los datos  $\phi_0^\pm$ se localizan en los extremos $x_\pm$ del intervalo finito
$I$\footnote{Nótese las similitud con la geometría estudiada en \cite{maldamaoz}. }. Nótese que en el contexto Euclídeo la solución
en el interior queda completamente especificada por los datos del borde, existen soluciones no normalizables y rotar el contorno de Feynman al eje
temporal imaginario conduce a soluciones no singulares. La expresión explícita para el lado derecho de (\ref{eucZ}) es
\be\label{eucl-AdSCFT}
\mathcal{Z}_{Egravity}\left[\phi(\phi_0^+,\,\phi_0^-)\right]\sim\, e^{
-\frac12\int d\mathbf{y}\,\, d\mathbf{y}'\phi_0^i(\mathbf{y})\,\tilde\Delta_{ij}(\mathbf{y},\mathbf{y}')\,\phi_0^j(\mathbf{y}')},
\ee
en donde $\tilde{\Delta}_{ij}$ denota el propagador volume-borde (\ref{two bdy deltas}) rotado de manera Euclídea.
Finalmente, a partir de (\ref{eucZ}) y (\ref{eucl-AdSCFT}) se obtiene la prescripción gauge/gravedad para una agujero de gusano a temperatura finita
\be
\label{thermalAdSCFT2}
\mathsf{Tr} \,\left[\, e^{-\beta H}\, \,e^{-\int d {\bf y}\,\phi_0^+({\bf y}) {\cal O^+({\bf y})} -
\int d {\bf y}\,\phi_0^-({\bf y}){\cal O}^-({\bf y})} \,\right]\sim \, e^{
-\frac12\int d\mathbf{y}\,\, d\mathbf{y}'\phi_0^i(\mathbf{y})\,\tilde\Delta_{ij}(\mathbf{y},\mathbf{y}')\,\phi_0^j(\mathbf{y}')}\,.
\ee
Esta es la fórmula central del argumento debido a que si uno asume que los grados de libertad $\Psi_+,\Psi_-$ son desacoplados, esto es
\be
\label{decoupled}
H[\Psi_+ ,\Psi_-]= H_+[\Psi_+] + H_-[\Psi_-]\,,
\ee
luego, $\rho_{_\beta}= e^{-\beta H_+}\,e^{-\beta H_-}$, y el lado izquierdo de (\ref{thermalAdSCFT2}) se factoriza en un producto de dos cantidades: una
que depende de $\phi_0^+$ y otra que depende de $\phi_0^-$\footnote{Nótese que en ausencia de interacción $[{\cal O}_+,{\cal O}_-]=0$ sin importar si
los puntos en donde están insertados los operadores están espaciados temporal/espacialmente. }.
Sin embargo, el cálculo de gravedad no se factoriza debido a los términos no nulos $\tilde{\Delta}_{\pm\mp}$. Se interpreta este resultado
como la manifestación de existencia de un acoplamiento no trivial $H_{int}$ entre los grados de libertad duales $\Psi_\pm$:  la teoría de campos dual
a la geometría de agujero de gusano contiene un término de acoplamiento no trivial entre los grados de libertad de sus bordes $\Psi_+$ y $\Psi_-$.

~

\subsection*{Aplicaciones}

El resultado de las observaciones anteriores es que la geometría de agujero de gusano codifica la descripción de una teoría
de campos dual consistente de dos copias de campos fundamentales en interacción. Además, el estado cuántico descripto por el
agujero de gusano está entrelazado. En particular, la continuación Euclídea puede ser vista como prescripción para \emph{separar},
en la teoría de campos dual, efectos de entrelazamiento y de interacción $H_{int}$.
Una función de correlación no nula entre operadores localizados en bordes diferentes debe interpretarse como originados a partir de
una interacción $H_{int}$, en lugar de ser vistos como producido por efectos de entrelazamiento. En lo que sigue se confrontarán estos
dos puntos de vista con otras dos geometrías relevantes en la literatura: AdS$_{1+1}$ y el agujero negro eterno de AdS \cite{Eternal}.

~

\noindent {\bf AdS$_{1+1}$:}

El análisis de AdS$_{1+1}$ (\ref{ads2}) es enteramente análogo al agujero de gusano estudiado. La sección Euclídea de la métrica global
(\ref{ads2}) tiene dos bordes antes de compactificar la dirección temporal y por lo tanto se requieren dos condiciones de borde $\phi_0^\pm$
en el límite semiclásico (ver (\ref{Z-2Bs})). Los argumentos que conducen a (\ref{thermalAdSCFT2}) se aplican con las funciones de correlación
Euclídeas $\tilde\Delta_{ij}$ inmediatamente obtenidas a partir de las fórmulas Lorentzianas (\ref{ads2++}). Se concluye entonces que el echo de tener
$\tilde \Delta_{\pm\mp}\ne0$ nos dice que AdS$_{1+1}$ es dual a una mecánica cuántica conforme compuesta de dos sectores interactuantes.

~

\noindent {\bf Agujero negro eterno de AdS:}

La diferencia crucial entre el agujero de gusano (\ref{wh}) y la solución de agujero negro de AdS máximamente extendida \cite{Eternal} es la siguiente:
al realizar la continuación Euclídea  de la solución de agujero negro, la existencia de un horizonte en signatura Lorentziana genera una singularidad
cónica en la variedad Euclídea, que solo puede ser evitada pidiendo una periodicidad precisa en el tiempo Euclídeo.
La geometría Riemanniana resultante tiene inevitablemente solo un borde asintótico, y entonces requiere imponer solo un dato en el borde, $\phi_0$,
esto indica la existencia de un único conjunto de grados de libertad $\Psi$ y un único propagador volumen-borde  $\tilde\Delta$.
El sistema en equilibrio con un baño térmico de temperatura fija (determinada por la masa del agujero negro) tiene una funcional generatriz
que se escribe
\be
\label{thermalBH}
\mathsf{Tr} \,\left[e^{-\beta H}  e^{-\int d{\bf y}\, \phi_0({\bf y})\, {\cal O}({\bf y})}\right] \sim \,
e^{-\frac{1}{2}\int d{\bf y}  d{\bf y}'  \,\phi_0 ({\bf y}) \,\tilde{\Delta} ({\bf y},{\bf y}')\, \phi_0 ({\bf y}')  } ,
\ee
La solución en tiempo real (agujero negro máximamente extendido) fue analizada e interpretada en el contexto de AdS/CFT en \cite{Eternal}.
El segundo borde {\it causalmente desconectado}, presente en signatura Lorentziana, fue entendido como soportando los acompañantes
de TFD (dinámica de campos térmicos) necesarios para obtener un estado térmico a partir de su integración y conduce a una duplicación del espacio de Hilbert $\cal H= H_+\otimes H_-$. El segundo conjunto de grados de libertad está causalmente desconectado del conjunto original a temperatura cero; aunque ellos aparezcan
en signatura Lorentziana y den lugar a funciones $\Delta_{ij}$ no triviales, ellos son ficticios desde el punto de vista físico.
La desconexión causal entre los bordes está relacionada con que la función de partición de dos puntos $\Delta_{\pm\mp}$ nunca es singular.

\section{Conclusiones}

En este capítulo se resumió la prescripción GKPW en el contexto Lorentziano relacionando la ambigüedad de agregar modos normalizables a  (\ref{phi})
con el contorno de integración $\cal C$ requerido para evitar las singularidades en el integrando de (\ref{deltatt}), (\ref{w++})
proviene de la existencia de modos normalizables. Para calcular cantidades Lorentzianas se necesita fijar un contorno de referencia
${\cal C}_{\sf ref}$ y las dos elecciones sensibles son la retardada o la de Feynman. Estas elecciones son duales, en el lado de la QFT, a calcular
funciones de respuesta o de correlación. Se eligió el camino de Feynman como contorno de referencia, cualquier otro contorno $\cal C$
difiere de ${\cal C}^{ F}$ en contribuciones que provienen de enlazar polos, estos polos enlazados fijan los estados inicial y final
en la función de correlación (ver (\ref{2ptf})).

En la sección \ref{gkpww} se extendió la prescripción de GKPW a espacio-tiempos con más de un borde asintótico, en particular se estudió
el caso de dos bordes. Se propuso escribir el campo en el volumen del espacio en términos de dos valores de borde asintóticos y como ejemplo de
 juguete se aplicó esta construcción al caso AdS$_{1+1}$ reobteniendo resultados previos, después se calcularon las funciones de correlación de
 dos puntos para el agujero de gusano de Einstein-Gauss-Bonnet (\ref{wh}). En este punto debo enfatizar  que la diferencia principal entre este método
y el utilizado en \cite{balasu}, para un orbifolio de AdS$_3$, consiste en que hemos considerado los valores de borde  $\phi_0^\pm$ del campo escalar en el volumen como cantidades independientes, además, se mostró explícitamente la posibilidad de construir dos propagadores volumen-borde  ${\sf K}^\pm$
(su condición de borde fue dada en (\ref{condition2boundaries})). La construcción realizada en \cite{balasu} utilizó una relación entre los valores de borde $\phi_0^\pm$ y esto fue entendido como un indicativo de que las fuentes $\pm$ para la teoría de campos dual están prendidas de manera correlacionada. Una cuestión que permanece pendiente es cuándo los dos conjuntos de datos son independientes o redundantes en la formulación dual, por otro lado otra cuestión es el origen de valores no nulos para operadores ubicados en bordes diferentes ($\Delta_{\pm\mp}$) esto puede ser producido por entrelazamiento o por interacciones o ambos casos simultáneamente.

En la sección \ref{holentanglement} se aplicaron las ideas de entrelazamiento holográfico desarrolladas en \cite{takaya} a la geometría de agujero de gusano.
El resultado no nulo obtenido para ${\sf S}_{_{\tilde \Sigma}}$ de los grados de libertad que viven sobre bordes opuestos sugieren que el agujero de gusano
debe ser entendido como representando a un estado entrelazado en $\cal H=\cal H_+\otimes\,\cal H_-$. Por otro lado,
la conexión causal entre los bordes sugiere que también puede existir un acoplamiento. Para atacar este problema se consideró al sistema del
agujero de gusano en contacto con un baño térmico, después de la Euclideanización, la geometría resultante aún tiene dos bordes,
esto indica, para evitar una contradicción, que la QFT dual consiste de dos copias ${\cal H}_{\pm}$ en interacción. En resumen,
el número de bordes desconectados de la sección Euclídea determina la cantidad de grados de libertad físicos.

Me gustaría enfatizar finalmente las implicaciones de este trabajo sobre la gravedad cuántica, la cual puede ser vista como una de las motivaciones. Este tema a sido discutido en diferentes contextos en los últimos años y es referido como espacio-tiempo emergente \cite{seiberg}.
En este sentido, se mostró cómo se decodifican propiedades topológicas y causales (conectividad) en la acción de la QFT, y esta parte de la información no está solamente en el estado fundamental sino en su estructura de interacciones. Espero que esta conclusión pueda contribuir a la construcción de reglas hacia una ingeniería de geometrías.

Debo mencionar que en presencia de interacciones se necesita decir cómo están identificados los puntos sobre bordes opuestos. Un primer paso hacia el entendimiento de este problema es identificar los puntos $\bf y,y'$ en el espacio de configuración en los cuales $\Delta_{\pm\mp}$ diverge y estudiar la consistencia de identificarlos, esto esta siendo investigado actualmente.

\def\baselinestretch{1.66}
\newpage
\def\baselinestretch{1}
\chapter{Estabilidad de lazos de Wilson}\label{Wh}

En los últimos años, la correspondencia gauge/gravedad
\cite{malda, GKP, Ewitten} nos permitió estudiar nuevos aspectos referidos al
fenómeno de confinamiento. El punto crucial es que el aspecto de
cuerda que presenta el tubo de flujo cromoeléctrico se manifiesta en la teoría
dual mediante la aparición de una dimensión holográfica
\cite{polyakov, polyakov2}. En \cite{rey, maldawilson} se propuso
desde la teoría de cuerdas una prescripción para calcular el valor
de expectación del lazo de Wilson: el lazo de la teoría de gauge
se puede pensar localizado en infinito en la coordenada radial
holográfica y el lazo de Wilson para quarks fundamentales queda
definido por una cuerda abierta cuyos extremos están sobre él.
En el límite de $N_c$ grande (límite de ´t Hooft) las
auto-interacciones de la cuerda de QCD se anulan \cite{thooft} y
la prescripción gravitatoria dual para calcular la energía
potencial entre quarks en una dada teoría de gauge queda reducida a
encontrar una superficie mínima en la teoría de gravedad.

Como se explicó en la sección \ref{Wilint}, el cálculo canónico de la energía potencial entre un par de quarks
fundamentales (para un lazo rectangular), involucra una cuerda con
forma de U extendida en la dirección holográfica, en donde la
separación, $L$, de los quarks en la teoría de gauge se traduce en
la separación entre los extremos fijos de la cuerda ubicados en
infinito. A medida que varía la separación entre los quarks la
cuerda explora la dirección holográfica, luego, la posición radial
mínima alcanzada por la cuerda, $r_0$, depende de la distancia de
separación entre los extremos L. Este procedimiento ha sido
aplicado a un gran número de ejemplos paradigmáticos y a dado
resultados consistentes con lo que es esperable obtener de la
teoría de gauge, en particular, existe un teorema que establece
condiciones suficientes para obtener fondos gravitatorios confinantes
\cite{sonnen} (ver \cite{sonnwilson} para un resumen). Extensiones
a representaciones más altas del grupo de gauge y lazos de ´t
Hooft han sido analizadas y propuestas en \cite{groo, hk, gomis,
passerini}. Estas, involucran D-branas de dimensiones más altas
con y sin campos de gauge prendidos respectivamente.

En muchas aplicaciones a fondos gravitatorios duales, la
prescripción \cite{rey, maldawilson} ha sido aplicada a orden cero
para establecer confinamiento, transiciones de fase o propiedades
de transporte \cite{sonnwilson, gubser}, solo recientemente ha
sido estudiada la estabilidad de algunas formas de embeber las
cuerdas clásicas en el fondo de gravedad \cite{kmt}-\cite{Siampos}
(ver también \cite{cg}-\cite{dgt}). Una de las motivaciones para
el análisis de la estabilidad, en general, ha sido la existencia
de multiples formas de embeber la cuerda en el espacio de fondo
para unas dadas condiciones de contorno
\cite{pufu}-\cite{Siampos},\cite{argyres} (ver también
\cite{Brandh}) señaladas por la presencia de un extremo en la
función longitud $L(r_0)$ (ver \cite{Bigazzi}, \cite{Bigazzi2}).
La presencia de una longitud de separación máxima fue interpretada
como el fenómeno dual a la presencia de apantallamiento en la
teoría de gauge. El motivo del presente trabajo es mostrar que
aquellas ramas de la solución con $L'(r_0)>0$ son inestables. Se
confirmará esta afirmación mostrando explícitamente la
existencia, en fondos de gravedad particulares, de modos
inestables ($\omega^2<0$) para la rama con $L'(r_0)>0$. Es
oportuno mencionar que este es un resultado satisfactorio, ya que
la forma físicamente aceptable de la función $L(r_0)$ calculada
desde la correspondencia gauge/gravedad esta dada por $L'(r_0)<0$.
Voy a mencionar brevemente cálculos de lazos de ´t Hooft en duales
gravitatorios con dichas inestabilidades.

En el camino, se discutirán varios aspectos de las diferentes formas de
embeber una cuerda en el espacio de fondo: el primer análisis estará
relacionado a la interpretación física de la configuración
empleada para obtener una energía potencial finita entre los
quarks en duales gravitatorios suaves, la segunda trata sobre la
invarianza ante difeomorfismos de la acción de la cuerda y su
relación con las posibles elecciones de gauge para la orientación
de las fluctuaciones en un plano, la tercera es la relación entre
las inestabilidades producida por la forma de embeber la cuerda y $L(r_0)$,
es decir, la relación entre la separación de los extremos de la
cuerda como función de la profundidad máxima alcanzada por la
misma en la dirección holográfica. En \cite{avramis, avramis2,
Siampos} se demostró que la presencia de un extremo en $L(r_0)$ es
un indicativo de la existencia de un modo cero en las
fluctuaciones longitudinales, lo cual señala la presencia de una inestabilidad. Voy a
confirmar esta echo calculando explícitamente los modos de
fluctuación más bajos en diferentes duales gravitatorios.

En este capítulo se estudiará la estabilidad de soluciones clásicas para la hoja de mundo de una cuerda, la cual se emplea para calcular la energía potencial
entre quarks fundamentales en duales gravitatorios confinantes y no confinantes. Se discutirá el fijado ante difeomorfismos de la acción de la cuerda,
su relación con la orientación de la fluctuación y la interpretación de la sustracción de la masa de los quarks necesaria para calcular la energía
potencial en fondos de gravedad (confinantes) suaves. Voy a estudiar varios duales gravitatorios y mostraré mediante análisis numérico la existencia de
inestabilidades ante fluctuaciones lineales para soluciones clásicas de cuerdas que tienen derivada positiva de su función longitud $L'(r_0)>0$.
Finalmente se discutirán brevemente los lazos de ´t Hooft en fondos gravitatorios no conformes.

El capítulo estará organizado de la siguiente manera: en la
sección \ref{wil} voy a resumir la prescripción para calcular valores de
expectación de lazos de Wilson desde el punto de vista del dual
gravitatorio. En la sección \ref{bkg} se describirán los fondos de
gravedad bajo estudio y se calcularán las funciones de longitud y
energía. En la sección \ref{stab} se van a estudiar fluctuaciones a orden
cuadrático y se calcularán numéricamente los modos mas bajos de
fluctuación. En la sección \ref{scho} se analizarán los resultados
obtenidos en la sección \ref{stab}  llevando las ecuaciones de movimiento
para las fluctuaciones a ecuaciones de Schrodinger. En la sección
\ref{thooftloop} se va a discutir brevemente el lazo de ´t Hooft y en las sección \ref{higherreps}
se estudiarán fluctuaciones bosónicas y fermiónicas en el dual gravitatorio de un operador de Wilson-'t Hooft
evaluado en la representación simétrica del grupo de gauge.

\section{Soluciones de Cuerda y Lazos de Wilson}
\label{wil}
\subsection{Cuerdas estáticas}

El punto de partida para realizar cálculos de valores de
expectación de lazos de Wilson, en el límite de N$_c,\lambda$
grandes (límite de ´t Hooft), a partir de duales gravitatorios con
métrica $g_{\mu\nu}$ es la acción de Nambu-Goto \footnote{En
general uno debería tener en cuenta contribuciones de los campos
de fondo $B_2$, no obstante como propuesta de solución se va a
considerar que no contribuye a (\ref{NG}).}

\be
S=\frac\eta{2\pi\alpha'}\int d\tau d\sigma\sqrt{\eta\, h}\,.
\label{NG}
\ee

donde $h=\det h_{\alpha\beta}$,
$h_{\alpha\beta}=g_{\mu\nu}\partial_\alpha X^\mu
\partial_\beta X^\nu$ es la métrica inducida sobre la hoja de mundo de la cuerda,
$\partial_\alpha=\partial/\partial\xi^\alpha$ con
$\xi^\alpha=\{\tau,\sigma\}$ son las coordenadas sobre la hoja de
mundo y $X^\mu$ tomando valores sobre el espacio de fondo
(target). El signo $\eta$ da cuenta de posibles configuraciones
Euclídeas ($\eta=+$) o Lorentzianas (tipo tiempo) ($\eta=-$). Las
clases de métricas que se van a considerar son de la forma

\be
ds^2=-g_{t}(r)dt^2+g_{x}(r)dx_i^2+g_{r}(r)dr^2+g_{ab}(r,\theta)d\theta^a
d\theta^b\,. \label{gravity}
\ee

Las coordenadas $t,x_i\,(i=1,2,3)$ representan las coordenadas
sobre la teoría de gauge, $r$ es la coordenada holográfica y los
$\theta_a\,(a,b=1,..,5)$ son coordenadas angulares adicionales que
parametrizan un espacio compacto 5-dimensional $\Sigma_5$. Para
conocer la energía potencial entre los quarks se debe resolver la
acción de NG para hojas de mundo (tipo tiempo) correspondientes a
cuerdas cuyos extremos en infinito se localizan sobre el lazo a
calcular, típicamente los extremos están separados por una
distancia $L$ en una de las coordenadas $x_i$ a la cual llamaré
$x$ (ver \cite{sonnwilson}). Voy a comenzar analizando formas
estáticas de embeber la cuerda $t(\tau),x(\sigma),r(\sigma)$, con
todas las otras coordenadas fijadas a constantes \footnote{
Generalizaciones que consideran quarks en movimiento en el borde,
relevantes para aplicaciones a plasmas de quarks y gluones (QGP)
son refraseadas en términos de componentes no diagonales en la
métrica \cite{pufu},\cite{argyres}.}. Las ecuaciones de movimiento
correctas se obtienen de insertar la propuesta de solución (ansatz)
en la acción, se tiene entonces

\bea
S&=&-\frac1{2\pi\alpha'}\int d\tau d\sigma \sqrt{g_{t}(r)\,\dot t^2\left( g_{x}(r)\,\acute x^2+g_{r}(r)\,\acute r^2\right)}\nn\\
 &=&-\frac1{2\pi\alpha'}\int dt d\sigma \sqrt{g_{t}(r)\left( g_{x}(r)\,\acute x^2+g_{r}(r)\,\acute r^2\right)}\nn\\
 &=&-\frac{\cal T}{2\pi\alpha'}\int d\sigma \sqrt{ f^2(r)\,\acute x^2+g^2(r)\,\acute r^2}\,.
 \label{effec}
\eea

donde $g^2(r)=g_{t}(r)g_{r}(r)$ y $f^2(r)=g_{t}(r)g_{x}(r)$. La
invarianza ante reparametrizaciones de (\ref{NG}) factoriza la
extensión temporal del lazo $\cal T$ y reduce el cálculo del lazo
de Wilson a encontrar una geodésica en la geometría efectiva 2-
dimensional

\be
ds^2_{eff}=f^2(r)dx^2+g^2(r)dr^2\,. \label{geff}
\ee

La carga conservada asociada a traslaciones en $x$ en (\ref{effec}) es

\be
\frac{f^2(r)\, \acute x(\sigma)}{\sqrt{f^2(r)\acute x(\sigma)^2+g^2(r)\acute r(\sigma)^2}}=A
\ee

de donde se obtiene

\be
\acute x(\sigma)=\pm A{\frac{g(r)}{f(r)}}\frac1{\sqrt{f^2(r)-A^2}}\,\acute r(\sigma)\,.
\label{eqxr}
\ee

La invarianza ante reparametrizaciones garantiza que (\ref{eqxr})
resuelve la ecuación de movimiento en la dirección $r$. Nombrando $r_0$
al punto dado por $f(r_0)=A$, (\ref{eqxr}) puede reescribirse como \cite{sonnen}

\be
\frac{dx}{dr}=\pm {\frac{g(r)}{f(r)}}\frac{f(r_0)}{\sqrt{f^2(r)-f^2(r_0)}}\,.
\label{sig}
\ee

Las condiciones de borde en infinito para la separación de los extremos
de la cuerda son $\Delta x|_{r=\infty}=L$. A partir de (\ref{sig}) se nota
que la cuerda alcanza el borde de manera ortogonal. Hay dos elecciones de
gauge naturales que se encuentran en la literatura: $x(\sigma)=\sigma$
($x$-gauge) o $r(\sigma)=\sigma$ ($r$-gauge). La primera elección ($x$-gauge) tiene el
beneficio de proveernos una parametrización completa de $r(x)$  cuando se imponen
$x\in[-L/2,L/2]$ y $r(\pm L/2)=\infty$ (la localización de la punta (tip)  de la cuerda se escoge en
$(r,x)=(r_0,0)$). Haciendo $x\to t$, la ecuación (\ref{sig}) puede ser entendida como un movimiento
de energía cero en un potencial $U(r)$ dado por

\be
U(r)=\frac{f^2(r)(f^2(r)-f^2(r_0))}{g^2(r)f^2(r_0)},
\ee
en donde el punto $r_0$ puede verse como el valor mínimo alcanzado por la cuerda
en la coordenada holográfica. La segunda elección (r-gauge) da una función
$x(r)$ bivaluada cuando se impone $r\in[r_0,\infty)$ y $x(\infty)=\pm L/2$.
Sin embargo, existen una gran cantidad de ejemplos que conducen a una expresión analítica
cerrada para $x(r)$. Además, cuando se calculan fluctuaciones alrededor
de soluciones estáticas, se simplifican drásticamente las ecuaciones de movimiento
debido a que no contribuyen las fluctuaciones provenientes de las componentes de la métrica $g_{\mu\nu}$
(ver \cite{sonnen},\cite{avramis,avramis2, Siampos}). Se debe tener en cuenta
que el tip de la cuerda es un punto especial en el gauge $r$ ya que
en dicho punto deben pegarse correctamente las dos ramas correspondientes a los signos
$\pm$ en (\ref{sig}) (ver en la sección siguiente).

Integrando (\ref{eqxr}) se puede escribir la longitud $L(r_0)$ como,
\be
L(r_0)=2\int_{r_0}^\infty
\frac{g(r)}{f(r)}\frac{f(r_0)}{\sqrt{f^2(r)-f^2({r_0})}}\,dr\,.
\label{generalL}
\ee

Asumiendo que $f(r),f'(r)>0$ el límite de integración inferior $r_0$ en (\ref{generalL})
es, en general, integrable \footnote{ Para $f'(r_0)=0$ hay una divergencia logarítmica
dependiente de $r_0$ en (\ref{generalL}) (en \cite{piai}
hay una discusión general sobre divergencias en la función longitud).}. Nótese que
$L(r_0)$ finito en el lado izquierdo de (\ref{generalL}) demanda que $g/f^2$
decaiga en infinito más rápidamente que $\frac1r$. En las siguientes secciones voy
a analizar más profundamente la relación (\ref{generalL}). La derivada $L'(r_0)$
puede calcularse de la siguiente manera \cite{avramis, avramis2, Siampos} (Ver también el trabajo
\cite{piai}),

\bea
\frac{L'(r_0)}2&=&-\left.\frac{g(r)}{\sqrt{f^2(r)-f^2({r_0})}}\right|_{r\to r_0}
+f'(r_0)\int_{r_0}^\infty  \frac{f(r)g(r)}{(f^2(r)-f^2(r_0))^{\frac32}}\,dr\nn\\
&=&-\left.\frac{g(r)}{\sqrt{f^2(r)-f^2({r_0})}}\right|_{r\to r_0}+f'(r_0)\int_{r_0}^\infty dr \frac{g(r)}{f'(r)} \frac d{dr}
\left(-\frac1{\sqrt{f^2(r)-f^2(r_0)}}\right)\nn\\
&=&-\left.\frac{f'(r_0)g(r)}{f'(r)\sqrt{f^2(r)-f^2({r_0})}}\right|_{r\to\infty}+\int_{r_0}^\infty dr \frac{f'(r_0)}{\sqrt{f^2(r)-f^2(r_0)}}\frac d{dr}
\left(\frac{g(r)}{f'(r)}\right)\,,\nn
\eea
donde se ha integrado por partes al pasar de la segunda a la tercera linea.
Como el primer término del lado derecho de la tercera linea es nulo en todos
los fondos de gravedad estudiados en este capítulo se tiene

\be
{L'(r_0)}=2\int_{r_0}^\infty dr \frac{f'(r_0)}{\sqrt{f^2(r)-f^2(r_0)}}\frac d{dr}
\left(\frac{g(r)}{f'(r)}\right)\,.\label{Lprima}
\ee

La energía de la configuración $q\bar q$ fue propuesta en  \cite{rey, maldawilson}
como la longitud de la solución de cuerda (\ref{eqxr}) en la métrica efectiva 2-dimensional
(\ref{geff}),

\be
E=\frac 1{2\pi\alpha'}\int d\sigma\sqrt{f^2(r)\acute
x(\sigma)^2+g^2(r)\acute r(\sigma)^2}
\label{energy}
\ee

Usando (\ref{eqxr}) se puede escribir la energía en el gauge $x$ y en el $r$
de la siguiente forma

\ba
E(r_0)&=&\frac 1{2\pi\alpha'}\int_{-L/2}^{L/2} dx\,\frac{f^2(r(x))}{f(r_0)}\\
&=&\frac 1{\pi\alpha'}\int_{r_0}^\infty
dr\frac{g(r)f(r)}{\sqrt{f^2(r)-f^2(r_0)}}\,.
\label{er}
\ea

La energía calculada con la expresión (\ref{energy}) diverge debido a la
extensión infinita de la cuerda \footnote{En general, la divergencia en (\ref{energy})
es independiente de $r_0$. Adicionalmente puede aparecer una divergencia
que sí depende de $r_0$ en las ecuaciones (\ref{generalL}) y (\ref{energy}) cuando la
cuerda explora regiones en donde $f'(r_0)=0$.}. La interpretación de esta divergencia
proviene de que la ecuación (\ref{energy}) involucra, además de la energía potencial entre los quarks, la autoenergía
(masa) de los mismos \cite{rey, maldawilson}. Con el fin de obtener una cantidad con sentido físico
y encontrar el potencial entre quarks, debemos comparar (\ref{energy}) con respecto
a un estado de referencia teniendo la precaución de sustraer una cantidad independiente
de $r_0$. Es costumbre tomar la longitud de una cuerda recta a lo largo de
la coordenada $r$ que sale de infinito y explora el interior del espacio-tiempo, con todas
las otras coordenadas fijadas a constante, como la masa desnuda del quark. Nombrando $r=r_{min}$
al valor mínimo permitido por la geometría (\ref{gravity}) para la coordenada radial, sea por
la existencia de un horizonte (por ejemplo, fondos de AdS en coordenadas de Poincaré o
agujeros negros térmicos ), o porque realmente el espacio-tiempo se acaba suavente
(ej. el Soliton AdS de Witten, la solución de Maldacena-Nuñez y la de Klebanov-Strassler), la masa del quark
toma la forma \footnote{ Véase la discusión al final de las secciones \ref{adsglobal} y \ref{mnsol}
para aspectos relacionados con la interpretación del estado de referencia en fondos
gravitatorios suaves.}

\be
m_q=\frac1{2\pi\alpha'}\int_{r_{min}}^\infty g(r)\, dr\, .
\label{se}
\ee

~

La energía potencial entre el par quark-antiquark obtenida de
(\ref{energy}) luego de sustraer la autoenergía de los mismos (\ref{se})
es
\bea
E_{q\bar q}(r_0)&=&E(r_0)-2\,m_q\nn\\
&=&\frac 1{\pi\alpha'}\left[\int_{r_0}^\infty \frac{g(r)f(r)}{\sqrt{f^2(r)-f^2(r_0)}}\,dr-\int_{r_{min}}^\infty g(r)\, dr\right]\,.
\label{enerren}
\eea

Eliminando $r_0$ de (\ref{generalL}) y (\ref{enerren}) se obtiene
la propuesta que nos ofrece la correspondencia AdS/CFT para la energía potencial
entre quarks, $V_{\sf string}(L)$, en el límite de grandes valores de la constante de ´t Hooft.
En las secciones que siguen se graficarán estas relaciones en varios ejemplos y se
analizará su forma funcional. Por completitud, la derivada de (\ref{enerren}) es

\be
E'_{q\bar q}(r_0)=\frac1{\pi\alpha'}\left[-\left.\frac{g(r)f(r)}{\sqrt{f^2(r)-f^2({r_0})}}\right|_{r=r_0}+\int_{r_0}^\infty dr
 \frac{f(r)g(r)f(r_0)f'(r_0)}{(f^2(r)-f^2(r_0))^{\frac32}}\right]\nn\,.
\ee
Usando la primera linea de (\ref{Lprima}) se obtiene \cite{avramis, avramis2, Siampos}
\be
E'_{q\bar q}(r_0)=\frac1{2\pi\alpha'}f(r_0)\,L'(r_0)\Rightarrow\frac{dE_{q\bar q}}{dL}=\frac1{2\pi\alpha'}f(r_0)\,,
\label{eprima}
\ee
donde $r_0$ en la última expresión debe entenderse como la función $r_0(L)$ obtenida de invertir (\ref{generalL}).

Voy a finalizar esta sección describiendo algunas condiciones que debe satisfacer cualquier potencial que pretenda
describir la interacción entre quarks. Las llamadas condiciones de "\,concavidad"\\ demostradas en
\cite{bachas} son

\be
\frac{dV}{dL}>0,~~~~~~\frac{d^2V}{dL^2}\leq0\,.
\label{convexity}
\ee

Estas condiciones son válidas independientemente del grupo de gauge y los detalles del
sector de materia. La interpretación física de (\ref{convexity}) nos dice que la fuerza entre
el par quark-antiquark es:(i) siempre atractiva y (ii) una función no creciente de su distancia de separación.
De (\ref{generalL}),(\ref{enerren}),(\ref{eprima})
se encuentra que para la función propuesta por la conjetura $V_{\sf string}(L)$ se tiene \cite{Brandh}
\be
\frac{dV_{\sf string}}{dL}=\frac{dE_{q\bar q}}{dr_0}\frac{dr_0}{dL}=\frac{1}{2\pi\alpha'} f(r_0)
,~~~~~~~~\frac{d^2V_{\sf string}}{dL^2}=\frac{1}{2\pi\alpha'}\left(\frac{dL}{dr_0}\right)^{-1}{f'(r_0)}\,.
\label{concavity}
\ee

La primera condición siempre se satisface en fondos de gravedad duales, debido a que por definición $f(r)>0$.
Aunque en cada uno de los ejemplos estudiados en esta tesis se tiene $f'(r)>0$, la segunda condición
no se va a satisfacer cuando $L'(r_0)$ sea positiva. Voy a presentar casos en donde aparecen estas manifestaciones
no física y mostraré que en estas circunstancias la solución para la forma de embeber la cuerda (\ref{sig})-(\ref{generalL})
es inestable ante pequeñas perturbaciones. Esta última afirmación es el motivo del presente capítulo.

\subsection{Análisis de Estabilidad de embeddings de cuerdas clásicas}
\label{stability}

Se estudiará en esta sección la estabilidad de la solución clásica $(r_{\sf cl}(\sigma),x_{\sf cl}(\sigma))$
dada por (\ref{sig})-(\ref{generalL}) ante pequeñas (y lineales) perturbaciones. Una fluctuación general alrededor
de la solución puede escribirse como
\be
X^\mu=(\tau,x_{\sf cl}(\sigma)+\delta x_1(\tau,\sigma),\delta x_2(\tau,\sigma),\delta x_3(\tau,\sigma),
r_{\sf cl}(\sigma)+\delta r(\tau,\sigma),\theta^a+\delta\theta^a(\tau,\sigma))\, .\label{embedding}
\ee

Es posible usar la invarianza ante difeomorfismos de la acción para fijar $t=\tau$ y olvidarnos
de la ecuación de movimiento en la dirección $t$. Para las clases de métricas consideradas en (\ref{gravity}),
las fluctuaciones  $\delta x_2$ y $\delta x_3$ se desacoplan y satisfacen, como era de esperar,
la misma ecuación de movimiento, las fluctuaciones en los ángulos $\delta\theta^a$ se mezclan entre ellas
(en el caso general de variedades compactas) y dan lugar a 5 ecuaciones de movimiento (no voy a analizar fluctuaciones
angulares en el presente trabajo y pueden fijarse a 0 de forma consistente). Finalmente las fluctuaciones
$\delta x_1$ y $\delta r$ se mezclan dando lugar a 2 ecuaciones acopladas.

Es fácil de ver que las ecuaciones de movimiento obtenidas en el gauge $r$ son proporcionales
a las obtenidas en el gauge $x$. El difeomorfismo remanente debe ser usado para fijar la orientación
del vector $(\delta r,\delta x_1)$ (fluctuaciones en el plano) en cada punto de la solución
 $(r_{\sf cl}(\sigma),x_{\sf cl}(\sigma))$. Después de imponer un vínculo de gauge una ecuación describe
 las fluctuaciones en el plano $(r,x_1)$ y tenemos un sistema de ecuaciones diferenciales bien
 definido. La elección física del gauge (n-gauge) es el que orienta la fluctuación a lo largo
 de la dirección normal a $(r_{\sf cl}(\sigma),x_{\sf cl}(\sigma))$, en general esto significa
 que se están teniendo en cuenta ambas fluctuaciones, en la coordenada $r$ y en la $x_1$.

Hay otras dos
 posibilidades naturales consideradas en la literatura correspondientes a fijar
$\delta x_1(\tau,\sigma)=0$ ($x$-gauge) o $\delta r(\tau,\sigma)=0$ ($r$-gauge). Los fijados de
gauge $n$ y $x$, como se ha mencionado en la sección previa, parametrizan las fluctuaciones a lo
largo de toda la solución clásica pero las ecuaciones de movimiento resultantes son tediosas debido a que
las fluctuaciones $\delta r$ dan contribuciones adicionales provenientes de fluctuaciones de la métrica
\cite{avramis, avramis2,Siampos,sonnkin}.

\begin{figure}[h]
\begin{minipage}{7cm}
\vspace{3cm}
\hspace{2.5cm}
\pscircle[linewidth=.3pt](2.5,0){.3}
\psline[linewidth=.3pt]{->}(0,-3)(0,3)
\psline[linewidth=.3pt]{<-}(0,0)(3.5,0)
\psarc[linewidth=.5pt,linestyle=dashed](0,0){2.5}{270}{90}
\pscurve(0,2.5)(0.05,2.505)(0.3,2.9)
(0.7,1.8)(1.2,2.7)(1.6,1.2)(2.1,2.3)(2.48,.8)
(2.5,0)(2.48,-.8)(2.1,-2.3)(1.6,-1.2)
(1.2,-2.7)(0.7,-1.8)(0.3,-2.9)(0.05,-2.505)(0,-2.5)
\rput[bI](-0.3,2.9){$x$}
\rput[bI](3.7,-0.3){$r$}
\rput[bI](3,-0.4){$r_0$}
\rput[bI](2.9,0.3){$?$}
\rput[bI](-.2,2.2){$\frac L2$}
\rput[bI](-.4,-2.7){$-\frac L2$}
\vspace{3cm}
\caption{{\sf  $r$-gauge}: La línea discontinua representa la forma clásica de embeber la
cuerda que uno perturba. En el tip $r_0$, la fluctuación esta orientada a lo largo de la cuerda,
y por lo tanto no es física.}
\label{r-gauge}
\end{minipage}
\   \  \
\hfill
\begin{minipage}{7cm}
\vspace{3cm}
\hspace{2cm}
\psline[linewidth=.3pt]{->}(0,-3)(0,3)
\psline{->}(2.5,0)(3.2,0)
\psline[linewidth=.3pt]{<-}(0,0)(4,0)
\psarc[linewidth=.5pt,linestyle=dashed](0,0){2.5}{270}{90}
\pscurve(0,2.5)(0.5,2.5)(1.3,2.3)(0.5,2.1)(2.4,1.7)(1,1.3)(2.9,0.8)
(1.5,0.5)(3.1,0.2)(3.2,0)(3.1,-0.2)(1.5,-0.5)(2.9,-0.8)(1,-1.3)(2.4,-1.7)(0.5,-2.1)(1.3,-2.3)
(0.5,-2.5)(0,-2.5)
\rput[bI](-0.3,2.9){$x$}
\rput[bI](3.4, 0.2){$\delta r_0$}
\rput[bI](3.9,-0.3){$r$}
\rput[bI](-.2,2.2){$\frac L2$}
\rput[bI](-.4,-2.7){$-\frac L2$}
\vspace{3cm}
\caption{{\sf $x$-gauge}: La línea discontinua representa la forma clásica de embeber la
cuerda que uno perturba. Esta es una elección con sentido físico bien definido en toda la
solución, las fluctuaciones gráficadas (pares) cambian la posición del tip.}
\label{xgauge}
\end{minipage}
\end{figure}
 Nótese que a primera vista los fijados de gauge $n$ y $x$
aparentemente permiten oscilaciones del tip, mientras que el fijado de gauge $r$ no lo permite.
Voy a elegir trabajar en el gauge $r$, lo que significa tener $\delta r(\tau,\sigma)=0$ y trabajar con
ecuaciones de movimiento más simples definidas en la mitad de la solución. Con el fin de obtener
soluciones con sentido será entonces necesario
analizar las condiciones de borde en el tip $(r_0,0)$ del embedding.
Además, como se están considerando fluctuaciones a lo largo de la coordenada $x_1$, precisamente
en el tip la fluctuación $\delta x_1$ se orienta a lo largo del volumen de mundo de la cuerda y no transversalmente
al mismo, es decir:\emph{ la fluctuación $\delta x_1$ en el gauge $r$ no es física en el tip del embedding}. El análisis
requerido ha sido desarrollado en \cite{pufu},\cite{avramis, avramis2, Siampos} y será discutido en secciones
posteriores.
Por último, otra ventaja de usar el gauge $r$ es que nos quedan expresiones cerradas de las ecuaciones de
movimiento linealizadas para las fluctuaciones (ver (\ref{SL})-(\ref{spp})), la isometría a lo largo
de $x_1$ implica que no se necesita la solución analítica explicita de $x_{cl}(r)$, lo que contribuye a las
ecuaciones de movimiento es su derivada (\ref{sig}) (como ejemplo comparar (\ref{xg}) con (\ref{cg})).
Me gustaría hacer referencia al trabajo \cite{pufu} en donde se consideraron fluctuaciones $\delta r$
en el gauge $\delta x_1=0$ sobre la mitad de la solución clásica parametrizada en el gauge $r$ (ver debajo ecuaciones
(\ref{x1}),(\ref{rr})).

En lo que sigue voy  estudiar las ecuaciones de movimiento para las fluctuaciones $\delta x_i$, estas se obtienen
usando el ansatz
\be
t=\tau,~~~~x_1=x_{\sf cl}(r)+\delta x_1(t,r),~~~~x_2=\delta
x_2(t,r),~~~~x_3=\delta x_3(t,r),~~~~r=\sigma \, ,
\label{Avramis}
\ee

en la acción (\ref{NG}). Expandiendo a segundo orden en fluctuaciones se obtiene
\bea
2\pi\alpha'{\cal
L}^{(2)}&=&\frac1{g(r)f(r)\sqrt{f^2(r)-f^2(r_0)}}
\left[\phantom{\sum}\!\!\!\!\!\!\!\!h^2(r)\,(f^2(r)-f^2(r_0))\,(\delta \dot x_1)^2-(f^2(r)-f^2(r_0))^2(\delta \acute x_1)^2\right.\nn\\
&&\left.+f^2(r)h^2(r)((\delta \dot x_2)^2+(\delta \dot x_3)^2)-f^2(r)\,
(f^2(r)-f^2(r_0))\,((\delta \acute x_2)^2+(\delta \acute x_3)^2)\phantom{\sum}\!\!\!\!\!\!\!\!\right]\,,
\label{lagr}
\eea
donde $h^2(r)=g_{x}(r)g_{r}(r)$. La ecuación de Euler-Lagrange para la fluctuación $\delta x_1$ es
\be
\left[\frac{d}{dr}\left(\frac{(f^2(r)-f^2(r_0))^{\frac32}}{g(r)f(r)}\frac{d}{dr}\right)+
\omega^2\frac{h^2(r) \sqrt{f^2(r) -f^2(r_0) }}{g(r)f(r)}\right]\delta x_1(r)=0\,,
\label{SL}
\ee
en donde se ha factorizado la dependencia temporal de las fluctuaciones como $\delta x(t,r)=\delta x(r)\,e^{-i\omega t}$.
Las ecuaciones para las fluctuaciones transversales al plano $(r,x_1)$ obtenidas de (\ref{lagr}) son
\bea
\left[\frac{d}{dr}\left(\frac{f(r)\sqrt{f^2(r)-f^2(r_0)}}{g(r)}\frac{d}{dr}\right)+
\omega^2\frac{h^2(r)f(r)}{g(r)\sqrt{f^2(r)-f^2(r_0)}}\right]\delta x_m(r)=0\,,~~~~m=2,3\, .
\label{spp}
\eea

Las ecuaciones (\ref{SL})-(\ref{spp}) son ecuaciones diferenciales del tipo Sturm-Liouville definidas en
la semi-recta $r_{min}\leq r_0 \leq r<\infty$ y estoy interesado en analizar la existencia de inestabilidades,
en particular, determinando el rango de valores de $r_0$ para el cual $\omega^2<0$.

Las condiciones de borde a imponer en el problema son del tipo Dirichlet, esto se traduce en fluctuaciones
que dejan los extremos de la cuerda fijos en el borde $\delta x_i(\tau,\sigma)|_{r=\infty}=0$. Pero,
como el gauge $r$ parametriza solo la mitad de $(r_{\sf cl}(\sigma),x_{\sf cl}(\sigma))$ se requiere un estudio
adicional de las condiciones de borde en el tip $r=r_0$ (punto singular de (\ref{SL})-(\ref{spp})). Voy a comenzar
analizando (\ref{spp}), la expansión alrededor del tip nos da\footnote{Considero a $f(r)$ como una función creciente
de $r$ que no tiene ceros excepto, quizá, en el fondo de la geometría $r=r_{min}$.}
\be
\frac{d}{dr}\left(\!\sqrt{r-r_0}\;\frac{d\delta x_m(r)}{dr}\right)+
\frac{\omega^2h^2(r_0)}{2f(r_0)f'(r_0)}\frac1{\sqrt{r-r_0}}\,\delta x_m(r)\approx0~~~\Longrightarrow ~~~
\delta x_m(r)\approx C_0+C_1\sqrt{r-r_0}+O({r-r_0})  \,.
\label{xperp}
\ee
Aquí $C_{0,1}$ son constantes arbitrarias correspondientes a las dos soluciones independientes de la ecuación
diferencial (\ref{spp}), las cuales una vez elegidas, determinan en su totalidad la expansión
en serie para $\delta x_m(r)$. Físicamente estas se corresponden con fluctuaciones pares e impares alrededor del tip
respectivamente (una vez que las pegamos con las fluctuaciones alrededor de la otra mitad de la solución, las cuales
obviamente satisfacen las ecuaciones de movimiento). Un conjunto discreto de valores $\omega^2$ es esperable si existen
soluciones no normalizables en el límite de $r$ grande, también se espera que la solución par tenga el menor
autovalor $\omega^2$ (ver apéndice \ref{kmt}).

Ahora analizemos las fluctuaciones sobre el plano $\delta x_1$. Expandiendo (\ref{SL}) alrededor de $r=r_0$
se encuentra
\be
\frac{d}{dr}\left((r-r_0)^{\frac32}\frac{d\delta
x_1(r)}{dr}\right)+\frac{\omega^2h^2(r_0)}{2f(r_0)f'(r_0)}\sqrt{r-r_0}\,\delta
x_1(r)\approx0 ~~~\Longrightarrow ~~~ \delta x_1(r)\approx
C'_0+C'_1\frac1{\sqrt{r-r_0}}+O(\sqrt{r-r_0}) \,.
\label{asymr0}
\ee

Aparece un desarrollo singular para $\delta x_1$ en el tip y uno esta tentado a cancelarlo
imponiendo $C'_1=0$. Sin embargo, como se mencionó anteriormente se debe tener en cuenta que
el gauge $r$ implica que, en el tip, $\delta x_1$ esta dirigido a lo largo del volumen de mundo
de la cuerda y no transversal a él, entonces, en el gauge $r$ el desplazamiento $\delta x_1$
(en el tip) no es físico. Con motivo de dar una interpretación física a las constantes $C'_{0,1}$ en
(\ref{asymr0}) voy a cambiarme del gauge $r$ al gauge $x$ (confrontar con \cite{gubser},\cite{avramis,avramis2,Siampos}).
El cambio de gauge en el ansatz (\ref{Avramis}) puede implementarse mediante un cambio de variables sobre
la solución (\ref{SL}) desde $r$ a una nueva variable que llamaré $u$. Esto puede ser implementado perturbativamente y
a primer orden en las fluctuaciones la relación es \cite{avramis,avramis2,Siampos},
\be u=r+\Delta
(t,r)~~~~~\mathrm {donde}~~~\Delta (t,r)=\frac{\delta
x_1(t,r)}{x'_{\sf cl}(r)}\, .
\label{change}
\ee
Esta transformación nos permite realizar el cambio de gauge deseado ya que
\bea
x_1&=&x_{\sf cl}(r)+\delta x_1(t,r)=x_{\sf cl}(u-\Delta (t,r))+\delta
x_1(t,r)\nn\\&\approx& x_{\sf cl}(u)-x'_{\sf cl}(r)
\frac{\delta x_1(t,r)}{x'_{\sf cl}(r)}+\delta x_1(t,r)=x_{\sf cl}(u)\label{x1}\\
r&\approx&u-\frac{\delta x_1(t,u)}{x'_{\sf cl}(u)}\,,
\label{rr}
\eea
aquí $r_0\le u<\infty$. Este es precisamente el gauge empleado en \cite{pufu}. El segundo término
en el lado derecho de (\ref{rr}) es interpretado como la fluctuación en la dirección $r$
inducida por la (gauge $r$) fluctuación en $x_1$. Ahora es fácil de ver que es finita.
El desarrollo asintótico (\ref{asymr0}) y la expansión alrededor del tip de (\ref{sig}),
$x'_{\sf cl}(r)\sim(r-r_0)^{-1/2}$, insertada en (\ref{rr}) es
\be
r\approx r_0- \alpha(C'_0{\sqrt{u-r_0}}+C'_1)+O(u-r_0)\,,
\label{rgauge}
\ee
con $\alpha$ una constante finita. El resultado (\ref{rgauge}) muestra que la fluctuación
física $\delta r$ originada por la fluctuación no física $\delta x_1$ en el tip es manifiestamente
finita. Entonces podemos interpretar a ($C'_0$) $C'_1$ en (\ref{asymr0}) como las fluctuaciones que
(no) mueven la posición del tip.

En las secciones siguientes voy a estudiar soluciones numéricas de (\ref{SL}) para varios
fondos gravitatorios determinando los autovalores más bajos que dan soluciones normalizables. Voy
a resolver (\ref{SL}) mediante el método de shooting integrando numéricamente desde $r_0$ hasta un valor
grande $r_{\infty}$. Los valores permitidos para $\omega^2$ se obtendrán de imponer que la solución numérica
se anule en $r=r_{\infty}$. Las condiciones de borde en el tip correspondientes a soluciones pares
son $C'_0=0$ y $C'_1$ arbitrario, por propósitos numéricos se utiliza $C'_1=1$, este valor fija
la normalización de la fluctuación. Una solución par alrededor del tip satisface
\be
 \left.\frac{d\delta r(t,r)}{dx_1}\right|_{r=r_0}=0~~~~~\mathrm {where}~~~\delta r(t,r)=-\frac{\delta
x_1(t,r)}{x'_{\sf cl}(r)}\, .
\label{bcc}
\ee
Usando (\ref{x1}) se puede escribir (\ref{bcc}) en términos de $\delta x_1(r)$. Las condiciones de borde para
soluciones pares de (\ref{SL}) se implementan numéricamente como
\be
     \delta x_1(r)+2(r-r_0)\frac{d\delta x_1(r)}{dr}=0, ~~~~r\rightarrow
     r_0\nn
\ee
\be
~~~~~~~~~~~~~~~~~\sqrt{r-r_0}\,\delta x_1(r)=1\,
,~~~~r\rightarrow r_0\label{bc}\,.
\ee
Para soluciones impares $C'_0=1$ y $C'_1=0$ se implementan como
\be
     \delta x_1(r)+2(r-r_0)\frac{d\delta x_1(r)}{dr}=1,~~~~r\rightarrow
     r_0\nn
\ee
\be
~~~~~~~~~~~~~~~~~\sqrt{r-r_0}\,\delta x_1(r)=0\,
,~~~~r\rightarrow r_0\label{bc2}\,.
\ee

Resumiendo, en fondos gravitatorios generales, la relación funcional de la solución
clásica (\ref{sig}) entre las coordenadas $x_1$ y $r$ en el tip toman la forma
$x_{\sf cl}^2(r)\approx r-r_0$, y el desarrollo asintótico de las fluctuaciones
en $x_1$, en el gauge r, es de la forma (\ref{asymr0}). Aunque aparezca una parte divergente
en (\ref{asymr0}), un cambio de gauge apropiado muestra que las partes divergentes y no divergentes
corresponden respectivamente a fluctuaciones (física) pares e impares alrededor del tip.

\section{Fondos Gravitatorios}
\label{bkg}

En esta sección voy a calcular los embeddings de cuerdas (\ref{eqxr}) duales a lazos de Wilson
rectangulares para un número de casos paradigmáticos. Se resumirán los casos de $AdS_5\times S^5$ \cite{rey,maldawilson}
y $AdS_5$-Schwarzschild$\times S^5$ \cite{Brandsonnen}. Luego, se realizará el análisis numérico de las ecuaciones
(\ref{generalL}) y (\ref{enerren}) para los fondos de Maldacena-Nuñez \cite{mn}, Klebanov-Strassler \cite{ks} y la
generalización de Maldacena-Nuñez \cite{cnp,cnp2}. En todos los caso la geometría es soportada por $p$-flujos
no triviales, pero estos no son relevantes en los cálculos.

\subsection{$AdS_5\times S^5$}

Esta solución de gravedad es dual a $\cal N$ $=4$ SYM con grupo de gauge $G=SU(N)$ en la fase de Coulomb.
La curvatura de AdS $R$ se relaciona con la teoría de gauge mediante la constante de ´t Hooft $\lambda$
como $R^4=\alpha'^2\lambda$ y el flujo de la 5-forma que soporta la geometría $N=\int_{S^5} F_5$ se relaciona
con el rango del grupo de gauge a través de $N={\sf Rank}(G)$ \cite{malda,GKP,Ewitten}. La invarianza conforme
de la teoría de gauge implica un desarrollo tipo Coulomb para el potencial entre quarks $V(L)\sim 1/L$. Lo novedoso del
cálculo en el lado gravitatorio es que permite obtener el coeficiente de proporcionalidad y su dependencia en el acoplamiento de gauge.

\subsubsection{Coordenadas de Poincaré \cite{rey, maldawilson}}

Este sistema de coordenadas describe la teoría de gauge formulada en $\mathbb R^{3,1}$.
La métrica se escribe como (compárese con la coordenadas utilizadas en la sección \ref{Wilint} en donde $z=\frac1r$)
\be
ds^2=\frac{r^2}{R^2}(-dt^2+dx_idx_i)+R^2\frac{dr^2}{r^2}+R^2d\Omega_5^2\,.
\label{ads5}
\ee
Se encuentra que $f^2(r)={r^4}/{R^4},~g^2(r)=1$. El rango de la coordenada radial es $0<r<\infty$,
en $r=0$ hay un horizonte de Killing. La ecuación (\ref{sig}) puede ser resuelta analíticamente en el gauge $r$,
se obtiene
\be
x_{\sf cl}(r)=\pm\left\{ cte-\frac{R^2}{4r_0}\mathsf{B}\left(\left(\frac{r_0}r\right)^4;\frac34,\frac12\right)\right\},~~~~~~r_0\le r<\infty
\label{xads}
\ee
donde $\mathsf{B}(z;a,b)$ is la función beta incompleta $\mathsf{B}(z;a,b)=\int_0^zt^{a-1}(1-t)^{b-1}dt$. Las condiciones
de borde fijan la constante en (\ref{xads}) y relaciona los parámetros $r_0$ y $L$, fijando $x_{\sf cl}(r_0)=0$ y $x_{\sf cl}(\infty)=\pm L/2$
se obtiene \cite{rey, maldawilson},
\be
L(r_0)=\frac{R^2}{2r_0}\mathsf{B}\left(\frac34,\frac12\right)=\frac{R^2}{r_0}\frac{(2\pi)^{\frac32}}{\Gamma[\frac14]^2}\,.
\label{lads}
\ee
La energía (\ref{enerren}) toma la forma \cite{rey, maldawilson}
\be
E_{q\bar q}(r_0)=\frac{r_0}{\pi\alpha'}(K(-1)-E(-1))=-\frac{r_0}{2\pi\alpha'}\frac{(2\pi)^{\frac32}}{\Gamma[\frac14]^2}\,,
\label{eads}
\ee
aquí, $K(m),E(m)$ son las integrales elípticas completas de primer y segundo tipo respectivamente. Eliminando $r_0$ de las expresiones
(\ref{lads})-(\ref{eads}) la propuesta de la conjetura AdS/CFT para la energía de interacción entre quarks en la representación
fundamental del grupo de gauge en el límite $\lambda$ grande para la teoría ${\cal N}=4$ SYM es \cite{rey, maldawilson}
\be
V_{\mathsf{string}}(L)=-\frac{(2\pi)^{2}}{\Gamma[\frac14]^4}\frac{R^2/\alpha'}{L}\sim -\frac{\sqrt\lambda} L\,.
\label{Vads}
\ee

Como es de esperar debido a la invarianza conforme se obtuvo un potencial de Coulomb atractivo. El resultado importante es la dependencia
en $\sqrt\lambda=(g_{YM}^2N)^{\frac12}$ (recordar que en el régimen perturbativo el factor de proporcionalidad
es $\lambda$). Esto sugiere que en el regimen de acoplamiento fuerte existe algún tipo de renormalización de las cargas \cite{rey, maldawilson}.
Nótese que para obtener la cantidad negativa (\ref{Vads}) habiendo comenzado de una cantidad definida positiva (ver ec.(\ref{energy})) fue crucial
sustraer la masa de los quarks (\ref{se}) .

\subsubsection{Coordenadas Globales}
\label{adsglobal}

Este ejemplo va a ser discutido debido a que clarifica conceptos involucrados en el proceso de sustracción de la masa (\ref{enerren}) en
duales gravitatorios suaves y completos (ver secciones \ref{mnsol}, \ref{kssol}, \ref{gmnsol}).

\begin{figure}[h]
\hspace{-0.25cm}\centering
\begin{minipage}{13cm} 
\centering
\includegraphics[width=7cm]{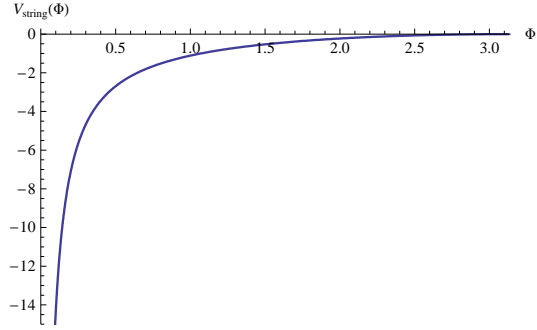}
\caption{$V_{\mathsf{string}}(\Phi)$ obtenido a partir de (\ref{fi})
y (\ref{enadsren}). Para ángulos de separación pequeños
entre los extremos de la cuerda $\Phi\ll1$ se encuentra el comportamiento de Coulomb esperable
$V\sim\sqrt\lambda/(R\Phi)$. Para separaciones grandes la solución se desvía debido a que $S^3$ es compacta.}
\label{Vadsglobal}
\end{minipage}
\end{figure}

Cálculos realizados en coordenadas globales representan una teoría de gauge ${\cal N}=4$ SYM definida sobre $S^3\times \mathbb R$.
La métrica de AdS se escribe
\be
ds^2=R^2[-\cosh^2\!\rho\,dt^2+d\rho^2+\sinh^2\!\rho\, d\Omega_3^2]\,.
\ee
Todas las coordenadas son adimensionales, con el radio de AdS, $R$, fijando la escala. Se escribe la métrica de
$S^3$ como $d\Omega_3^2=d\theta_1^2+\sin^2\!\theta_1(d\theta_2^2+\sin^2\!\theta_2d\varphi^2)$. Siendo $\varphi$ una coordenada
cíclica, el ansatz apropiado para la cuerda es $t=\tau,~\rho=\rho(\sigma),~\varphi=\sigma $, esto conduce a
$f^2(\rho)=\frac14\sinh^2\!2\rho,~g^2(\rho)=\cosh^2\!\rho$. Las coordenadas angulares remanentes deben ser fijadas a $\theta_i=\frac\pi2$
(el ecuador de $S^3$) en orden de satisfacer las ecuaciones de movimiento. La carga conservada debida a la coordenada cíclica $\varphi$ conduce
a una ecuación de movimiento unidimensional de energía cero
\be
\acute \rho^2+U(\rho)=0\,,
\label{global}
\ee
donde $\acute \rho=d\rho/d\varphi$ y el potencial $U(\rho)=\sinh^2\!\rho\left(1-\frac{\sinh^2\!2\rho}{\sinh^2\!2\rho_0} \right)$.
$\rho_0$ es la posición radial mínima alcanzada por la cuerda cuando los extremos de la cuerda en infinito se separan
en $\Delta\varphi=\Phi$. La relación (\ref{generalL}) para $\Phi(\rho_0)$ se calcula trivialmente
\be
\Phi(\rho_0)=2\int_{r_0}^\infty \frac{\sinh 2\rho_0}{\sinh \rho\sqrt{\sinh^2\!2\rho-\sinh^2\!2\rho_0}}\,d\rho\,.
\label{fi}
\ee
Debido a que la teoría de gauge esta definida sobre $S^3$ existe una separación máxima para los quarks la cual corresponde
a ubicarlos en las antípodas sobre el ecuador de la $S^3$. Esto resulta en una cuerda que alcanza el origen $\Phi(0)=\pi$
y que da lugar a una hoja de mundo recta y suave (extendida a lo largo de la coordenada $r$) parametrizada por dos mitades
en $\varphi=\varphi_0$ y $\varphi=\pi+\varphi_0$. La energía (divergente) (\ref{er}) de la configuración (\ref{global}) es
\be
E(\rho_0)=\frac R{2\pi\alpha'}\int_{\rho_0}^\infty  \frac{\sinh^2\!2\rho}{\sinh \rho \sqrt{\sinh^2\! 2\rho-\sinh^2\! 2\rho_0}}\,d\rho\,.
\label{diven}
\ee
Substrayendo la masa de los quarks (\ref{enerren}) se obtiene,
\be
E_{q\bar q}(\rho_0)=\frac R{2\pi\alpha'}\left[\int_{\rho_0}^\infty \left(\frac{2\cosh \rho}{\sqrt{1-\frac{\sinh^2\!2\rho_0}{\sinh^2\!2\rho}}}
-2\cosh \rho\right)\,d\rho-2\sinh \rho_0\right]\,,
\label{enadsren}
\ee
la cual es finita y definida negativa ( ver figura \ref{Vadsglobal}). El resultado finito (\ref{enadsren})
debe entenderse como resultante de comparar (\ref{diven}) con respecto al estado de referencia proveniente de la
configuración de la cuerda recta con sus extremos en infinito y en las antípodas del ecuador de la $S^3$.
Se interpreta esta última configuración como la única correspondiente a quarks "\,infinitamente" separados en la esfera.
Nótese que el estado de referencia y la configuración que se esta analizando satisfacen diferentes condiciones de borde.

\subsection{$AdS_5$-Schwarzschild$\times S^5$}
\label{adssch}

\begin{figure}[h]
\begin{minipage}{7.5cm}
\begin{center}
\includegraphics[width=7.5cm]{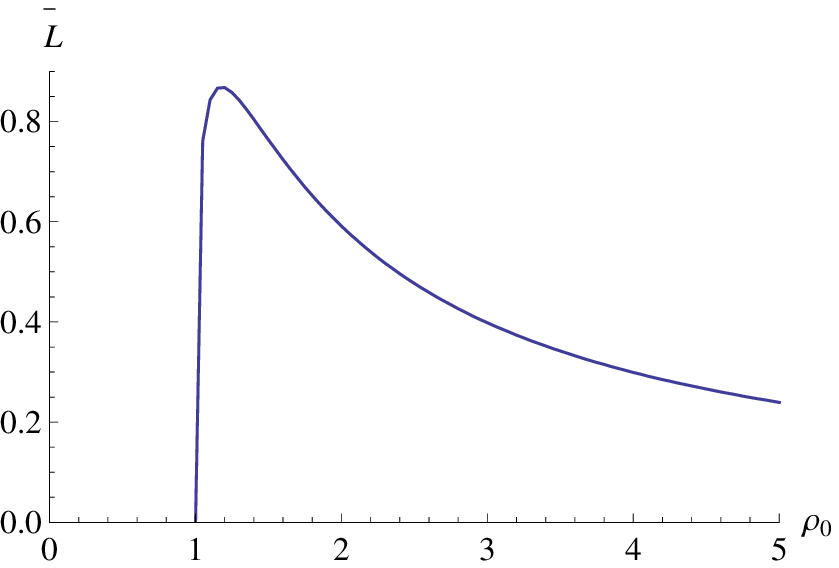}
\caption{$L(\rho_0)$ (\ref{ltads}): Se observa un máximo para$\rho_0\simeq1.177$. Las
dos ramas de la izquierda/derecha del máximo en $\rho_{0c}\simeq1.177$
dan lugar a una relación bivaluada para $V_{\sf string} (L)$ en la fig. \ref{E(L)}.}
\label{L(r_0)}
\end{center}
\end{minipage}
\   \
\hfill \begin{minipage}{7cm}
\begin{center}
\includegraphics[width=7.5cm]{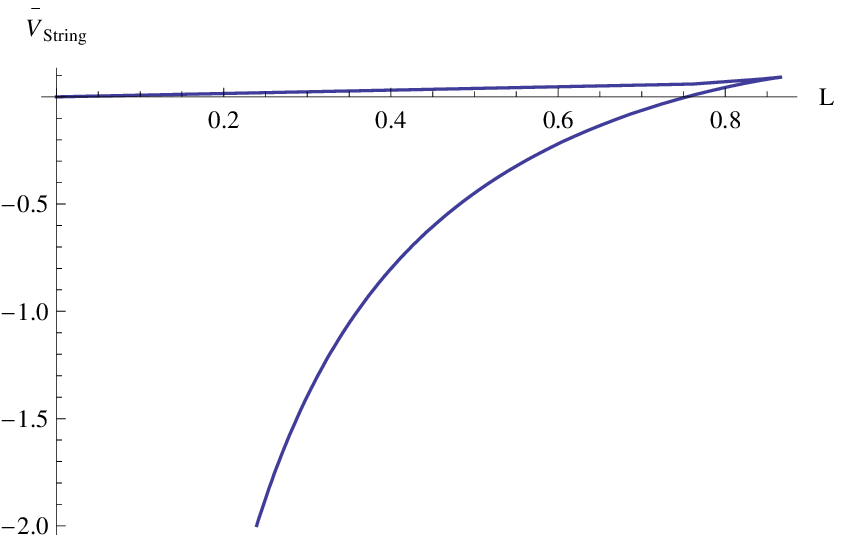}
\caption{$V_{\sf string}(L)$ bivaluada obtenida desde (\ref{ltads}) y (\ref{etads})
eliminando numéricamente $\rho_0$. La curva superior correspondiente a la rama izquierda en la fig.
\ref{L(r_0)} no satisface las condiciones (\ref{convexity}).}
\label{E(L)}
\end{center}
\end{minipage}
\begin{center}
\begin{minipage}{15cm}
\begin{center}
\includegraphics[width=7.5cm]{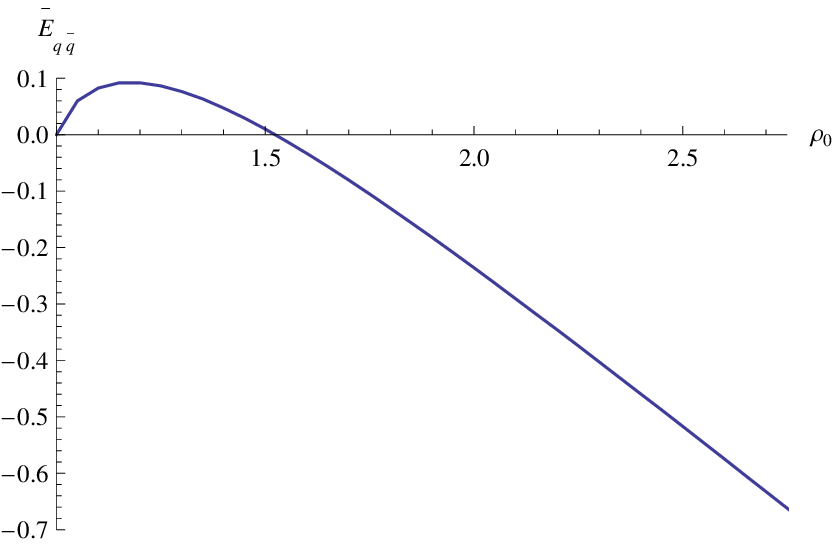}
\caption{Función $\bar E_{q\bar q}(\rho_0)$ para ${\cal N}=4$ SYM térmico. La ecuación (\ref{eprima}) garantiza la existencia de un extremo
en $E_{q\bar q}(r_0)$ y $L(r_0)$ para el mismo valor de $r_0$. La curva corta el eje en $\rho_{0m}\simeq1.524$, para $\rho_0>\rho_{0m}$
la solución en forma de U es un mínimo absoluto, para $\rho_0<\rho_{0m}$ la solución de dos líneas rectas (referencia) es el mínimo
absoluto.}
\label{e(r_0)}
\end{center}
\end{minipage}
\end{center}
\end{figure}

Teorías de gauge a temperatura finita se describen considerando soluciones de agujero negro en duales gravitatorios \cite{wittenT}.
La geometría de horizonte cercano de $N$ $D3$-branas no extremales (negras) se conjetura dual a ${\cal N}=4$ SYM a
temperatura finita. Como se explica en \cite{wittenT}, el fondo de agujero negro apropiado para describir
la teoría de gauge sobre $\mathbb R^{3,1}$ involucra un límite de masa infinita delicado del agujero negro $AdS_5$-Schwarzschild
resultante en,\footnote{El límite proviene de que la métrica es dependiente de solo una escala, al contrario
del caso de masa finita en donde hay dos parámetros: la temperatura (masa del agujero negro) y el radio de AdS. Esta
última geometría exhibe una transición de fase (Hawking-Page), la cual fue interpretada como dual a la transición de fase
confinante/deconfinante en ${\cal N}=4$ sobre $S^3$ \cite{wittenT}.}
\be
ds^2=\frac{r^2}{R^2}\left[-(1-\frac{\mu^4}{r^4})dt^2+d{x_i}dx_i\right]
+\frac{R^2}{r^2}\frac1{1-\frac{\mu^4}{r^4}}dr^2+R^2d\Omega_5^2\,.
\label{thermalads}
\ee
El horizonte del agujero negro esta localizado en $r=\mu$ y su temperatura es $T=\frac\mu{R^2}\pi$. Es conveniente trabajar
con coordenadas adimensionales, escaleando $r=\mu\,\rho$, $t=R^2/\mu\, \bar
t$ y $x=R^2/\mu\, y$ se obtiene
\be
ds^2=R^2\left[-\left({\rho^2}-\frac{1}{\rho^2}\right)d\bar
t^2+{\rho^2}d{y_i}dy_i+\frac1{{\rho^2}-\frac{1}{\rho^2}}d\rho^2+d\Omega_5^2\right].
\label{adimTads}
\ee
La escala de la teoría de gauge adimensional $\bar t,y_i$ en (\ref{adimTads}) esta fijada por $R^2/\mu$ y
se tiene $f^2(\rho)=\rho^4-1,~g^2(\rho)=1$ y $\rho=1$ para la posición del horizonte. Las expresiones
para la longitud de separación adimensional (\ref{generalL}) y la energía potencial (\ref{enerren}) para el par
$q\bar{q}$ pueden calcularse analíticamente \cite{avramis, avramis2, Siampos,Brandsonnen, BrandIt,Brandh}
\be \bar
L(\rho_0)=\frac{(2\pi)^{\frac32}}{\Gamma[\frac14]^2}\frac{\sqrt{\rho_0^4-1}}{\rho_0^3}\,\,_2F_1\!
\left(\frac34,\frac12,\frac54;\frac{1}{\rho_0^4}\right)
\label{ltads}
\ee
\be \bar
E_{q\bar{q}}(\rho_0)=\frac{R^2}{\pi\alpha'}\left[1-
\frac{(2\pi)^{\frac32}}{2\Gamma[\frac14]^2}\,\rho_0
\,_2F_1\!\left(-\frac12,-\frac14,\frac14;\frac{1}{\rho_0^4}\right)\right]
\label{etads}
\ee
Aquí $\rho_0\ge1$ es la posición radial mínima alcanzada por la cuerda, el valor radial mínimo $r_{min}$ en
(\ref{enerren}) ha sido tomado en la ubicación del horizonte $r_{min}=\mu$. Se puede chequear fácilmente que en el
límite de temperaturas bajas $LT\ll1$ (correspondiente a $\rho_0\gg1$) se recupera el comportamiento a temperatura cero
(\ref{lads})-(\ref{Vads}). Se grafican en las figuras \ref{L(r_0)} y \ref{e(r_0)} la forma de la longitud
(\ref{ltads}) y la energía (\ref{etads}) como función de $\rho_0$. En la figura \ref{E(L)} se grafica la
relación $V_{\sf string}(L)$ obtenida a partir de (\ref{ltads})-(\ref{etads}) eliminando $\rho_0$, el resultado
es una función bivaluada.

La figura \ref{L(r_0)} muestra un máximo $\bar L_c\simeq0.869$ en $\rho_{0c}\simeq1.177$ lo cual implica que no existe
solución suave que conecte al par de quarks para $\bar L>\bar L_c$. La única solución existente para $\bar L>\bar L_c$ corresponde
a dos cuerdas rectas que alcanzan el horizonte. Esta configuración, usada para la sustracción en (\ref{enerren}), es interpretada
como la correspondiente a un par de quarks libres. La existencia de un máximo en la función $L(r_0)$ para el fondo de agujero
negro ha sugerido interpretarlo como el dual gravitatorio de un baño térmico de apantallamiento  \cite{Brandsonnen,BrandIt}.
La figura \ref{L(r_0)} también muestra la existencia de dos ramas de soluciones para cada distancia de separación $L<L_c$
entre los extremos de la cuerda. La rama izquierda ($L'>0$) da lugar a un potencial $V_{\sf string}(L)$ que no satisface las condiciones
(\ref{convexity}) y por lo tanto no es física (curva superior en la figura \ref{E(L)}).
El resultado, como veremos, es que la teoría de cuerdas es sabia: la rama izquierda no debería ser
válida debido a que es inestable ante pequeñas perturbaciones (ver sección \ref{nond3}).

El último punto para comentar es que aunque, como vimos, toda la región para la cual $L'<0$ ($1.177<\rho_0<\infty $)
es estable ante pequeñas perturbaciones, se espera que la curva de abajo en la figura \ref{E(L)} sea metaestable
cuando $E_{q\bar q}>0$ ($0.754<\bar L<0.869$ o $1.177<\rho_0<1.524$). La razón de esto es que el estado de referencia,
las solución correspondiente a dos líneas rectas tiene $E_{q\bar q}=0$ y entonces es el mínimo absoluto en $0.754<\bar L<0.869$
(Ver fig. \ref{e(r_0)}). Teniendo en cuentas estas dos últimas cosas en \cite{Brandsonnen} se sugiere tomar
$\bar L_{max}\simeq0.754$ como la longitud de apantallamiento.

\subsection{Maldacena-Núñez}
\label{mnsol}

La región $r\approx0$ de este fondo gravitatorio es dual al regimen IR de
una teoría $\cal N$ $=1$ SYM en $d=4$ más una torre de estados de Kaluza-Klein. La configuración de branas de prueba que
permiten obtener esta solución consiste en $N$ $D5$-branas enroscadas
sobre un 2-ciclo finito en el origen de un conifold resuelto.  Cuando se tiene en
cuenta la acción de las $D5$ sobre la geometría ocurre una transición (ver \cite{vafa})
que nos conduce a un espacio-tiempo dotado de una $S^2$ que colapsa y una $S^3$ finita en el origen
(como sucede en el conifold deformado, ver \cite{paredes}). Esta solución
fue encontrada independientemente en el contexto de supergravedades con
simetrías de gauge (gauged sugra) en \cite{Cham, Cham2}.

La métrica se escribe \cite{mn}
\beq
ds^2= \alpha'  Ne^{{\phi}}\,\,\Big[-dt^2+dx_idx_i+
dr^2+e^{2h}\,(d\theta^2+\sin^2\theta d\varphi^2)+ {\frac1
4}\,(w^i-A^i)^2\Big]\,, \label{metric}
\eeq
en donde $w^i$ ($i=1,2,3$) son las formas invariantes a derecha de $su(2)$
\ba
 \label{su2}
 w^1+iw^2= e^{-i\psi}(d\tilde\theta+i\sin\tilde\theta\, d\tilde\varphi),~~~~
 w^3=d\psi\,+\,\cos\tilde\theta\, d\tilde\varphi\,,
\ea
y $A^i,\phi, h$  esta dada por
\ba
A^1&=&-a(r)\, d\theta\,, \,\,\,\,\,\,\,\,\, A^2\,=\,a(r) \sin\theta\,
d\varphi\,, \,\,\,\,\,\,\,\,\, A^3\,=\,- \cos\theta\, d\varphi\,\nn\\
 a(r)&=&\frac{2r}{\sinh 2r}\nn\\
 e^{2h}&=&r\coth 2r\,-\,\frac{r^2}{\sinh^2 2r}\,-\, {\frac1 4}\nn\\
e^{2\phi}&=&e^{2\phi_0}\, \frac{\sinh 2r}{ 2e^h}\,.
\label{oneform}
\ea
$\phi_0$ es una constante de integración que fija el valor del dilatón
$\phi$ en el origen (algunos autores escriben $g_s=e^{\phi_0}$).

Las coordenadas $t,x_i,r$ son adimensionales y su escala queda fijada por $(\alpha'  N)^{\frac12}$.
La métrica (\ref{metric}) es regular en el límite $r\to 0$ y la 2-esfera ($\theta,\varphi$) colapsa
suavemente, la topología del espacio-tiempo resultante es de la forma ${\cal M}_7\times S^3$, contrariamente
a los casos discutidos anteriormente en donde la topología tomaba la forma ${\cal M}_5\times X^5$ con $X^5$ compacto.
La razón de esto es que la métrica de fondo (\ref{metric}) modela una teoría de gauge
5+1 dimensional sobre la $D5$-brana enroscada. Sin embargo, se espera tener una teoría efectiva
3+1 dimensional a energías $E<1/R_{S^2}$, donde $R_{S^2}$ es el radio de la esfera que envuelve
la $D5$.

\begin{figure}[h]
\begin{minipage}{7cm}
\begin{center}
\includegraphics[width=7.5cm]{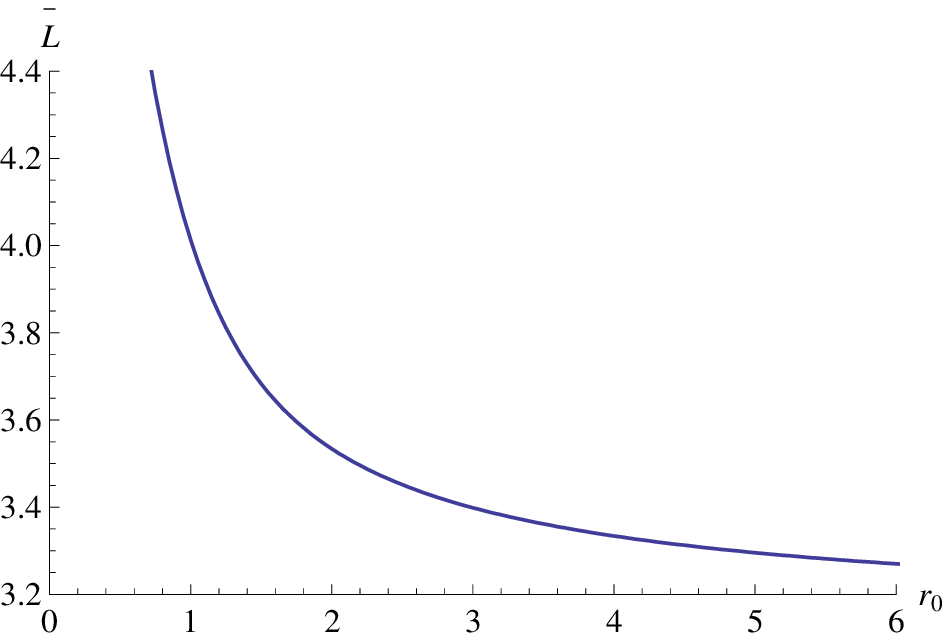}
\caption{$\bar L(r_0)$ (\ref{lmn}). Solución de MN.}
\label{lvsRmin MN}
\end{center}
\end{minipage}
\   \
\hfill
\begin{minipage}{7cm}
\begin{center}
\includegraphics[width=7.5cm]{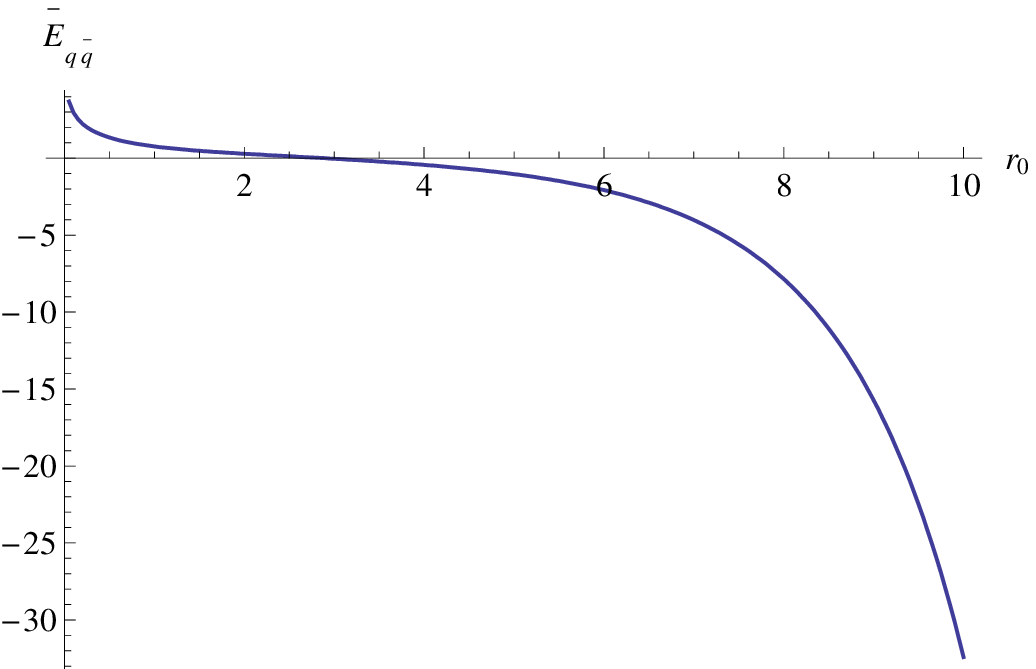}
\caption{$E(r_0)$ (\ref{emn}). Solución de MN.}
\label{EvsRmin MN}
\end{center}
\end{minipage}
\end{figure}

El ansatz de la cuerda estática con forma de U nos conduce a $f^2(r)=g^2(r)=e^{2\phi}$, con lo cual
la longitud de separación (\ref{generalL}) y la energía (\ref{enerren}) toman la forma
\be
\bar L(r_0)=2\int_{r_{0}}^{\infty}{\frac{e^{\phi(r_{0})}}{\sqrt{e^{2\phi(r)}-e^{2\phi(r_{0})}}}}\,dr
\label{lmn}
\ee
\be
\bar E_{q\bar q}(r_0)=\frac {N }{\pi}\left[\int_{r_{0}}^{\infty}\frac{e^{2\phi(r)}}{\sqrt{e^{2\phi(r)}-e^{2\phi(r_{0})}}}\,dr-
\int_{0}^{\infty}e^{\phi(r)}dr\right]\,.
\label{emn}
\ee
La expresión para la energía puede ser reescrita de forma conveniente \cite{sonnen}
\bea
\bar E_{q\bar q}(r_0)&=&\frac {N}{\pi}\left[\int_{r_{0}}^{\infty}\left(\frac{e^{2\phi(r)}+e^{2\phi(r_0)}-e^{2\phi(r_0)}}
{\sqrt{e^{2\phi(r)}-e^{2\phi(r_{0})}}}-e^{\phi(r)}\right)\,dr-
\int_{0}^{r_0}e^{\phi(r)}dr\right]\nn\\
&=&\frac {N}{\pi}\left[e^{\phi(r_{0})}\frac{\bar L(r_0)}2+\int_{r_{0}}^{\infty}{dr(\sqrt{e^{2\phi(r)}-e^{2\phi(r_{0})}}-e^{\phi(r)})}
 -\int_0^{r_{0}}{e^{\phi(r)}}dr\right]\,.
 \label{energMN}
\eea
En el límite de $L$ grande la cuerda alcanza el fin del espacio ($r_0\to0$, ver figura \ref{lvsRmin MN})
y los últimos dos términos en (\ref{energMN}) no contribuyen (see \cite{sonnen}). En el límite
de $\bar L\gg1$ se obtiene, explicitando nuevamente las dimensiones  \cite{mn}
\be
V_{\sf string}(L)\approx\frac {e^{\phi_0}}{2\pi\alpha'} L \quad\Rightarrow\quad T_{\sf string}
=\frac {e^{\phi_0}}{2\pi\alpha'}\,.
\label{tension}
\ee

La métrica de fondo (\ref{metric}) predice, por lo tanto, confinamiento lineal para la separación entre
quarks en perfecto acuerdo con las propiedades de su teoría de gauge dual $d=4$, ${\cal N}=1$ SYM.
A partir de la tensión de la cuerda (\ref{tension}) se puede ver que el valor del dilatón en el origen
$\phi_0$ se relaciona con la escala dinámica generada de la teoría de gauge dual.
\\
En las figuras \ref{lvsRmin MN} y \ref{EvsRmin MN} se grafican las funciones $\bar L(r_0)$ y $\bar E(r_0)$
provenientes de las ecuaciones (\ref{lmn}) y (\ref{emn}). La divergencia $L\to \infty$ para $r_0\to 0$
en la figura  \ref{lvsRmin MN} se debe a que $\left.\frac d{dr}(e^{2\phi(r)})\right|_{r=0}=0$ (ver \cite{piai} para
una discusión sobre este echo).
\begin{figure}[h]
\begin{center}
\begin{minipage}{15cm}
\begin{center}\includegraphics[width=7.5cm]{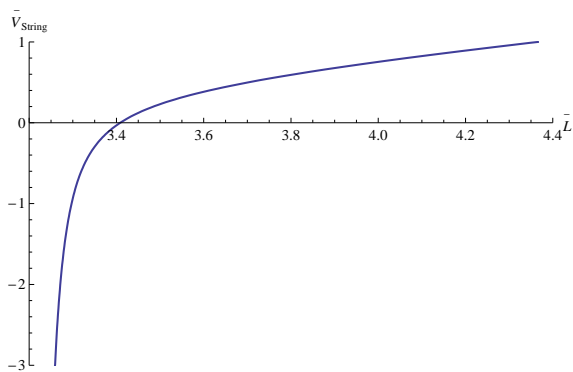}
\caption{Energía potencial $V_{\sf string}(L)$ para el par $q\bar q$ obtenida a
partir de la solución de Maldacena-Nu{\~n}ez eliminando numéricamente
$r_0$ de (\ref{lmn})-(\ref{emn}). Nótese el cambio en el comportamiento de la función
pasando de ser tipo Coulomb a lineal.}
\label{EwilsonMN}
\end{center}
\end{minipage}
\end{center}
\end{figure}\\

El resultado para el potencial $V_{\sf string}(L)$ en la figura \ref{EwilsonMN} es gratificante, pero el
comportamiento lineal ocurre para configuraciones que tienen energías por encima de cero. Una pregunta que surge
es si debemos confiarle a la solución cuando $V_{\sf string}>0$ (ver el último párrafo de la sección \ref{adssch}).
La sustracción en (\ref{emn}) corresponde a un par de cuerdas rectas expandidas a lo largo de la dirección radial
con las demás coordenadas espaciales fijas. Al ser regular el fondo gravitatorio, las cuerdas no pueden terminar
en ningún punto en el interior y la única posibilidad para obtener una solución de referencia suave es
ubicar los extremos de la cuerda antipodalmente en la coordenada $\varphi$ y teniendo ambos la misma
coordenada $x_1$ (ver la linea azul horizontal en la fig.\ref{substraction} y la discusión relacionada
en la sección \ref{adsglobal}). Se concluye que aunque el confinamiento lineal ocurre para configuraciones
con $E_{q\bar q}>0$, la solución es estable y no puede decaer al estado de referencia $E_{q\bar q}=0$ ya que
las dos configuraciones a comparar en (\ref{emn}) satisfacen diferentes condiciones de borde.
En la figura \ref{substraction} se dibujan las hojas de mundo relevantes para el calculo
del valor de expectación del lazo de Wilson rectangular en fondos de gravedad suaves.

\vspace{1cm}

\begin{figure}[h]
\begin{center}
\hspace{-2cm}
\psline[linewidth=.3pt]{->}(1,-.7)(1,0.3)
\psline[linewidth=.3pt]{->}(1,-.7)(2,-.7)
\rput[bI](0.8,.5){$x_1$}
\rput[bI](2.2,-.8){$r$}
\psbezier[linewidth=.3pt]{->}(.7,-.2)(.8,-.4)(1.2,-.4)(1.3,-.2)
\rput[bI](1.6,-.3){$\varphi$}
\psellipse[linewidth=.5pt](1,-1.5)(2,.3)
\psbezier[linewidth=.5pt](-1,-6)(-.90,-6.5)(2.9,-6.5)(3,-6)
\psbezier[linewidth=.5pt,linestyle=dashed](-1,-6)(-.90,-5.5)(2.9,-5.5)(3,-6)
\psline[linewidth=.5pt](-1,-6.)(-1,-1.5)
\psline(-1,-6)(3,-6)
\psline[linecolor=blue,showpoints=true](-1,-6)(3,-6)
\psline[linewidth=.5pt](3,-6.)(3,-1.5)
\psline[linewidth=.3pt]{<->}(3.3,-6.)(3.3,-2)
\rput[bI](4,-4){$L$}
\psline[linewidth=.3pt,linestyle=dashed](1,-1)(1,-6.7)
\pscurve(3,-6)(1.4,-5)(1.3,-4)(1.4,-3)(3,-2)
\psdots[dotsize=0.2](3,-2)(-1,-6)(3,-6)
\vspace{7cm}
\caption{ Geodésicas empleadas para el cálculo de lazos de Wilson rectangulares en fondos
gravitatorios suaves. La curva negra vertical muestra la solución en forma de U, la linea
azul horizontal es el estado de referencia con respecto al cual comparar la energía de la
linea negra. Las configuraciones negra y azul satisfacen diferentes condiciones de borde.}
\label{substraction}
\end{center}
\end{figure}

\subsection{Klebanov-Strassler}
\label{kssol}

\begin{figure}[h]
\begin{minipage}{7cm} 
\begin{center}
\includegraphics[width=7.5cm]{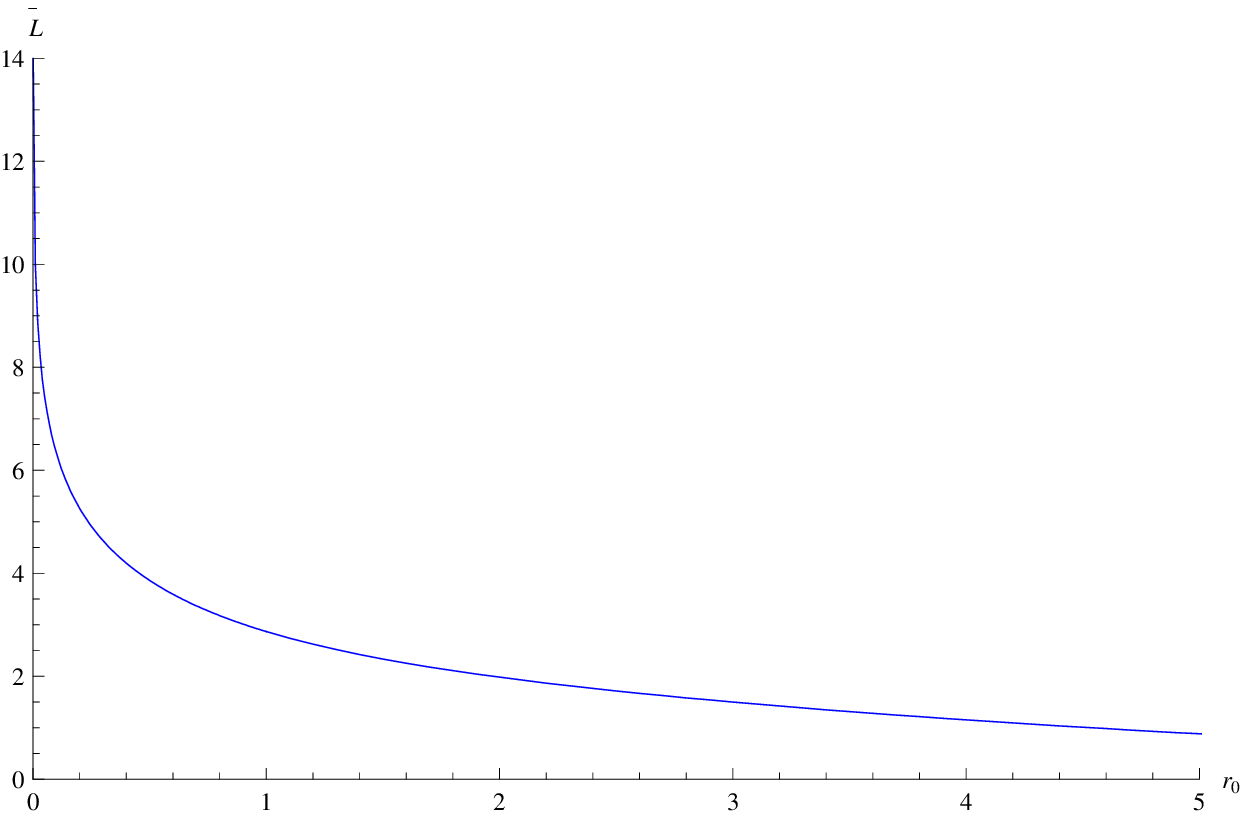}
\caption{$\bar L(r_0)$ (\ref{lks}). Solución de KS.}
\label{LvsrminKS}
\end{center}
\end{minipage}
\   \
\hfill
\begin{minipage}{7cm}
\begin{center}
\includegraphics[width=7.5cm]{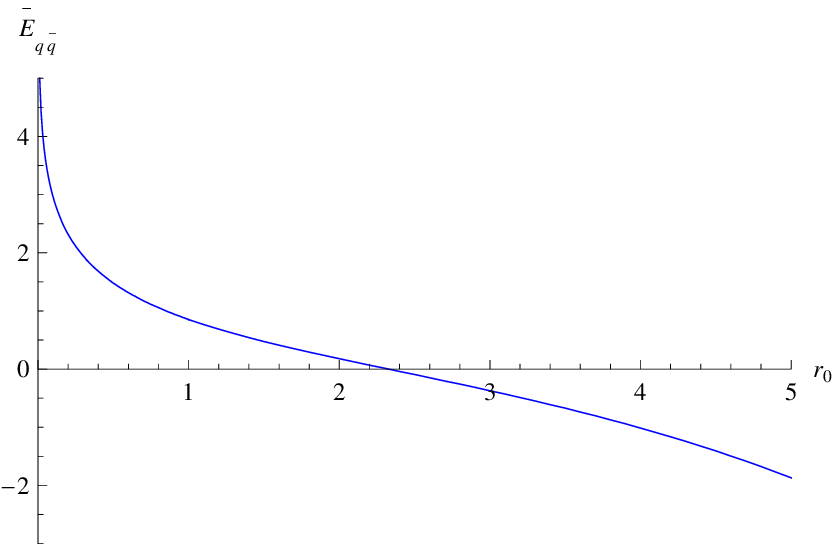}
\caption{$\bar E_{q\bar q}(r_0)$ (\ref{eks}). Solución de KS.}
\label{EvsRminKS}
\end{center}
\end{minipage}
\end{figure}

Este fondo gravitatorio describe una (quiver) teoría de gauge $\cal N$ $=1$ con campos
de materia en la representación fundamental del grupo de gauge $SU(N+M)\times SU(N)$.
La configuración de branas de prueba que conducen a esta geometría es: $N~D3$ branas
en el vértice del conifold singular más $M~D5$-branas enrolladas en la $S^2$
del conifold y compartiendo las restantes 3 dimensiones con la $D3$. La solución esta
soportada por un dilatón constante $\phi=\phi_0$, el cual puede ser fijado para tener
$g_s\ll1$ en todo el espacio (al contrario que en la solución de MN). El flujo $\cal N$ $=1$
$SU(N+M)\times SU(N)\rightarrow SU(M)$ que se realiza a través de una cascada de dualidades de
Seiberg  en la teoría de gauge (ver \cite{Strassler}) se manifiesta en la geometría por medio de la
variación del flujo de la 5-forma.

\begin{figure}[h]
\begin{center}
 \includegraphics[width=7.5cm]{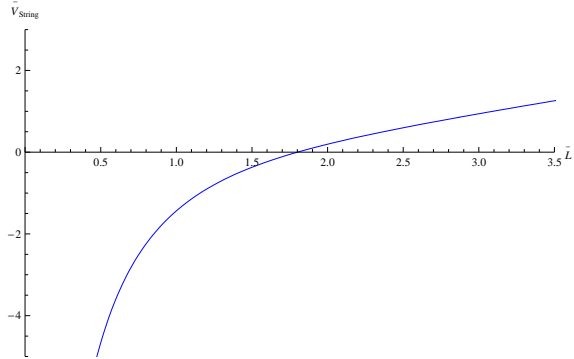}
\caption{$V_{\sf string}(L)$ para un lazo de Wilson rectangular sobre el fondo de gravedad de KS.}
\label{EwilsonKS}
 \end{center}
\end{figure}

La métrica es \cite{ks}
\be
ds^2=g_s \alpha'\! M\, [h^{-\frac12}(r)(-dt^2+dx_{i}dx_{i})+h^{\frac12}(r)ds_{6}^2] \label{ks}
\ee
La geometría del conifold deformado $ds_6$ se puede escribir como
\bea ds_6^2& =& \frac{1}{2} K(r)\left[\frac{(dr^2 +
(g^5)^2) }{3 K^3(r)} +
\cosh^2\frac{r}{2}\, ((g^3)^2 +(g^4)^2) + \sinh^2 \frac{r}{2} \,
((g^1)^2 + (g^2)^2) \right],
\label{defconif}
\eea
en donde
\be
K(r)=\frac{[\sinh (2r) - 2r]^{\frac13}}{2^{\frac13}\sinh r}
\label{kks}
\ee
y las $g_i$ están definidas por
\bea
g^1 = \frac{e^1 - e^3}{\sqrt2}, \quad g^2 = \frac{e^2 -
e^4}{\sqrt2},\quad g^3 = \frac{e^1 + e^3}{\sqrt2}, \quad g^4 =
\frac{e^2 + e^4}{\sqrt2},\quad g^5 = e^5, \label{fbasis}
\eea
con
\begin{eqnarray}
e^1= -\sin\theta_1\,d\phi_1, ~~~e^2= d\theta_1, \qquad
e^3= - \sin\psi\, d\theta_2+\cos\psi\,\sin\theta_2\,d\phi_2 , \nonumber \\
e^4=\cos\psi\, d\theta_2+\sin\psi\,\sin\theta_2\,d\phi_2 , \qquad e^5=
d\psi + \cos\theta_1\,d\phi_1 + \cos\theta_2\,d\phi_2.
\end{eqnarray}
Las coordenadas en (\ref{ks}) son adimensionales, las coordenadas de la teoría de gauge
$t,x_i$ escalean como $\frac{g_s M \alpha'}{\ell_{cf}}$, y la escala de la coordenada
holográfica $r$ esta fijada por $\ell_{cf}$. El factor $h(r)$ toma la forma
\bea
h(r) = 2^{\frac23}
\int_r^\infty d x \frac{x\coth x-1}{
\sinh^2 x}\, (\sinh 2x - 2x)^{\frac13}.
\label{hks}
\eea
El fondo gravitatorio (\ref{ks}) esta soportado en parte por un $B_{\mu\nu}$ no trivial
pero la forma de embeber la cuerda que consideramos hace que este no contribuya en los
cálculos  \cite{loewy}. Las funciones en (\ref{geff}) están dadas por $f^2(r)=\frac{1}{h(r)},~g^2(r)=\frac{1}{6K^2(r)}$.
Las expresiones adimensionales para la longitud (\ref{generalL}) y la energía (\ref{enerren}) son
\be
\bar {L}(r_0)=2\int_{r_0}^\infty\frac{dr}{\sqrt6K(r)}\frac{{h(r)}}{\sqrt{h(r_0)-h(r)}}\,.
\label{lks}
\ee
\be
\bar{E}_{q\bar q}(r_0)=\frac{g_sM}{\pi }\left[
\int_{r_0}^\infty\frac{dr}{\sqrt6K(r)}\frac{\sqrt{h(r_0)}}{\sqrt{h(r_0)-h(r)}}-\int_0^{r_0}\frac{dr}{\sqrt6K(r)}\right]\,.
\label{eks}
\ee
En las figuras \ref{LvsrminKS} y \ref{EvsRminKS} se graficaron estas dos últimas expresiones. Al igual que en el
caso de MN, se espera encontrar una divergencia en $r_0=0$ debido a que $\left.\frac {dh(r)}{dr}\right|_{r=0}=0$.
Eliminando numéricamente $r_0$ de las ecuaciones (\ref{lks})-(\ref{eks}) se grafica en la figura \ref{EwilsonKS}
la función $V_{\sf string}(L)$. Como se puede observar fácilmente, existe una relación lineal para el
potencial de interacción para valores de $L\gg1$. Procediendo como en (\ref{energMN}) encontramos que
la tensión de la cuerda confinante es
\be
T_{\sf string}=\frac{1}{2\pi\alpha'}\frac{\ell_{cf}^2}{g_s \alpha'\! M\sqrt{h_{0}}}\,,
\label{ts}
\ee
en donde $h_0=h(0)\simeq1.1398$. Como en el caso de MN, la contribución dominante al area mínima (\ref{NG}),
en el límite de $L$ grande, proviene de la región $r\approx0$. De nuevo, hay un problema relacionado
a cuándo se le debe creer a configuraciones con $E_{q\bar q}>0$, como se discutió al final de la sección
anterior y se dibujó en la fig. \ref{substraction} las soluciones con $E_{q\bar q}>0$ son clásicamente
estables.

\subsection{ Maldacena-N\'u\~nez generalizado}
\label{gmnsol}

Este tipo de fondos gravitatorios fueron obtenidos en \cite{cnp} generalizando la solución
descripta en la sección \ref{mnsol}. Estas soluciones fueron ampliamente discutidas en
\cite{hn} y se interpretan como duales a teorías de gauge mínimamente supersimétricas
que contienen operadores irrelevantes de dimensión 6. El operador cambia drásticamente
el desarrollo ultravioleta (UV) de las teorías tomando la solución "\,afuera " del límite
de horizonte cercano de las $D5$-branas que generan la geometría. El análisis en \cite{hn}
muestra que asintóticamente, $r$ grande, la solución es el producto de un espacio de Minkowski en
4 dimensiones con el conifold deformado.

\begin{figure}[h]
\begin{minipage}{7cm}
\begin{center}
\includegraphics[width=7.5cm]{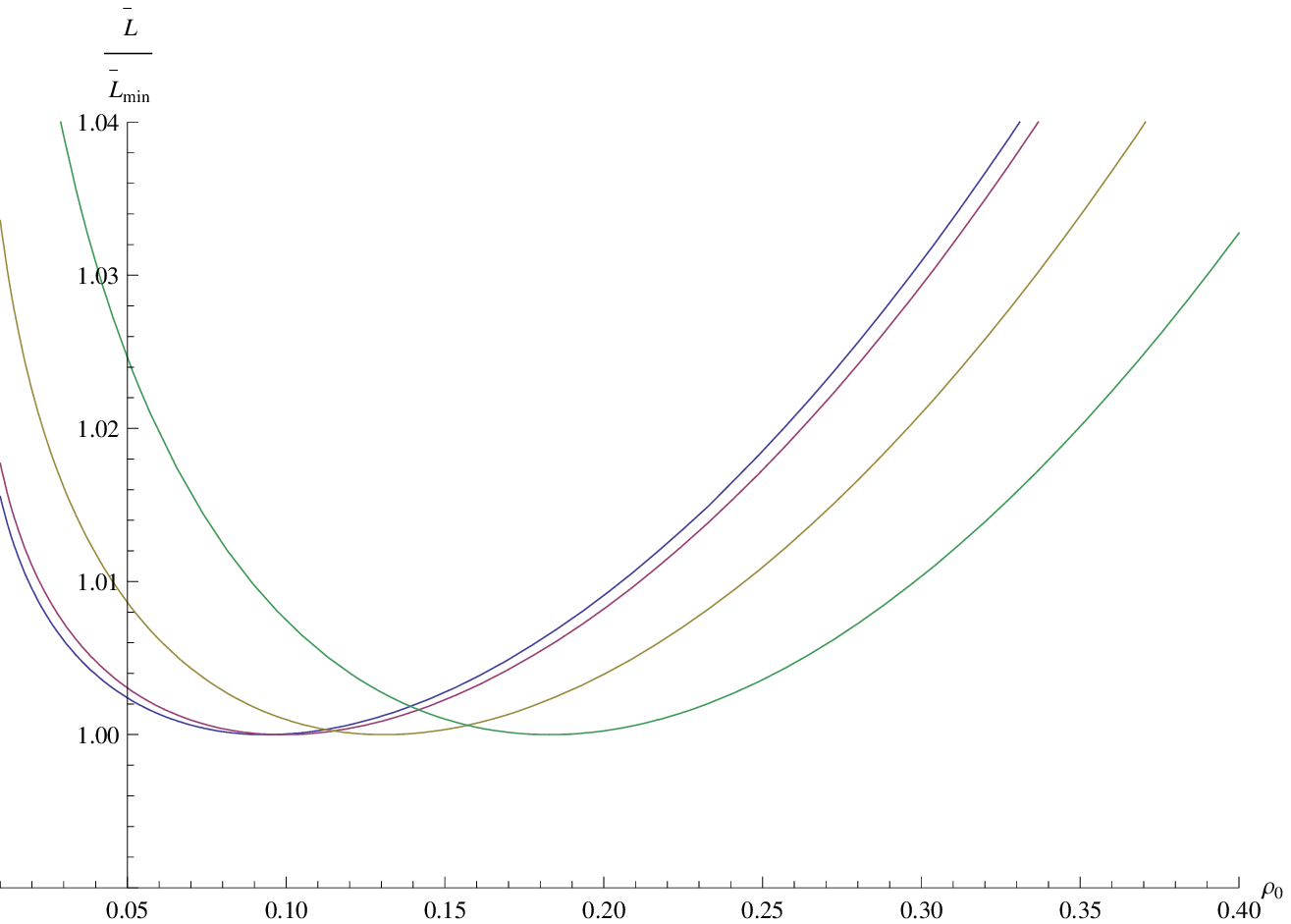}
\caption{$L(\rho_0)$ (\ref{LMNgen}). Las curvas azul, violeta, amarilla y verde se
corresponden con los valores
$\mu=-1.8,\,-1.5,\, -1,\,-.8$
($\rho_\infty=7$). Se observa una distancia de separación mínima para los quarks.}
\label{LMNgmumenosuno}
\end{center}
\end{minipage}
\   \
\hfill \begin{minipage}{7cm}
\begin{center}
\includegraphics[width=7.5cm]{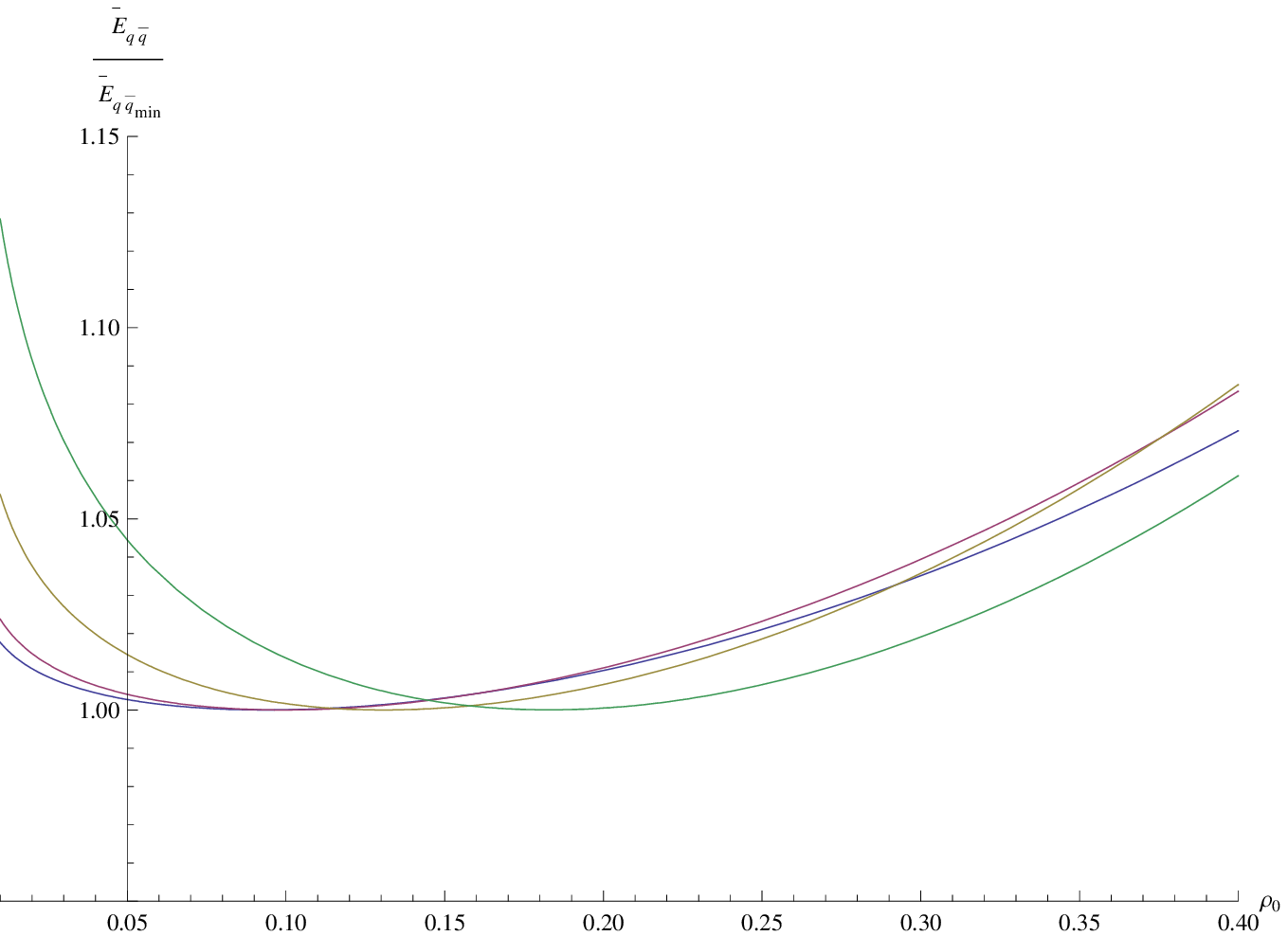}
\caption{$E(r_0)$ (\ref{EMNgen}). Los colores se corresponden con los valores de $\mu$ de la fig. \ref{LMNgmumenosuno}.}
\label{EMNgmumenosuno}
\end{center}
\end{minipage}
\end{figure}

~

La métrica es \cite{cnp}
\bea
ds^2&=&g_s \alpha'\! N\, e^{{4f(r)}}  \Big[-dt^2+dx_{i}dx_{i}+ dr^2+e^{2h(r)}\,(d\theta^2+\sin^2\theta d\varphi^2)\nn\\
&&+ \frac{e^{2g(r)}}{4}\left((w_1+a(r)d\theta)^2+(w_2-a(r)\sin\theta d\varphi)^2\right)
+\frac{e^{2k(r)}}{4}(w_3+\cos\theta d\varphi)^2\Big]\,.
\label{gMN}
\eea
Realizando el cambio de coordenadas $d\rho\equiv e^{-k(r)}dr$, las funciones $a$, $k$ y $f$ satisfacen
\bea
\partial_\rho a &=&\frac{-2}{-1+2\rho\coth2\rho}\left[{e^{2k}} \frac{(a\cosh2\rho-1)^2}{\sinh2\rho}+a\,(2\rho
-a\sinh2\rho)\right]\nn\\
\partial_\rho k &=&\frac{2(1+a^2-2a\cosh2\rho)^{-1}}{-1+2\rho\coth2\rho}\left[ {e^{2k}} a\sinh2\rho\,(
a\cosh2\rho-1)+(2\rho-4a\,\rho\cosh2\rho+\frac{a^2}{2}\sinh4\rho)\right]\nn\\
\partial_\rho f &=&-\frac1{4\sinh^{2}2\rho}\left[\frac{(1-a \cosh 2\rho )^2(-4\rho+\sinh4\rho)}{(1+a^2-2a\cosh2\rho)(-1+2\rho\coth2\rho)}\right],
\label{eqdf}
\eea
y las funciones  $g(\rho),h(\rho)$ en (\ref{gMN}) están dadas por
\be
e^{2g }=\frac{b \cosh 2\rho -1}{a \cosh 2\rho -1},~~~~
e^{2h }=\frac{e^{2g }}{4}(2a \cosh 2\rho -1-a^2),~~~\mathrm{con}~~b(\rho)=\frac{2\rho}{\sinh2\rho}\,.\label{bghMNg}
\ee
Las primeras dos ecuaciones diferenciales en (\ref{eqdf}) tienen una familia de soluciones regulares dependientes de un solo parámetro.
Para pequeños valores de $r$ se encuentra \cite{cnp}
\be
a(\rho)=1+\mu\rho^2+...,~~~~e^{2k(\rho)}= \frac{4}{6+3\mu}-\frac{20+36\mu+9\mu^2}{15(2+\mu)}\rho^2+....
\label{akMN}
\ee
con $\mu$ tomando valores en el intervalo $(-2,-\frac23)$. Insertando (\ref{akMN}) en la tercera ecuación
de (\ref{eqdf}) y en (\ref{bghMNg}) se obtiene
\be
e^{2g(\rho)}= \frac{4}{6+3\mu}+...,~~~~e^{2h(\rho)}= \frac{4\rho^2}{6+3\mu}+....,~~~~e^{2f(\rho)}=1+\frac{(2+\mu)^2}{8}\rho^2+...
\ee
La constante arbitraria para $f$ proveniente de (\ref{eqdf}) fue factorizada como $g_s$ in (\ref{gMN}).
Los valores límites para $\mu$ dan resultados conocidos: el caso con $\mu=-\frac23$ reproduce la solución
de MN estudiada en la sección \ref{mnsol} con $\phi=4f$  ($k=const.$), y el caso con $\mu=-2$
nos conduce al producto del espacio de Minkowski en 4 dimensiones con el conifold deformado ($\phi$ en este caso
es constante). Finalmente, el límite $\rho\rightarrow\infty$ de todas las soluciones (excepto $\mu=-\frac23$ )
tiene al conifold deformado como límite asintótico (ver \cite{cnp} para más detalles).

La expresiones para la longitud (\ref{generalL}) y la energía (\ref{enerren}) quedan
\be
\bar
L(\rho_0)=2\int_{\rho_0}^{\rho_\infty}\frac{e^{4f(\rho_0)}}{\sqrt{e^{8f(\rho)}-e^{8f(\rho_0)}}}\,e^{k(\rho)}d\rho
\label{LMNgen}
\ee
\be
\bar
E_{q\bar{q}}(\rho_0)=\frac{Ng_s}\pi\left[\int_{\rho_0}^{\rho_\infty}
\frac{e^{8f(\rho)}}{\sqrt{e^{8f(\rho)}-e^{8f(\rho_0)}} }\,
e^{k(\rho)}d\rho -\int_{0}^{\rho_\infty}
e^{4f(\rho)}e^{k(\rho)}d\rho\right]\,.
\label{EMNgen}
\ee
Nótese que la integración a lo largo de la coordenada radial en ambas expresiones se extiende
hasta una distancia finita $\rho_\infty$. La razón, estudiada en \cite{cnp}, proviene de notar
que (\ref{LMNgen}) es divergente \footnote{ Ver \cite{piai} para encontrar una discusión sobre las
divergencias cuando se calcula la función longitud $L(r_0)$.}. Se debe tener en cuenta que el cálculo
del lado de cuerdas se corresponde con estudiar la teoría de gauge dual con quarks muy masivos
para $\rho_\infty\gg1$ (pero no infinitamente masivos). Además, como los extremos de la cuerda están fijos
a una distancia radial finita, se puede chequear que la cuerda de prueba no alcanza a
lo largo de la dirección normal a la teoría de gauge definida sobre la brana.
\begin{figure}
\begin{minipage}{7cm} 
\begin{center}
\includegraphics[width=7.5cm]{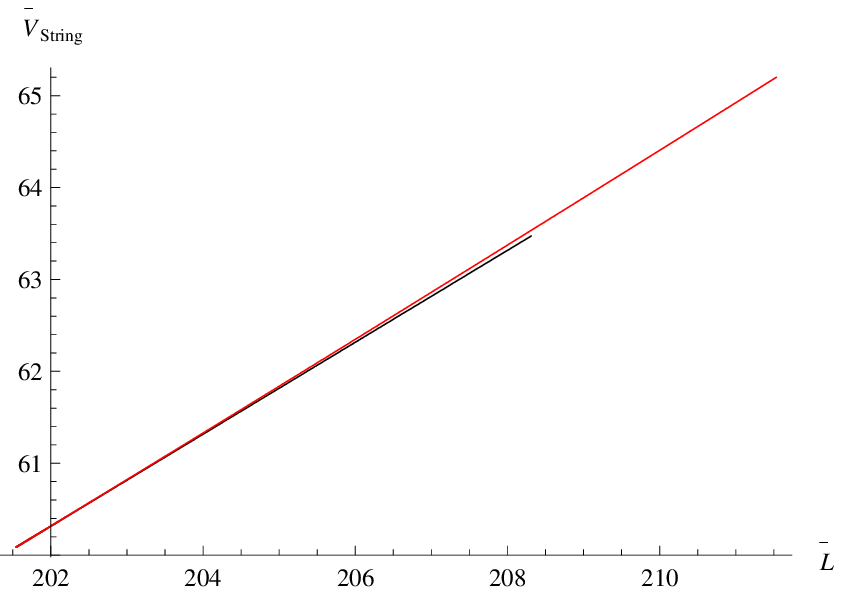}
\caption{$V_{\sf string}(L)$ bivaluada para el caso de gMN con $\mu=-1$. La curva superior
es la rama no física correspondiente a soluciones de cuerda a la derecha del mínimo en la figura \ref{LMNgmumenosuno}.
La curva no alcanza el origen, manifestando así la existencia de una longitud separación mínima entre quarks.}
\label{gancho}
\end{center}
\end{minipage}\  \
\hfill
\begin{minipage} {7cm} 
\begin{center}
\includegraphics[width=7.5cm]{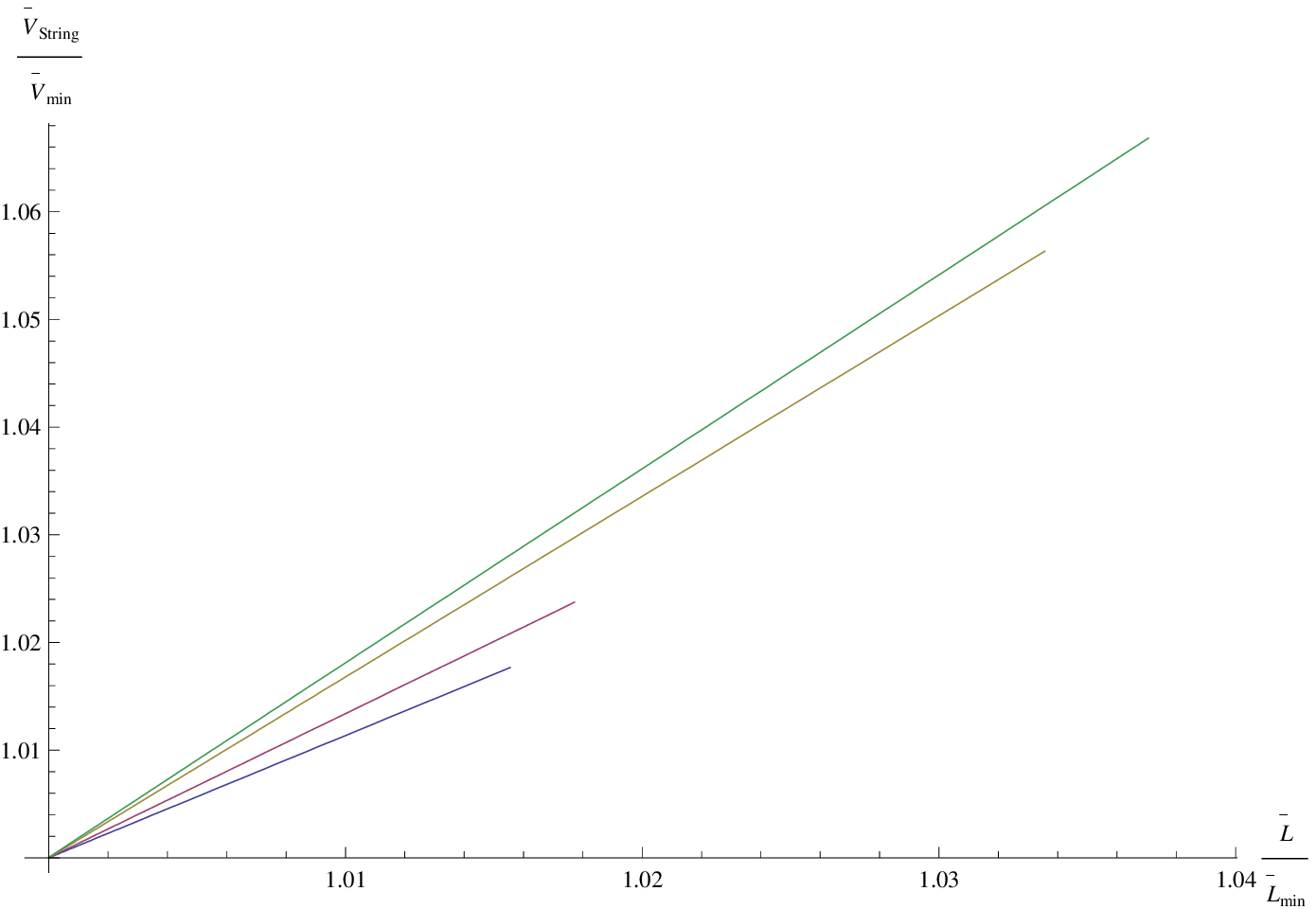}
\caption{ $\bar V_{\sf string}(L)$ normalizado y lineal para la rama física (izquierda) de la figura
\ref{LMNgmumenosuno}. Los colores se corresponden con los de la fig. \ref{LMNgmumenosuno}. }
\label{elvarios}
\end{center}
\end{minipage}
\end{figure}

En las figuras \ref{LMNgmumenosuno} y \ref{EMNgmumenosuno} se grafican las soluciones numéricas de (\ref{LMNgen}) y (\ref{EMNgen})
para varios valores de $\mu$. La figura (\ref{LMNgmumenosuno}) muestra que $L(\rho_0)$ presenta un mínimo global
para todos los valores de $\mu$ (excepto para $\mu\ne-\frac23$). En otras palabras, no existen soluciones
para distancias de separación $L<L_{min}$. Es interesante notar que el valor mínimo de $L$ se encuentra para todos los valores
de $\mu$ en una región pequeña en la coordenada $\rho$ cerca del origen. Basándonos en las consideraciones de concavidad
discutidas al final de la sección \ref{wil} se espera que la hoja de mundo de la cuerda
a la izquierda del mínimo $\rho_c$ tenga un sentido físico bien definido (que sea estable) y que la rama
de la derecha de $\rho_c$ sea no física (inestable). En la sección siguiente voy a mostrar que esto es lo que
sucede cuando se analizan fluctuaciones cuadráticas alrededor de las soluciones. Se encuentran autovalores
negativos para la rama de soluciones con $L'(\rho_0)>0$.

En las figuras (\ref{gancho}) y (\ref{elvarios}) se grafica la relación $V_{\sf string}(L)$ para este fondo gravitatorio en
donde se puede observar un confinamiento lineal. La figura (\ref{gancho}) muestra la función bivaluada
$V_{\sf string}(L)$ para $\mu=-1$, la rama superior (no física, en rojo) corresponde a configuraciones de cuerda
a la derecha del mínimo de la fig. (\ref{LMNgmumenosuno}). En la figura \ref{elvarios} se muestra la función
$V_{\sf string}(L)$ para las ramas físicas de la fig.\ref{LMNgmumenosuno} y para diferentes valores de $\mu$.
Procediendo como en (\ref{energMN}) se encuentra que todas las soluciones conducen a una tensión de la cuerda
independiente de $\mu$
\be
T_{\sf string}
=\frac {g_s}{2\pi\alpha'}\,.
\ee

\section{Análisis de Estabilidad}
\label{stab}

En esta sección se va a estudiar, para los fondos gravitatorios presentados en la sección previa,
el problema de autovalores definido por la ecuación de movimiento (\ref{SL}) para fluctuaciones en el plano
en el gauge $r$. Estoy interesado en encontrar los modos inestables. Por las razones discutidas en el
final de la sección \ref{wil}, el objetivo es mostrar que dichos autovalores negativos
($\omega^2<0$) se producen para soluciones de cuerda en regiones en donde $L'(r_0)>0$. Se estudiarán
soluciones pares, esto significa que en (\ref{asymr0}) las condiciones iniciales en el tip serán $C'_1=1$ y $C'_0=0$
\footnote{ La soluciones pares corresponden a elegir arbitrariamente $C'_1$ en el tip, su valor
fija la normalización de la solución. }. Estas condiciones van a ser implementadas numéricamente
como (ver (\ref{bc}))
\be
     \delta x_1(r)+2(r-r_0)\frac{d\delta x_1(r)}{dr}=0, ~~~~r\rightarrow
     r_0\nn
\ee
\be
~~~~~~~~~~~~~~~~~\sqrt{r-r_0}\,\delta x_1(r)=1\,
,~~~~r\rightarrow r_0\,.
\label{bcs}
\ee
Resolviendo numéricamente, los autovalores $\omega^2$ permitidos para (\ref{SL}) se obtienen pidiendo que
$\delta x_1(r)$ sea una solución normalizable
\be
\delta x_1(r)= 0\,, ~~~~~~r\to\infty.
\ee
Por completitud se van a escribir las relaciones entre los modos cero y los puntos
críticos de la función $L(r_0)$ \cite{avramis}. La solución de modo cero de (\ref{SL}) puede inmediatamente
escribirse como
\be
\delta x_1^{(0)}(r)=C\int_r^\infty d\bar r\frac{g(\bar r)f(\bar r)}{(f^2(\bar r)-f^2(r_0))^{\frac32}}+C'
\label{zero}
\ee
en donde $C'$ y $C$ son las constantes  de integración, se fija $C'=0$ para obtener una solución normalizable y se escoge $C'=1$.
Integrando por partes  en (\ref{zero}) y usando (\ref{Lprima}) se obtiene
\bea
\delta x_1^{(0)}(r)&=&-\int_r^\infty d\bar r\frac{g(\bar r)}{f'(\bar r)}\frac{d}{d\bar r}\left(\frac1{\sqrt{f^2(\bar r)-f^2(r_0)}}\right)\nn\\
&=&\frac{g(r)}{f'(r)\sqrt{f^2(r)-f^2(r_0)}}+\frac{L'(r)}{2f'(r)} \,.
\eea
Expandiendo esta última expresión alrededor del tip $r=r_0$ se tiene
\be
\delta x_1^{(0)}(r)=\frac{g(r_0)}{\sqrt2(f'(r_0))^{\frac32}}\frac1{\sqrt{r-r_0}}+\frac{L'(r_0)}{2f'(r_0)}+{\cal O}\left(\sqrt{r-r_0}\right)\,.
\label{expzero}
\ee
Por lo general, el primer factor en el lado derecho de (\ref{expzero}) es no nulo, luego, una condición
necesaria y suficiente para obtener una solución par con modo cero (ver (\ref{bcs})) requiere que el segundo
termino en (\ref{expzero}) se anule, ó, equivalentemente, $r_0$ debe ser un punto crítico de la función
longitud $L(r_0)$ \cite{avramis}.

\subsection{$AdS_5$}

\begin{figure}
\begin{minipage}{7cm} 
\begin{center}
\includegraphics[width=7.5cm]{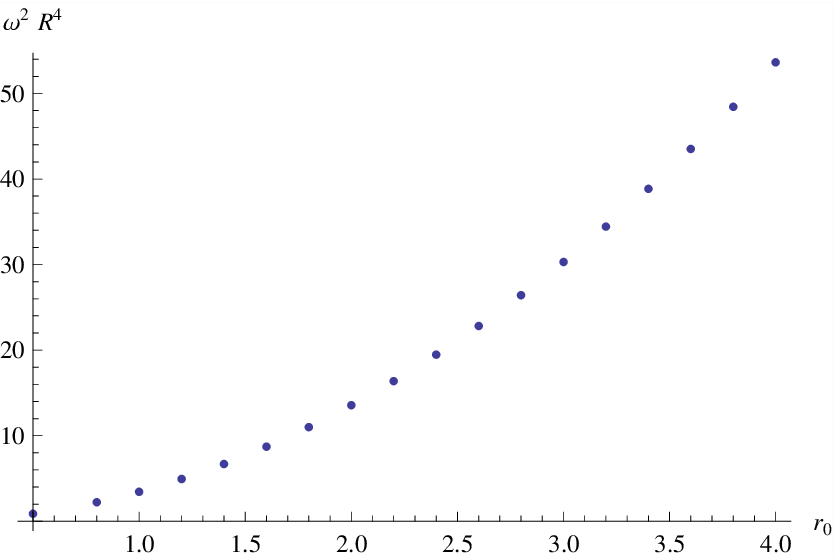}
\caption{Autovalor mínimo $\omega^2$ de (\ref{fluctads5})
que conduce a una solución normalizable como función de $r_0$.}
\label{wvsrminAdS}
\end{center}
\end{minipage}\  \
\hfill
\begin{minipage}{7cm} 
\begin{center}
\includegraphics[width=7.5cm]{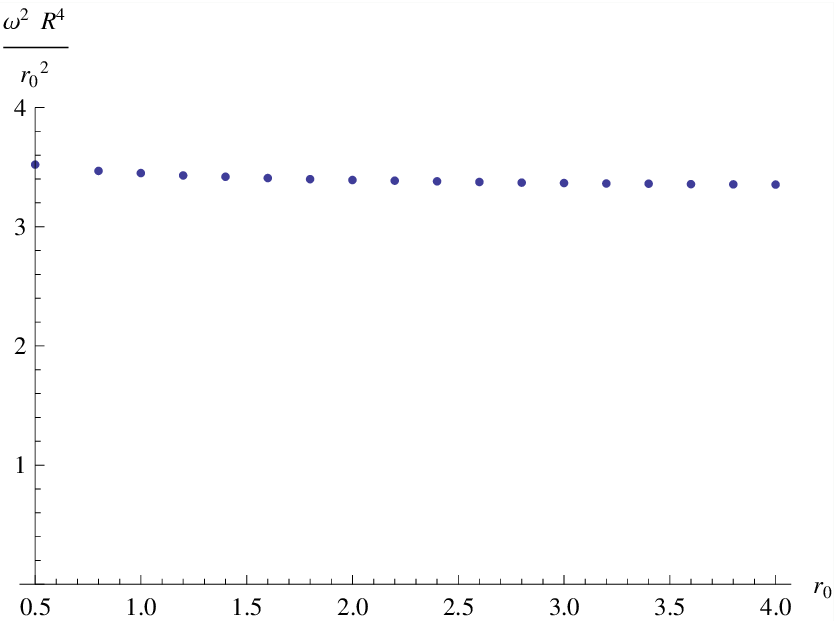}
\caption{Autovalor mínimo propiamente normalizado de la fig. \ref{wvsrminAdS}  como
función de $r_0$ (ver (\ref{fluctnormalized})).}
\label{wvsrminAdS2}
\end{center}
\end{minipage}
\end{figure}

Las ecuaciones de movimiento (\ref{SL}) para la fluctuaciones en el caso de AdS
en coordenadas de Poincaré (\ref{ads5}) toma la forma
\be
\left[\frac{d}{dr}\left(\frac{(r^4-r_0^4)^{\frac32}}{r^2}\frac{d}{dr}\right)+
\omega^2R^4\frac{\sqrt{r^4-r_0^4}}{r^2}\right]\delta x_1(r)=0\,~~~~0<r_0\le r<\infty\,.
\label{fluctads5}
\ee
La invarianza ante dilataciones nos permite factorizar la dependencia en $r_0$.
Haciendo $r=r_0\, \rho$ se obtiene
\be
\left[\frac{d}{d\rho}\left(\frac{(\rho^4-1)^{\frac32}}{\rho^2}\frac{d}{d\rho}\right)+
\frac{\omega^2R^4}{r_0^2}\frac{\sqrt{\rho^4-1}}{\rho^2}\right]\delta x_1(\rho)=0\,.
\label{fluctnormalized}
\ee
El desarrollo asintótico ($\rho\rightarrow\infty$) de (\ref{fluctnormalized}) es
\be
\left[\frac{d}{d\rho}\left(\rho^4\frac{d}{d\rho}\right)+\frac{\omega^2R^4}{r_0^2}\right]\delta x_1(\rho)\approx0\,,
~~~~~~\rho\gg 1\,,
\label{asymAds}
\ee
cuya solución se escribe
\be
\delta x_1(\rho)\approx\alpha_0+\frac{\alpha_1}{\rho^3}\,,~~~~~~\rho\gg 1\,,
\label{asymads2}
\ee
en donde $\alpha_0,\,\alpha_1$ son constantes de integración. El comportamiento
(\ref{asymads2}) implica que existirán soluciones normalizables ($\alpha_0=0$)
para valores particulares (discretos) de $\omega_n^2$.

Con el objeto de testear el método de shooting empleado, se integrará numéricamente
(\ref{fluctads5}) para diferentes valores de $r_0$ y se determinará el autovalor mínimo
$\omega^2$ compatible con un solución normalizable. En la figura \ref{wvsrminAdS} se grafican
estos $\omega^2$ como función de $r_0$. Los autovalores son positivos para todo valor de $r_0$, lo cual
señala la estabilidad de la configuración de cuerda. En la figura \ref{wvsrminAdS2} se muestra la
esperada independencia en $r_0$ de los modos apropiadamente normalizados (ver (\ref{fluctnormalized})).
En la siguiente tabla se muestran los primeros autovalores correspondientes a condiciones de borde pares en
el tip.
\begin{center}
\begin{tabular}{|r||c|l|}
 \hline
    &   ${\omega_n^2R^4}/{r_0^2}$\\
     \hline
  $n=1$  & ~3.450\\
  $n=3$  & ~22.113\\
  $n=5$  & ~52.325\\
  $n=7$  & ~94.558\\
  $n=9$  & ~148.845\\  \hline
\end{tabular}
\end{center}
En la sección \ref{scho} se probará la estabilidad de la configuración, transformando la
ecuación diferencial (\ref{fluctads5}) en una ecuación tipo Schrodinger (ver apéndice \ref{sl2sc}).

\subsection{$D3$-branas no extremales}
\label{nond3}

\begin{figure}[h]
\begin{center}
\includegraphics[width=7.5cm]{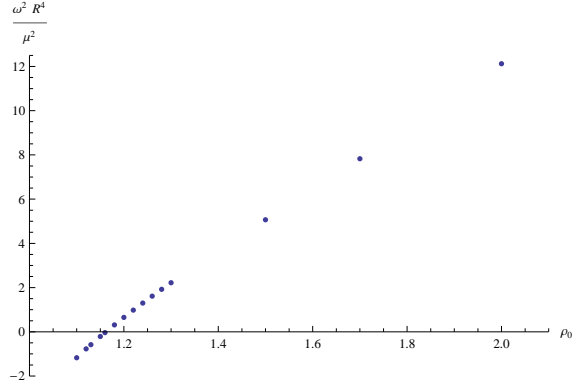}
\caption{Minimo autovalor $\omega^2$ de (\ref{eqdiff}) que conduce a una solución normalizable como función de $\rho_0$.
Existe un modo cero para $\rho_0\simeq1.177$. La solución clásica con $\rho_0<1.177$ es inestable ante perturbaciones lineales.}
\label{wvsrminnonextremal}
\end{center}
\end{figure}

La ecuación de movimiento para la fluctuación (\ref{SL}) en el fondo gravitatorio (\ref{thermalads}) toma la forma
\be
\left[\frac{d}{d\rho}\left(\frac{(\rho^4-\rho_0^4)^{\frac32}}{\sqrt{\rho^4-1}}\frac{d}{d\rho}\right)+
\frac{\omega^2R^4}{\mu^2}\frac{\rho^4\sqrt{\rho^4-\rho_0^4}}{(\rho^4-1)^{\frac32}}\right]\delta x_1(\rho)=0\,,~~~~1<\rho_0\le \rho<\infty\,.
\label{eqdiff}
\ee
El fondo de gravedad (\ref{thermalads}) tiende asintóticamente a AdS y por lo tanto el
comportamiento asintótico ($\rho\rightarrow\infty$) de la solución de (\ref{eqdiff}) esta dado por
(\ref{asymads2}). Como en el caso anterior, se espera encontrar un conjunto de autovalores que conducen
a soluciones normalizables.

Se grafica en la figura \ref{wvsrminnonextremal} el mínimo autovalor encontrado numéricamente resolviendo
(\ref{eqdiff}) y que permite soluciones normalizables. Aparece un modo cero precisamente en el punto crítico de
la función longitud $L(\rho_0)$, esto es para $\rho_0\simeq1.177$ (ver fig.\ref{L(r_0)}), en acuerdo con (\ref{expzero}).

Se concluye que la rama (izquierda) en la figura \ref{L(r_0)} que posee $L'(\rho_0)>0$ es inestable ante perturbaciones
lineales. Finalmente, nótese que el análisis numérico indica que las soluciones de la rama derecha de la figura \ref{L(r_0)}
, para las cuales $L'(\rho_0)<0$, son estables ante perturbaciones lineales. Sin embargo, como se discutió
al final de la sección \ref{adssch} se espera que la solución con $1.177<\rho_0<1.524$ sea metaestable, decayendo
a un par de quarks libres.

\subsection{Maldacena-N\'u{\~n}ez}
\label{mnbkg}

La ecuación de movimiento para la fluctuación en el plano en la geometría de Maldacena-Nuñez
(\ref{metric}) toma la forma (\ref{SL})
\be
\left[\frac{d}{dr}\left(\frac{(e^{2\phi(r)}-e^{2\phi(r_0)})^{\frac32}}{e^{2\phi(r)}}
\frac{d}{dr}\right)+\bar \omega^2\sqrt{e^{2\phi(r)}-e^{2\phi(r_0)}}\right]\delta
x_1(r)=0\,,~~~~0<r_0\le r<\infty\,.
\label{SLMN}
\ee
en donde $\bar \omega^2=\omega^2\alpha' N$. Se analizará ahora el desarrollo asintótico de (\ref{SLMN}) para
ver cuando debiese uno esperar un espectro discreto o no. En el límite $r\rightarrow \infty$ la ecuación
de movimiento para $\delta x_1(r)$ queda
\be
\left[\frac{d}{dr}\left(e^rr^{-\frac14}\frac{d}{dr}\right)+\bar\omega^2e^rr^{-\frac14}\right]\delta
x_1(r)=0\,,~~~~~~r\gg 1\,,
\label{asymMNinfinity}
\ee
en donde se usó que $e^{2\phi(r)}\rightarrow e^{2r}r^{-\frac12}$ para $r\gg1$. Esta última ecuación
puede escribirse como
$$
\left[\frac{d^2}{dr^2}+\left(1-\frac{1}{4r}\right)\frac{d}{dr}+\bar \omega^2\right]\delta x_1(r)=0\,,~~~~~~r\gg 1\,.
$$
El término $r^{-1}$ puede omitirse en el límite de $r$ grande y la solución asintótica para (\ref{SLMN}) es
\be
\delta x_1(r)\simeq e^{-\frac12r}(\beta_0\, e^{r\alpha}+\beta_1\,e^{-r\alpha})\,,~~~~~~r\gg 1\,,
\label{asymMN}
\ee
en donde $\alpha=\frac{\sqrt{1-4\bar\omega^2}}{2}$. A partir de (\ref{asymMN}) se sigue que cualquier
$\bar \omega^2>0$ conduce a soluciones normalizables, el espectro de fluctuaciones es por lo tanto continuo.
En el caso $\bar\omega^2\leq0$ ($\alpha\ge\frac12$), $\beta_0$ debe ser fijada a cero y se tiene la posibilidad
encontrar un espectro discreto de autovalores negativos. El análisis numérico no encontró ningún autovalor
normalizable negativo, sugiriendo la estabilidad de la configuración clásica y en acuerdo con la
condición de concavidad (\ref{convexity}).

En la sección \ref{scho} se mostrará que no existen autovalores negativos a partir del estudio de
la ecuación de Schr\"{o}dinger de (\ref{SLMN}) (ver apéndice \ref{sl2sc}).

\subsection{Klebanov-Strassler}
\label{ksbkgd}

En este caso, la ecuación de movimiento para la fluctuación toma la forma
\be
\left[\frac{d}{dr}\left(\frac{K(r)}{h(r)}\left(1-\frac{h(r)}{h(r_0)}\right)^{\frac32}\frac{d}{dr}\right)+
\bar \omega^2\frac1{6K(r)}\sqrt{1-\frac{h(r)}{h(r_0)}}\right]\delta x_1(r)=0\,,~~~~0<r_0\le r<\infty\,,
\label{slks}
\ee
con $\bar \omega^2$ adimensional y $K(r)$ y $h(r)$ dadas por (\ref{kks}) y (\ref{hks}) respectivamente. Del límite
$r\rightarrow\infty$ de $K(r)$ y $h(r)$ se obtiene
\be
\left[\frac{d}{dr}\left(\frac{e^{r}}{r}\frac{d}{dr}\right)+\bar\omega^2\frac{e^{\frac{r}{3}}}{2^{\frac43}}\right]\delta x_1(r)=0\,,~~~~r\gg 1\,,
\ee
con lo cual da
\be
\left[\frac{d^2}{dr^2}+\left(1-\frac1r\right) \frac{d}{dr}+\bar \omega^2 \frac{re^{-\frac23r}}{2^{\frac43}}\right]\delta x_1(r)=0\,,~~~~r\gg 1\,.
\label{asymKS}
\ee
En el límite de valores grandes para $r$ el término con $r^{-1}$ y el último término en (\ref{asymKS}) pueden ser omitidos
y la fluctuación asintóticamente tiende a
\be
 \delta x_1(r)\simeq \alpha_0+\alpha_1e^{-r}\,,~~~~~~r\gg 1\,.
\ee
La constante de integración $\alpha_0$ debe ser fijada a cero para obtener soluciones normalizables y por lo tanto
se espera obtener un conjunto discreto de autovalores.
\begin{figure}[h]
\begin{center}
\includegraphics[width=7.5cm]{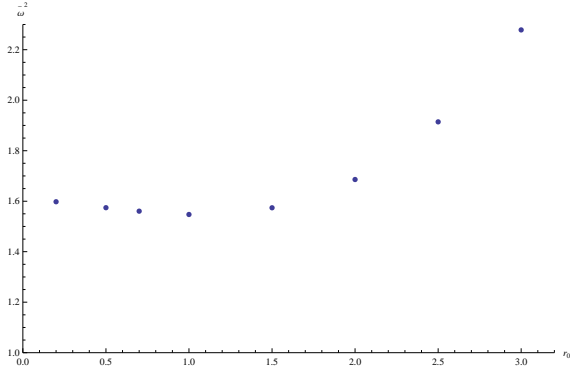}
\caption{Soluciones numéricas para los menores $\omega^2$ que permiten encontrar soluciones normalizables
como función de $r_0$ para el fondo gravitatorio de Klebanov-Strassler. No se han encontrado modos negativos.}
\label{wvsrminKS}
\end{center}
\end{figure}
En la figura (\ref{wvsrminKS}) se grafican (como función de $r_0$) los autovalores más bajos de (\ref{slks}) tales que
que conducen a soluciones normalizables . No se han encontrado numéricamente autovalores negativos.
En la sección  \ref{scho} se probará la estabilidad de la solución clásica transformando (\ref{slks}) en una ecuación del tipo
Schr\"{o}dinger mostrando que no pueden existir modos negativos (ver apéndice \ref{sl2sc}).

\subsection{Maldacena-N\'u\~nez generalizada}
\label{gMNbkgd}

La ecuación de movimiento para la fluctuación $\delta x_1(\rho)$ en la geometría (\ref{gMN}) es
\be
\left[\frac{d}{d\rho}\left(\frac{(e^{8f(\rho)}-e^{8f(\rho_0)})^{\frac32}}{e^{8f(\rho)+k(\rho)}}\frac{d}{d\rho}\right)+
\bar\omega^2e^{k(\rho)}\sqrt{e^{8f(\rho)}-e^{8f(\rho_0)}}\right]\delta x_1(\rho)=0\,,~~~~0<\rho_0\le \rho<\infty\,.
\label{SLMNgen}
\ee
En el caso $\mu=-\frac23$ ($k(\rho)=const.$) se recupera la ecuación (\ref{SLMN}) para el fondo de Maldacena-Nuñez
(a partir de ahora se considerará $\mu\ne-\frac{2}{3}$). En el límite de $\rho$ grande, la solución de gMN
tiende al conifold deformado y la función $f$ toma un valor constante $f_\infty$, el desarrollo asintótico
queda dado por
\be
\left[e^{-k(\rho)}\frac{d}{d\rho}\left(e^{-k(\rho)}\frac{d}{d\rho}\right)+\bar\omega^2\frac{e^{8f_\infty}}
{e^{8f_\infty}-e^{8f(\rho_0)}}\right]\delta x_1(\rho)=0\,,~~~~\rho\gg 1\,.
\label{asymgMN}
\ee
Volviendo a la variable original $r$ en (\ref{gMN}) ($dr=e^{k(\rho)}d\rho$) se obtiene
\be
\left[\frac{d^2}{dr^2}+\tilde\omega^2\right]\delta x_1(r)=0\,,~~~~r\gg 1\,.
\ee
cuya solución son ondas planas $e^{\pm i\tilde\omega r}$ para $\bar\omega^2>0$ y exponenciales reales
$e^{\pm\tilde\omega r}$ para $\bar\omega^2<0$. Una palabra de precaución, como se discutió en la sección \ref{gmnsol}
la teoría de gauge que vive sobre la brana debe localizarse a una distancia finita $\rho_\infty$,
luego, para (\ref{SLMNgen}) definida sobre $\rho_0\le\rho\le \rho_\infty$ existirán autovalores positivos.
En el caso $\bar\omega^2<0$ existe la posibilidad de obtener autovalores negativos, y de echo se encuentran soluciones normalizables
con modos negativos precisamente para las región en la que la solución clásica no cumple con las condiciones de
convexidad (\ref{convexity}).
\begin{figure}
\begin{minipage}{7cm} 
\begin{center}
\includegraphics[width=7.5cm]{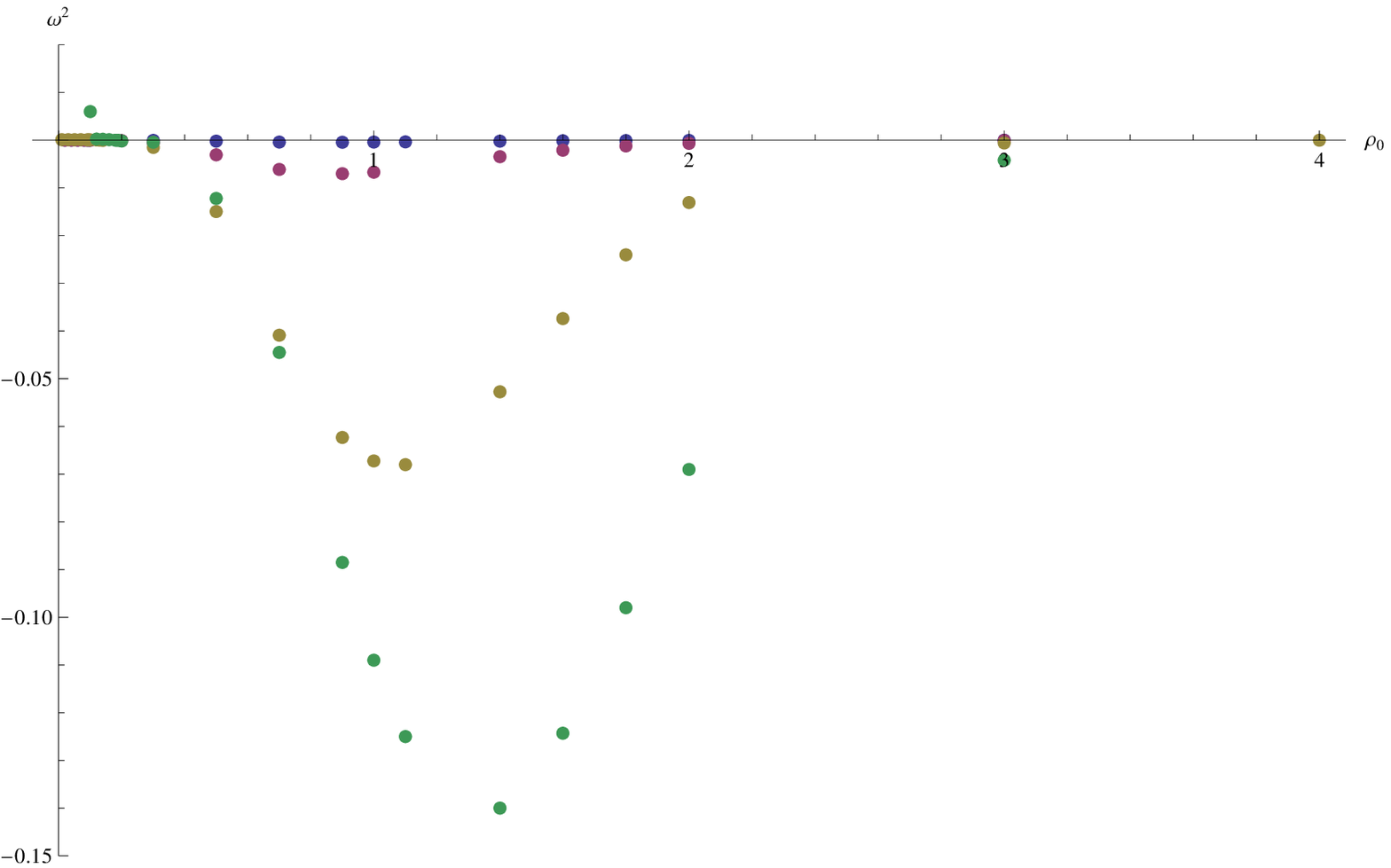}
\caption{Autovalores más bajos $\omega^2$ de (\ref{SLMNgen}) que conducen a soluciones normalizables como función de $\rho_0$.
Modos negativos (inestables) se encuentran precisamente para las soluciones clásicas que satisfacen $L'(\rho_0)>0$. Colores como en la fig. \ref{LMNgmumenosuno}.}
\label{wvsrminMNgenmu}
\end{center}
\end{minipage}\  \
\hfill
\begin{minipage}{7cm} 
\begin{center}
\includegraphics[width=7.5cm]{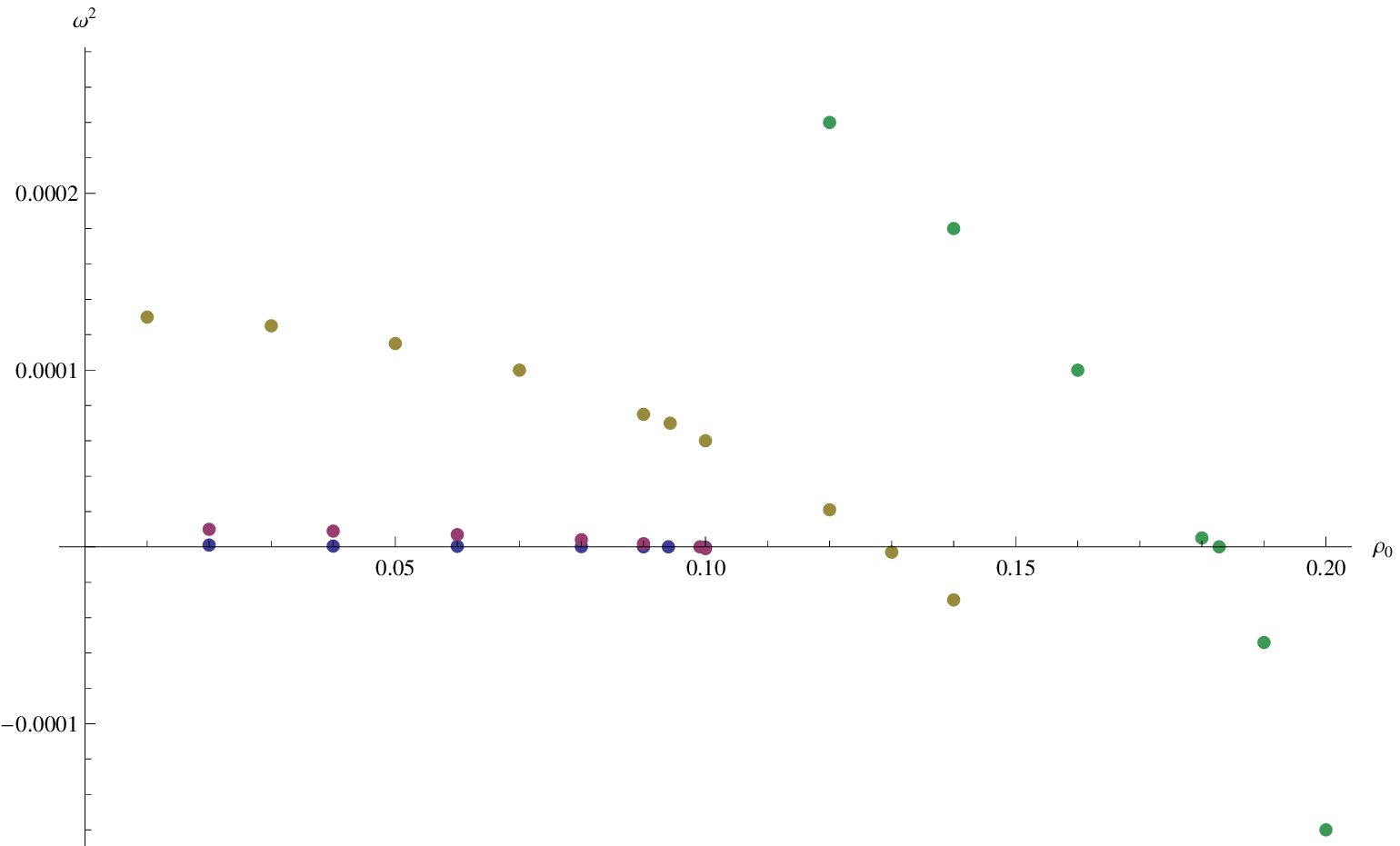}
\caption{Ampliación de la figura \ref{wvsrminMNgenmu} cerca del origen. Los autovalores más bajos son positivos para las soluciones que satisfacen $L'(\rho_0)<0$. Aparece un modo cero precisamente para el valor critico de la función
$L(\rho_0)$ (ver fig. \ref{LMNgmumenosuno}).}
\label{wvsrminMNgenzoommu}
\end{center}
\end{minipage}
\end{figure}
En las figuras \ref{wvsrminMNgenmu} y \ref{wvsrminMNgenzoommu} se grafican los autovalores mínimos que permiten soluciones
normalizables obtenidos numéricamente como función de $r_0$. Se encuentra un acuerdo completo con la figura
\ref{LMNgmumenosuno}: no se encuentran inestabilidades para la solución clásica que satisface $L'(r_0)<0$, por otro
lado, se encuentran modos negativos (inestables) para la rama derecha (soluciones con $L'(r_0)>0$) en la figura \ref{LMNgmumenosuno}.
Estos resultados son gratificantes debido a que los modos inestables se encuentran precisamente para las soluciones
clásicas de cuerda que no satisfacen las condiciones (\ref{convexity}). En la sección siguiente
voy a resumir estos resultado transformando las ecuaciones de movimiento en ecuaciones del tipo Schr\"{o}dinger.

\section{Análisis de Potenciales de Schr\"{o}dinger}
\label{scho}

En esta sección se analizarán la ecuación de movimiento para la fluctuación (\ref{SL})
llevándola a una expresión de tipo Schr\"{o}dinger (ver apéndice \ref{sl2sc}). A partir de
la forma del potencial es posible mostrar la existencia, o no, de autovalores negativos y
por lo tanto probar la estabilidad de la configuración de cuerda correspondiente.

\subsection{$AdS_5$}

\begin{figure}
\begin{minipage}{7cm} 
\begin{center}
\includegraphics[width=7.5cm]{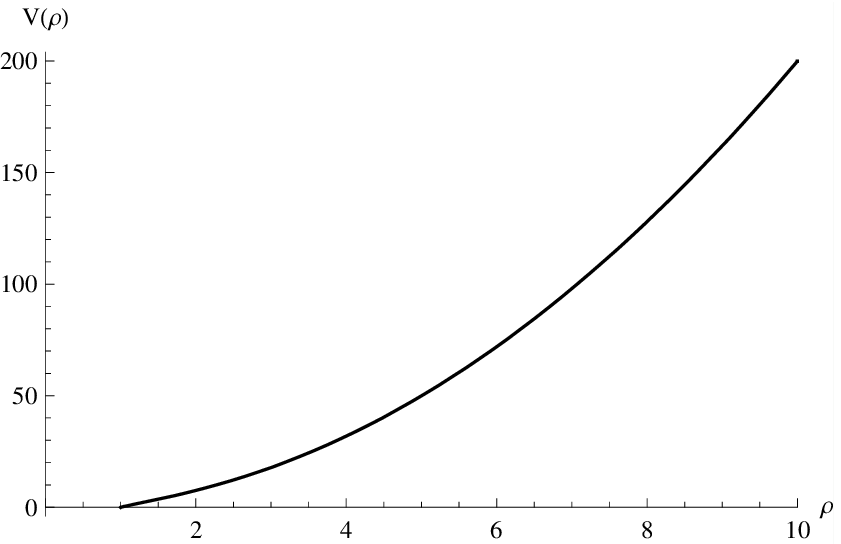}
\caption{Potencial de Schrodinger (\ref{potads}) para fluctuaciones en
 $AdS$. Al ser definido positivo se garantiza la ausencia de modos negativos.}
\label{potschrads5s5}
\end{center}
\end{minipage}\  \
\hfill
\begin{minipage}{7cm} 
\begin{center}
\includegraphics[width=7.5cm]{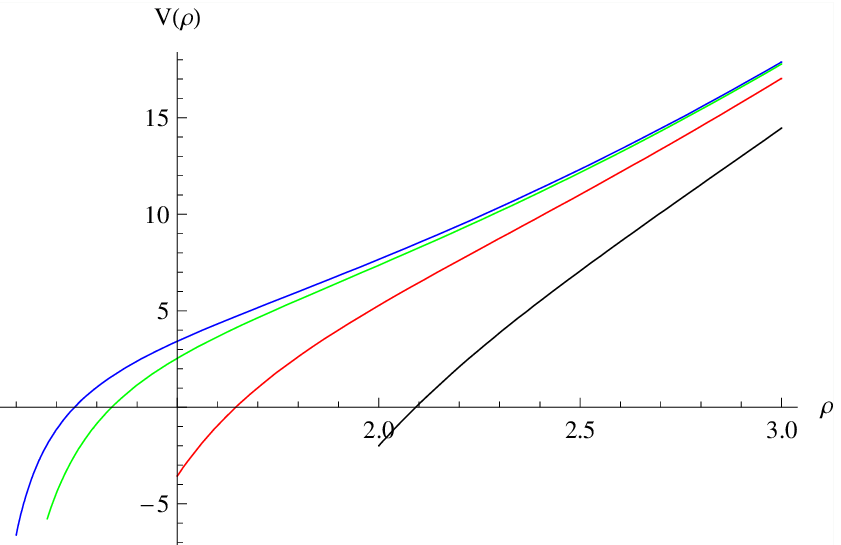}
\caption{$V(\rho,\rho_0)$ (ec.(\ref{pottads})) para diferentes valores de  $\rho_0$ .
La línea azul corresponde a $\rho_0=1.1$, la verde al valor crítico
$\rho_0=1.177$ y las líneas roja y negra a $\rho_0=1,\,2$ respectivamente.
La región en donde el potencial es negativo disminuye a medida que $\rho_0$ aumenta.
Los autovalores negativos dejan de existir para $\rho_0\ge1.177$.}
\label{potschrnonextremal}
\end{center}
\end{minipage}
\end{figure}

El potencial de Schr\"{o}dinger (\ref{schrodinger}) para la ecuación (\ref{fluctnormalized}) toma la forma \cite{avramis}
\be
V(\rho)=2\,\frac{\rho^4-1}{\rho^2}\,,~~~~\rho\in[1,\infty)\,,
\label{potads}
\ee
aquí $\rho$ debe ser entendido como $\rho=\rho(y)$. El cambio de variables (\ref{chv}) que nos conduce a la ecuación de Schrodinger
(\ref{sch}) puede ser calculado analíticamente
\be
y(\rho)=y_0-\frac14\, \mathsf{B}\left(\frac1{\rho^4};\frac14,\frac12\right)\,,
\label{chvads}
\ee
con $y_0=\frac{\Gamma[\frac14]^2}{4\sqrt{2\pi}}$. La semi-recta $\rho\in[1,\infty)$ del problema original
de Sturm-Liouville, ante el cambio de variables (\ref{chvads}), se mapea en el intervalo infinito $y\in[0,y_0]$.
El potencial (\ref{potads}) es divergente en $y_0$. Se ha definido un problema de Schrodinger
sobre un intervalo finito con condiciones de borde canónicas (ver (\ref{bcsch})-(\ref{bcsschrod})) y por lo tanto,
dado que el potencial (\ref{potads}) es definido positivo, se obtendrá un espectro discreto. Argumentos
estándar de mecánica cuántica permiten afirmar que no existirán soluciones con autovalores negativos.
Se concluye que el fondo de $AdS$ dado por (\ref{xads}) es estable ante perturbaciones lineales. La figura
\ref{potschrads5s5} muestra el potencial (\ref{potads}) como función de $\rho$, la verdadera variable
del problema de Schrodinger es $y$ dada por (\ref{chvads}), la cual es solo un reescaleo del eje horizontal
de la fig. \ref{potschrads5s5} mapeando $\rho=\infty$ a una distancia finita.

\subsection{$AdS_5$-Schwarzschild}

El potencial de Schr\"{o}dinger (\ref{schrodinger}) para la ecuación (\ref{eqdiff}) toma la forma \cite{avramis}
\be
V(\rho,\rho_0)=2\,\frac{\rho^8(\rho^{4}-\rho_0^{4})-\rho_0^4(4\rho^4-1)-3\rho^4}{\rho^6(\rho^4-1)}\,,~~~~1<\rho_0\le \rho<\infty\,.
\label{pottads}
\ee
El comportamiento de este potencial para diferentes valores de $r_0$ a sido graficado en la figura \ref{potschrnonextremal}.
A diferencia del caso de $AdS_5\times S^5$, existen regiones en donde el potencial toma valores negativos, lo cual
esta en acuerdo con los resultados de la sección \ref{nond3} en donde se habían encontrado autovalores negativos.
El potencial comienza tomando un valor negativo en $\rho_0$ dado por $V(\rho_0,\rho_0)=-8/\rho_0^2$.
A medida que $\rho_0$ se incrementa la región negativa disminuye y los modos negativos dejan de existir en el
valor crítico, encontrado numéricamente en la sección \ref{nond3},  $\rho_{0c}\simeq1.177$ el cual precisamente
coincide con el valor crítico de la función longitud $L(\rho_0)$. Se concluye que la solución clásica  que satisface
$L'(\rho_0)>0$ es inestable ante perturbaciones lineales (ver también \cite{avramis} para un análisis perturbativo de los
autovalores). La solución inestable $L'(\rho_0)>0$  tiene energía finita $E_{q\bar{q}}$ positiva (ver fig. \ref{e(r_0)}),
debido a que la configuración de referencia satisface las mismas condiciones de borde, el candidato natural para
el proceso de decaimiento es el estado de referencia (quarks libres).

Por completitud se mencionará que debido a que el desarrollo asintótico en esta caso coincide con el de el ejemplo
previo, la ecuación de Schr\"{o}dinger para las fluctuaciones queda definida en un intervalo finito. El espectro, por lo tanto,
es discreto.

\subsection{Maldacena-N\'u\~nez }
\label{schMN}

\begin{figure}
\begin{minipage}{7cm} 
\begin{center}
\includegraphics[width=7.5cm]{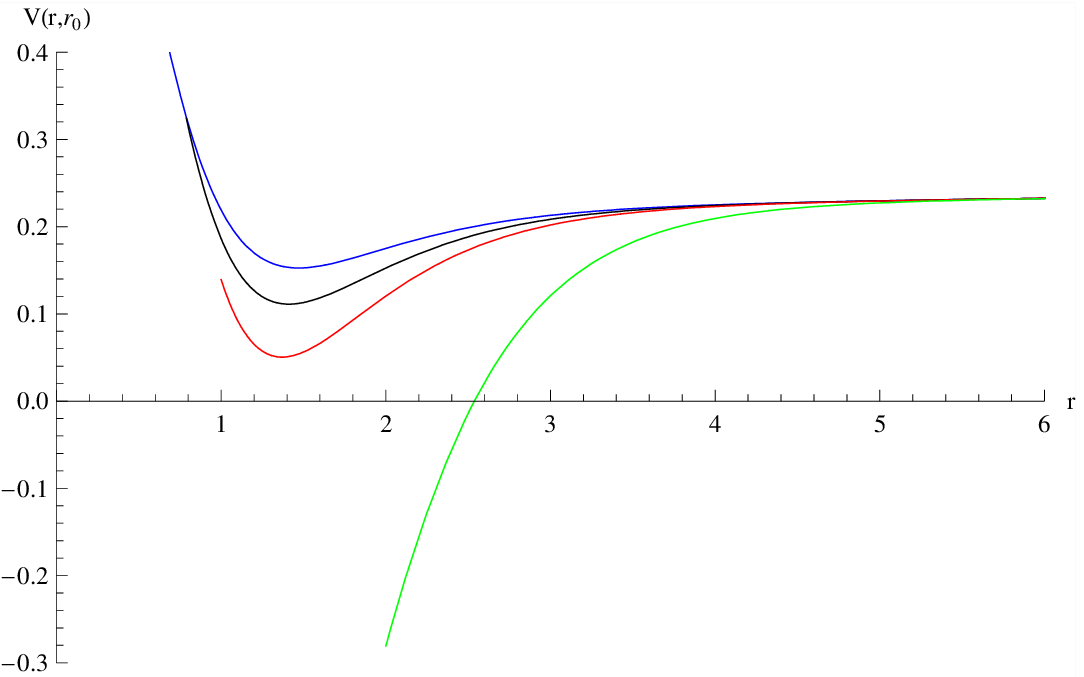}
\caption{Potencial de Schr\"{o}dinger (\ref{potMN}). Las líneas azul, negra, roja y verde corresponden con
$r_0=0.2,\,0.7,\,1,\,2$.
el mínimo de potencial se hace negativo para $r_0\ge1.1605$.}
\label{potschrMN}
\end{center}
\end{minipage}\  \
\hfill
\begin{minipage}{7cm} 
\begin{center}
\includegraphics[width=7.5cm]{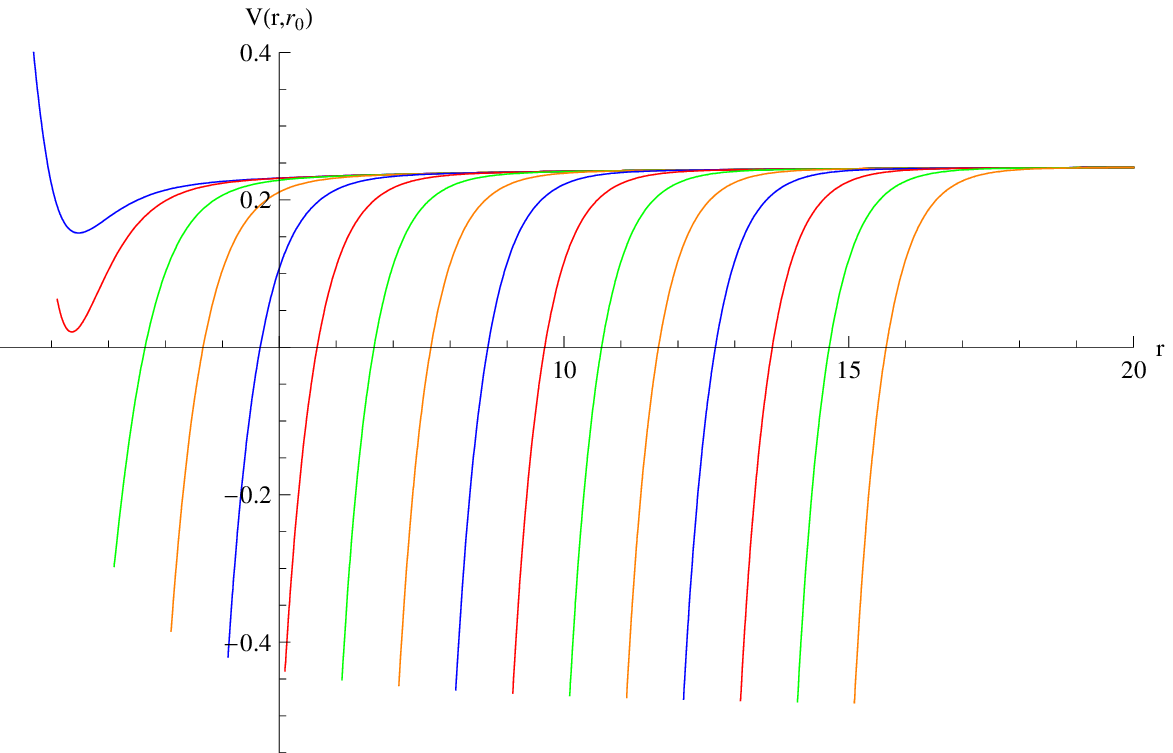}
\caption{Potencial (\ref{potMN}) como función de $r_0$. El potencial asintóticamente tiende a
$V_\infty=\frac14$ en concordancia con (\ref{asymMN}). Para $r_0<1.1605$ es positivo y
 para $r_0\ge 1.1605$ contiene regiones negativas.
 Cerca del tip el potencial puede aproximarse asintóticamente (para $r_0$ grande) por
$V(r)\simeq-\frac12+\frac32 (r-r_0)$}
\label{potenMN}
\end{center}
\end{minipage}
\end{figure}

El potencial de Schr\"{o}dinger para (\ref{SLMN}) toma la forma
\be
V(r,r_0)=\frac{e^{-2\phi(r)}}{4}\left((e^{2\phi(r)}-3e^{2\phi(r_0)})\phi'^2(r)
+2(e^{2\phi(r)}+e^{2\phi(r_0)})\phi''(r)\right),~~~~0<r_0\le r<\infty\,.
\label{potMN}
\ee
Como antes, $r$ debe ser entendida como $r=r(y)$ y al contrario que los últimos dos casos con el cambio de variables
(\ref{chv}) se obtiene un problema de Schr\"{o}dinger en la coordenada $y$ definida sobre la semi-recta $y\in[0,\infty)$.
Las figuras \ref{potschrMN} y \ref{potenMN} muestran el potencial de Schr\"{o}dinger (\ref{potMN}) para diferentes
valores de $r_0$. Se deben confrontar estas figuras con los resultados obtenidos en la sección \ref{mnbkg}.
Lo concluido en esa sección fue que, para todo valor de $r_0$, existe un espectro continuo para
$\omega^2>0$ y numéricamente no se encontraron modos normalizables negativos. En primer lugar, analizemos el
espectro continuo para $\omega^2>0$. La Figura \ref{potschrMN} muestra que el potencial es definido positivo
para $r_0\le1.1605$ y asintóticamente toma el valor $V_{\infty}=\frac14$. Luego, se podría concluir  que
no existen soluciones para $0<\bar\omega^2<V_{min}$ y que se encuentra un espectro discreto para
$V_{min}<\bar \omega^2<\frac14$ (en caso de ser posible) y un espectro continuo, pero no normalizable, para
$\bar\omega^2>\frac14$, todo en contradicción con los resultados mencionados. La concordancia se obtiene
cuando se tiene en cuenta el factor $(PQ)^{-\frac14}$ que relaciona la solución de la ecuación de Schr\"{o}dinger
$\Psi$ con la fluctuación $\delta x_1$ (ver apéndice \ref{sl2sc} eqn. (\ref{chv}))
\be
\delta x_1=\frac{e^{\frac{\phi(r)}2}}{(e^{2\phi(r)}-e^{2\phi(r_0)})^{\frac12}}\Psi\simeq e^{-\frac{r}2}\Psi\,,~~~~r\to\infty\,.
\ee
El factor $e^{-\frac{r}2}$ hace que todas las soluciones con $\bar\omega^2>0$ del problema de Schr\"{o}dinger satisfagan
$\delta x_1|_{r=\infty}=0$ independientemente de si $\Psi(y)$ es normalizable o no (asintóticamente se tiene $y\simeq r$).
Sin embargo, para soluciones con $\bar\omega^2<0$ el factor no es suficiente para hacer que las soluciones (divergentes) satisfagan
las condiciones de borde. Se concluye que para todo valor de $r_0$ se tiene un espectro continuo para $\bar\omega^2>0$.

El segundo punto a tener en cuenta es la posibilidad de encontrar estados ligados para $-\frac12<\bar\omega^2<0$
en el límite de $r_0$ grande. Como se ha visto en la figura \ref{potenMN}, asintóticamente, el potencial comienza
tomando el valor $V(r_0)\simeq-\frac12$ y en la aproximación lineal se obtiene $V(r)\simeq-\frac12+\frac32 (r-r_0)$.
La relación entre las coordenadas $r$ e $y$ (\ref{chv}) en el mismo límite es $(r-r_0)\simeq y^2/2$. Todo esto
conduce a un oscilador armónico en las coordenadas $y$ con energía de ligadura positiva. Se concluye, por lo tanto,
que no existen estados ligados y se encuentra una perfecta concordancia entre el análisis de Schr\"{o}dinger
y los resultados numéricos de la sección \ref{mnbkg}.

El resultado es atractivo debido a que si se hubiesen encontrado inestabilidades, no hay candidato
disponible para el decaimiento (confrontar con sección \ref{schgMN}).

\subsection{Klebanov-Strassler}
\label{schKS}

\begin{figure}[h]
\hspace{-0.25cm}
\begin{minipage}[b]{1\linewidth} 
\centering
\includegraphics[width=7.5cm]{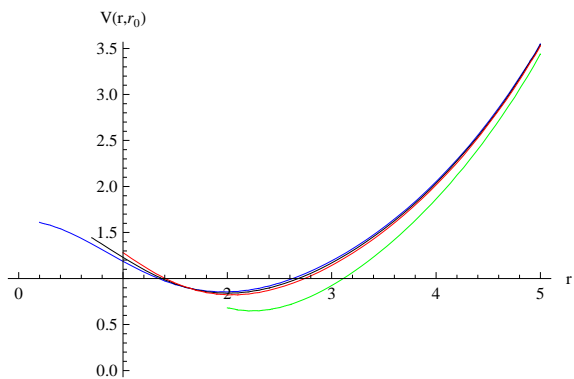}
\caption{Potencial de Schrodinger (\ref{sKS}). Las líneas azul, negra, roja y verde corresponden
a $r_0=0.2,\,0.7,\,1,\,2$. El potencial es definido positivo y por lo tanto se obtienen
autovalores positivos.}
\label{potschrKS}
\end{minipage}
\end{figure}
El potencial para la fluctuación en el caso de Klebanov-Strassler (\ref{slks}) es
\bea
V(r,r_0)&=&-\frac{3K(r)}{8{h^{3}(r)h(r_0)}}\left[{4h(r)(h(r)+h(r_0))h'(r)k'(r)}
\right.\nn\\
&&\left.-k(r)(3h(r)+7h(r_0))h'^2(r)+4h(r)(h(r)+h(r_0))h''(r))\right]
\label{sKS}
\eea
El comportamiento asintótico de las funciones $P,Q$ en este caso conduce a un problema de Schr\"{o}dinger
definido en un intervalo finito en la coordenada $y$ (ver apéndice \ref{sl2sc}). La figura (\ref{potschrKS})
muestra la forma del potencial para varios valores de $r_0$. El potencial es definido positivo y, por lo tanto, no
existen soluciones con $\omega^2<0$. El intervalo finito en el cual el problema de Schr\"{o}dinger esta definido
implica un conjunto discreto de autovalores. Se encuentra perfecta concordancia con los resultados de
la sección \ref{ksbkgd}.

\subsection{Maldacena-N\'u\~nez Generalizado}
\label{schgMN}
\begin{figure}
\begin{minipage}{7cm} 
\begin{center}
\includegraphics[width=7.5cm]{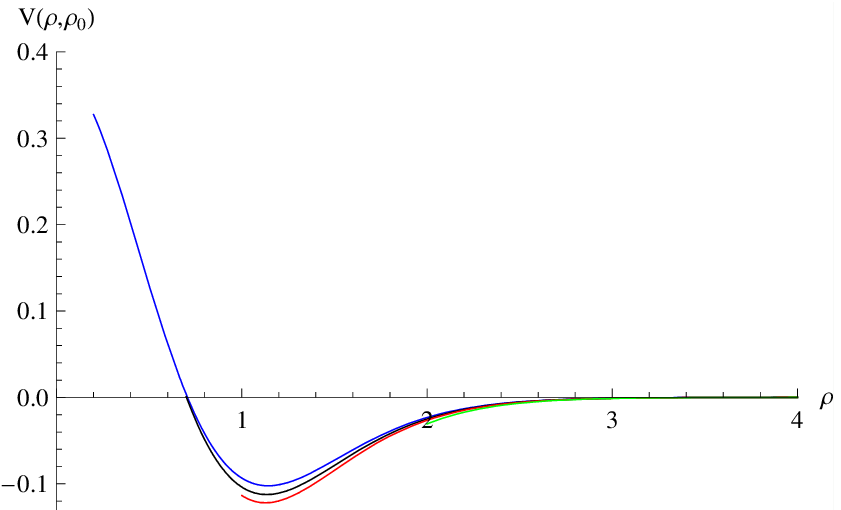}
\caption{Potencial de Schr\"{o}dinger (\ref{VgMN}) para
diferentes valores de $\rho_0$ y $\mu=-1$. Las líneas azul, negra, roja y verde se corresponden con
$\rho_0=0.2,\,0.7,\,1,\,2$ respectivamente.}
\label{potschrMNgen}
\end{center}
\end{minipage}\  \
\hfill
\begin{minipage}{7cm} 
\begin{center}
\includegraphics[width=7.5cm]{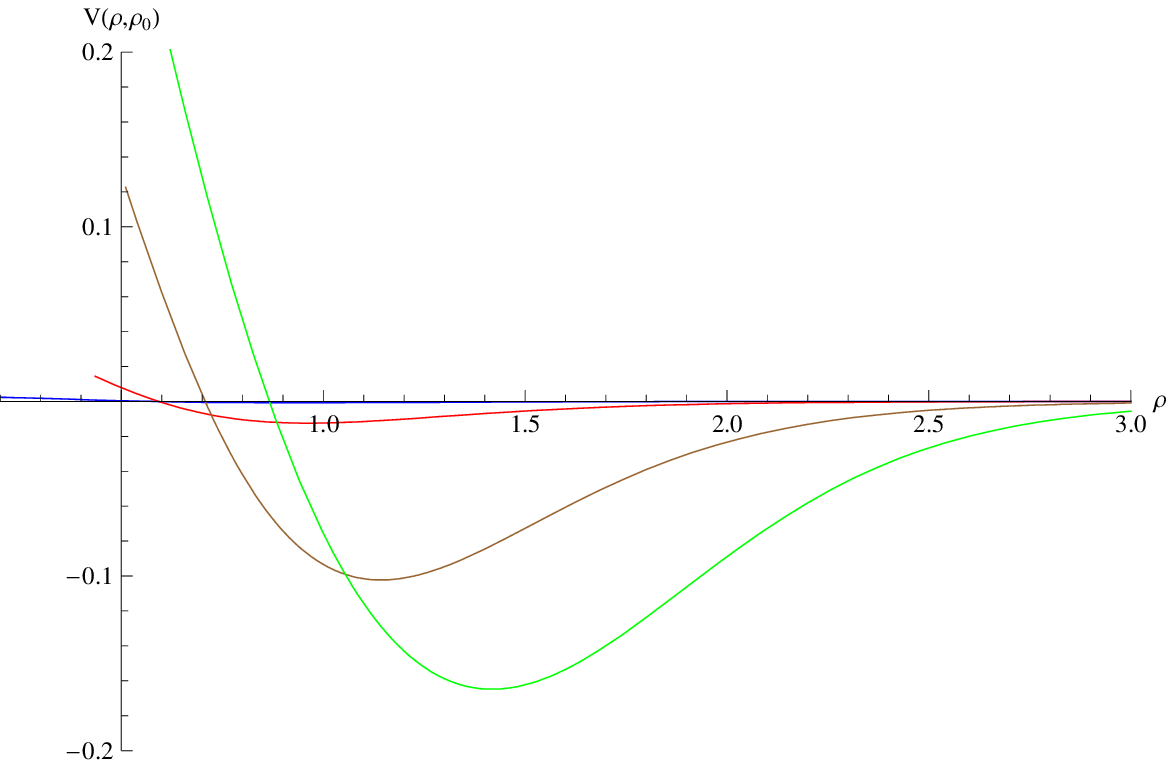}
\caption{Potencial de Schr\"{o}dinger para $\rho_0=0.2$ y para diferentes valores
del parámetro $\mu$.  Las curvas azul, roja, marrón y verde se corresponden
con $\mu=-1.8,\,-1.5,\,-1,\,-0.8$.}
\label{potschrgMNvariosmu}
\end{center}
\end{minipage}
\end{figure}

El potencial de Schr\"{o}dinger para fluctuaciones en el caso generalizado de Maldacena-Nuñez
puede escribirse como
\be
V(\rho,\rho_0)=\frac2{e^{8f(\rho)+2k(\rho)}}\left(2 (e^{8f(\rho)}-e^{8f(\rho_0)})f'^2(\rho)
+(e^{8f(\rho)}+e^{8f(\rho_0)})(f''(\rho)-k'(\rho)f'(\rho))\right)
\label{VgMN}
\ee
El comportamiento asintótico de las funciones $P,Q$ (ver apéndice \ref{sl2sc} y la ec. (\ref{asymgMN})) conduce a
un problema de Schr\"{o}dinger formulado en la semi-recta $y\in[0,\infty)$. La figura (\ref{potschrMNgen}) muestra el
comportamiento del potencial para diferentes valores de $r_0$ y $\mu=-1$. El potencial se hace negativo por sobre
el valor crítico $\rho_*$ y asintóticamente $V_\infty=0$ en concordancia con (\ref{asymgMN}) y con la existencia
de modos negativos encontrados numéricamente en la sección \ref{gMNbkgd}. El factor $(PQ)^{-\frac14}$ que relaciona
la función de onda de Schr\"{o}dinger $\Psi$ con la función $\delta x_1$ tiende a una constante en infinito, lo cual
no cambia el desarrollo asintótico de las soluciones $\Psi$ (confrontar con la sección \ref{schMN}).
La figura \ref{potschrgMNvariosmu} muestra el potencial de Schr\"{o}dinger para un valor fijo $\rho_0=0.2$ y para
diferentes valores de $\mu$. El mínimo de potencial disminuye a medida que $\mu$ se acerca al valor $-\frac23$.
Como se ha mencionado, la solución original de Maldacena-Nuñez (\ref{metric}) no se conecta continuamente con la
clase generalizada de soluciones (\ref{gMN}). Se encontró perfecta concordancia con los resultados numéricos de la sección
\ref{schgMN} pero no es claro cuál es el estado final de decaimiento.

\section{Lazo de 't Hooft}
\label{thooftloop}

El dual electromagnético de las líneas de Wilson en teorías de Yang-Mills son las líneas de ´t Hooft \cite{thooftline}.
En cuatro dimensiones, se supone que el mecanismo para confinamiento se debe a la condensación de monopolos
magnéticos (efecto Meissner dual), el análisis en \cite{thooftline} concluye que se debería observar
un potencial apantallado entre un par $m\bar m$ de monopolos cuando se observa confinamiento entre el par $q\bar q$.
El confinamiento oblicuo es una generalización de esta idea en la cual se estudia el confinamiento de diones.

La prescripción dada por la conjetura para calcular lazos de ´t Hooft en las soluciones de MN y KS ha sido propuesta
en los mismos trabajos \cite{mn}-\cite{ks} (ver \cite{BM} y también \cite{paredes}, en donde se analiza una cuestión técnica,
corrigiendo el 2-ciclo propuesto en \cite{mn}) y consiste en enroscar una $D3$-brana de prueba en el mismo 2-ciclo sobre
el cual están enroscadas las $D5$-branas que generan la geometría (ver también \cite{groo}). El resultado de esta construcción
es una D1-brana efectiva (cuerda) la cual se analiza en completa analogía a la cuerda fundamental que ha sido discutida
en secciones previas. La diferencia con respecto al lazo de Wilson es que el lazo de ´t Hooft en teorías genéricas
que no son duales ante simetría S es sensible a la variedad interna de 5 dimensiones.

\subsection{Maldacena-N\'u\~nez}

\begin{figure}[h]
\begin{minipage}{7cm}
\begin{center}
\includegraphics[width=7.5cm]{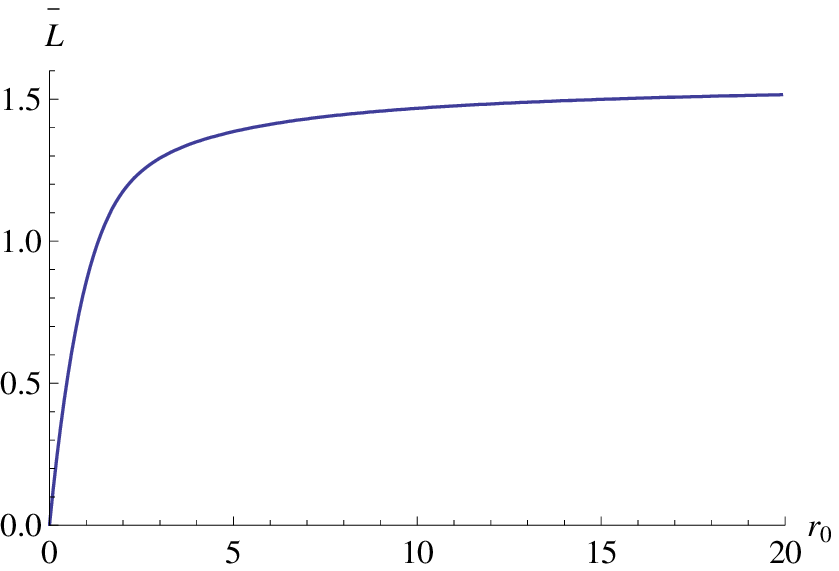}
\caption{Función longitud para la cuerda efectiva como función de $r_0$ para el lazo de Wilson
 en el caso de Maldacena-N\'u\~nez.}
\label{lvsRminMNthooft}
\end{center}
\end{minipage}
\   \
\hfill \begin{minipage}{7cm}
\begin{center}
\includegraphics[width=7.5cm]{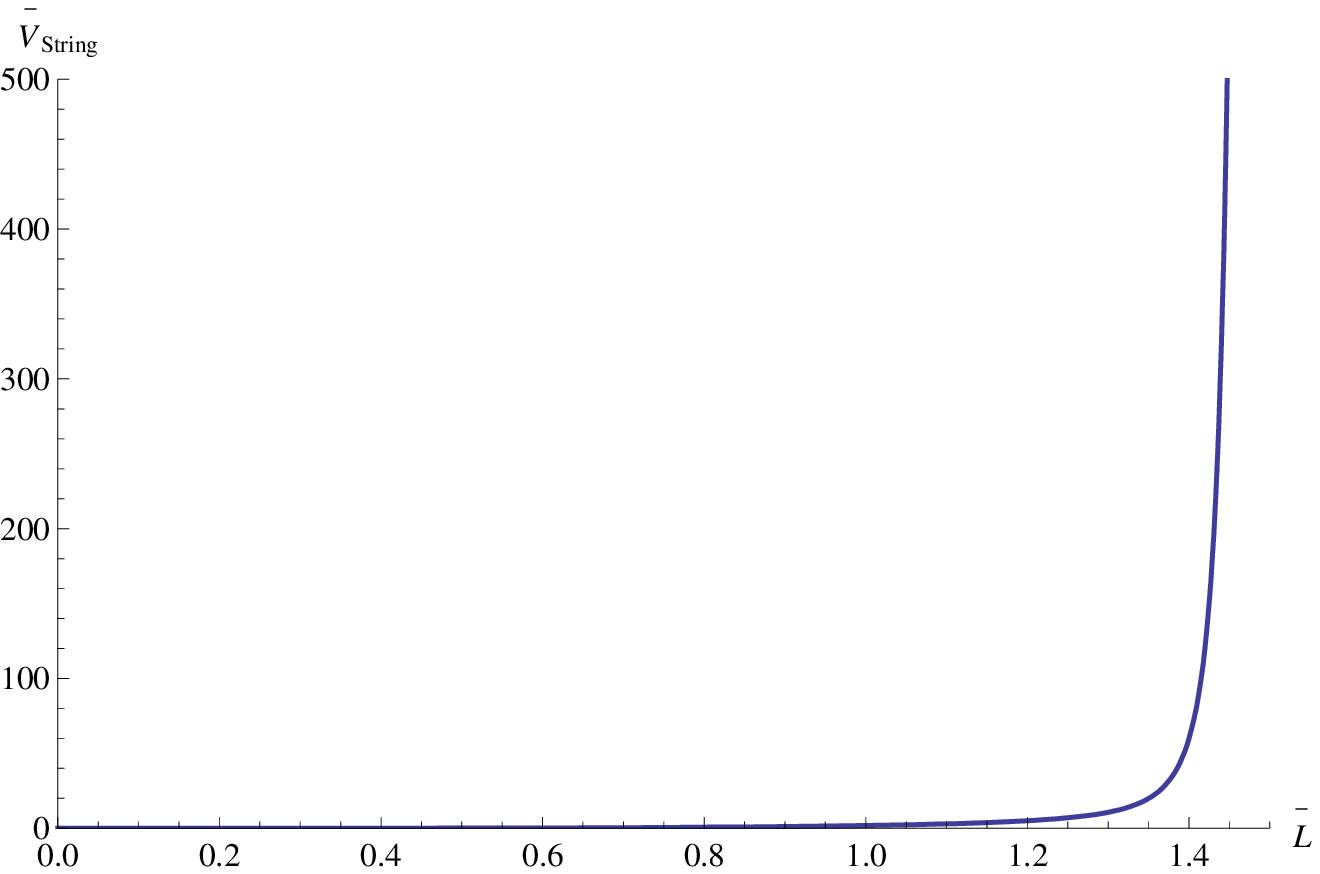}
\caption{Energía del par monopolo-antimonopolo como función de la longitud de separación.}
\label{EvsLMNthooft}
\end{center}
\end{minipage}
\end{figure}

La variedad en la que se embebe la $D3$ en la métrica (\ref{metric}) es \cite{BM}\footnote{ Las coordenadas restantes
se fijan a constantes. El valor de la coordenada $\psi$ se fija pidiendo que la $S^2$ tenga volumen mínimo.}
\be
{\cal M}_4=[t,x,r(x),\theta=\tilde\theta,\varphi=2\pi-\tilde\varphi,\psi=\pi]\,.
\label{ansatz}
\ee
La métrica inducida sobre ${\cal M}_4$ es
\be
ds^2_{ind}= \alpha'  Ne^{{\phi}}\,\,\Big[-dt^2+(1+\acute r^2)dx^2+(e^{2h}+\frac14(1-a)^2)\,(d\theta^2+\sin^2\theta d\varphi^2)\Big]\,,
\ee
y de las expresiones (\ref{oneform}) para $a,h$ queda
\be
V_{S^2}(r)\equiv \frac14(1-a(r))^2+e^{2h(r)} =r\tanh{r}\,.
\label{s2vol}
\ee
Nótese que la $S^2$ colapsa suavemente en el origen. Integrando la acción DBI \footnote{ Usando el ansatz
para el fijado de gauge (\ref{ansatz}) en la acción (\ref{dbi}) se obtiene la ecuación de movimiento correcta
para $r(x)$ lo cual coincide con (\ref{sig}).}
\be
S_{DBI}=-T_{D3}\int d^4\sigma e^{-\phi} \sqrt{g_{ind}}
\label{dbi}
\ee
sobre la variedad interna ($S^2_{\theta\varphi}$) y la coordenada temporal se tiene
\be
S_{eff}= 4\pi T_{D3}{\cal T}{ (\alpha'N)^2}\int{ e^\phi\,r\tanh r \sqrt{1+\acute r^2}}\,dx\,.
\ee
La diferencia importante con respecto al cálculo del lazo de Wilson en el fondo de MN
(ver sección \ref{mnsol}) reside en las funciones $f(r)$ y $g(r)$ que en (\ref{lmn})-(\ref{emn}) están
multiplicadas por el volumen de la 2-esfera (\ref{s2vol}).

\begin{figure}[h]
\begin{minipage}{7cm}
\begin{center}
\includegraphics[width=7.5cm]{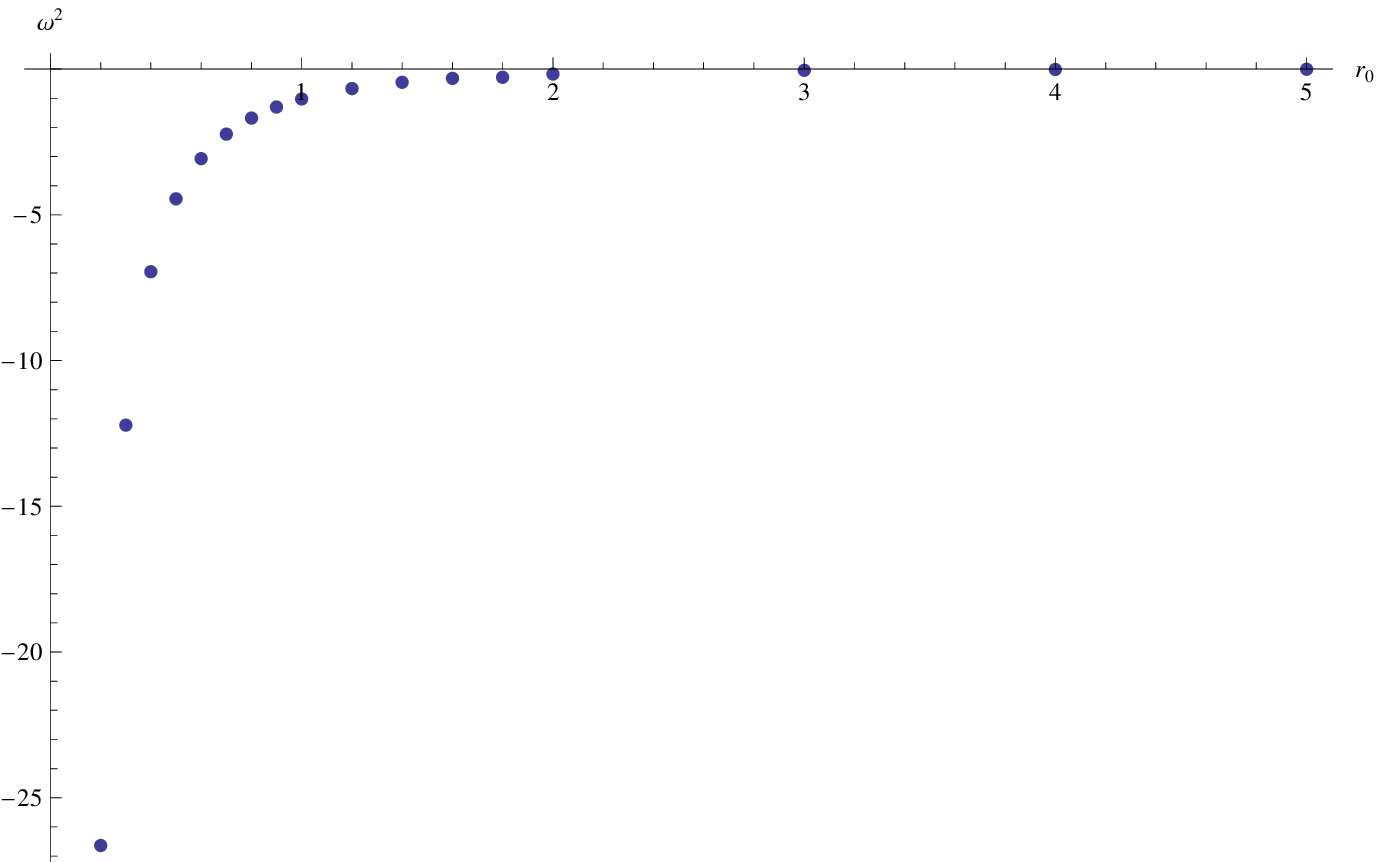}
\caption{Autovalores más bajos $\omega^2$ para fluctuaciones que solo dependen de las coordenadas $t,r$
como función de $r_0$ en el caso del lazo de ´t Hooft para el fondo de Maldacena-N\'u\~nez.}
\label{wvsrminMNthooft}
\end{center}
\end{minipage}
\   \
\hfill \begin{minipage}{7cm}
\begin{center}
\includegraphics[width=7.5cm]{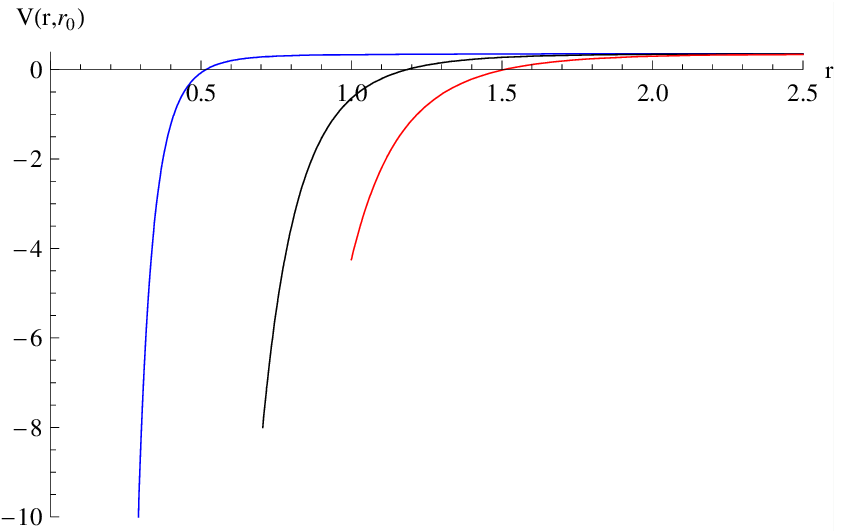}
\caption{ Potencial de Schrodinger para fluctuaciones que dependen de las coordenadas $t,r$
como función de $r_0$. Las curvas azul, negra y roja corresponden a
$r_0=0.2,\,0.7, 1$.}
\label{potschrMNthooft}
\end{center}
\end{minipage}
\end{figure}

En las figuras \ref{lvsRminMNthooft} se gráfica el comportamiento de la función longitud (\ref{generalL}) como función de $r_0$.
La función longitud es una función creciente en $r_0$ y a partir de las discusiones previas
se espera que las soluciones sean inestables.
Puede verse la inestabilidad de la forma de embeber la D1 efectiva en el caso de MN debido a que una fluctuación
a lo largo de la dirección $x_1$ dependiente solamente de $t,r$ está desacoplada de las fluctuaciones angulares
(fluctuación consistente). La ecuación de movimiento para $\delta x_1$ es (\ref{SL}) con $f(r)=g(r)= h(r)=r\tanh r\, e^{\phi(r)}$.
El comportamiento asintótico de la fluctuación es el mismo que en el caso del lazo de Wilson, sin embargo,
este comportamiento cambia drásticamente cerca del origen debido a que $f(r)$ tiende a cero.
Como se ve en la figura \ref{wvsrminMNthooft} existen autovalores negativos para todo valor de $r_0$.
En la figura \ref{potschrMNthooft} se muestra el comportamiento del potencial de Schr\"{o}dinger
asociado con la ecuación de movimiento para la fluctuación.

En la figura \ref{EvsLMNthooft} se gráfica la energía como función de la longitud de separación $L$.
La energía de la configuración es positiva para todo $L$, este echo y la inestabilidad de la configuración
sugiere que la configuración estable para dadas condiciones de borde, es la correspondiente a dos
'' líneas rectas ''. Contrariamente al caso del lazo de Wilson, las '' líneas rectas '' (usadas como estado de referencia
para regularizar la energía) pueden terminar en el origen debido a que estas corresponden a $D3$ enroscadas en una
$S^2$ topológica de (\ref{metric}) que colapsa suavemente en el origen.

\subsection{Klebanov-Strassler}

\begin{figure}[ht]
\begin{minipage}{7cm}
\begin{center}
\includegraphics[width=7.5cm]{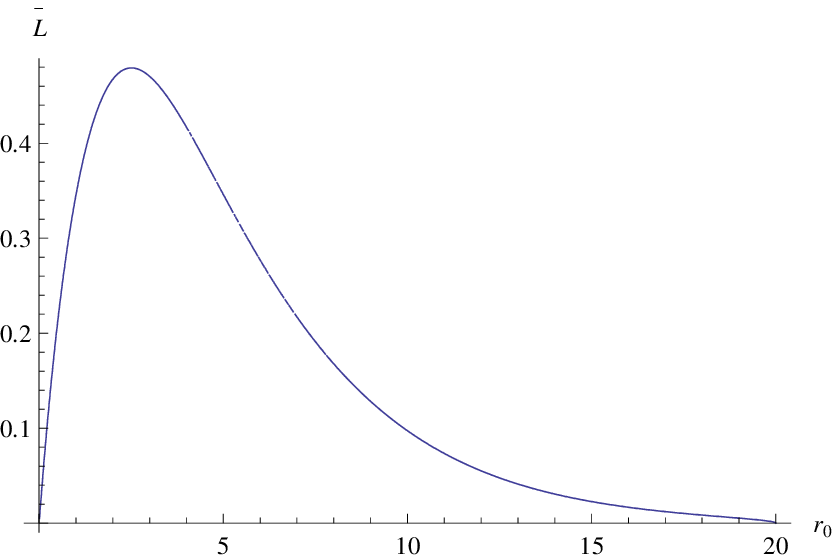}
\caption{$L(r_0)$ para la cuerda efectiva en el fondo de KS.}
\label{lvsRminKSthooft}
\end{center}
\end{minipage}
\   \
\hfill \begin{minipage}{7cm}
\begin{center}
\includegraphics[width=7.5cm]{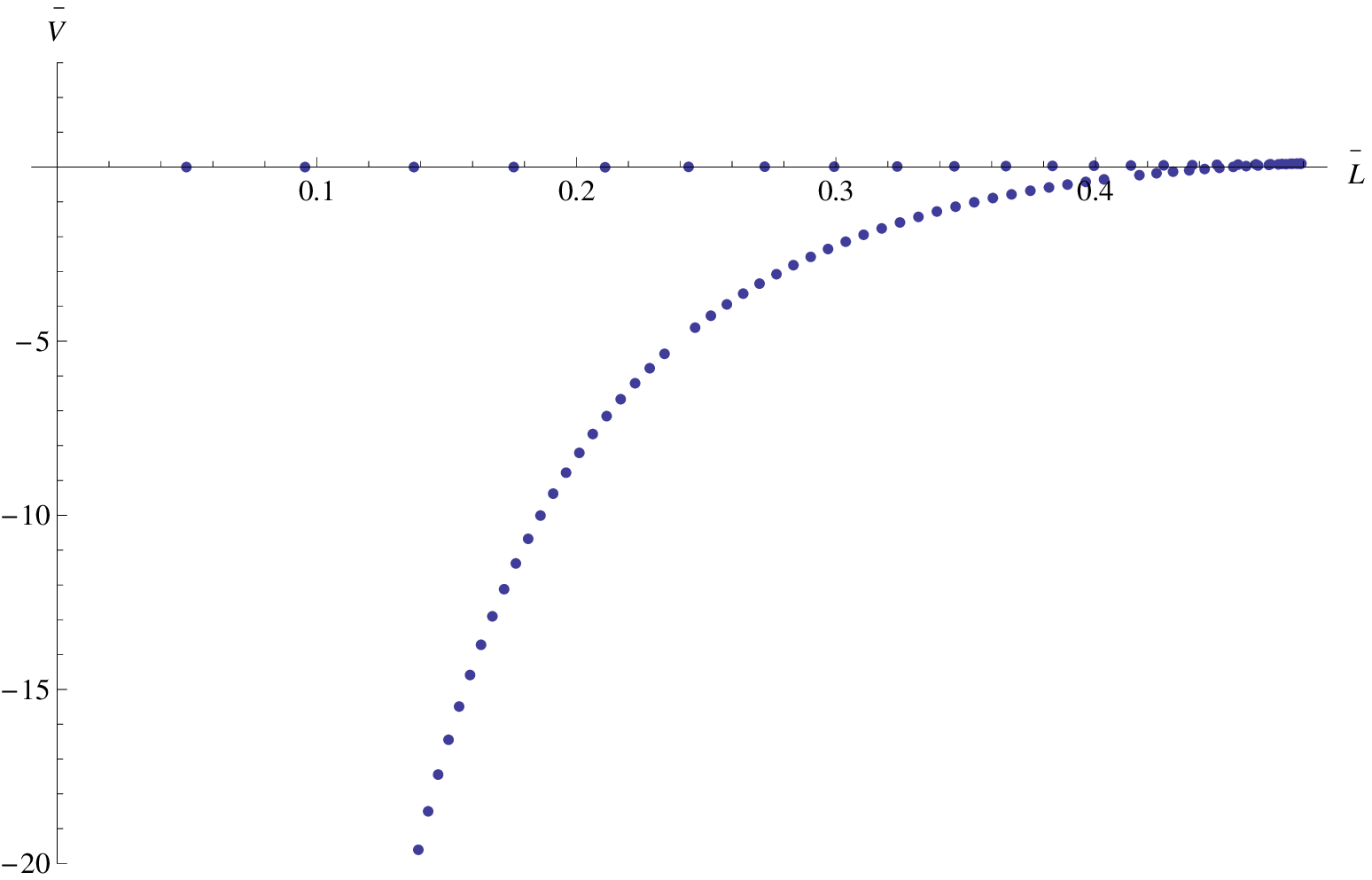}
\caption{ La energía del par monopolo-antimonopolo como función de la longitud de separación.}
\label{EvsrminthooftKS}
\end{center}
\end{minipage}
\end{figure}

Este caso de nuevo involucra enroscar una $D3$ sobre una $S^2$ topológica dentro de (\ref{ks}) (para su parametrización ver
el apéndice A de \cite{ho}). Hay diferencias importantes con respecto al caso de MN, en esta caso la $H_3$ que soporta
la geometría de KS contribuye a la acción de la cuerda (\ref{NG}), y además conduce a un entrelazamiento entre las
fluctuaciones de las coordenadas angulares y las del plano. El comportamiento diferente en el UV con respecto al
caso de MN es la razón por la cual la función $L(r_0)$ tiene, a priori, regímenes estables (ver figura \ref{lvsRminKSthooft}).
El comportamiento de $\bar V(\bar L)$ muestra que el potencial esta apantallado para valores grandes de $L$ lo cual
concuerda con el potencial lineal para el caso del lazo de Wilson (ver \cite{groo} para estudiar un ejemplo relacionado).
El análisis del presente caso es análogo al realizado en la sección \ref{adssch}, en resumen, cuando la configuración
de energía se hace positiva, la solución de dos líneas rectas es favorecida. Como para el caso de MN esto es posible
sin un horizonte en este caso debido a que la $D3$ esta enroscada sobre una $S^2$ que colapsa suavemente en el origen.
No se realizó el análisis de las ecuaciones de movimiento para las fluctuaciones acopladas para chequear las inestabilidades
de la rama izquierda de la fig. \ref{lvsRminKSthooft}.

\section{Fluctuaciones del Lazo de Wilson en representaciones de orden más alto}\label{higherreps}

En esta sección se estudiarán las fluctuaciones bosónicas y fermiónicas de la configuración de branas dual a un lazo de Wilson en la representación simétrica mencionada en la sección \ref{Wilint}.

En particular se analizarán las  fluctuaciones correspondientes a un lazo de Wilson- 't Hooft supersimétrico \cite{Kapustin, Chen}, para un camino ${\cal C}$ recto en la representación simétrica del grupo. A partir de este operador se puede obtener el resultado tanto para el lazo de Wilson como para el de 't Hooft.

\subsection{Acción general para D-branas}

\subsubsection{Acción Bosónica}

La parte bosónica de la acción de un Dp-brana, llamada acción de Dirac-Born-Infeld se escribe
\be
 S_B=-T_{Dp}\int d^{p+1}\sigma\,e^{-\varphi}\sqrt{-\textrm{det}\left(g+2\pi\alpha'F\right)}+T_{Dp}\int P[C_{p+1}]\label{DBIWZ},
\ee
en donde $T_{Dp}=(2\pi)^{-p}\alpha'^{-\frac{p+1}{2}}$ es la tensión de la brana, $F$ es el campo de gauge que vive en su volumen de mundo,
$\varphi$ es el dilatón y $g$ y $P[C_{p+1}]$ son el pullback de la métrica y de la $p+1$ forma a la cual se acopla respectivamente. El segundo término
 de la acción es el denominado término de Wess-Zumino-Witten y es responsable de la existencia de gravitones gigantes \cite{Grisaru, Toumbas} y el efecto Myers \cite{Myersdielectric}.

Físicamente la expresión \eqref{DBIWZ} es la acción efectiva de baja energía para los modos no masivos de una brana con campos de fondo ($g_{\mu\nu}$, $\varphi$ y formas de Ramond-Ramond $C_{p+1}$) \cite{Polchobook,braneprimer}. En el caso del lazo de Wilson en la representación simétrica, que corresponde a una $D3$-brana, solo tendremos acoplamiento al $C_4$ de fondo.

\subsubsection{Acción Fermiónica}

La acción fermiónica cuadrática para nuestro caso ($D3$-brana  en un fondo AdS$_5\times S^5$) toma la forma \cite{Martucci}

\be
S_{Dp}^{(F)}=\frac{T_{Dp}}{2}\int d^{p+1}\sigma
\sqrt{-Det(M)}\bar\Theta(1-\Gamma_{Dp})(\tilde
M^{-1})^{\alpha\beta}\Gamma_\beta D_\alpha\Theta\label{FermionicAction},
\ee
en donde $M=g+F$. El campo fermiónico $\Theta$ es un fermion de Weyl con quiralidad positiva en $d=9+1$, esto significa que posee 32 componentes reales y es reducible frente a $SO(9,1)$ a 2 fermiones MW. En lo que sigue consideraremos a $\Theta$ como un
fermion de 64 componentes, las partes superiores e inferiores corresponden a los dos fermiones MW, las matrices $\Gamma$ se representaran entonces como matrices de $64\times 64$ dimensiones. La derivada covariante $D_\alpha$ se escribe

\be
D_\alpha=(\partial_\alpha
x^m)\left(\nabla_m+\frac{1}{16\cdot5!}F_{npqrt}\Gamma^{npqrt}(i\sigma_2)\Gamma_m\right),
~~~~\nabla_m=\partial_m+\frac{1}{4}\omega_m^{\underline{np}}\Gamma_{\underline{np}}\label{covariantder}.
\ee
Letras griegas denotan índices sobre el volumen de mundo de la brana, letras latinas corresponden a coordenadas del espacio 10-dimensional y los indices subrayados corresponden a coordenadas del espacio tangente (planos). $\tilde M_{\alpha\beta}=g_{\alpha\beta}+\tilde\Gamma F_{\alpha\beta}$ con $\tilde\Gamma=\gamma^{11}\otimes \sigma_3$\footnote{$\gamma^{11}$ es la matriz quiral 32 x 32 en $d=9+1$}. La matriz
$\Gamma_{Dp}$ en \eqref{FermionicAction} es

\bea
\Gamma_{Dp}&=& \frac{\sqrt{-Det (g)}}{\sqrt{-Det(g+F)}}\Gamma_{Dp}^{(0)}\otimes (\sigma_3)^{\frac{p+1}{2}}(-i\sigma_2)\sum_{q}\Gamma^{\alpha_1\ldots\alpha_{2q}}
F_{\alpha_1\alpha_2}\ldots F_{\alpha_{2q-1}\alpha_{2q}}\otimes\frac{\sigma_3^q}{2^qq!},\nn  \\
\Gamma_{Dp}^{(0)}&=&\frac{\epsilon^{\alpha_1\ldots\alpha_{p+1}}}{(p+1)!\sqrt{-Det (g)}}\Gamma_{\alpha_1\ldots\alpha_{p+1}}
\eea

La acción \eqref{FermionicAction} más la parte bosónica \eqref{DBIWZ} preserva la simetría $\kappa$ y proyecta al volumen de mundo la simetría del target. Es un generalización de la acción de Green y Schwarz para la cuerda \cite{GreenS, Simon}.

\subsection{Lazo de Wilson-'t Hooft y sus fluctuaciones}\label{wilsonthooft}

El valor de expectación del lazo de Wilson en la representación simétrica
puede obtenerse en la teoría dual gravitatoria estudiando soluciones de $D3$-branas enroscadas de forma tal que su volumen de mundo sea AdS$_2\times S^2\subset$ AdS$_5 $, con campo eléctrico encendido y ubicadas en
un punto de $S^5$ \cite{drukker}.

\subsubsection{Solución}
El fondo en el que embebemos la brana tiene encendidos $g_{\mu\nu}$, $F_5$ y $\varphi$. El dilaton (que aparece en \eqref{DBIWZ}) es constante para el fondo AdS, y ya fue factorizado en la definición de $T_p$.
Escribimos la geometría AdS$_5\times S^5$ como
\be\label{bkgd}
ds^2=L^2\left(\frac{\cosh^2(u)}{r^2}\left(-dt^2+dr^2\right)+du^2+\sinh^2(u)(d\theta^2+\sin^2\theta d\phi^2)\right)+L^2(d\alpha_1^2+\sin^2\alpha_1d\Omega_4^2).
\ee
En estas coordenadas $F_5$ toma la forma
\be
F_{5}=4L^4\frac{\sinh^2(u)\cosh^2(u)}{r^2}dt\wedge dr\wedge du\wedge d\theta \wedge d\phi\equiv dC_4,
\ee
y su potencial puede ser escrito como
\be
C_4=4L^4\frac{f(u)}{r^2}dt\wedge dr\wedge d\theta \wedge d\phi,
\ee
con $f(u)=\frac{1}{32}\sinh(4u)-\frac{u}{8}$.

Agreguemos una D3-brana en esta geometría con un campo electromagnético viviendo en su volumen de mundo. Debemos ahora escoger un ansatz que de cuenta de un lazo recto en la teoría de gauge. La elección  apropiada es
$X^\mu(\sigma)=(t,r,u,\theta,\phi,\alpha_i)=(\sigma_0,\sigma_1, u(\sigma_1),\sigma_2,\sigma_3,ctes\ldots$). Geométricamente esta elección corresponde a enrollar la $D3$ en la $S^2$ parametrizada por $\theta,\phi$ en \eqref{bkgd} y a extenderla en las coordenadas $t, r$. El ansatz para el campo electromagnético en el volumen de mundo es
\be
F=\frac{q}{r^2}\,dt\wedge dr+k\sin\theta d\theta\wedge d\phi
\ee
donde $q$ y $k$ dan cuenta de las cargas eléctricas y magnéticas. Desde el punto de vista dual la solución con ambas cargas encendidas
se correlata con el Wilson loop de un dyon (lazo de Wilson-t'Hoooft).

La acción \eqref{DBIWZ} queda
\be
S_B=-T_{D3}\int d^4\sigma\,\sqrt{-\textrm{det}\left(g+F\right)}+T_{D3}\int P[C_4]\label{DBIWZD3},
\ee
en donde sin pérdida de generalidad se utilizarán unidades en las cuales $2\pi\alpha'=1$. Para obtener la solución correspondiente a un lazo de Wilson se debe fijar $k=0$ \cite{Faraggi} mientras que para dar cuenta del lazo de 't Hooft se debe tomar $q=0$.

Al insertar el ansatz en las ecuaciones de movimiento obtenemos que $u=u_{eq}$ es constante y su valor queda fijado por las cargas ($q,k$) según
\be
q= \coth(u_{eq})\sqrt{L^4\sinh^2(u_{eq})-k^2},\label{q vs k}
\ee
El resultado obtenido generaliza el de \cite{Faraggi} que se obtiene para $k=0$.

\subsubsection{Fluctuaciones}

En esta sección se calcularán las acciones bosónica y fermiónica para pequeñas fluctuaciones alrededor de la solución \eqref{q vs k}.

~

~

~

\begin{itemize}
\item {\bf Bosónicas}
\end{itemize}

El objetivo es expandir la acción \eqref{DBIWZD3} a segundo orden en fluctuaciones. Las perturbaciones a la solución se escriben
\bea
    u=u_{eq}+\delta u,~~~~
    \theta^{\hat{i}}=\theta^{\hat{i}}_0+\delta\theta^{\hat{i}},~~~~
    F=\frac{q}{r^2}\,dt\wedge dr+k\sin\theta\, d\theta\wedge d\phi+f.
\eea
En lo que sigue se ignorarán las fluctuaciones $\delta\theta^{\hat{i}}$ debido a que las mismas se desacoplan. La acción bosónica para las fluctuaciones relevantes se puede escribir
\bea\label{dionic fluct B}
    S_B^{(2)}&=&\frac{T_{D3}}{4}\int d^4\sigma\left[\frac{1}{r^2} \left(\frac{1}{L^4\sinh^4(u_{eq})+
   k^2}\left[2k r^2 \sqrt{\sinh^2(u_{eq})L^4-k^2} \left(f_{t\phi}f_{r\theta}-f_{t\theta}f_{r\phi}\right.\right.\right.\right.\nn\\&&\left.\left.\left.\left.+
   f_{tr}f_{\theta\phi}\right) + L^4 \cosh (u_{eq}) \csc (\theta ) \sinh ^3(u_{eq})
   \left(\left(-f_{\theta\phi}^2+r^2\left(f_{t\phi}^2-f_{r\phi}^2\right)\right)\right.\right.\right.\right.\nn\\&&
   \left.\left.\left.\left.+r^2\sin\theta^2\left(f_{t\theta}^2-f_{r\theta}^2+r^2f_{tr}^2\right)\right)\right]\right)\right.\nn\\&&\left.-\frac{L^4\cosh ( u_{eq})
    \sinh ( u_{eq})}{r^2} \left(\csc (\theta) \partial_\phi\delta u^2
   +\sin (\theta ) \left(\left(\partial_r\delta u^2-\partial_t\delta u^2\right) r^2+\partial_\theta\delta u^2\right)\right)\right]\nn\\
\eea

Nótese que el primer término se anula tanto para el operador de Wilson ($k=0$) como para el de 't Hooft ($q=0$) de lo
 cual se sigue que este es un termino enteramente originado por el campo diónico. Introduciendo la métrica deformada $\hat g$ (métrica de cuerda abierta)
\bea\label{deformed metric}
    \hat{g}_{\alpha\beta}=g_{\alpha\beta}-F_{\alpha\gamma}g^{\gamma\delta}F_{\delta\beta}
    &=\frac{\sinh ^4(u_{eq}) L^4+k^2}{L^2\sinh^2(u_{eq})}\left(
                                            \begin{array}{cccc}
                                              {\displaystyle-\frac{1}{r^2}} & 0 & 0 & 0 \\
                                              0 & {\displaystyle\frac{1}{r^2}} & 0 & 0 \\
                                              0 & 0 & 1 & 0 \\
                                              0 & 0 & 0 & \sin^2\theta \\
                                            \end{array}
                                          \right),
\eea
la ecuación \eqref{dionic fluct B} toma la forma
\bea\label{Sb fluct dyon}
    S_B^{(2)}&=&\frac{T_{D3}}{2}\left(\frac{L^4\cosh(u_{eq})\sinh^3(u_{eq})}{L^4\sinh^4(u_{eq})+k^2}\right)
    \int d^4\sigma\left[\sqrt{-\hat{g}}\left(L^2\hat{g}^{\alpha\beta}\partial_{\alpha}\delta u\partial_{\beta}\delta u+\frac12\,\,\
    \hat{g}^{\alpha\beta}\hat{g}^{\gamma\delta}f_{\alpha\gamma}f_{\beta\delta}\right)\right.\nn\\&&\left.+
    \frac{k \sqrt{\sinh^2(u_{eq})L^4-k^2}}{L^4\cosh(u_{eq})\sinh^3(u_{eq})}\left(f_{t\phi}f_{r\theta}-f_{t\theta}f_{r\phi}+
   f_{tr}f_{\theta\phi}\right)\right]\nn\\
    &=&\frac{T_{D3}}{2}\left(\frac{L^4\cosh(u_{eq})\sinh^3(u_{eq})}{L^4\sinh^4(u_{eq})+k^2}\right)
    \int d^4\sigma\sqrt{-\hat{g}}\left(L^2\hat{g}^{\alpha\beta}\partial_{\alpha}\delta u\partial_{\beta}\delta u+\frac12\,\,\
    \hat{g}^{\alpha\beta}\hat{g}^{\gamma\delta}f_{\alpha\gamma}f_{\beta\delta}\right.\nn\\&&\left.+
    \frac{k\, q}{2L^4\sinh^2(2u_{eq})}\hat{\epsilon}^{\alpha\beta\gamma\delta}f_{\alpha\beta}f_{\gamma\delta}\right),
\eea
en donde $\hat{\epsilon}^{\alpha\beta\gamma\delta}=\frac{\epsilon^{\alpha\beta\gamma\delta}}{\sqrt{-\hat{g}}}$. Nótese que como consecuencia de lo comentado luego de \eqref{dionic fluct B} se obtiene un término proporcional a $\tilde f f$ cuya constante de proporcionalidad es $ \frac{k\, q}{2L^4\sinh^2(2u_{eq})}$ de lo cual se observa que es una contribución propia de este operador de Wilson-'t Hooft.

~

\begin{itemize}
\item {\bf Fermiónicas}
\end{itemize}

A continuación escribiremos el resultado para las fluctuaciones de la acción fermiónica
para la D3 brana,  su derivación se relegará a la sección \ref{fluctfermD3} del apéndice.

Puesto que en la solución clásica \eqref{q vs k} los campos fermiónicos están apagados, $\Theta$ en \eqref{FermionicAction} representa la fluctuación
propiamente dicha, el resultado final es
\be
S_F^{(2)}=\frac{T_{D3}}{2}\left(\frac{L^4\cosh(u_{eq})\sinh^3(u_{eq})}{L^4\sinh^4(u_{eq})+k^2}\right)\int d^4\sigma\sqrt{\hat{g}}\,\overline{\Theta}\,
\hat{\Gamma}^{\alpha}\hat{\nabla}_{\alpha}\Theta,
\ee
en donde $\hat g$ se definió en \eqref{deformed metric} y en esta ultima expresión el espinor $\Theta$ tiene 16 componentes reales debido a que el fijado de gauge de la simetría $\kappa$, $\tilde\Gamma\Theta=\Theta$ anula la mitad de las componentes. Sorprendentemente una series de redefiniciones y rotaciones del campo fermiónico permiten reescribir la acción resultante como una acción para un fermion libre en espacio curvo.

\section{Conclusiones}

En el capitulo \ref{Wh} se ha analizado la propuesta de la conjetura de Maldacena para calcular lazos de Wilson
via teoría de cuerdas en fondos gravitatorios y se ha estudiado su estabilidad ante perturbaciones lineales.

La prescripción del lado de cuerdas involucra calcular el área mínima
para la hoja de  mundo de una cuerda cuyos extremos están sobre el lazo
y ubicados en un valor fijo de la coordenada holográfica. Cuando los extremos
se ubican en infinito se obtiene un area divergente y es necesario
realizar una regularización para obtener un resultado que tenga sentido.
En la sección \ref{wil} se ha resumido esta prescripción y se ha mostrado cómo
obtener un resultado finito. Se ha elegido regularizar la acción mediante
el procedimiento original propuesto en  \cite{rey, maldawilson}. Este proceso
tiene en cuenta que en el cálculo del área mínima de la acción de Nambu-Goto hay
contribuciones de la auto-energía (masa) de los quarks externos. Con esta
interpretación se reproducen resultado conocidos para $AdS$ y $AdS$ térmico.
De echo, la regularización es responsable de cambiar el área positiva en una energía
potencial negativa y atractiva. Cuando se estudian fondos gravitatorios que terminan
suavemente ($AdS$ en coordenadas globales, MN y KS) nos encontramos con un rompecabezas
debido a que las cuerdas rectas que se extienden a lo largo de la dirección
redial usadas para la substracción deben terminar en algún lugar en el volumen del
espacio.
Se concluye que la interpretación correcta para el proceso de substracción
es que se está comparando la hoja de mundo de la cuerda con respecto a un estado
de referencia que consiste de la hoja de mundo de dos cuerdas rectas cuyos
extremos están en las antípodas de una dirección compacta (representada gráficamente
en la fig.\ref{substraction}). Se dice entonces que el estado de referencia, en general,
satisface diferentes condiciones de borde que la hoja de mundo con la que se calcula
el valor de expectación del lazo de Wilson. Esta última observación es bienvenida para los
casos de MN y KS en donde ocurre confinamiento lineal para hojas de mundo que poseen
energía regularizada positiva (ver figs. \ref{EwilsonMN} and \ref{EwilsonKS}): si el estado
de referencia satisficiera las mismas condiciones de borde que la hoja de mundo
del lazo de Wilson, el comportamiento lineal obtenido no debería ser tomado en cuenta debido
a que el estado de referencia  ($E_{q\bar q}=0$) sería el de menor energía ( confrontar
con el último párrafo de la sección \ref{adssch}), pero del análisis previo se ve que este
no es el caso. Me gustaría recordar una observación de \cite{groo} que establece que la relación
entre valores de expectación de lazos de Wilson y cuerdas en duales gravitatorios (al nivel semiclásico)
\be
\langle W\rangle\simeq e^{-A}
\ee
es esquemático debido a que la adición de términos de borde a la acción de Nambu-Goto
no cambia el área mínima de las soluciones pero cambia el valor de la acción clásica en
algo diferente al area. En \cite{groo} esta arbitrariedad fue usada para realizar una transformación de
Legendre de la acción de Nambu-Goto mostrando que la cantidad resultante, para el caso de lazos
en $AdS$, está libre de divergencias lineales que provienen del comportamiento de la
hoja de mundo cerca del borde de $AdS$.

También se han discutido las condiciones de concavidad (\ref{convexity}) que deben ser satisfechas
por cualquier potencial que pretenda describir la interacción entre quarks físicos. Duales
gravitatorios genéricos tienen funciones $f(r)$ crecientes y positivos, por lo tanto las condiciones
de concavidad no se satisfacen cuando la función longitud es una función creciente de la posición radial
mínima, $r_0$, alcanzada por la cuerda. En la sección \ref{bkg} se analizaron las funciones longitud y potencial,
$L(r_0)$ y $V_{\sf string}(L)$, para diferentes fondos de gravedad y se mostró que algunos de ellos
conducen a soluciones del embedding de la cuerda que no satisfacen las condiciones de concavidad.

Basándome en los trabajos previos \cite{pufu}-\cite{avramis} he estudiado fluctuaciones lineales
alrededor del embedding para testear su estabilidad. Se concluyó que cuando la solución conduce a un
potencial no físico, que no satisface las condiciones (\ref{convexity}), existen modos inestables ante
fluctuaciones lineales. Durante el análisis se han discutido los diferentes fijados de gauge que pueden
imponerse y su relación con el difeomorfismo de la acción de Nambu-Goto. Se estudiaron tres fijados de gauge
naturales y se eligió trabajar en el gauge $r$ debido a que se obtienen expresiones simples y cerradas para
las ecuaciones de movimiento (ver ec. (\ref{lagr})). El gauge $r$ nos conduce a un comportamiento singular
para las fluctuaciones en el tip de la solución, pero refiriendo a \cite{avramis} se ha mostrado
que dichas fluctuaciones son físicas.

En la sección \ref{stab} se estudió la estabilidad de las soluciones analizadas en la sección \ref{bkg}.
Se mostró mediante un análisis numérico que los fondos de gravedad $AdS_5\times S^5$, Maldacena-N\'u\~nez y Klebanov-Strassler
son estables. Por otro lado para $AdS$ térmico y la generalización de Maldacena-N\'u\~nez de la sección
\ref{gmnsol} se encontraron modos inestables, en concordancia con el comportamiento de $L(r_0)$. Este
último caso es patológico porque el lazo no puede estar ubicado en infinito y además, se encontró
que existe un mínimo de separación más allá del cual no existen soluciones que conecten los extremos
de la cuerda.
En la sección \ref{scho} se han transformado las ecuaciones de movimiento para las fluctuaciones ( que son del tipo Sturm-Liouville),
en ecuaciones de Schr\"{o}dinger con el objetivo de reanalizar el problema de estabilidad. Este análisis se
encontró en perfecta concordancia con los resultados obtenidos en \ref{stab}.
Se concluye que las regiones en donde se encuentran modos inestables coincide con las regiones en
donde la condición de concavidad no se satisface.

\vspace{1.5mm}

En la sección \ref{thooftloop} se estudio el caso de la interacción monopolo-antimonopolo en los casos de
Maldacena-Nuñez y Klebanov-Strassler. Se discutió la prescripción de la conjetura
para calcular lazos de ´t Hooft, la cual consiste en enroscar una $D3$ sobre una $S^2$ topológica presente
en la geometría. El caso de MN mostró ser inestable para todo valore de $r_0$. Fue factible
un análisis de la fluctuación en el plano debido a que su ecuación de movimiento
estaba desacoplada de las fluctuaciones angulares, se encontraron, mediante un análisis numérico,
los modos inestables. El comportamiento del caso de KS, similar al $AdS$ térmico, tiene presumiblemente
regiones estables e inestables, pero un análisis de fluctuaciones nos conduce a ecuaciones
de movimiento acopladas que aún no se han analizado.

Se concluye que el análisis de lazos de Wilson/´t Hooft calculados en duales gravitatorios estudiando
el valor de $f^2$ en el origen debe ser acompañado de un análisis de la función $L(r_0)$.

En la última sección se calcularon fluctuaciones tanto bosónicas como fermiónicas para el dual gravitatorio al valor de expectación de un lazo de Wilson- 't Hooft evaluado en la representación simétrica del grupo de gauge. En este caso aún hay trabajo por hacer, ya que puede continuarse el analisis estudiando en que representaciones del grupo de simetría transforman dichas fluctuaciones.

\def\baselinestretch{1.66}

\def\baselinestretch{1}
\chapter{Superconductores Holográficos}\label{SC}
\def\baselinestretch{1.66}

Luego de los trabajos \cite{Hartnoll, Horowitz} la conjetura AdS/CFT se convirtió en una herramienta útil para estudiar sistemas de materia condensada.
En particular, por medio del análisis de la teoría semiclásica de gravedad se han estudiado desde un punto de vista teórico sistemas de materia condensada
fuertemente correlacionados (ver \cite{lecturesadssmt}).

En este capítulo se analizará la deformación de la geometría del espacio-tiempo dual a un superconductor\footnote{La diferencia sutíl
entre un superconductor y un superfluído proviene de lo siguiente: en ambos se produce la ruptura espontanea de una simetría pero en el primer
caso esta es una simetría local mientras que en el segundo caso es global. En esta tesis no se va a hacer diferencia entre uno y otro debido a que
para los fenómenos considerados esta distinción no produce ninguna diferencia.} tipo $p$ en 3+1 dimensiones \cite{Gubserp} (véase \cite{Ammon, Zhang} para
un tratamiento similar en 4+1 dimensiones). En el camino se reproducirá la deformación del dual gravitatorio al superconductor tipo $p+ip$ estudiado
en \cite{Gubserpip}. También se usará la prescripción dada en \cite{Ryu,rt,Takayanagi} para calcular la entropía de entrelazamiento desde el punto de vista
holográfico para ambos duales gravitatorios. Cálculos similares de la entropía de entrelazamiento en pueden encontrarse en \cite{Albash, Cai1, Cai2, Cai3}.

La superconductividad tipo $p$  es una fase de la materia que se produce cuando los electrones con momento angular relativo $j=1$ condensan formando pares
de Cooper. En otras palabras, existe un vector cargado ante una simetría $U(1)$ que condensa. Este tipo de superconductores se originan a partir de
electrones fuertemente correlacionados y por lo tanto la utilización de la teoría de Bardeen, Cooper y Schrieffer (BCS)
 no es la manera correcta de describir su dinámica microscópica.
Este fenómeno es un desafío para la física teórica que, debido a la propiedad fundamental de la conjetura de describir teorías de campos fuertemente acopladas,
puede ser estudiado a partir de su dual gravitatorio. Se introducirán ahora los ingredientes mínimos necesarios para reproducir
la dinámica del superconductor. Este tipo de enfoque nos permite reproducir las propiedades del sistema de materia condensada sin explicar su origen
microscópico.

En los trabajos \cite{Erdmenger2,Basu, Erdmenger3, Peeters, Ammon2, Kaminski} fueron estudiados superconductores tipo $p$ a partir de una teoría de
cuerdas dual. Los mínimos ingredientes necesarios en la teoría gravitatoria para tener temperatura finita, potencial químico y ruptura espontánea
de simetría (SSB) son: una geometría de agujero negro y un campo de gauge no Abeliano  \cite{Hartnoll2, Hartnoll3}. Las soluciones a ser consideradas
son geometrías asintóticamente AdS con un campo de gauge $SU(2)$. La SSB se realiza mediante una condición asintótica no trivial (pelo)
sobre este campo de gauge. El potencial químico y la SSB provienen de prender dos direcciones independientes dentro del grupo de gauge no Abeliano.
La ruptura espontánea de simetría ocurre en el lado gravitatorio a través de la formación de un condensado fuera del horizonte.

La entropía de entrelazamiento (EE) entre un subsistema ${\cal A}$ y su complemento ${\cal B}$ está definida por la entropía de von Neumann
\be
{\cal S_{\cal A}}=-Tr_{{\cal A}}(\rho_{\cal A}\ln\rho_{\cal A}).
\ee
Aquí $\rho_{\cal A}=Tr_{\cal B}(\rho)$ es la matriz densidad obtenida de tomar la traza sobre los grados de libertad del subsistema ${\cal B}$
en la matrix densidad del sistema completo $\rho$. Inocentemente $S_{\cal A}$ mide la cantidad de información que está oculta dentro de ${\cal B}$
cuando subdividimos el sistema. Desde el punto de vista de la teoría de gravedad la EE fue conjeturada \cite{Ryu} proporcional al valor del area mínima
para una superficie, $\gamma_{\cal A}$, en el volumen cuyo borde en infinito coincide con el borde de la región  ${\cal A}$ (véase \cite{Takayanagi})
\be
{\cal S_{\cal A}}= \frac{2\pi Area(\gamma_{\cal A})}{\kappa^2}.
\ee
Aquí $\kappa$ es la constante gravitatoria. Nótese que la entropía térmica estándar se obtiene como un caso particular de la EE cuando la región ${\cal A}$
es el sistema entero.
En \cite{Casini} los autores demostraron cómo esta técnica holográfica para calcular la EE arroja el resultado correcto al estudiarse el caso particular
de una superficie esférica y a temperatura cero en teorías de campos conformes. Al final de este capítulo se calculará la EE para una geometría con forma de
banda en los duales gravitatorios correspondientes a los superconductores tipo $p$ y $p+ip$.

\section{Superconductores tipo $p$ y $p+ip$}\label{holsup}

Como ya ha sido mencionado, el dual gravitatorio a un superconductor tipo $p$ se modela con una teoría de Einstein-Yang-Mills (EYM). En \cite{Gubserp, Gango}
fue estudiada dicha teoría de gravedad en 3+1 dimensiones y en su límite de prueba, es decir, despreciando la deformación en la geometría debida
al campo de gauge. Además, estos autores mostraron que la teoría dual al superconductor tipo $p+ip$ estudiada en \cite{Gubserpip} es inestable
ante pequeñas fluctuaciones siendo la configuración estable aquella para el superconductor tipo $p$. En esta sección se tendrán en cuenta las deformaciones
de la geometría 3+1 dimensional dual al superconductor $p$ producidas por el campo de gauge no Abeliano y se compararán los resultados con los obtenidos para
el caso del $p+ip$.

\begin{figure}[htbp]
\centering
\includegraphics[width=0.6\textwidth]{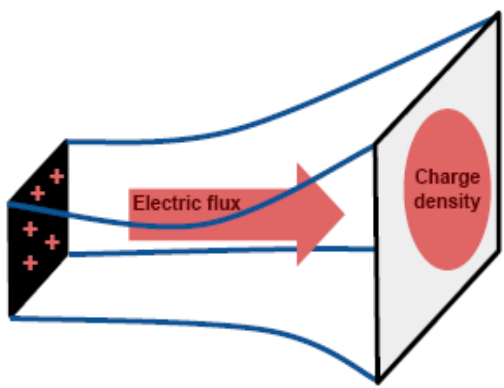}
\caption{El agujero negro AdSRN plano. La fuente de la densidad de carga se encuentra en el horizonte del agujero negro.}
\label{RNSC}
\end{figure}

Se trabajará con un grupo de gauge $SU(2)$. En el caso del $p+ip$, se propondrá un campo de gauge tal que rompa el subgrupo $U(1)$ del $SU(2)$ y la simetría
rotacional $SO(3)$ del espacio pero que deje invariante un subgrupo diagonal de ellos. Por otro lado, en el dual al superconductor tipo $p$, el campo de gauge
romperá ambas simetrías $U(1)$ por completo. La solución gravitatoria que describe la dinámica de acoplamiento fuerte en ambos tipos de superconductores
se describe de la siguiente manera (ver figura \ref{RNSC}): se forma una capa superconductora cargada fuera del horizonte producida por la competencia entre la repulsión
eléctrica (con el agujero negro cargado) y el potencial gravitatorio de la geometría asintóticamente AdS. A temperaturas suficientemente altas no se produce pelo
fuera del agujero negro y la solución es AdS-Reissner-Nordström (AdSRN). Por debajo de una temperatura crítica $T_c$ se genera un campo de gauge no trivial
con potencial químico no nulo sobre el borde de la geometría y aparece un condesado sin necesidad de tener una fuente para el mismo lo cual origina un
rompimiento espontáneo de la simetría de gauge $SU(2)$.

\subsection{Superconductor tipo $p$ en 2+1 dimensiones}\label{pwave}
\subsubsection{Solución}

Comencemos a partir de una teoría de Yang-Mills en $3+1$ dimensiones y con grupo de gauge
 $SU(2)$ en un espacio AdS (ver \cite{Winst}), su densidad Lagrangiana es

\be
\label{action}
\kappa_{(4)}^2{\cal L} =R -2\Lambda-\frac14Tr(F_{\mu\nu}F^{\mu\nu})
\ee
en donde $\Lambda=-\frac{3}{{\hat R}^2}$, $\kappa_{(4)}$ es la constante gravitacional en cuatro dimensiones y el $F_{\mu\nu}$ para el campo de gauge es
\begin{equation}
\label{energymomentumtensor}
F^a_{\mu \nu}=\partial_\mu A^a_\nu-\partial_\nu A^a_\mu+g_{_{YM}}\epsilon^{abc}A^b_\mu A^c_\nu
\end{equation}
con $g_{_{YM}}=\frac{\hat g_{_{YM}}}{\kappa_{(4)}}$ denotando al parámetro que mide la deformación de la geometría y
$\hat g_{_{YM}}$ el acoplamiento de Yang-Mills usual. Se usarán letras latinas como índice sobre el grupo $SU(2)$ y letras griegas para denotar
coordenadas del espacio-tiempo. Escaleando el campo de gauge como $\tilde A= \frac{A}{g_{_{YM}}}$ se puede ver que el límite de $g_{_{YM}}$ grande
corresponde al límite de prueba del campo de gauge. Aproximadamente uno puede pensar que $\frac{1}{\hat g_{_{YM}}^2}$ cuenta el número de grados
de libertad cargados ante el $SU(2)$ de la teoría dual. Además, $\frac{1}{\kappa_{(4)}^2}$ cuenta el número total de grados de libertad. Considerar
la deformación producida por el campo de gauge significa que el número de grados de libertad de los estados cargados es del mismo orden
que el número de grados de libertad del sistema.

Las ecuaciones de movimiento obtenidas a partir de minimizar la acción son
\bea\label{EYM}
G_{\mu\nu}= R_{\mu\nu}-\frac12g_{\mu\nu} R&=&\frac{3}{R^2} g_{\mu\nu}+\frac12Tr[F_{\mu\gamma}F_\nu^\gamma]-\frac{g_{\mu\nu}}{8}Tr[F_{\gamma\rho}F^{\gamma\rho}]\\
\nonumber
\\
D_\mu F^{\mu\nu}&=&0\label{Max}
\eea
y con el objetivo de resolverlas se propondrá la siguiente solución \cite{Ammon, Manvelyan}
\begin{equation}
\label{metricansatz}
ds^2 = -M(r)\sigma(r)^2dt^2 + \frac{1}{M(r)}dr^2 +r^2 h(r)^{2}dx^2 + r^2h(r)^{-2} dy^2 \,,
\end{equation}
para la geometría de fondo, y
\begin{equation}
\label{gaugefieldansatz}
A=\phi(r)\tau^3 dt+\omega(r)\tau^1d x\,,
\end{equation}
para el campo de gauge. Aquí se ha usado la notación matricial $A=A_\mu^a\tau^adx^\mu$ con $\tau^a=\frac{\sigma^a}{2i}$ y $\sigma^a$ las matrices de Pauli usuales, los generadores de $SU(2)$ satisfacen $[\tau^a,\tau^b]=\epsilon^{abc}\tau^c$. Un solución que satisfaga que $\omega\neq0$ en el campo
de gauge \eqref{gaugefieldansatz} romperá la simetría $U(1)$ asociada con rotaciones alrededor de $\tau^3$ (usualmente denominada $U(1)_3$). Además si $h\neq0$ se romperá la simetría $U(1)_{xy}$ asociada a rotaciones en el plano $xy$. A temperaturas suficientemente altas se espera que no haya pelo fuera del agujero negro y la solución con condensado nulo es AdSRN con
\bea
\omega(r)&=&0,\,\,\,\nn\\  h(r)&=&1 \nn\\\sigma(r)&=&1,\,\,\, \nn\\ \phi(r)&=&\mu\left(1-\frac{r_h}{r}\right),\nn\\ M(r)&=&r^2+\frac{\mu^2r_h^2}{r^2}-\left(\frac{\mu^2}{8}+r_h^2\right)\frac{r_h}{r}.
\eea

Reemplazando la solución propuesta en las ecuaciones de movimiento para la acción de EYM se obtienen 5 ecuaciones, tres de ellas son ecuaciones diferenciales de segundo orden y las dos restantes son ligaduras de primer orden
\bea
M'&=&\frac{3r}{{\hat R}^2}-\frac{1}{8\sigma^2}\left(\frac{g_{_{YM}}^2\phi^2\omega^2}{rh^2M}+r\phi'^2\right)-M
\left(\frac{1}{r}+\frac{rh'^2}{h^2}+\frac{\omega'^2}{8rh^2}\right)\nn\\
\sigma'&=&\frac{\sigma}{h^2}\left(r
h'^2+\frac{\omega'^2}{8r}\right)+\frac{g_{_{YM}}^2\phi^2\omega^2}{8r
M^2h^2\sigma};\nn\\
h''&=&\frac{1}{8r^2h}\left(-\omega'^2+\frac{g_{_{YM}}^2\phi^2\omega^2}{M^2\sigma^2}\right)-h'\left(\frac{2}{r}-\frac{h'}{h}+\frac{M'}{M}+\frac{\sigma'}{\sigma}
\right);\nn\\
\omega''&=&-\frac{g_{_{YM}}^2\phi^2\omega}{M^2\sigma^2}+\omega'\left(\frac{2h'}{h}-\frac{M'}{M}-\frac{\sigma'}{\sigma}\right);\nn\\
\phi''&=&\frac{g_{_{YM}}^2\phi\,\omega^2}{r^2h^2M}-\phi'\left(\frac{2}{r}-\frac{\sigma'}{\sigma}\right).
\label{eomsp}
\eea
Este sistema de ecuaciones goza de 4 simetrías de escala que serán útiles cuando intentemos resolverlas numéricamente, ellas son
\begin{enumerate}
    \item $\sigma\rightarrow\lambda\sigma,~~~~\phi\rightarrow\lambda\phi$
    \item $\omega\rightarrow\lambda\omega,~~~~h\rightarrow\lambda h$
    \item $M\rightarrow\lambda^{-2} M,~~~~\sigma\rightarrow\lambda \sigma,~~~~g_{_{YM}}\rightarrow\lambda^{-1}g_{_{YM}},~~~~{\hat R}\rightarrow\lambda {\hat R}$
    \item $M\rightarrow\lambda^{2} M,~~~~ r\rightarrow \lambda r,~~~~\phi\rightarrow\lambda\phi,~~~~\omega\rightarrow\lambda\omega $
\end{enumerate}
Usando estas simetrías de escala podemos elegir $\hat R=r_h=1$ y fijar el valor en el borde de las funciones de la métrica $\sigma(\infty)=h(\infty)=1$. La geometría y el campo de gauge deben ser regulares en el horizonte, esto motiva la siguiente expansión en el IR ($r$ pequeño)
\bea\label{IRp}
M&=&M_1(r-r_h)+M_2(r-r_h)^2+\ldots\nn\\
h&=&h_0+h_2(r-r_h)^2+\ldots\nn\\
\sigma&=&\sigma_0+\sigma_1(r-r_h)+\sigma_2(r-r_h)^2+\ldots\nn\\
\omega&=&\omega_0+\omega_2(r-r_h)^2+\omega_3(r-r_h)^3+\ldots\nn\\
\phi&=&\phi_1(r-r_h)+\phi_2(r-r_h)^2+\ldots
\eea
Por otro lado, en el UV ($r$ grande) el desarrollo deseado es
\bea
M&=& r^2+\frac{M_1^b}{r}+\frac{(\omega_1^b)^2+\rho^2}{8r^2}+\ldots\nn\\
h&=&1+\frac{h_3^b}{r^3}-\frac{(\omega_1^b)^2}{32r^4}+\ldots\nn\\
\sigma&=&1-\frac{(\omega_1^b)^2}{32r^4}+\ldots\nn\\
\omega&=&\omega_0^b+\frac{\omega_1^b}{r}-\frac{g_{_{YM}}^2\mu^2\omega_1^b}{6r^3}+\ldots\nn\\
\phi&=&\mu+\frac{\rho}{r}+\frac{g_{_{YM}}^2\mu^2\omega_1^b}{12r^4}+\ldots\label{bcp}
\eea
Para dar cuenta de la SSB, se buscan soluciones en donde la componente no normalizable se anule $\omega_0^b=0$. El diccionario estándar de AdS/CFT nos permite interpretar al valor de borde y al sub-leading (segundo termino más relevante en la expansión) de $\phi$ como el potencial químico $\mu$ y la densidad de carga $\rho$ de la teoría de campos dual \cite{KW}. Además, el coeficiente sub-leading $M_1^b$ en la expansión en el borde de $g_{tt}$ coincide con la acción Euclídea regularizada evaluada en la solución de las ecuaciones de movimiento \cite{Hartnoll}. El coeficiente normalizable en $\omega$ es dual al valor de expectación de vació de la corriente  $\langle J_x^1\rangle\propto \omega_1^b$ y hace las veces de parámetro de orden para el sistema.

\begin{figure}[ht]
\begin{minipage}{8cm}
\begin{center}
\includegraphics[width=8cm]{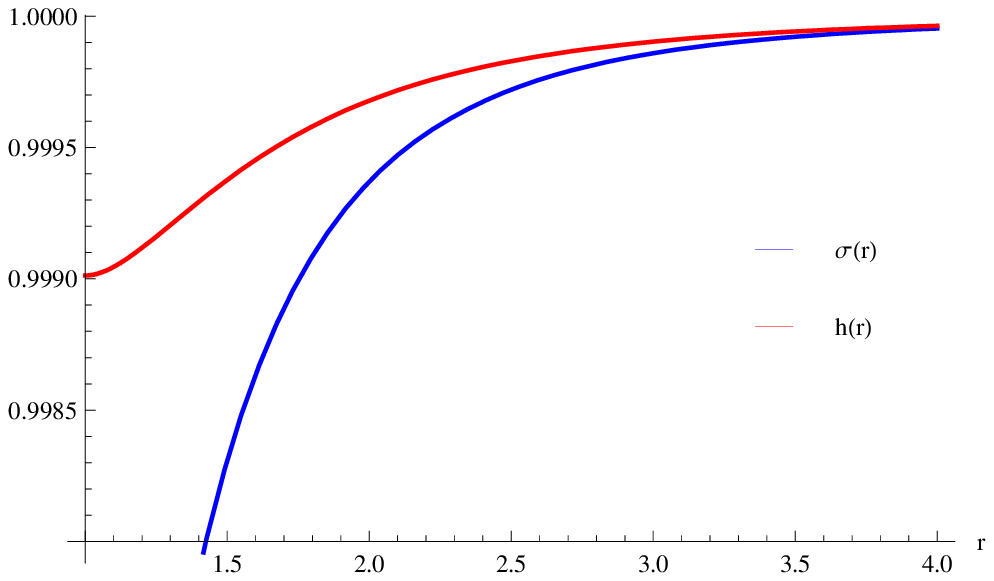}
\caption{Las funciones adimensionales de la métrica $\sigma(r)$ y $h(r)$ para $g_{_{YM}}=2$ y $T=0.2312\mu$.}
\label{hysigmap}
\end{center}
\end{minipage}
\  \
\hfill \begin{minipage}{8cm}
\begin{center}
\includegraphics[width=8cm]{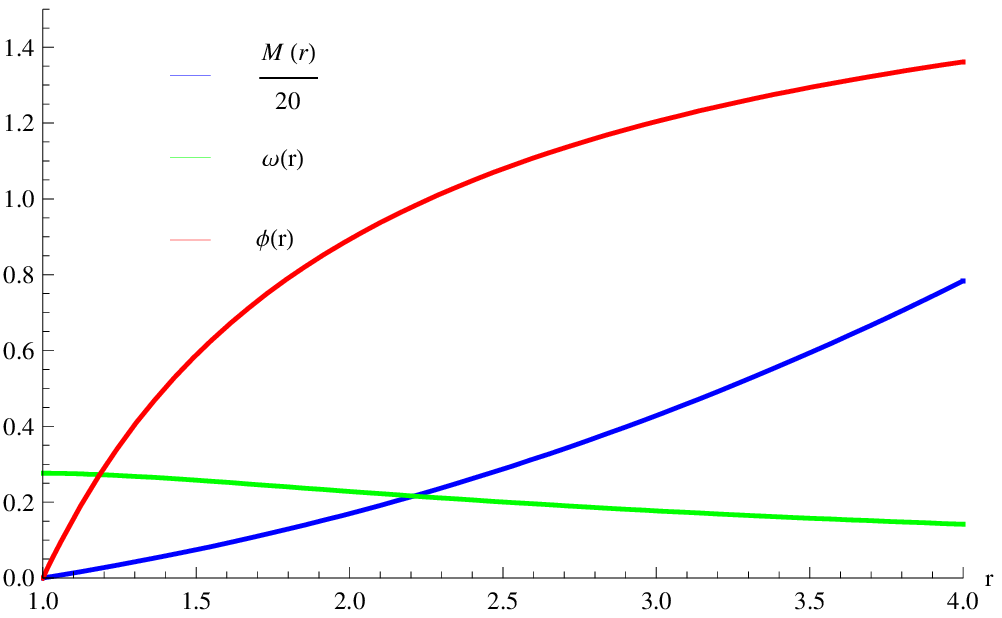}
\caption{Las funciones adimensionales de la métrica $M(r)$ y las funciones del campo de gauge $\omega(r)$ y $\phi(r)$ para $g_{_{YM}}=2, T=0.2312\mu$.}
\label{funcionesp}
\end{center}
\end{minipage}
\end{figure}

Las soluciones de las ecuaciones \eqref{eomsp} dependen de cuatro coeficientes IR $\phi_1,\omega_0,h_0,\sigma_0$
y del parámetro de deformación $g_{_{YM}}$. Todos los otros coeficientes en \eqref{IRp} pueden ser escritos en términos de estos.
Se procederá a integrar las ecuaciones de movimiento numéricamente desde el horizonte hacia el borde usando un método de shooting con
el objetivo de tener el comportamiento asintótico deseado. Se exploró el rango $g_{_{YM}}\in [0.85,24]$ y se observó que el comportamiento de
 las funciones no cambia cualitativamente al variar $g_{_{YM}}$. En las figuras \ref{hysigmap} y \ref{funcionesp} se muestran las
 soluciones de \eqref{eomsp} con condiciones de borde \eqref{bcp}. Se ha usado el potencial químico $\mu$ para adimensionalizar. Esto
 significa que se esta trabajando en el conjunto gran canónico.

\subsubsection{Termodinámica}

En esta sección se calcularán las cantidades termodinámicas asociadas con las soluciones. A partir del estudio de la función potencial en
el conjunto\footnote{Para pasar del conjunto canónico ($\mu$ fijo) con energía libre $\Omega$, al conjunto gran canónico ($\rho$ fijo) con energía libre $F$,
 se debe agregar un termino de borde a la acción Euclídea. Esto cambia el problema variacional e implica la relación de Gibbs $F=\Omega+\mu\rho$.} gran
 canónico se observará una transición de fase de segundo orden entre una fase superconductora y una normal.

\begin{figure}[h]
\begin{center}
\includegraphics[width=7.7cm]{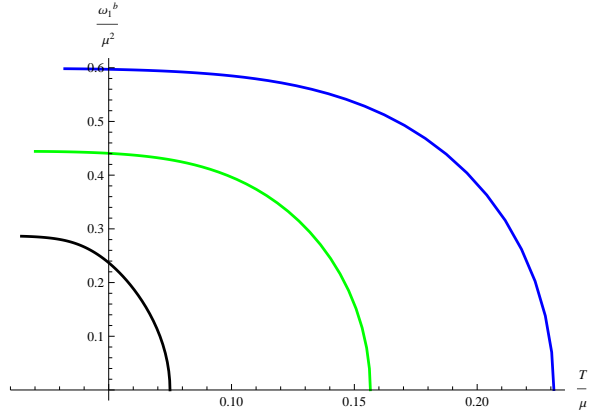}
\caption{ La figura muestra el comportamiento del coeficiente normalizable de la función $\omega$, el cual es proporcional al condensado $\langle J_x^1\rangle$. Las líneas negra, verde y azul refieren a soluciones con $g_{_{YM}}=1, 1.5, 2$ y $T_c=0.0749, 0.1565, 0.2312$ respectivamente. Nótese que el condensado se anula para $T>T_c$.}
\label{condp}
\end{center}
\end{figure}

\begin{figure}[h]
\begin{minipage}{7.7cm}
\begin{center}
\includegraphics[width=7.7cm]{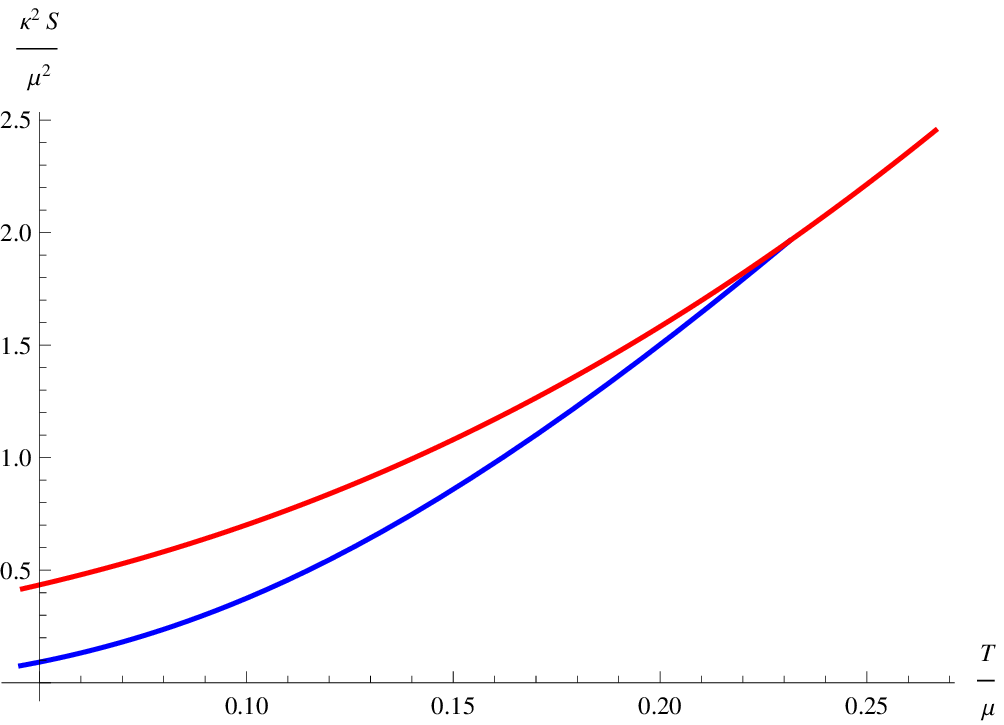}
\caption{ Entropía como función de la temperatura. La línea azul es para la fase superconductora con $g_{_{YM}}=2$ y la línea roja refiere a la fase normal (geometría AdSRN). Hay una transición de fase de segundo orden para $T=T_c=0.2312$.}
\label{EvsTp}
\end{center}
\end{minipage}
\ \
\hfill\begin{minipage}{7.7cm}
\begin{center}
\includegraphics[width=7.7cm]{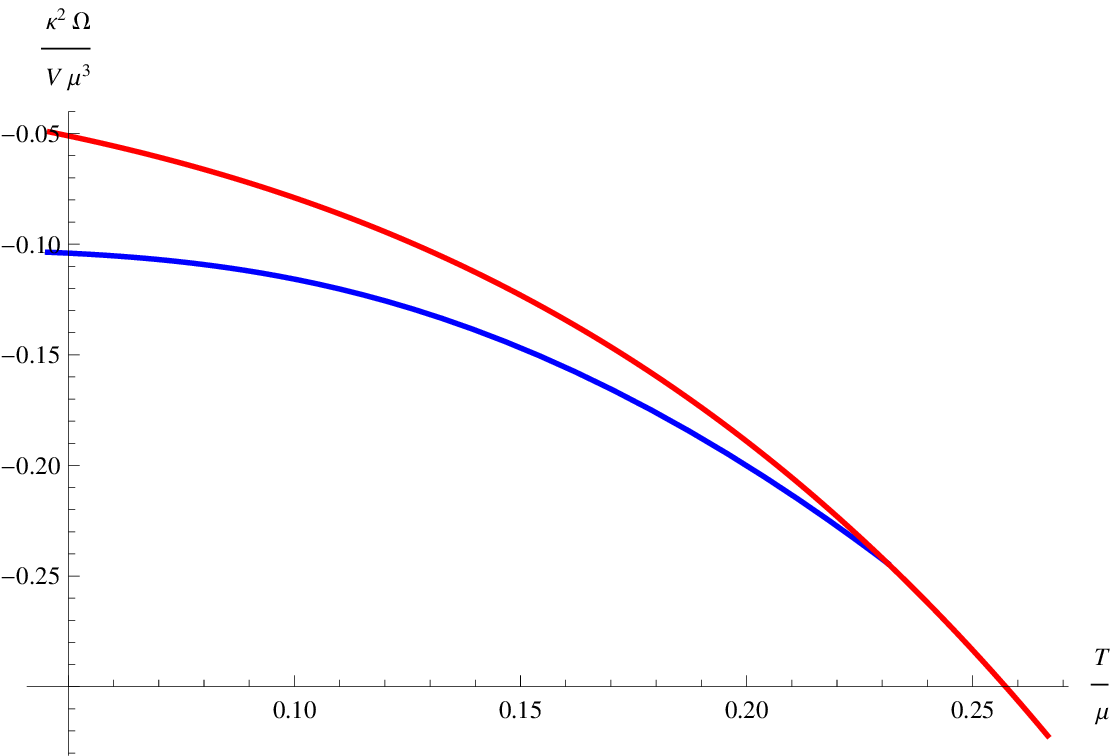}
\caption{Función potencial $\Omega$ calculada a partir de \eqref{omega} como unción de la temperatura $T$ para $g_{_{YM}}=2$. La línea roja es el potencial para la solución de RN y la azul para la fase superconductora.}
\label{freep}
\end{center}
\end{minipage}
\end{figure}

La temperatura de la teoría de campos dual está dada por la temperatura de Hawking del agujero negro
\be
T=\frac{M_1\sigma_0}{2\pi}=\frac{1}{16\pi}\left(24\sigma_0^2-\phi_1^2\right)r_h
\ee
en donde la segunda igualdad proviene de la consistencia de la expansión en serie\eqref{IRp} que
relaciona el coeficiente $M_1$ con $\sigma_0$ y $\phi_1$. El área del horizonte, $A_h$, conduce a la entropía
\be
S=\frac{2\pi}{\kappa^2_{(4)}} A_h
=\frac{2\pi^2VT^2}{\kappa^2_{(4)}}\frac{12^2}{\left(24\sigma_0^2-\phi_1^2\right)^2}\label{Sp}
\ee
en donde $V=\int dx\,dy$. En la figura \ref{condp} se muestra el parámetro de orden $\omega_1^b$ (es decir, el valor de expectación de la corriente $\langle J_x^1\rangle$) como función de la temperatura. Nótese que a $T=T_c$ el condensado se anula mostrando la desaparición de el estado superconductor para $T>T_c$. A partir de los resultados numéricos se encuentra que cerca de la temperatura crítica $T_c$ la corriente satisface $\langle J_1^x\rangle\propto(1-\frac{T}{T_c})^{1/2}$ y por lo tanto el valor del exponente crítico es $1/2$. La figura \ref{EvsTp} muestra el comportamiento de la entropía de Bekenstein-Hawking \eqref{Sp} como función de la temperatura para mi solución y la del agujero negro AdSRN.

La correspondencia AdS/CFT identifica la acción Euclídea on-shell de la teoría de gravedad $S_E$ multiplicada por la
temperatura $T$ como la función potencial $\Omega$ del sistema en el conjunto gran canónico. Para calcularla, se hará una continuación a
signatura Euclídea y se compactificará el tiempo con periodo $\frac{1}{T}$ para evitar singularidades. La acción on-shell tiene un factor $\frac{1}{T}$ proveniente de la integración en la coordenada temporal, escribiendo $ S_{on-shell}=\frac{\tilde S_{bulk}}{T}$ se tiene
\be\label{onshellp}
\tilde S_{bulk}=-\int dx\,dy\,dr\sqrt{-g}\,{\cal L}
\ee
en donde la densidad Lagrangiana está escrita en \eqref{action}. La componente $yy$ del tensor energía-momento es proporcional a la métrica y por lo tanto las ecuaciones de Einstein \eqref{EYM} implican que
\be
G_{yy}=\frac{r^2}{2h^2}\left(\kappa^2_{(4)}{\cal L}-R\right)
\ee
Luego, tenemos
\be
G_\mu^\mu=-R=G_r^r+G_t^t+G_x^x+\frac{1}{2}\left(\kappa^2_{(4)}{\cal L}-R\right)
\ee
a partir de lo que se obtiene
\be
{\cal L}=\frac{2}{r^2\sigma\kappa^2_{(4)}}\left[\frac{r^3M\sigma}{h}\left(\frac{h}{r}\right)'\right]'
\ee
en donde $'$ denota derivada con respecto a la coordenada holográfica $r$. Luego, la contribución a la acción on-shell \eqref{onshellp} proveniente del volumen del espacio-tiempo se puede escribir como
\be
\tilde S_{bulk}=-\int dx\,dy\,dr\sqrt{-g}{\cal L}=-\frac{2V}{\kappa^2_{(4)}} \left[\frac{r^3 M\sigma}{h}\left(\frac{h}{r}\right)'\right]_{r=r_\infty}
\ee
en donde $r_{\infty}$ es el borde del espacio. Como es usual, para tener un problema variacional bien definido cuando se imponen
condiciones de borde Dirichlet en la métrica necesitamos agregar a la acción el término de Gibbons-Hawking
\be
\tilde S_{GH}=-\frac{1}{\kappa^2_{(4)}}\int dx\,dy\sqrt{-g_{\infty}}\,\nabla_\mu n^\mu=-\frac{V}{\kappa^2_{(4)}}r^2\sigma\left[\frac{M'}{2}+M\left(\frac{\sigma'}
{\sigma}+\frac{2}{r}\right)\right]_{r=r_\infty},\label{GH}
\ee
siendo $n^\mu dx_\mu=\sqrt{M}dr$ el vector unitario normal al borde y $g_\infty$ el determinante de la métrica inducida sobre el borde. Precisamente en $r=r_{\infty}$ \eqref{GH} diverge y por lo tanto debe ser regularizada mediante la suma de contra-términos de borde
\be
\tilde S_{ct}=\frac{1}{\kappa^2_{(4)}}\int dx\,dy\sqrt{-g_{\infty}}=\frac{V}{\kappa^2_{(4)}}\left[r^2\sqrt{M}\sigma\right]_{r=r_\infty}
\ee
Finalmente, el potencial termodinámico dual $\Omega$ resulta
\bea\label{omega}
\Omega&=&\lim_{r_\infty\rightarrow\infty}\tilde S_{on-shell}\nn\\
&=&\lim_{r_\infty\rightarrow\infty}(\tilde S_{bulk}+\tilde S_{GH}+\tilde S_{ct})
\eea

Luego de regularizar la acción el potencial $\Omega$ coincide con el valor sub-leading de la componente $g_{tt}$ de la métrica
 de fondo, es decir, $\Omega=M_1^b$ \cite{Hartnoll}. Se ha verificado la solución numérica calculando $\Omega$ de ambas maneras
  y se encontró una concordancia excelente. En la figura \ref{freep} se muestra el potencial \eqref{omega} como función de la temperatura.
  Como se mencionó anteriormente se produce una transición de fase de segundo orden en $T=T_c$: el gran potencial y la entropía son continuas pero
  $S$ no es diferenciable. Por debajo de $T_c$ el sistema se encuentra en la fase superconductora, al incrementar la temperatura por sobre $T_c$ la
   geometría de AdSRN domina la energía libre, lo que modela la transición de la fase superconductora a la normal.

\subsection{Superconductores tipo $p+ip$}\label{p+ip}

En esta sección se recopilarán los resultados de \cite{Gubserpip} y se compararán con los resultados obtenidos en la sección anterior.
Se encontró que a $T=T_c$ el sistema tiene una transición de fase de segundo orden y que en todo el rango de temperatura el gran
potencial del superconductor tipo $p$ es menor que el del $p+ip$. Esto implica que la fase estable del sistema es la tipo $p$, lo cual concuerda
con el análisis de estabilidad desarrollado en \cite{Gubserp}.

\subsubsection{Solución}

La geometría y el campo de gauge para modelar la solución son
\bea
ds^2&=&-M(r)dt^2+r^2h(r)^2(dx^2+dy^2)+\frac{dr^2}{M(r)}\\
A&=&\phi(r)\tau^3dt+\omega(r)(\tau^1dx+\tau^2dy).
\eea
Aquí hay un diferencia importante con el superconductor tipo $p$ de la sección previa. Ahora el campo de gauge rompe el grupo $U(1)_3\times U(1)_{xy}$
 dejando invariante una combinación diagonal. En la sección anterior el campo de gauge rompía por completo el grupo $U(1)_3\times U(1)_{xy}$.
 Esto nos permite proponer una métrica totalmente simétrica en el plano $xy$.

\begin{figure}[ht]
\begin{minipage}{7.8cm}
\begin{center}
\includegraphics[width=7.8cm]{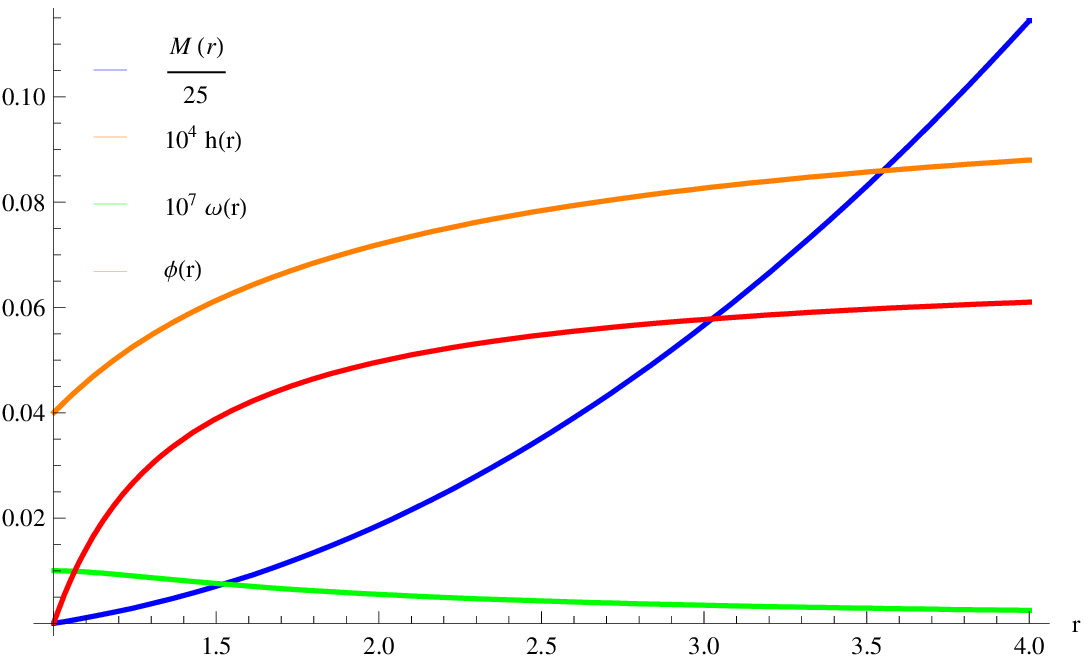}
\caption{Comportamiento de las funciones adimensionales $M(r),h(r),\omega(r)$ y $\phi(r)$, para $g_{_{YM}}=2, T=0.2312\mu$.}
\label{funcionesp+ip}
\end{center}
\end{minipage}
\ \ \ \
\hfill\begin{minipage}{8cm}
\begin{center}
\includegraphics[width=8cm]{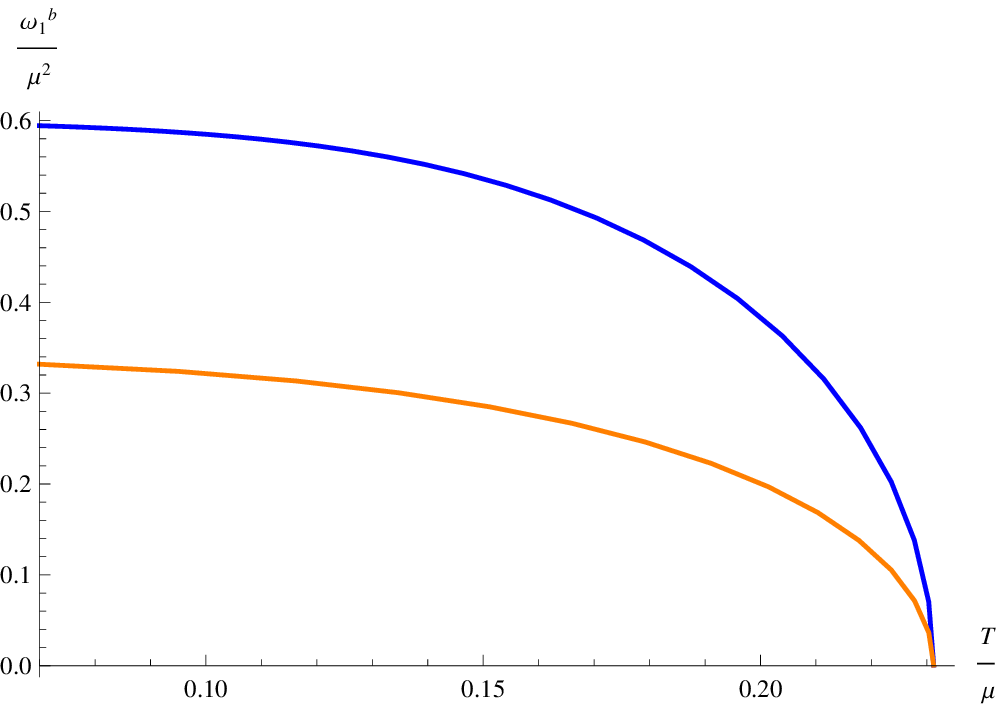}
\caption{ El valor de expectación $\langle J_x^1\rangle\propto \omega_1^b$ de la teoría de campos dual como función de la temperatura para el superconductor tipo $p$ (línea azul) y tipo $p+ip$ (línea naranja) para $g_{_{YM}}=2$. Se anula para $T>T_c=0.2312$, lo cual sugiere la existencia de una transición de fase entre el estado superconductor y el normal.}
\label{condpypip}
\end{center}
\end{minipage}
\end{figure}

Las 5 ecuaciones de movimiento obtenidas se dividen en 4 ecuaciones diferenciales de segundo orden y una ligadura de primer orden proveniente de la componente $rr$ de la ecuaciones de Einstein
\bea
h''&=&-\frac{h}{2}\left[\frac{1}{r^2}-\frac{3}{{\hat R}^2M}+\frac{M'}{r M}+\frac{\phi'^2}{8M}+\frac{\omega'^2}{4r^2h^2}\right]-\frac{h'}{2}\left[\frac{6}{r}+\frac{h'}{h}
+\frac{M'}{M}\right]-\frac{g_{_{YM}}^2\omega^2}{8r^2hM}\left[\frac{\phi^2}{M}+\frac{\omega^2}{2r^2h^2}\right]\nn\\
M''&=&\frac{3}{{\hat R}^2}+\frac{M}{r}\left[-\frac{M'}{M}+\frac1r+\frac{\omega'^2}{4rh^2}\right]-\frac{h'}{h}\left[M'-\frac{h'}{h}-\frac2r\right]+\frac38\phi'^2+
\frac{g_{_{YM}}^2\omega^2}{4r^2h^2}\left[\frac{\phi^2}{M}+\frac{3\omega^2}{2r^2h^2}\right]\nn\\
\omega''&=&\frac{g_{_{YM}}^2\omega}{M}\left[\frac{\omega^2}{r^2h^2}-\frac{\phi^2}{M}\right]-\frac{M'\omega'}{M}\nn\\
\phi''&=&\frac{2g_{_{YM}}^2\phi\omega^2}{r^2h^2M}-2\phi'\left[\frac1r+\frac{h'}{h}\right]\nn\\
0&=&-\frac{3}{{\hat R}^2}+\frac{M}{r^2}\left[1-\frac{\omega'^2}{4h^2}+\frac{M'}{M}r\right]+\frac{h'}{h}\left[M\left(\frac2r+\frac{h'}{h}\right)
+M'\right]+\frac18\phi'^2.\nn
\eea\label{pip}
Las ecuaciones tienen tres simetrías de escala que nos ayudarán a resolver el sistema numéricamente. Estas son
\begin{enumerate}
\item $\omega\rightarrow\lambda\omega,~~~~h\rightarrow\lambda h$
\item $M\rightarrow\lambda^{-2}M,~~~~\phi\rightarrow\frac{\phi}{\lambda},~~~~{\hat R}\rightarrow\lambda {\hat R},~~~~g_{_{YM}}\rightarrow\frac{g_{_{YM}}}{\lambda}$
\item $M\rightarrow\lambda^2M,~~~~h\rightarrow\frac{h}{\lambda},~~~~\phi\rightarrow\lambda\phi,~~~~r\rightarrow\lambda r$\label{scalpip}
\end{enumerate}
y permiten escoger $R=r_h=1$ y fijar el valor de $h(r)$ en el borde a $h(\infty)=1$. El comportamiento en el IR de estas ecuaciones es el mismo que para un agujero negro cargado
\bea\label{IRpip}
M&=&M_1(r-r_h)+M_2(r-r_h)^2+\ldots\nn\\
h&=&h_0+h_1(r-r_h)+h_2(r-r_h)^2+\ldots\nn\\
\omega&=&\omega_0+\omega_1(r-r_h)+\omega_2(r-r_h)^2+\ldots\nn\\
\phi&=&\phi_1(r-r_h)+\phi_2(r-r_h)^2+\ldots
\eea
en donde, como antes, se impuso que el potencial de Maxwell $\phi$ se anule en el horizonte con el objeto de obtener un campo de gauge bien definido cuando se realize la continuación Euclídea.
En el UV se pedirá
\bea
M&=&r^2+2 h_1^b r+(h_1^b)^2+\frac{M_1^b}{r}+\frac{-8h_1^b M_1^b+\rho^2+2(\omega_1^b)/3}{8r^2}+\ldots\nn\\
h&=&1+\frac{h_1^b}{r}-\frac{(\omega_1^b)^2}{48 r^4}+\ldots\nn\\
\omega&=&\frac{\omega_1^b}{r}-\frac{h_1^b\omega_1^b}{r^2}+\ldots\nn\\
\phi&=&\mu+\frac{\rho}{r}-\frac{\rho h_1^b}{r^2}+\ldots
\eea
Nótese que para que ocurra una ruptura espontánea de simetría no se permite un término no normalizable en $\omega$. Como antes, las simetrías de escala \eqref{scalpip} permiten fijar $h_0^b=1$. En la figura \ref{funcionesp+ip} se muestra el comportamiento de las solución y en la figura \ref{condpypip} se grafica el parámetro de orden $\langle J_x^1\rangle>\propto \omega_1^b$ como una función de la temperatura. Para $T=T_c$ ambos condensados se anulan y ocurre una transición de fase de segundo orden. Nótese que los valores del condensado para el caso del $p+ip$ son menores que aquellos para el superconductor tipo $p$.

\subsubsection{Termodinámica}

\begin{figure}[ht]
\begin{minipage}{8cm}
\begin{center}
\includegraphics[width=7.5cm]{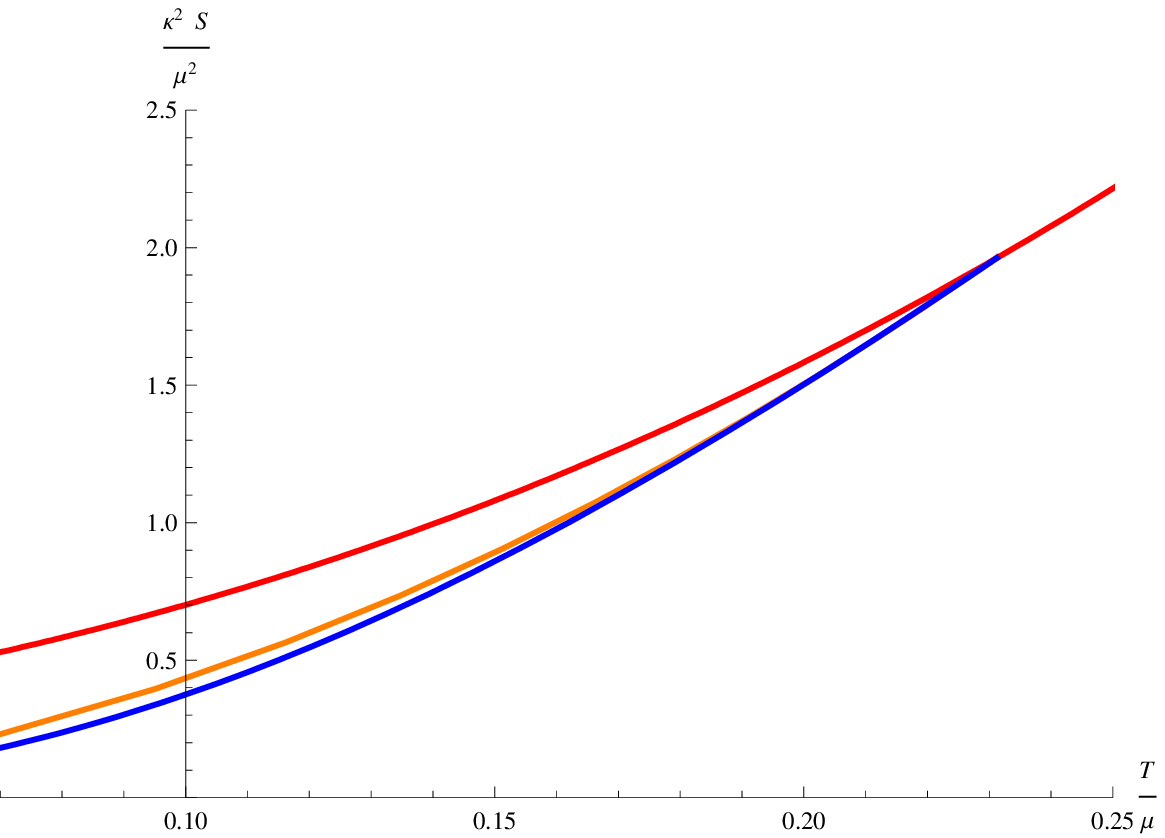}
\caption{ Entropía de Bekenstein-Hawking para el agujero negro de RN (rojo) y las soluciones de superconductro tipo $p$ (azul) y $p+ip$ (naranja) con  $g_{_{YM}}=2$ y $T_c=0.2312$.  Hay una transición de fase de segundo orden para ambos superconductores en $T=T_c$.}
\label{SvsT}
\end{center}
\end{minipage}
\ \
\hfill\begin{minipage}{8cm}
\begin{center}
\includegraphics[width=7.5cm]{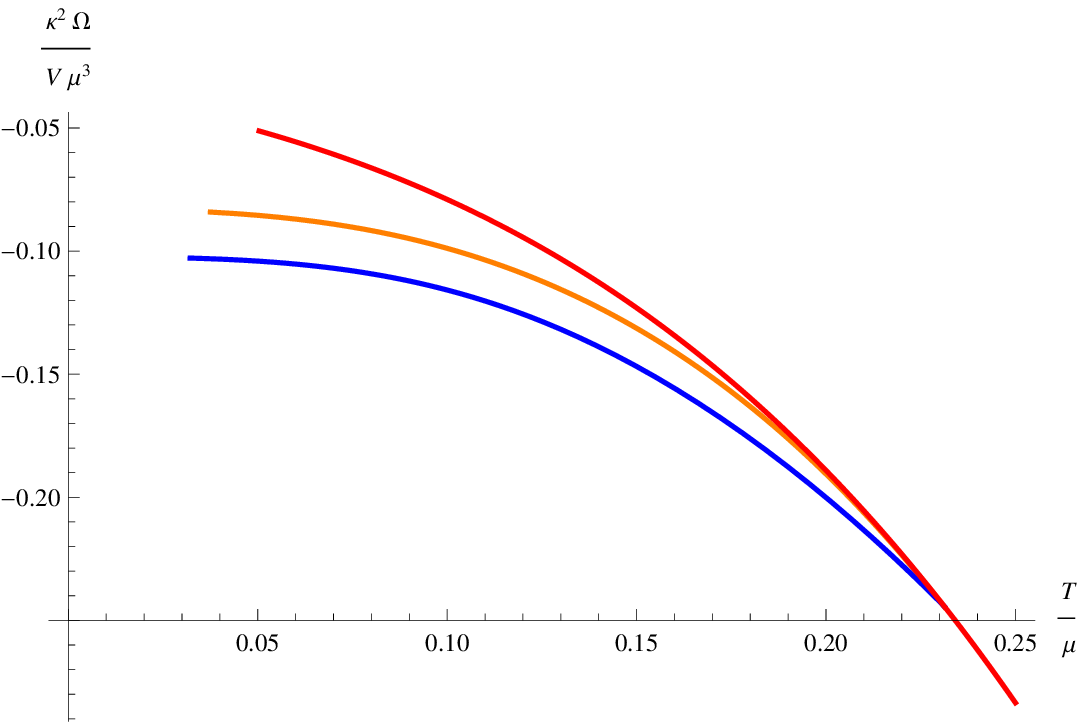}
\caption{ Potencial gran canónico como función de la temperatura. Solución de RN (rojo), $p+ip$ (naranja) y $p$ (azul). Para todo rango de temperaturas por debajo de $T_c$ la solución de superconductor tipo $p$ es preferida por sobre la $p+ip$ ($g_{_{YM}}=2$ y $T_c=0.2312$).}
\label{FvsT}
\end{center}
\end{minipage}
\end{figure}

La temperatura asociada a la solución de fondo es proporcional a la derivada de la componente $g_{tt}$ de la métrica evaluada en el horizonte. En este caso se tiene
\be
T=\frac{M_1}{2\pi}
\ee
La formula de Bekenstein-Hawking, que relaciona la entropía con el area del horizonte del agujero negro, en este caso es
\be
S=\frac{2\pi}{\kappa^2_{(4)}} A_h=\frac{2\pi}{\kappa^2_{(4)}} r_h^2h_0^2.
\ee
En la figura \ref{SvsT} se muestra la entropía en el caso del superconductor tipo $p+ip$ (linea naranja), del tipo $p$ (línea azul) y del agujero negro de RN (línea roja).

El grand potencial $\Omega$ está dado por el coeficiente del término sub-leading de la función $g_{tt}$ de la métrica,
\be
\Omega=\frac{V M_1^b}{\kappa^2_{(4)}}
\ee
Este potencial se dibuja en la figura \ref{FvsT}, en donde claramente se muestra que para cualquier valor de la temperatura la solución tipo $p$ (azul) es preferida por sobre la $p+ip$ (naranja) debido a su menor energía libre. Para $T>T_c$ el sistema se encuentra en su fase normal (en rojo) y el condensado se anula (ver figura \ref{condpypip}).

\section{Entropía de entrelazamiento}\label{Entanglement}

\begin{figure}[ht]
\begin{center}
\includegraphics[width=10cm]{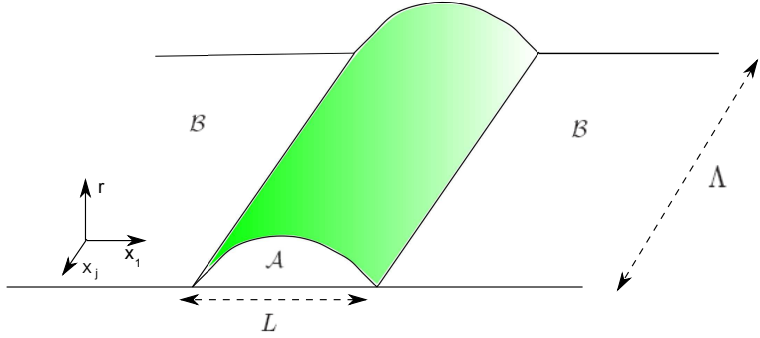}
\caption{ Diagrama de la región en forma de banda ${\cal A}$ usada para calcular la entropía de entrelazamiento.}
\label{EEpicture}
\end{center}
\end{figure}

Una prescripción holográfica para calcular la entropía de entrelazamiento (EE) en el espacio AdS$_{d+1}$ dual a una CFT$_d$ fue propuesta en \cite{Ryu} en términos de superficies mínimas. La prescripción involucra subdividir el sistema en dos regiones, ${\cal A}$ y su complemento ${\cal B}$, y encontrar las superficie estática mínima $d-1$ dimensional (a tiempo constante) $\gamma_{\cal A}$ tal que su borde coincide con el borde del subsistema ${\cal A}$ (ver figura \ref{EEpicture}).

La entropía de entrelazamiento entre las dos regiones es el area clásica de $\gamma_{\cal A}$,
\be
{\cal S}_{\cal A}=\frac{2\pi}{\kappa_{(d+1)}^2}\int_{\gamma_{\cal A}} d^{(d-1)}\sigma\sqrt{g^{(d-1)}_{ind}},\label{HEE}
\ee
en donde $g^{(d-1)}_{ind}$ es la métrica inducida sobre la superficie y $\kappa_{(d+1)}^2$ es la constante gravitatoria en $d+1$ dimensiones.

En \cite{KlebanovEE} y \cite{PZ} la EE fue calculada para fondos gravitatorios duales a teorías de gauge confinantes y a geometrías de agujero negro respectivamente. En esta sección se realizará un cálculo similar para un fondo de gravedad general y se aplicará la prescripción a las soluciones de superconductores tipo $p$ y $p+ip$ encontrados en la sección \ref{holsup}. Nótese que la prescripción es muy similar a la realizada en \cite{rey,maldawilson} para calcular valores de expectación de lazos de Wilson (ver capítulo \ref{Wh}).

Escribamos la métrica de fondo $d+1$ dimensional como
\be
ds_{d+1}^2=-g_{tt}(r)dt^2+ g_{x_i x_i}(r)dx_i^2+g_{rr}(r)dr^2,~~~~~~ i=1\ldots d-1,
\ee
en donde $r$ es la coordenada holográfica. La región de interés será una banda recta en la dirección $x_j$ con ancho $L$ en la dirección $x_1$. La forma de embeber la superficie estáticamente es $x_1=x_1(\zeta),\, x_j=\zeta_j, r=r(\zeta)$, con $j=2,\ldots, d-1$. Aún se tiene invarianza ante difeomorfismos en ${\cal S}_{\cal A}$, dependiendo del contexto esta puede ser fijada como $x_1=\zeta$  (embedding global) o $r=\zeta$.
La entropía \eqref{HEE} es
\be
{\cal S}_{\cal A}=\frac{2\pi\Lambda}{\kappa_{(d+1)}^2}\int d\zeta \sqrt{g_{x_2x_2}(r)\ldots g_{x_{d-1} x_{d-1}}(r)}\sqrt{g_{rr}(r)r'^2+g_{x_1x_1}(r)x'^2_1},\label{SA}
\ee
en donde $\Lambda=\int d\zeta_2\ldots d\zeta_{d-1}$ y el símbolo $'$ denota derivadas en $\chi$.

Definiendo $g_{\chi\chi}(r)=g_{x_2x_2}(r)\ldots g_{x_{d-1} x_{d-1}}(r)$ y las funciones
\be
f^2(r)=g_{\chi\chi}(r)g_{x_1x_1}(r),~~~~~~ \eta^2(r)=g_{\chi\chi}(r)g_{rr}(r)
\ee
la entropía de entrelazamiento se puede escribir
\be
{\cal S}_{\cal A}=\frac{2\pi\Lambda}{\kappa_{(d+1)}^2}\int d\zeta \sqrt{\eta^2(r)r'^2+f^2(r)x'^2_1}.\label{onshellEE}
\ee
Minimizando \eqref{onshellEE} se obtiene
\be
x'_1(\zeta)=\pm \frac{f(r_0)\eta(r)}{f(r)}\frac{r'(\zeta)}{\sqrt{f^2(r)-f^2(r_0)}}\label{rprima},
\ee
en donde $r=r_0$ es el valor mínimo en la coordenada holográfica alcanzada por la superficie. Dependiendo del fondo gravitatorio estudiado este punto podría ser el radio del horizonte o el fin del espacio-tiempo. Invirtiendo esta relación se puede leer la longitud de la banda en la dirección $x_1$
\be
L=2\int_{r_0}^\infty dr \frac{dx_1}{dr}=2\int_{r_0}^\infty dr \frac{\eta(r)}{f(r)}\frac{f(r_0)}{\sqrt{f^2(r)-f^2(r_0)}}.\label{Lenght}
\ee
Ahora se fijará la invarianza ante difeomorfismo como $x_1(\zeta)=\zeta$, esta elección tiene la ventaja de proveer una parametrización completa de
$r(x_1)$, ($x_1\in[-\frac{L}{2},\frac{L}{2}]$ y las condiciones de borde son $r(\pm\frac{L}{2})=\infty$). A partir de \eqref{rprima} en \eqref{onshellEE} la entropía de entrelazamiento queda
\be
{\cal S}_{\cal A}(r_0)=2\frac{2\pi\Lambda}{\kappa_{(d+1)}^2}\int_{r_0}^\infty dr \frac{f(r)\eta(r)}{\sqrt{f(r)^2-f(r_0)^2}}.\label{connectedsurface}
\ee
la expresión \eqref{connectedsurface} diverge en $r=\infty$ debido a la extensión infinita de la superficie. Esta divergencia se entiende a partir de la existencia de otra solució, con las mismas condiciones de borde, consistente de dos superficies desconectadas que se expanden todo a lo largo de la dirección radial. Su area es
\be
{\cal S}_{{\cal A}_{disc}}=2\frac{2\pi\Lambda}{\kappa_{(d+1)}^2}\int_{r_{min}}^\infty dr\,\, \eta(r),\label{disconnected}
\ee
aquí $r_{min}$ es el valor mínimo de $r$ permitido por la geometría. La EE queda definida con respecto al estado de referencia \eqref{disconnected}
\be
\Delta {\cal S}_{{\cal A}}=\frac{4\pi\Lambda}{\kappa_{(d+1)}^2}\left(\int_{r_0}^\infty dr \frac{f(r)\eta(r)}{\sqrt{f(r)^2-f(r_0)^2}}-\int_{r_{min}}^\infty dr\,\, \eta(r)\right).\label{DeltaS}
\ee
En lo que sigue vamos a estudiar la EE para las soluciones de \eqref{eomsp} y \eqref{pip}.

En el caso del superconductor tipo $p$ las funciones relevantes son
\be
f^2_p(r)=g_{yy}g_{xx}=r^4,~~~~~~\eta^2_p(r)=g_{yy}g_{rr}=\frac{r^2}{h^2 N}
\ee
en donde el subíndice $p$ nos recuerda que corresponden al tipo $p$. Con esto, podemos calcular explícitamente la cantidad \eqref{DeltaS}
With this, we can compute explicitly the quantity \eqref{DeltaS}
\be
\Delta {\cal S}_{\cal A}=\frac{4\pi\Lambda}{\kappa_{(4)}^2}\left(\int_{r_0}^\infty dr \frac{r^3}{h\sqrt{N}\sqrt{r^4-r_0^4}}-\int_{r_{min}}^\infty dr\frac{r}{h\sqrt{N}}\right)
\ee

\begin{figure}[ht]
\begin{minipage}{7cm}
\begin{center}
\includegraphics[width=7.5cm]{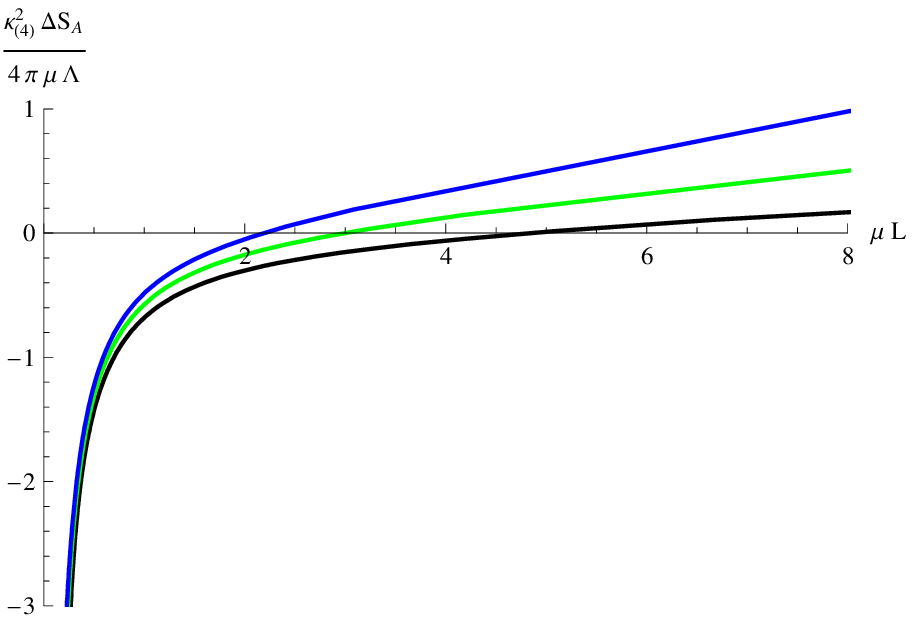}
\caption{ Entropía de entrelazamiento como función del tamaño de la banda para la solución tipo $p$. Las curvas negra, verde y azul corresponden a $g_{_{YM}}=1, T=0.0749\mu$, $g_{_{YM}}=1.5,
T=0.1565\mu$ y $g_{_{YM}}=2, T=0.2312\mu$ respectivamente.}
\label{EEvsLp}
\end{center}
\end{minipage}
\ \
\hfill\begin{minipage}{7cm}
\begin{center}
\includegraphics[width=7.5cm]{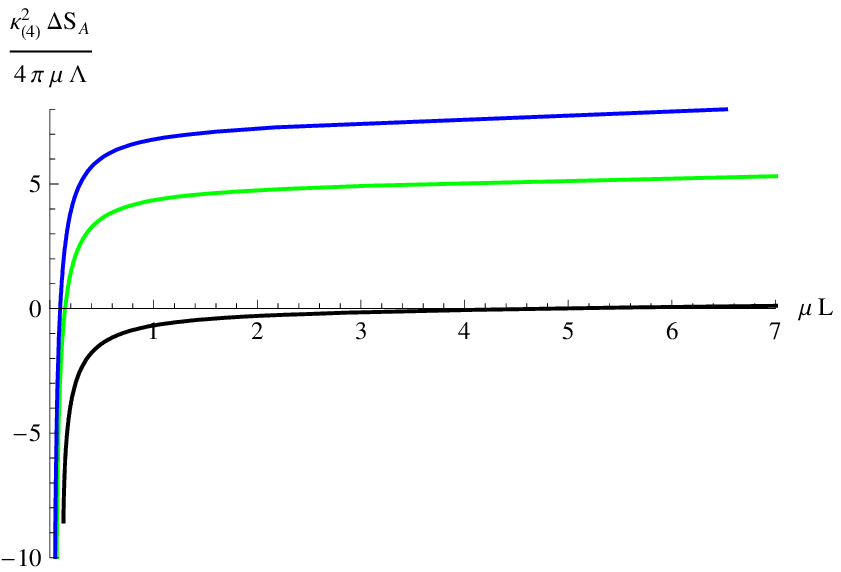}
\caption{Entropía de entrelazamiento como función del tamaño de la banda para la solución tipo $p+ip$. Las curvas negra, verde y azul corresponden a $g_{_{YM}}=1, T=0.0749\mu$, $g_{_{YM}}=1.5,
T=0.1565\mu$ y $g_{_{YM}}=2, T=0.2312\mu$ respectivamente.}
\label{EEvsLpip}
\end{center}
\end{minipage}
\end{figure}

\begin{figure}[ht]
\begin{center}
\includegraphics[width=7.5cm]{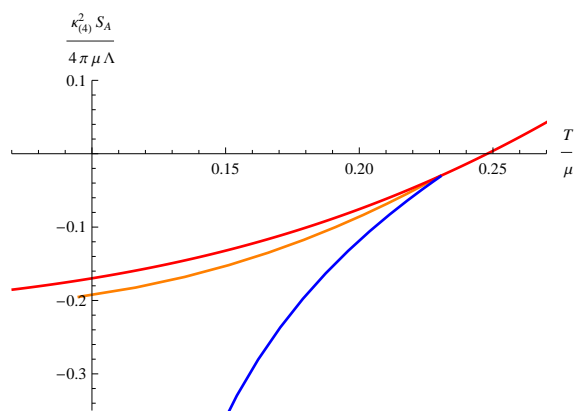}
\caption{ Entropía de entrelazamiento como función de la temperatura y a $\mu L=3$ fijo. La curva azul es para la solución tipo $p$, la naranja para la tipo $p+ip$ y la roja para RN. Se utilizó $g_{_{YM}}=2$ y $T_c=0.2312$.}
\label{EEvsT}
\end{center}
\end{figure}

En la figura \ref{EEvsLp} se muestra $\Delta {\cal S}_{\cal A}$ como función de la longitud de la banda $L$. Se graficó para diferentes valores del parámetro de deformación y diferentes valores de la temperatura. Como era esperado, la curva inferior es la que tiene menor temperatura debido a que al bajar la temperatura hay más grados de libertad condensados. El comportamiento lineal para valores grandes de $\mu L$ es una manifestación de la ley de area propuesta en \eqref{HEE}. La figura \ref{EEvsT} muestra la EE de las fases condensadas (línea azul) y normal (linea roja) como función de la temperatura y para un valor constante de $L$.

Con el objetivo de obtener una entropía finita sin sustraer la solución desconectada, podemos escribir la EE de la siguiente manera
\be
{\cal S}_{\cal A}(r_0)=\frac{4\pi\Lambda}{\kappa_{(4)}^2}\int_{r_0}^{\cal R} dr \frac{r^3}{h\sqrt{N}\sqrt{r^4-r_0^4}}=S_{\cal A}+\frac{4\pi\Lambda}{\kappa_{(d+1)}^2}{\cal R}
\ee
en donde $S_{\cal A}$ tiene dimensiones de longitud y no presenta divergencias. La figura \ref{EEvsT} muestra que la EE para el superconductor (linea azul) es menor que para la solución de RN (linea roja). Como se mencionó anteriormente esto es esperable debido que la solución superconductora tiene grados de libertad condensados.

Realizando el mismo análisis para la solución de \eqref{pip} se obtiene
\be
f^2_{p+ip}(r)=g_{yy}g_{xx}=r^4h^4,~~~~~~\eta^2_{p+ip}(r)=g_{yy}g_{rr}=\frac{r^2h^2}{M},
\ee
y la siguiente EE
\be
\Delta {\cal S}_{\cal A}=\frac{4\pi\Lambda}{\kappa_{(4)}^2}\left(\int_{r_0}^\infty dr \frac{r^3h^3}{\sqrt{M}\sqrt{r^4h^4-r_0^4h(r_0)^4}}-\int_{r_{min}}^\infty dr\frac{r h}{\sqrt{M}}\right)
\ee
La figura \ref{EEvsLpip} muestra el comportamiento de $\Delta {\cal S}_{\cal A}$ para este caso, y podemos realizar el mismo análisis que para el caso del superconductor $p$. De nuevo, podemos utilizar una aproximación diferente para obtener la entropía no divergente sin realizar la sustracción de las soluciones desconectadas. Esto consiste en separar la parte divergente de la integral \eqref{connectedsurface} y tener en cuenta su parte finita, $S_{\cal A}$. En este caso
\be
{\cal S}_{\cal A}(r_0)=\frac{4\pi\Lambda}{\kappa_{(4)}^2}\int_{r_0}^{\cal R} dr \frac{r^3h^3}{\sqrt{M}\sqrt{r^4h^4-r_0^4h(r_0)^4}}=S_{\cal A}+\frac{4\pi\Lambda}{\kappa_{(4)}^2}{\cal R}.
\ee
La figura \ref{EEvsT} muestra esta parte finita (linea naranja) para $g_{_{YM}}=2$ y $T_c=0.2312$, de donde se ve que, como se esperaba, es menor que la EE de la solución de RN. Además la figura muestra que la EE en el caso del superconductor tipo $p$ es menor que para el $p+ip$. Esto sugiere que a una dada temperatura hay más grados de libertad condensados en el superconductor $p$.

\subsection{Conclusiones}\label{summary}

En este capítulo se estudiaron los duales holográficos a superconductores tipo $p$ y $p+ip$ en 3+1 dimensiones. Se calculó la deformación de la geometría dual al tipo $p$ y sus propiedades termodinámicas. Como era esperable, y en contraste a la solución en 4+1 dimensiones de \cite{Ammon}, se encontró una transición de fase de segundo orden entre las fases normal y superconductora. Luego, se resumió la deformación producida por el campo de gauge sobre la geometría del $p+ip$ y se comparó con nuestra solución.
A partir del estudio de las cantidades termodinámicas y en particular de sus potencial gran canónico, se notó que para un valor fijo de la temperatura la solución tipo $p$ tiene menor potencial y por lo tanto es la preferida por el sistema. Esto se relaciona con el echo de que la solución tipo $p+ip$ es inestable frente a pequeñas fluctuaciones y decae a la solución tipo $p$.

Finalmente, usando la propuesta holográfica estudiada en \cite{Ryu}, se calculó la entropía de entrelazamiento de la teoría cuántica de campos usando
 su dual gravitatorio, para una geometría de banda y como función de la temperatura y el tamaño de la banda. Se observó que en ambos casos la EE se
 comporta lineal para valores grandes de $L$, lo cual confirma la ley de area propuesta. La EE vs L tiende a valores mas grandes de $\Delta {\cal S_{A}}$
  a medida que se incrementa su temperatura. Como función de la temperatura se observa que la mayor EE es para la solución de RN. Esto era esperable debido
  a que el superconductor tiene grados de libertad condensados. Además la solución tipo $p$ presenta más grados de libertad condensados que el $p+ip$. Esto podría explicar el echo de que el valor del condensado para el tipo $p$ es más grande que para el $p+ip$ a una dada temperatura.

\newpage

\def\baselinestretch{1}
\chapter{Conclusiones}\label{Conclusiones}
\def\baselinestretch{1.66}

En esta tesis se estudiaron varias aplicaciones de la conjetura de Maldacena. En particular, a lo largo del trabajo se han analizado desde el punto de vista holográfico (gravitatorio) teorías de campos conformes, no conformes y sistemas de materia condensada.
A lo largo de la tesis se desarrollaron distintos aspectos técnicos de la conjetura de Maldacena mostrando cómo se obtienen las funciones de correlación y los valores de expectación del lazo de Wilson de la teoría de gauge dual a partir de cálculos en gravedad. También se estudiaron las recientes aplicaciones de la conjetura al estudio de un sistema superconductor. Los aspéctos estudiados corresponden a cálculos originales que condujeron a publicaciones en revistas internacionales con referato.

En el capítulo \ref{Wormhole} se ha extendido la prescripción GKPW para calcular funciones de correlación de teorías de campos duales a fondos de
gravedad que poseen dos bordes asintóticamente AdS. En el camino se analizaron las caracterísiticas particulares de la prescripción de GKPW en
espacio-tiempos de signatura Lorentziana. Se propuso también un método que nos permite concluir si las teorías de campos en dichos bordes interactúan entre sí o se entrelazan.

En el capítulo \ref{Wh} se estudió el valor de expectación del operador lazo de Wilson para diferentes representaciones del grupo de gauge
desde el punto de vista dual gravitatorio. En particular, se analizaron fluctuaciones para diferentes fondos de gravedad duales a teorías de campos no conformes y con supersimetría no maximal. La estabilidad se estudió para el caso del lazo de Wilson rectangular y en la representación fundamental. Asimismo la conjetura provee un marco para estudiar el potencial monopolo-antimonopolo: estos se estudiarón para las geometrías de Maldacena-Nuñez y Klebanov-Strassler, soluciones de gravedad que modelan el régimen IR de ${\cal N}=1$ SYM, con el objeto de chequear las ideas de ' t Hooft \cite{thooft}. Se analizaron fluctuaciones tanto bosónicas como fermiónicas al fondo gravitatorio dual a una línea de Wilson- 't Hooft.

Por último en el capítulo \ref{SC} se estudió la termodinámica de un superconductor tipo $p$ en 2+1 dimensiones, en el límite en el que el número de
grados de libertad cargados y el número de grados de libertad totales son comparables, mediante su descripción gravitatoria dual. Se observó que,
al igual que en el caso del $p+ip$, existe un transición de fase de segundo orden entre un estado normal y uno superconductor al variar la temperatura.
Además se concluyó que por debajo de una temperatura crítica la configuración más estable es la solución dual correspondiente al superconductor tipo $p$.
 Con el objeto de probar la solución, se calculó la entropía de entrelazamiento utilizando la propuesta de Ryu - Takayanagi y se mostró que esta satisface
 la ley de área.

\newpage

\appendix
\addcontentsline{toc}{chapter}{Apéndices}
\def\baselinestretch{1}
\chapter{Lazos de Wilson}
\def\baselinestretch{1.66}
\section{De Sturm-Liouville a Schr\"odinger}
\label{sl2sc}

Las ecuaciones (\ref{SL})-(\ref{spp}) son del tipo de Sturm-Liouville
\be
\Big[-\frac{d}{dr}\left(P(r,r_0)\frac{d}{dr}\right)+U(r,r_0)\Big]
\Phi(r)=\omega^2Q(r,r_0)\Phi(r),~~~ r_0 \leq r<\infty
\label{sliouv2}
\ee
las funciones $P(r,r_0)$ y $Q(r,r_0)$ pueden leerse de (\ref{SL})-(\ref{spp}), en ambos casos $U(r,r_0)=0$
. El cambio de variables
\be
y=\int_{r_0}^r \sqrt{\frac{Q }{P }}\;dr,~~~~
\Phi(r)=\left(PQ\right)^{-\frac14}\Psi(y)
\label{chv}
\ee
transforma (\ref{sliouv2}) en una ecuación del tipo Schr\"{o}dinger
\be
\left[-\frac{d^2}{dy^2}+V \right]\Psi =\omega^2\Psi ,~~~ 0 \leq y\leq y_0\,.
\label{sch}
\ee
Aquí $y_0=\int_{r_0}^\infty dr\sqrt{\frac{Q}{P}}$ el cual puede ser infinito o finito dependiendo de la naturaleza de $Q,P$
y se puede chequear que (\ref{chv}) es integrable en el límite inferior dado por $y\sim\sqrt{r-r_0}$. El potencial
$V$ esta dado por
\bea
V&=&\frac U Q+\left[(P Q )^{-\frac14}\frac{d ^2}{dy^2}\right](P Q )^{\frac14}\nn\\
&=&\frac U Q+\left[\frac{P ^{\frac14}}{Q ^{\frac34}}\frac{d}{dr}\left(\sqrt{\frac{P}{Q}}\frac{d}{dr}\right)\right]
(PQ)^{\frac14}
\label{schrodinger}
\eea
Los puntos $r=r_0$ y $r=\infty$ se mapean a $y=0$ e $y=y_0$ respectivamente. Las condiciones de borde que deben imponerse
en la solución de (\ref{sch}) son \cite{avramis}:
\begin{itemize}
\item Infinito: extremos de la cuerda fijos\footnote{Ver, sin embargo, un detalle en el contexto de Maldacena-N\'u\~nez
(secc. \ref{schMN}) cuando se impone (\ref{bcsch}).}
\be
\delta x|_{r=\infty}=0 \Rightarrow\Psi|_{y=y_0}=0
\label{bcsch}
\ee

\item Tip $r=r_0$: para fluctuaciones en el plano, $\delta x_1$, y fluctuaciones transversales
$\delta x_m$  se obtiene a partir de (\ref{xperp}),(\ref{bc}),(\ref{bc2})
\be
\begin{array}{c}
  ~\mathrm{Soluciones~~pares}:\left.\frac{d\Psi}{dy}\right|_{y=0}=0 \\
  \mathrm{Soluciones~~impares}: ~~\left. \Psi \right|_{y=0}=0\,.
\end{array}\label{bcsschrod}
\ee
 \end{itemize}

\section{Espectro para fluctuaciones transversales en $AdS_5\times S^5$}
\label{kmt}

En esta sección se repetirá la solución de \cite{kmt} para el espectro exacto de las
fluctuaciones longitudinales en el fondo de gravedad $AdS_5\times S^5$ y se comparará con
los valores obtenidos numéricamente en esta tesis descriptos en el final de la sección \ref{stability}.
La métrica de $AdS$ en coordenadas de Poincare se escribe
\be
ds^2=\frac{R^2}{z^2}(-dt^2+dx_{i}dx_{i}+dz^2)+R^2d\Omega_5^2\, .
\ee
El fijado de gauge $x$ $t=\tau, x=x_{\sf cl}, z=z_{\sf cl}(x)$ conduce a
\be
\left(\frac {dz_{\sf cl}}{dx} \right)^2=\frac{z_0^4-(z_{\sf cl})^4}{(z_{\sf cl})^4}\,.
\label{zcl}
\ee
La solución a (\ref{zcl}) con los extremos de la cuerda separados por una distancia $L$ es (\ref{xads})-(\ref{lads}).
\be
x_{\sf cl}(z)=\pm z_0\left[\frac{(2\pi)^{\frac32}}{2\Gamma[\frac14]^2}
-\frac14\mathsf{B}\left( \frac{z^4}{z_0^4} ;\frac34,\frac12\right)\right]
\label{adssol}
\ee
en donde $z_0=z_{\sf cl}(0)=({\Gamma[\frac14]^2}/{(2\pi)^{\frac32}})L$ es la distancia radial máxima alcanzada por la
cuerda (tip de la cuerda) y $\mathsf{B}(z;a,b)$ es la función beta incompleta $\mathsf{B}(z;a,b)=\int_0^zt^{a-1}(1-t)^{b-1}dt$.

Las fluctuaciones alrededor de la solución (\ref{adssol}) en las coordenadas transversales $x_m$ ($m=2,3$) son
desacopladas, escribiendo $X^\mu=(t,x_{\sf cl}(\sigma),\delta x_m(t,\sigma),z_{\sf cl}(\sigma))$ las ecuaciones a
orden lineal son \cite{kmt},\cite{cg}
\bea
x\mathrm{ -gauge}:&&\left[\partial_t^2-\frac{z_{\sf cl}^4(x)}{z_0^4}\,\partial_x^2\right]\delta x_m(t,x)=0\label{xg}\\
r\mathrm{ -gauge}:&&\left[\partial_t^2-(1-\frac {z^4}{z_0^4})\, \partial_z^2+\frac2z\,\partial_z\right]\delta x_m(t,z)=0~~~~~~~m=2,3\,.
\label{cg}
\eea
Como se ha mencionado en la sección \ref{wil}, nótese que la ecuación de movimiento en el gauge $x$ (\ref{xg}) depende
explícitamente de la solución clásica $z_{\sf cl}(x)$. Las ecuaciones se relacionan mediante el cambio de variables
en (\ref{adssol}). Escribiendo $\delta x_m=e^{-iwt} f(z)$ en (\ref{cg}) y nombrando $\tilde z=z/z_0$ se obtiene \cite{cg}
\be
\left[(1-\tilde z^4)\partial_{\tilde z}^2 -\frac2{\tilde z}+\xi^2\right]f(\tilde z)=0,~~~~~0\le \tilde z\le 1\,,
\label{xads2}
\ee
en donde $\xi=z_0\omega$. El cambio de variables  \cite{btt}
\bea
f(\tilde z)&=&\sqrt{1+\xi^2\tilde z^2}F(q)\nn\\
q(\tilde z)&=&\pm 2\int_{\tilde z}^1\frac {t^2}{(1+(\xi
t)^2)\sqrt{1-t^4}}\,dt
\eea
transforma la ecuación (\ref{xads2}) en un oscilador armónico simple
\be
 \frac
{d^2F}{dq}+\frac14\,\xi^2(\xi^4-1)F=0,~~~~q\in[-q_*,q_*]
\label{osci}
\ee
en donde $q_*=q(0)$. Las condiciones de borde en infinito $\delta x_m(t,0)=0$ se han mapeado en $F(q_*)=0$,
y cuantiza las frecuencias en (\ref{osci}) conduciendo a
 \be
\omega_nz_0\sqrt{\omega_n^4z_0^4-1}\int_0^1\frac{t^2dt}{(1+w_n^2z_0^2)\sqrt{1-t^4}}=\frac{n\pi}{2},~~~~n=1,2,...\label{exact}
\ee
La tabla siguiente muestra la comparación entre los autovalores exactos (\ref{exact}) y el análisis
numérico de autovalores de (\ref{xads2}) con $z_0=1$.
\begin{center}
\begin{tabular}{|r||c|l|}
 \hline
   & {Exact} & {Numeric} \\
     \hline
  $\omega_1$ & 2.203 & ~2.226\\
  $\omega_2$ & 3.467 & ~3.492\\
  $\omega_3$ & 4.697 & ~4.735\\
  $\omega_4$ & 5.914 & ~5.959\\
  $\omega_5$ & 7.125 & ~7.181\\
  $\omega_6$ & 8.332 & ~8.396\\
  $\omega_7$ & 9.537 & ~9.612\\
  $\omega_8$ & 10.741 & ~10.823\\  \hline
\end{tabular}
\end{center}
Los autovalores impares (pares) se obtuvieron resolviendo (\ref{xads2}) con condiciones de borde impares (pares)
discutidas después de (\ref{xperp}).

\section{Fluctuaciones fermiónicas del operador de Wilson-'t Hooft}\label{fluctfermD3}

 La matriz $\tilde{M}^{\alpha\beta}$ evaluada en la solución clásica \eqref{q vs k} toma la forma
\bea\label{tilde M Inverse}
    \tilde{M}^{\alpha\beta}&=\frac{L^2\sinh^2(u_{eq})}{k^2+L^4\sinh^4(u_{eq})}\left(
                                                        \begin{array}{cccc}
                                                          -r^2 & \frac{2r^2\sqrt{L^4\sinh^2(u_{eq})-k^2}}{L^2\sinh(2u_{eq})}\tilde\Gamma & 0 & 0 \\
                                                          -\frac{2r^2\sqrt{L^4\sinh^2(u_{eq})-k^2}}{L^2\sinh(2u_{eq})}\tilde\Gamma & r^2 & 0 & 0 \\
                                                          0 & 0 & 1 & -\frac{k}{L^2\sin\theta\sinh^2(u_{eq})}\tilde{\Gamma} \\
                                                          0 & 0 & \frac{k}{L^2\sin\theta\sinh^2(u_{eq})}\tilde{\Gamma} & \frac{1}{\sin^2\theta} \\
                                                        \end{array}
                                                      \right)\nn.
\eea
Nótese que los elementos ubicados en la diagonal de esta matriz son los mismos que lo elementos de la diagonal de la inversa de la métrica deformada \eqref{deformed metric}.

Ahora se procederá a calcular paso por paso los diferentes factores involucrados en la acción fermiónica \eqref{FermionicAction} para la solución de D3-brana analizada en la sección \ref{wilsonthooft}.

El pullback de la matrices de Dirac 10-dimensionales es
\bea
    \Gamma_t&=&\frac{L\cosh(u_k)}{r}\Gamma_{\underline{0}},~~~~
    \Gamma_r=\frac{L\cosh(u_k)}{r}\Gamma_{\underline{1}}, \nn \\
    \Gamma_{\theta}&=&L\sinh(u_k)\Gamma_{\underline{2}},~~~~
    \Gamma_{\phi}=L\sinh(u_k)\sin\theta\Gamma_{\underline{3}}\nn
\eea
y con estas expresiones se encuentra que
\bea
    \tilde{M}^{\alpha\beta}\Gamma_{\beta}D_{\alpha}=\frac{L\sinh(u_{eq})}{\sqrt{k^2+L^4\sinh^4(u_{eq})}}
    \left[-re^{2R_e\tilde\Gamma}\Gamma_{\underline{0}}D_t+re^{2R_e\tilde\Gamma}\Gamma_{\underline{1}}D_r
    +e^{2R_m\tilde{\Gamma}}\Gamma_{\underline{2}}D_{\theta}+\frac{1}{\sin\theta}e^{2R_m\tilde{\Gamma}}\Gamma_{\underline{3}}D_{\phi}\right]
    \label{mtildegamma},\nn\\
\eea
en donde han sido utilizadas las matrices de rotación $R_e$ y $R_m$ definidas por medio de
\be
    R_e=-\frac{1}{2}\sinh^{-1}\left(\sqrt{\frac{L^4\sinh^2(u_{eq})-k^2}{L^4\sinh^4(u_{eq})+k^2}}\right)\Gamma_{01},~~~~
    R_m=\frac{1}{2}\arcsin\left(\frac{k}{\sqrt{L^4\sinh^4(u_{eq})+k^2}}\right)\Gamma_{23}\nn.
\ee
Usando los siguientes vielbein para la métrica deformada $\hat g$
\bea
    \hat{e}^{\underline{0}}&=&\frac{\sqrt{\sinh^4(u_{eq})L^4+k^2}}{L \,r \sinh(u_{eq})}\,dt,~~~~
    \hat{e}^{\underline{1}}=\frac{\sqrt{\sinh^4(u_{eq})L^4+k^2}}{L \,r \sinh(u_{eq})}\,dr,\nn\\
    \hat{e}^{\underline{2}}&=&\frac{\sqrt{\sinh^4(u_{eq})L^4+k^2}}{L \sinh(u_{eq})}\,d\theta,~~~~
    \hat{e}^{\underline{3}}=\frac{\sqrt{\sinh^4(u_{eq})L^4+k^2}}{L \sinh(u_{eq})}\sin\theta\, d\phi\nn,
\eea
se puede simplificar la expresión \eqref{mtildegamma} a
\bea
    \tilde{M}^{\alpha\beta}\Gamma_{\beta}D_{\alpha}&=e^{{\cal{R}}\tilde{\Gamma}}\left[\hat{\Gamma}^{\alpha}e^{{\cal{R}}\tilde{\Gamma}}
    D_{\alpha}e^{-{\cal{R}}\tilde{\Gamma}}\right]e^{{\cal{R}}\tilde{\Gamma}},\nn
\eea
en donde se definió ${\cal{R}}=R_e+R_m$ y se usó que $\hat{\Gamma}_{\alpha}=\hat{e}^{\underline{\alpha}}_{\phantom{\underline{\alpha}}\alpha}\Gamma_{\underline{\alpha}}$ y $\hat{\Gamma}^{\alpha}=\hat{g}^{\alpha\beta}\hat{\Gamma}_{\beta}$.

El pullback de la derivada covariante 10-dimensional se puede escribir
\bea
    \nabla_{\alpha}d\sigma^{\alpha}&=d+\frac{1}{4}\omega^{\underline{\mu\nu}}\Gamma_{\underline{\mu\nu}}+\frac{1}{4}
    \omega^{\underline{ij}}\Gamma_{\underline{ij}}+\frac{1}{2}\sinh(u_k)e^{\underline{\mu}}\Gamma_{\underline{\mu4}}
    +\frac{1}{2}\cosh(u_k)e^{\underline{i}}\Gamma_{\underline{i4}}.\nn
\eea
Los primero 3 términos forman a la derivada covariante con respecto a la geometría deformada del volumen de mundo \eqref{deformed metric}. A esta derivada la llamaremos $\hat{\nabla}$ y se define como $\hat{\nabla}_\alpha=\partial_\alpha
+\frac14\hat{w}^{\underline{\beta\gamma}}\Gamma_{\underline{\beta\gamma}}$. Los dos términos restantes provienen de la curvatura extrínseca del volumen de mundo. Luego, se encuentra que
\bea
    e^{{\cal{R}}\tilde{\Gamma}}\nabla_{\alpha}d\sigma^{\alpha} e^{-{\cal{R}}\tilde{\Gamma}}&=\hat{\nabla}_{\alpha}d\sigma^{\alpha}+\frac{1}{2}\sinh(u_k)e^{\underline{\mu}}\Gamma_{\underline{\mu4}}
    e^{-2R_e\tilde{\Gamma}}+\frac{1}{2}\cosh(u_k)e^{\underline{i}}\Gamma_{\underline{i4}}e^{-2R_m\tilde{\Gamma}},\nn
\eea
y por lo tanto podemos escribir al primer termino de la derivada covariante \eqref{covariantder} como
\bea
    \tilde{M}^{\alpha\beta}\Gamma_{\beta}\nabla_{\alpha}&=e^{{\cal{R}}\tilde{\Gamma}}\left[\hat{\Gamma}^{\alpha}\hat{\nabla}_{\alpha}+\frac{L\sinh(u_{eq})}
    {\sqrt{L^4\sinh^4(u_{eq})+k^2}}
    \Gamma_{\underline{4}}\left(\sinh(u_{eq})e^{-2R_e\tilde{\Gamma}}+\cosh(u_{eq}) e^{-2R_m\tilde{\Gamma}}\right)\right]e^{{\cal{R}}\tilde{\Gamma}}\label{firstterm}.\nn\\
\eea

El segundo termino en \eqref{covariantder} proviene de la 5-forma de Ramond-Ramond y se puede ver fácilmente que
\bea
    {F}_5\Gamma_{\alpha}d\sigma^{\alpha}\otimes\left(i\sigma_2\right)&=-4\Gamma_{\underline{01234}}
    \otimes\left(i\sigma_2\right)\left(\cosh(u_{eq})e^{\underline{\mu}}\Gamma_{\underline{\mu}}+\sinh(u_{eq})e^{\underline{i}}
    \Gamma_{\underline{i}}\right)\left(1+\Gamma^{11}\right)\nn.
\eea
Como el espinor $\Theta$ satisface $\Gamma^{11}\Theta=\Theta$ se puede reemplazar $\Gamma^{11}=1$ en esta expresión y obtener
\bea
    \tilde{M}^{\alpha\beta}\Gamma_{\beta}{F}_5\Gamma_{\alpha}\otimes\left(i\sigma_2\right)&=-16e^{{\cal{R}}\tilde{\Gamma}}\Gamma_{\underline{01234}}
    \otimes\left(i\sigma_2\right)\frac{L\sinh(u_{eq})}{\sqrt{L^4\sinh^4(u_{eq})+k^2}}\left(\cosh(u_{eq})e^{-2R_m\tilde{\Gamma}}
    +\sinh(u_{eq})e^{-2R_e\tilde{\Gamma}}\right)e^{{\cal{R}}\tilde{\Gamma}}\label{secondterm}.\nn\\
\eea
A partir de \eqref{firstterm} y \eqref{secondterm} se tiene
\bea
    \tilde{M}^{\alpha\beta}\Gamma_{\beta}D_{\alpha}&=e^{{\cal{R}}\tilde{\Gamma}}\left[\hat{\Gamma}^{\alpha}\hat{\nabla}_{\alpha}+
    \left(1-\Gamma^{(0)}_{D3}\right)\frac{L\sinh(u_{eq})}{\sqrt{L^4\sinh^4(u_{eq})+k^2}}\Gamma_{\underline{4}}
    \left(\cosh(u_{eq})e^{-2R_m\tilde{\Gamma}}
    +\sinh(u_{eq})e^{-2R_e\tilde{\Gamma}}\right)\right]e^{{\cal{R}}\tilde{\Gamma}}\nn,
\eea
en donde
\bea
    \Gamma^{(0)}_{D3}&=\Gamma_{\underline{0123}}\otimes\left(i\sigma_2\right)\nn.
\eea

Ahora estudiemos al proyector de simetría $\kappa$ denominado $\Gamma_{D3}$ y definido en \cite{Martucci} como
\bea\label{Definition of Gamma_D3}
    \Gamma_{D3}&=-\frac{\epsilon^{\alpha_1\alpha_2\alpha_3\alpha_4}\Gamma_{\alpha_1\alpha_2\alpha_3\alpha_4}}{(p+1)!\sqrt{\textrm{det}\left(g+F\right)}}
    \otimes\left(i\sigma_2\right)\times\sum_{q}\Gamma^{\alpha_1\cdots\alpha_{2q}}F_{\alpha_1\alpha_2}\cdots F_{\alpha_{2q-1}\alpha_{2q}}\otimes\frac{\left(\sigma_3\right)^q}{q!2^q}\nn,
\eea
en donde $\epsilon^{0123}=1$. Para nuestro caso se obtiene
\be
\Gamma_{D3}=-\Gamma_{D3}^{(0)}\frac{L^4\cosh(u_{eq})\sinh^3(u_{eq})}{L^4\sinh^4(u_{eq})+k^2}\left[1+\left(\frac{k}{L^2\sinh^2(u_{eq})}
\Gamma_{\underline{23}}-\frac{\sqrt{L^4\sinh^2(u_{eq})-k^2}}{L^2\cosh(u_{eq})\sinh(u_{eq})}\Gamma_{\underline{01}}\right)\otimes\sigma_3\right].\nn
\ee
Actuando sobre el espinor $\Theta$ tenemos
\be
 \overline{\Theta}\Gamma_{D3}=-\overline{\Theta}e^{{\cal{R}}\tilde{\Gamma}}\Gamma^{(0)}_{D3}e^{-{\cal{R}}\tilde{\Gamma}}.\nn
\ee
para lo cual se ha usado que $\overline{\Theta}\Gamma^{11}=-\overline{\Theta}$.

Usando estas expresiones, el lagrangiano fermiónico para las fluctuaciones de la solución clásica \eqref{q vs k} se escribe
\bea
    \mathcal{L}_F&=&\overline{\Theta}e^{{\cal{R}}\tilde{\Gamma}}\left(1+\Gamma^{(0)}_{D3}\right)\left[\hat{\Gamma}^{\alpha}\hat{\nabla}_{\alpha}+
    \left(1-\Gamma^{(0)}_{D3}\right)\frac{L\sinh(u_{eq})}{\sqrt{L^4\sinh^4(u_{eq})+k^2}}\Gamma_{\underline{4}}
    \left(\cosh(u_{eq})e^{-2R_m\tilde{\Gamma}}
    +\sinh(u_{eq})e^{-2R_e\tilde{\Gamma}}\right)\right]e^{{\cal{R}}\tilde{\Gamma}}\Theta\nn.
\eea
Nótese que como $(\Gamma_{D3}^0)^2=1$ el segundo término dentro del paréntesis se anula. Redefiniendo $e^{{\cal{R}}\tilde{\Gamma}}\Theta=\Theta'\Rightarrow \overline{\Theta}e^{{\cal{R}}\tilde{\Gamma}}=\overline{\Theta}\,'$ se obtiene
\be
 \mathcal{L}_F=\overline{\Theta}\,'\left(1+\Gamma^{(0)}_{D3}\right)
    \hat{\Gamma}^{\alpha}\hat{\nabla}_{\alpha}\Theta'.\nn
\ee

El próximo paso es reducir las componentes del espinor usando la simetría $\kappa$, la cual demanda
\be
\tilde{\Gamma}\Theta'=\Theta',\nn
\ee
y nos dice que podemos fijar a cero la mitad de las componentes de $\Theta$. Renombrando $\Theta'=\Theta$ se obtiene la forma final para el Lagrangiano
\be
\overline{\Theta}\left(1-\Gamma_{D3}\right)\tilde{M}^{\alpha\beta}\Gamma_{\beta}D_{\alpha}\Theta=\overline{\Theta}\,
\hat{\Gamma}^{\alpha}\hat{\nabla}_{\alpha}\Theta.\nn
\ee

Con este cálculo se obtiene finalmente que a segundo orden en fluctuaciones, la acción fermiónica toma la siguiente forma
\be
S_F^{(2)}=\frac{T_{D3}}{2}\left(\frac{L^4\cosh(u_{eq})\sinh^3(u_{eq})}{L^4\sinh^4(u_{eq})+k^2}\right)\int d^4\sigma\sqrt{\hat{g}}\,\overline{\Theta}\,
\hat{\Gamma}^{\alpha}\hat{\nabla}_{\alpha}\Theta.
\ee


\end{document}